\documentclass[aps,prd,onecolumn,superscriptaddress,nofootinbib,floatfix,10pt]{revtex4-2}

\usepackage{amsmath,amssymb,amsthm,mathtools}
\usepackage{graphicx}
\usepackage{bm}

\usepackage[colorlinks=true,linkcolor=blue,citecolor=blue,urlcolor=blue]{hyperref}

\allowdisplaybreaks
\makeatletter
\renewcommand{\p@subsection}{}
\makeatother

\renewcommand{\thesection}{\arabic{section}}

\renewcommand{\thesubsection}{\thesection.\arabic{subsection}}

\numberwithin{equation}{section}

\newcommand{\dd}{\mathrm{d}}
\newcommand{\half}{\tfrac12}

\newcommand{\lhat}{\hat l}
\newcommand{\Om}{\Omega}

\begin{document}

\title{Polar-metric/axial-gauge perturbations of the Maldacena--Milekhin--Popov wormhole}

\author{Abhishake Sadhukhan}
\email{abhishake.physics@presiuniv.ac.in}
\affiliation{Department of Physics, Presidency University, Kolkata - 700073, India}

\begin{abstract}
We study linear perturbations of the Maldacena-Milekhin-Popov traversable wormhole in the sector where polar metric perturbations couple to axial gauge perturbations, and reduce it to two coupled master equations of Zerilli-Moncrief type. In the throat the Casimir energy of the lowest Landau level of charged fermions holds the wormhole open, and the linearised Bianchi identities hold only if the fermions respond. Once their state is fixed, their two-dimensional response follows from the conformal anomaly. We lift it to four dimensions on the field lines, assuming that the metric of each line scales with the local field strength. Through the trace anomaly this scaling couples the field strength to the worldsheet curvature, reminiscent of a Wen-Zee term. The response comes with a Hall current that sources the axial gauge mode. At zeroth order in the fermion backreaction $\alpha$ the throat is AdS$_2\times S^2$ and the modes are two Poschl-Teller problems with squared masses $l(l-1)$ and $(l+1)(l+2)$. We give the $O(\alpha)$ corrections in closed form. In the mouths the fermions act through the anomaly at the same order, and mouths and throat match. The higher Landau levels act through the one-loop Euler-Heisenberg action. We keep it to first order in $\eta_{\rm H}=g_c^2/12\pi^2$, which is not suppressed by $1/q$. It enlarges the AdS$_2$ radius and splits the levels shared by the two modes. On the closed-throat levels examined their $O(\alpha)$ mixing is of second order. The $O(\alpha)$ frequency shifts there, measured by the static observer at the throat centre, are real and depend on the assumed scaling, compared here with a lift on the coordinate planes. Region by region, at first order in $\alpha$ and $\eta_{\rm H}$ and on the levels examined, we find no growing mode at $l=2,3,4$, although the $O(\alpha)$ mouth rests on an estimate. Stability of the wormhole as a whole is not shown.
\end{abstract}

\maketitle
\raggedbottom

\tableofcontents

\section{Introduction}
\label{sec:intro}

Traversable wormholes need matter that violates the null energy condition \cite{Morris:1988cz}. Minimally coupled classical bosonic fields with canonical kinetic terms cannot supply it. A classical Dirac field can violate the null energy condition. Static wormholes supported by classical Dirac and Maxwell fields have been reported \cite{Blazquez-Salcedo:2020czn,Konoplya:2021hsm}. The construction of \cite{Blazquez-Salcedo:2020czn} does not solve the field equations at the throat \cite{Danielson:2021aor}, and \cite{Weinbaum:2026fbp} argues that a classical Dirac field that represents a single-particle state does not source a static spherically symmetric traversable wormhole. Quantum fields can supply the violation \cite{Gao:2016bin,Fu:2018oaq}. Gao, Jafferis and Wall \cite{Gao:2016bin} showed that a double-trace coupling between the two boundaries of an eternal AdS black hole makes it traversable. Maldacena and Qi \cite{Maldacena:2018lmt} built the matching eternal wormhole in nearly-AdS$_2$ gravity. Maldacena, Milekhin and Popov (MMP) \cite{Maldacena:2018gjk} then built a four-dimensional, asymptotically flat traversable wormhole in Einstein--Maxwell theory coupled to massless charged fermions. Versions of this construction large enough for a human to traverse were discussed in \cite{Maldacena:2020sxe}. Two magnetically charged, near-extremal Reissner-Nordstrom mouths of opposite charge are joined by a long throat of the form AdS$_2\times S^2$. The charged fermions in the monopole background fill a lowest Landau level of $q$ modes, one per unit of flux. These modes propagate along the magnetic field lines as two-dimensional fermions. Each field line threads the throat and closes through the ambient region, and each fermion therefore lives on a spatial circle. The Casimir energy of this two-dimensional conformal field theory (CFT) supplies the negative energy that holds the throat open. The construction is controlled by the dimensionless coupling $\alpha=g_c^2/4\pi^2 q$, which measures the backreaction of the fermions on the AdS$_2\times S^2$ throat. This coupling is small when the magnetic charge is large, and the throat, of length $\pi\ell$ with $\ell=4r_e/\pi\alpha$, is correspondingly long. The throat can also be viewed as a perturbation of the near-horizon geometry of a near-extremal Reissner-Nordstrom black hole. Kanai, Maeda and Yoshida took this view and found that no perturbative construction of this kind works when the only new ingredient is a higher-derivative correction \cite{Kanai:2025zgq}. Negative Casimir energy of the MMP type, or a background with less symmetry, is needed. Charged traversable throats can also form dynamically, by injecting negative-energy null dust into a black hole \cite{Koga:2025bqw}.

The MMP wormhole is a horizonless object with a long, nearly reflecting cavity between its two mouths. This makes it a natural candidate for the gravitational-wave echoes proposed as signatures of horizonless compact objects \cite{Cardoso:2016rao,Cardoso:2017cqb,Bueno:2017hyj,Konoplya:2016hmd}. The same signature is now being studied in dynamical settings, where accretion onto the reflecting surface compresses the echo train and removes it once a horizon forms \cite{Sharma:2026gvu}. The echo phenomenology of the MMP wormhole was studied for scalar and gauge test fields in \cite{Mondal:2025tht}. At the fiducial parameters used there and below, the light travel time through the throat is about $1.5\times10^{25}$\,s, and the interest of the echoes is structural rather than observational. The transmission analysis of throat spacetimes in \cite{Riley:2026avn} finds the stronger echoes at higher $l$ reported there consistent with its $l$-dependent barrier heights, and \cite{Mondal:2025tht} compares its transmission with the low-frequency scattering computed in \cite{Freivogel:2026ujn}. For wormholes supported by a phantom scalar field the quasinormal spectrum has been computed beyond test fields. The scalar and axial modes of the Ellis-Bronnikov wormhole tend to the Schwarzschild modes of the same mass when the mass is large at fixed $l$ \cite{Blazquez-Salcedo:2018ipc}. Its polar modes coincide with the axial modes for the massless wormhole and form two separate branches once the wormhole carries mass \cite{Azad:2022qqn}. A test-field treatment misses the coupled gravitational and electromagnetic perturbations, and it misses the response of the fermionic Casimir energy to a perturbation of the geometry. The present paper supplies the linearised Einstein--Maxwell--fermion system for the MMP wormhole in the sector that contains the polar gravitational degree of freedom, and reduces it to master equations of Zerilli-Moncrief type \cite{Regge:1957td,Zerilli:1970se,Zerilli:1974ai,Moncrief:1974ng,Moncrief:1974am,Chandrasekhar:1979iz,Chandra,Martel:2005ir}. Coupled metric and gauge-field perturbations of charged black holes remain an active area. Recent work has extended the master-function program to new branches, to new symmetries, and to a Hamiltonian formulation in which Darboux transformations act as canonical transformations between master functions \cite{Lenzi:2021wpc,Liu:2025reu,deCesare:2024csp,Lenzi:2026aay}. Quantum corrections to a charged background yield coupled Regge--Wheeler master equations and break the classical isospectrality of the two parity sectors \cite{delRio:2024kms}. Our system is of that kind. The main correction here comes from the Casimir stress tensor of the fermions. A local one-loop action for the gauge field, from the higher Landau levels, adds a second one (Sec.~\ref{sec:hll}).

Two features distinguish this problem from the textbook Reissner-Nordstrom one. The first is kinematic. The MMP mouths are magnetically charged, and on a magnetic background the parity sectors of gravitational and electromagnetic perturbations pair crosswise. An even-parity (polar) metric perturbation couples to an odd-parity (axial) gauge perturbation, and vice versa \cite{Pereniguez:2023wxf,DeFelice:2023rra}. This follows from the parity of the monopole field strength alone. Electric--magnetic duality maps the magnetic mouth onto the electric Reissner-Nordstrom black hole treated by Chandrasekhar \cite{Chandrasekhar:1979iz,Chandra}, and it flips the parity of the gauge perturbation while it leaves the metric perturbation alone \cite{Pereniguez:2023wxf,DeFelice:2023rra}. His polar equation therefore governs the axial-gauge sector of the magnetic mouth, and his potentials cannot be carried over labelled as before. Perturbations of magnetically charged black holes are under active study beyond Einstein--Maxwell theory, including four-derivative couplings between the curvature and the field strength \cite{Stashko:2026kja}. We treat the sector consisting of polar $h_{\mu\nu}$ and axial $a_\mu$ throughout. The other sector, axial metric perturbations with polar gauge perturbations, is treated in the companion paper \cite{Sadhukhan:2026mcu}.

The second feature is dynamical, and it is the main new physics of the paper. In the throat the background fermion stress tensor $T^Q_{\mu\nu}$ is nonzero, since it is the Casimir energy that supports the wormhole. The linearised Bianchi identity then contains terms in which the perturbed connection multiplies the background $T^Q$, and these terms do not vanish when the fermion response $\delta T^Q$ is set to zero. The fermion sector must therefore respond. The LLL fermions form a two-dimensional CFT, and in two dimensions any metric perturbation is a Weyl rescaling plus a diffeomorphism. The exact linear response is therefore fixed by the two-dimensional conformal anomaly \cite{Polyakov:1981rd,Birrell:1982ix,Christensen:1977jc,Davies:1976ei}, once the state of the fermions is given. We lift this response to four dimensions on the worldsheets of the field lines of the full field strength. An azimuthal electric field tilts these worldsheets in $\theta$, and a perturbed radial magnetic field changes the density of field lines. The lift then has mixed components, which are the drift of the field lines, and a density term. Once the density of the fermions follows the flux, conservation along the field lines holds only if the two-dimensional metric of each line is scaled by the local field strength. This scaling is an assumption, discussed in Sec.~\ref{sec:lift}. Conservation does not single it out, since a lift built on the coordinate planes, whose density and scale ignore the flux, passes the same test (Sec.~\ref{sec:response}). The first-order spectrum of the closed throat does depend on the choice (Sec.~\ref{sec:stab-normalmodes}). The $\theta$ component of the force balance $\nabla_\mu T_Q^{\mu\nu}=-F^{\nu\lambda}J_\lambda$ is not satisfied by the stress tensor alone. It requires an induced Hall current $J^\phi$, which sources the very axial gauge mode being solved for. With both the response and the Hall current included, the linearised Bianchi identities hold at $O(\alpha)$.

We also derive the Hall current a second way, with no reference to $T^Q_{\mu\nu}$. We vary the LLL effective action with respect to the gauge field. The action is a sum over field lines of the effective action of one two-dimensional fermion on the metric of its line. Varying the gauge field moves the field lines and changes the local field strength. This route agrees with the force-balance route for every term. Both routes follow from one functional, which also generates the lift, and the agreement tests the lift and its algebra rather than the scaling. Through the trace anomaly, the dependence of the line metric on the field strength gives a coupling $\tfrac{c_1}{48\pi}\int\sqrt{\mathsf g}\,R_{\mathsf g}\log|F|$ between the field strength $|F|$ and the curvature of the field-line worldsheet. Here $\mathsf g$ is the induced metric of the worldsheet, $R_{\mathsf g}$ its curvature scalar and $c_1=1$ the central charge of one two-dimensional fermion. A treatment on the coordinate planes must add the Euler density $\tfrac{c_1}{48\pi}\sqrt{\hat g}\hat R$ by hand to reproduce this piece. Here its coefficient is fixed by the anomaly, for a line metric that scales linearly with $|F|$. This structure resembles the Wen-Zee term of the quantum Hall literature \cite{Wen:1992ej,Gromov:2014gta,Bradlyn:2014wla,Abanov:2014ula}.

At $O(\alpha^0)$ the mouth reduces to a closed Einstein--Maxwell problem whose master equations we give in full. Its near-horizon limit at extremality reproduces the throat equations, which shows that the two regions match, in parallel with the recent matching between extreme Reissner-Nordstrom and Bertotti--Robinson perturbations \cite{deCesare:2024csp}. At $O(\alpha)$ the mouth is not free of fermions. Their Casimir energy is negligible there, but the anomaly part of their stress tensor is local and of order $\alpha$, as in the throat. It renormalises $r_e^2$ to $r_e^2(1+\alpha/3)$ at both ends of the mouth. MMP absorb this into the radii and discard the trace part in both regions, which is consistent for the background. Once the trace part is kept in the throat it must be kept in the mouth, and we extend the mouth to $O(\alpha)$. In the throat at $\alpha=0$ the master system decouples exactly into two Poschl-Teller problems with integer squared masses $m_-^2=l(l-1)$ and $m_+^2=(l+1)(l+2)$, the near-horizon limits of Chandrasekhar's potentials. The $O(\alpha)$ corrections are computed in closed form. The symmetric part of the throat potential is positive for $\alpha\rho\le0.1$ at $l=2,3,4$, with $\rho$ the radial coordinate of the throat, a range that contains the junction with the mouths. The higher Landau levels add the one-loop Euler-Heisenberg action of the gauge field, which makes the coupling run. We treat it as a second small parameter $\eta_{\rm H}=g_c^2/12\pi^2$, which is not suppressed by $1/q$ (Sec.~\ref{sec:hll}). It enlarges the AdS$_2$ radius and raises both throat masses in units of that radius. On the clock of the static observer at the centre of the throat it lowers every normal frequency, and it splits the levels that the two throat modes share at $\alpha=0$. In the mouth it turns the problem into one of nonlinear electrodynamics, and the master variables of Nomura, Yoshida and Soda \cite{Nomura:2020tpc} give it an energy that is positive in the cases we scanned. The results are collected in Sec.~\ref{sec:summary}.

Section~\ref{sec:parity} explains the parity classification on a magnetic background and fixes the ansatz. Section~\ref{sec:setup} describes the MMP background, the consistency condition that separates the mouth from the throat, the harmonic separation and the reduction algorithm used throughout. Section~\ref{sec:TQ} constructs the LLL stress tensor in the throat to $O(\alpha)$, with its physical origin, the lift to four dimensions, the corrected background and the exact linear response. Section~\ref{sec:hall} derives the Hall current from the force balance and from the effective action. Section~\ref{sec:throatmaster} gives the throat master equations, including the $O(\alpha)$ corrections and the exact $\alpha\to0$ limit. Section~\ref{sec:mouth} treats the mouth to $O(\alpha)$. Its $\alpha^0$ part is the Einstein--Maxwell mouth with Chandrasekhar's variables, and the section ends with the matching to the throat in the overlap. Section~\ref{sec:stability} analyses the null energy condition at first order and the linear stability. Section~\ref{sec:hll} adds the one-loop action of the higher Landau levels. Section~\ref{sec:summary} collects the results, the assumptions and the open questions. The lengthier expressions are collected in the appendices, and Appendix~\ref{app:numerics} describes the numerical methods.

\section{Parity sectors on a magnetically charged background}
\label{sec:parity}

\subsection{The parity map and the selection rule}
\label{sec:selection}

Perturbations on a static spherically symmetric background are classified by how they respond to the parity map
\begin{equation}
  \mathsf P:\quad (\theta,\phi)\;\longmapsto\;(\pi-\theta,\;\phi+\pi).
  \label{eq:parityop}
\end{equation}
The axisymmetric scalar harmonic transforms as $Y(\theta)=P_l(\cos\theta)\mapsto(-1)^l\,Y(\theta)$. A perturbation carrying the eigenvalue $(-1)^l$ is called even or polar, and one carrying $(-1)^{l+1}$ is called odd or axial. The background metric is invariant under \eqref{eq:parityop}. The two classes therefore cannot mix through the geometry alone, which is why the Regge--Wheeler and Zerilli problems are independent \cite{Regge:1957td,Zerilli:1970se}.

On the MMP background the gauge field breaks this. The background gauge field is that of a magnetic monopole, in the form used by MMP \cite{Maldacena:2018gjk},
\begin{equation}
  \bar A \;=\; \frac{q}{2}\cos\theta \,\dd\phi
  \qquad\Longrightarrow\qquad
  \bar F_{\theta\phi} \;=\; -\frac{q}{2}\sin\theta ,
  \qquad
  \frac{1}{2\pi}\int_{S^2}\bar F \;=\; -\,q .
  \label{eq:monopole}
\end{equation}
Following MMP we take $q$ to be a positive integer and interpret it as the number of flux quanta through the sphere, $q=\tfrac{1}{2\pi}\big|\!\int_{S^2}\bar F\big|$. This is the quantity the index theorem counts \cite{Atiyah:1963zz,Wu:1976ge} (Sec.~\ref{sec:throat-origin}), and every appearance of $q$ in the geometry and in the fermion stress tensor is through $q^{2}$ or through this count. With the orientation $\varepsilon_{\theta\phi}=+\sin\theta$ the potential \eqref{eq:monopole} gives the radial magnetic field $-q/2r^{2}$, pointing inward, and the opposite orientation is obtained by $q\to-q$. The orientation matters in one place only, the direction of the induced Hall current of Sec.~\ref{sec:hall}. Under \eqref{eq:parityop} we have $\dd\theta\mapsto -\dd\theta$, $\dd\phi\mapsto \dd\phi$ and $\sin\theta\mapsto\sin\theta$. The monopole field strength therefore obeys
\begin{equation}
  \bar F = -\tfrac{q}{2}\sin\theta\,\dd\theta\wedge \dd\phi
  \;\longmapsto\;
  +\tfrac{q}{2}\sin\theta\,\dd\theta\wedge \dd\phi = -\,\bar F ,
  \label{eq:Fbarparity}
\end{equation}
that is, $\bar F$ is parity odd. Parity acts on the monopole as $q\to-q$, a reversal of the orientation convention fixed below \eqref{eq:monopole}, and the selection rule derived below depends only on this sign flip. An electric background behaves oppositely. There $\bar F=(Q_{\rm e}/r^2)\,\dd t\wedge \dd r$ involves no angular one-forms, $\bar F\mapsto+\bar F$, and the background field strength is parity even. The background geometry is parity even in both cases, since the stress tensor is quadratic in $\bar F$ and the sign in \eqref{eq:Fbarparity} cancels. The distinction becomes visible only at linear order in the perturbations, where $\bar F$ appears once rather than twice.

Write the Maxwell stress tensor as $T_{\mu\nu}=g_c^{-2}\big[F_{\mu\sigma}F_\nu{}^\sigma-\tfrac14 g_{\mu\nu}F^2\big]$ and linearise about $(\bar g,\bar F)$. The terms first order in the gauge perturbation are the cross terms
\begin{equation}
  \delta T^{\rm mag}_{\mu\nu}\;\supset\;
  g_c^{-2}\Big[\,\bar F_{\mu\sigma}\,\delta F_\nu{}^\sigma + \delta F_{\mu\sigma}\bar F_\nu{}^\sigma\,\Big],
  \qquad \delta F_{\mu\nu}=\partial_\mu a_\nu-\partial_\nu a_\mu ,
  \label{eq:crossterm}
\end{equation}
each of which carries one factor of $\bar F$ and one factor of $\delta F$. Parity is multiplicative, hence
\begin{equation}
  \mathsf P\big[\delta T\big] \;=\; \mathsf P\big[\bar F\big]\times \mathsf P\big[\delta F\big].
  \label{eq:parityproduct}
\end{equation}
The linearised Einstein equation $\delta G_{\mu\nu}=\kappa\,\delta T_{\mu\nu}$ is consistent only if both sides carry the same parity eigenvalue, and $\delta G$ inherits its parity from the metric perturbation $h$. Combining this with \eqref{eq:parityproduct} gives the selection rule
\begin{equation}
  \mathsf P\big[h\big] \;=\; \mathsf P\big[\bar F\big]\times \mathsf P\big[\delta F\big] .
  \label{eq:selection}
\end{equation}
The same rule follows from the linearised Maxwell equation $\nabla_\mu F^{\mu\nu}=0$, whose perturbation contains a term linear in $\delta F$ and a term $\sim h\,\bar F$ coupling the metric perturbation to the background flux.

For an electric background $\mathsf P[\bar F]=+1$, and \eqref{eq:selection} reduces to $\mathsf P[h]=\mathsf P[\delta F]$. The sectors pair like with like, polar metric with polar gauge and axial metric with axial gauge. This is the Reissner-Nordstrom system of Zerilli, Moncrief and Chandrasekhar \cite{Zerilli:1974ai,Moncrief:1974ng,Moncrief:1974am,Chandrasekhar:1979iz}, in which each parity closes on itself. For the magnetic background $\mathsf P[\bar F]=-1$, and \eqref{eq:selection} becomes $\mathsf P[h]=-\mathsf P[\delta F]$. The pairing is crossed. An even-parity metric perturbation couples to an odd-parity gauge perturbation, and an odd-parity metric perturbation couples to an even-parity gauge perturbation \cite{Pereniguez:2023wxf,DeFelice:2023rra}. No assumption is involved, since \eqref{eq:Fbarparity} alone fixes it. This is why Chandrasekhar's equations for a given parity cannot be carried over to the MMP mouth with the parity labels unchanged, although the mouth metric is Reissner-Nordstrom, and it is why this paper pairs polar $h_{\mu\nu}$ with axial $a_\mu$. The relation between the two sectors and Chandrasekhar's electric problem is made explicit in Sec.~\ref{sec:mouth-RN}.

\subsection{The ansatz in Regge--Wheeler gauge}
\label{sec:ansatz}

We write the general static, spherically symmetric background as
\begin{equation}
  \dd s^2 \;=\; -A(r)\,\dd t^2 \;+\; B(r)\,\dd r^2 \;+\; R(r)^2\,\mathring g ,
  \qquad \mathring g = \dd\theta^2 + \sin^2\theta\,\dd\phi^2 ,
  \label{eq:bgmetric}
\end{equation}
so that
\begin{equation}
  \bar g_{\mu\nu} = \mathrm{diag}\!\left(-A,\;B,\;R^2,\;R^2\sin^2\theta\right),
  \qquad
  \sqrt{-\bar g} = \sqrt{AB}\,R^2\sin\theta ,
\end{equation}
and perturb with a bookkeeping parameter $\epsilon$,
\begin{equation}
  g_{\mu\nu} = \bar g_{\mu\nu} + \epsilon\, h_{\mu\nu},
  \qquad
  A_\mu = \bar A_\mu + \epsilon\, a_\mu ,
  \label{eq:pert}
\end{equation}
working to $O(\epsilon)$. To that order
\begin{equation}
  g^{\mu\nu} = \bar g^{\mu\nu} - \epsilon\, h^{\mu\nu},
  \qquad
  \sqrt{-g} = \sqrt{-\bar g}\left(1 + \tfrac{\epsilon}{2}\, h^\mu{}_\mu\right),
\end{equation}
with indices raised by $\bar g$. Sections~\ref{sec:parity} and \ref{sec:setup} use the generic coordinate names $(t,r)$. In the throat they are replaced by $(\tau,\rho)$ (Sec.~\ref{sec:system}).

We work in Regge--Wheeler gauge \cite{Regge:1957td} with axisymmetric ($m=0$) harmonics $Y(\theta)\equiv P_l(\cos\theta)$. The general even-parity metric perturbation on a spherically symmetric background is built from seven functions of $(t,r)$. Let $x^A=(\theta,\phi)$, with $A,B\in\{\theta,\phi\}$, denote coordinates on the unit two-sphere, whose round metric is
\begin{equation}
  \mathring g_{AB}\,\dd x^A \dd x^B \;=\; \dd\theta^2 + \sin^2\theta\,\dd\phi^2 ,
  \qquad
  \mathring g_{AB} = \mathrm{diag}\!\left(1,\;\sin^2\theta\right),
  \label{eq:gammaAB}
\end{equation}
and let $\mathring\nabla_A$ be its covariant derivative, with $\mathring\nabla^2\equiv\mathring g^{AB}\mathring\nabla_A\mathring\nabla_B$, $\mathring\nabla^2 Y=-\lhat\,Y$ and $\lhat\equiv l(l+1)$. Throughout the paper capital indices $A,B$ refer to the sphere, lowercase indices $a,b$ to the two-dimensional $(t,r)$ block, or $(\tau,\rho)$ in the throat, and Greek indices $\mu,\nu,\sigma,\lambda$ to four-dimensional spacetime. The seven functions are one for each of $h_{tt},h_{tr},h_{rr}$, two multiplying the even-parity vector harmonic $\partial_A Y$ in $h_{tA}$ and $h_{rA}$, and two multiplying the angular tensors $\mathring g_{AB}Y$ and the trace-free tensor harmonic $\mathring\nabla_A\mathring\nabla_B Y-\tfrac12\mathring g_{AB}\mathring\nabla^2 Y$ in $h_{AB}$. Even-parity gauge transformations are generated by a vector $v_\mu\,\dd x^\mu = \big(v_t\,\dd t + v_r\,\dd r\big)Y + v\,\partial_A Y\,\dd x^A$ with three independent components, which is enough to remove three of the seven. Regge--Wheeler gauge sets the two vector-harmonic amplitudes and the trace-free tensor amplitude to zero, leaving the four functions
\begin{align}
  h_{tt} &= A\,H_0(t,r)\,Y(\theta), &
  h_{tr} &= H_1(t,r)\,Y(\theta), &
  h_{rr} &= B\,H_2(t,r)\,Y(\theta),
  \label{eq:hansatz1}\\
  h_{\theta\theta} &= R^2 K(t,r)\,Y(\theta), &
  h_{\phi\phi} &= R^2\sin^2\theta\, K(t,r)\,Y(\theta), &
  a_\phi &= a(t,r)\,\sin\theta\, Y'(\theta),
  \label{eq:hansatz2}
\end{align}
with all other components zero. The factors of $A$ and $B$ in $h_{tt}$ and $h_{rr}$ make $H_0$ and $H_2$ dimensionless fractional perturbations of the background metric functions, and the field equations then come out with fewer explicit factors of $A,B$. That $h_{\theta\theta}$ and $h_{\phi\phi}$ are proportional to the same function $K$ in the ratio $1:\sin^2\theta$ is the statement that the trace-free tensor amplitude has been gauged away. The angular block is pure trace, $h_{AB}=R^2K\,\mathring g_{AB}Y$. The absence of $h_{t\theta}$ and $h_{r\theta}$ is the same gauge choice, since those components would carry the even-parity vector harmonic $\partial_\theta Y$.

The gauge perturbation has a single component for a different reason. The odd-parity vector harmonic on the sphere is $Y^{\rm odd}_A=\varepsilon_A{}^B\partial_B Y$, with $\varepsilon_{\theta\phi}=\sin\theta$. At $m=0$ the harmonic is $\phi$-independent, hence $\partial_\phi Y=0$ and $Y^{\rm odd}_\theta=0$, while $Y^{\rm odd}_\phi=\varepsilon_\phi{}^\theta\partial_\theta Y=-\sin\theta\,Y'(\theta)$, since $\varepsilon_\phi{}^\theta=-\sin\theta$. The axial gauge perturbation is therefore $a_A=-a(t,r)Y^{\rm odd}_A$, with the sign absorbed into $a$. It has only a $\phi$ component, with the angular profile written in \eqref{eq:hansatz2}. Its $t$ and $r$ components vanish by parity. $a_t$ and $a_r$ would multiply $Y$ itself and belong to the even-parity sector, which the selection rule \eqref{eq:selection} forbids from coupling to even-parity $h_{\mu\nu}$.

We rescale the gauge amplitude,
\begin{equation}
  w \;\equiv\; \frac{4a}{q},
  \label{eq:wdef}
\end{equation}
which removes $q$ from every radial Einstein--Maxwell equation. The charge enters the coupled system only through the cross terms \eqref{eq:crossterm}, which are linear in $\bar F\propto q$ and linear in $a$, together with the background relation $r_e^2=\kappa q^2/8g_c^2$ of Sec.~\ref{sec:system}. With the factor of $4$ and \eqref{eq:re} the combination $\kappa\bar F_{\theta\phi}\,a/g_c^2$ is $-r_e^2w\sin\theta$, with no numerical factor. The unknown functions are
\begin{equation}
  \{H_0,\;H_1,\;H_2,\;K,\;w\}(t,r) ,
\end{equation}
and $l$ is a free parameter throughout. The perturbed field strength is
\begin{equation}
  \delta F_{\mu\nu} = \partial_\mu a_\nu - \partial_\nu a_\mu
  \quad\Longrightarrow\quad
  \delta F_{t\phi} = \dot a \sin\theta\,Y',\qquad
  \delta F_{r\phi} = a' \sin\theta\,Y',\qquad
  \delta F_{\theta\phi} = a\,\partial_\theta(\sin\theta\,Y') .
  \label{eq:fpert}
\end{equation}

We take $l\ge2$ throughout. The restriction comes from the metric sector. The trace-free tensor harmonic $\mathring\nabla_A\mathring\nabla_B Y-\tfrac12\mathring g_{AB}\mathring\nabla^2 Y$ carries spin two on the sphere and vanishes identically for $l=0$ and $l=1$. For these multipoles there is no such amplitude to gauge away, and Regge--Wheeler gauge is not a complete gauge fixing. At $l=0$ the axial gauge amplitude also vanishes, since $a_\phi\propto\sin\theta\,Y'=0$. In the mouth the generalised Birkhoff theorem \cite{MTW} then reduces the perturbation to a change of mass plus gauge. The magnetic charge is quantised and cannot change \cite{Dirac:1931kp}, and an electric charge would belong to the other parity sector. In the throat the source $T^Q$ evades Birkhoff's theorem. There the $l=0$ sector contains the Jackiw--Teitelboim dilaton (Sec.~\ref{sec:alpha0}), which controls the throat length \cite{Almheiri:2014cka,Maldacena:2016upp,Maldacena:2018gjk}. It does not propagate. In wormholes supported by phantom scalar fields the corresponding mode is unstable \cite{Gonzalez:2008wd,Bronnikov:2012ch}. We do not treat it here, and its stability is open. At $l=1$ the polar metric perturbation carries no radiative content. In vacuum it is a displacement of the centre of mass \cite{Zerilli:1970wzz,Detweiler:2003ci}. The axial gauge amplitude does not vanish at $l=1$. It is a magnetic dipole potential, $a_\phi\propto\sin^2\theta$, and it carries an electromagnetic dipole mode, as in Reissner-Nordstrom \cite{Zerilli:1974ai,Moncrief:1974ng,Moncrief:1974am,Moncrief:1974gw}. The $l=1$ system therefore has one physical degree of freedom and not two, the electromagnetic dipole. In the $\alpha=0$ throat the mass $m_-^2=l(l-1)$ of Sec.~\ref{sec:alpha0} vanishes at $l=1$ while $m_+^2=6$ remains. Which of the two throat modes carries the dipole is decided only by a complete gauge fixing at $l=1$. The reduction of Sec.~\ref{sec:algorithm} assumes a complete gauge, and every master equation, potential and numerical result in this paper holds for $l\ge2$. The $l=1$ dipole needs a separate gauge fixing, which we do not carry out.

We set $m=0$ without loss of generality. The background is spherically symmetric and the radial equations cannot depend on $m$. Different values of $m$ within one multiplet describe the same physical mode seen along a rotated axis. With $m=0$ the angular harmonics are Legendre polynomials, and every angular dependence can be removed with Legendre's equation alone (Sec.~\ref{sec:separation}).

\section{The MMP wormhole and the reduction to master equations}
\label{sec:setup}

\subsection{The system and its backgrounds}
\label{sec:system}

The MMP wormhole \cite{Maldacena:2018gjk} is a solution of four-dimensional Einstein--Maxwell theory coupled to a massless charged Dirac fermion. The bosonic action is
\begin{equation}
 \mathcal I \;=\; \int \dd^4x\,\sqrt{-g}\left[\frac{{}^{(4)}\!R}{16\pi G} \;-\; \frac{F_{\mu\nu}F^{\mu\nu}}{4g_c^2}\right]
  \;+\; \mathcal I_{\rm fermion},
  \label{eq:action}
\end{equation}
with $\kappa \equiv 8\pi G$. Varying \eqref{eq:action} gives
\begin{align}
  G_{\mu\nu} &= \kappa\left(T^{\rm mag}_{\mu\nu} + T^{Q}_{\mu\nu}\right),
  \label{eq:einstein}\\[2pt]
  \nabla_\mu F^{\mu\nu} &= g_c^2\, J^\nu ,
  \label{eq:maxwell}
\end{align}
where
\begin{equation}
  T^{\rm mag}_{\mu\nu} \;=\; \frac{1}{g_c^{2}}\left(F_{\mu\sigma}F_{\nu}{}^{\sigma}
  - \tfrac14 g_{\mu\nu}F_{\sigma\lambda}F^{\sigma\lambda}\right),
  \label{eq:Tmag}
\end{equation}
and $T^Q_{\mu\nu}$, $J^\nu$ denote the expectation values of the fermion stress tensor and charge current in the relevant quantum state. In the throat they are the whole point, and Secs.~\ref{sec:TQ} and \ref{sec:hall} construct them. In the mouth they vanish at $O(\alpha^0)$ and are $O(\alpha)$, and Sec.~\ref{sec:mouth} constructs them there. The background field strength is the monopole \eqref{eq:monopole}, a closed but not exact two-form, and \eqref{eq:maxwell} is satisfied identically at zeroth order with $J^\nu = 0$. The extremal radius is fixed by
\begin{equation}
  r_e^2 \;=\; \frac{\kappa\, q^2}{8 g_c^2}
  \qquad\text{(MMP eq.~(2.3))},
  \label{eq:re}
\end{equation}
which is the relation that makes the $\epsilon^0$ Einstein equations hold on the extremal background. Equation numbers of MMP refer to the arXiv version of \cite{Maldacena:2018gjk}. We keep $r_e$, $GM$, $q$, $g_c$ and $\kappa$ as free parameters throughout and use \eqref{eq:re} as an identity. The dimensionless combination
\begin{equation}
  \alpha \;\equiv\; \frac{g_c^2}{4\pi^2 q} \qquad\text{(MMP eq.~(5.36))}
  \label{eq:alpha}
\end{equation}
organises the quantum corrections in the throat. Substituting $g_c^2 \to 4\pi^2\alpha q$ is a change of variables.

Both regions are covered by the form \eqref{eq:bgmetric}. The two cases are
\begin{align}
  \text{mouth:}&\qquad A = 1 - \frac{2GM}{r} + \frac{r_e^2}{r^2},\qquad B = \frac1A,\qquad R = r,
  \label{eq:bgmouth}\\
  \text{throat:}&\qquad A = r_e^2\!\left(1+\rho^2+\alpha\gamma(\rho)\right),\quad
  B = \frac{r_e^2}{1+\rho^2+\alpha\gamma(\rho)},\quad
  R^2 = r_e^2\!\left(1+\alpha\varphi(\rho)\right),
  \label{eq:bgthroat}
\end{align}
these being MMP eqs.~(5.32)--(5.33) and (5.34). In the mouth the coordinates are the asymptotic $(t,r)$. We write $GM$ as the extremal value of the mouth considered plus the mass defect $\delta_M$ of MMP's eq.~(5.42). The defect is of order $\alpha^2r_e$ and negative, and with it the mouth has no horizon. In the $\alpha^0$ mouth the extremal value is $r_e$, and $GM=r_e+\delta_M$. At $O(\alpha)$ the extremal value moves to $r_e(1-\alpha/18)$ (Sec.~\ref{sec:mouthTQ}). The higher Landau levels alone move it to $r_e(1+\eta_{\rm H})$ (Sec.~\ref{sec:hll-bg}), and the two corrections together to $r_e(1+\eta_{\rm H}-\alpha/18)$. In the throat the dimensionless radial coordinate is $\rho$ and the dimensionless time is $\tau$. They are related to the asymptotic coordinates by the MMP matching (their eq.~(5.19))
\begin{equation}
  \tau = \frac{t}{\ell},
  \qquad
  \rho = \frac{\ell\,(r - r_e)}{r_e^2},
  \qquad
  \ell \;=\; \frac{4r_e}{\pi\alpha},
  \label{eq:matching}
\end{equation}
in the overlap region $1\ll\rho$, $r-r_e\ll r_e$. Throughout, mouth quantities are functions of $(t,r)$ and throat quantities of $(\tau,\rho)$. The background functions $\gamma(\rho)$ and $\varphi(\rho)$ are the $O(\alpha)$ corrections to the AdS$_2$ and $S^2$ factors, determined in Sec.~\ref{sec:bgalpha}. The mouth receives its own $O(\alpha)$ correction in Sec.~\ref{sec:mouthTQ}.

\subsection{Expansion of the field equations and the consistency condition}
\label{sec:consistency}

Write the Christoffel symbols, Ricci tensor and Einstein tensor of \eqref{eq:pert} as
\begin{align}
  \Gamma^\lambda{}_{\mu\nu} &= \tfrac12 g^{\lambda\sigma}
  \left(\partial_\mu g_{\sigma\nu} + \partial_\nu g_{\sigma\mu} - \partial_\sigma g_{\mu\nu}\right),
  \\
  R_{\mu\nu} &= \partial_\lambda \Gamma^\lambda{}_{\mu\nu} - \partial_\nu\Gamma^\lambda{}_{\mu\lambda}
  + \Gamma^\lambda{}_{\lambda\sigma}\Gamma^\sigma{}_{\mu\nu}
  - \Gamma^\lambda{}_{\nu\sigma}\Gamma^\sigma{}_{\mu\lambda},
  \qquad
  G_{\mu\nu} = R_{\mu\nu} - \tfrac12 g_{\mu\nu}\,{}^{(4)}\!R ,
\end{align}
and expand each to first order in $\epsilon$,
\begin{equation}
  G_{\mu\nu} = \bar G_{\mu\nu} + \epsilon\,\delta G_{\mu\nu} + O(\epsilon^2),
  \qquad
  T_{\mu\nu} = \bar T_{\mu\nu} + \epsilon\,\delta T_{\mu\nu} + O(\epsilon^2).
\end{equation}
The field equations then split into
\begin{align}
  O(\epsilon^0):&\qquad \bar G_{\mu\nu} - \kappa\left(\bar T^{\rm mag}_{\mu\nu} + \bar T^{Q}_{\mu\nu}\right) = 0,
  \qquad \partial_\mu\!\left(\sqrt{-\bar g}\,\bar F^{\mu\nu}\right) = 0,
  \label{eq:order0}\\
  O(\epsilon^1):&\qquad \delta G_{\mu\nu} - \kappa\left(\delta T^{\rm mag}_{\mu\nu} + \delta T^{Q}_{\mu\nu}\right) = 0,
  \qquad
  \delta\!\left[\partial_\mu\!\left(\sqrt{-g}F^{\mu\nu}\right)\right] = g_c^2\,\delta\!\left(\sqrt{-g}J^\nu\right).
  \label{eq:order1}
\end{align}
The zeroth-order equations \eqref{eq:order0} say that the chosen background solves the theory. They are imposed first, and everything afterwards uses them. The first-order equations \eqref{eq:order1} are what we reduce.

The Bianchi identity gives $\nabla^\mu G_{\mu\nu}\equiv0$. The system \eqref{eq:order1} is therefore solvable only if the matter side is covariantly conserved in the perturbed geometry. The total stress tensor is
\begin{equation}
  T^{\mu\nu} \;=\; T_{\rm mag}^{\mu\nu} \;+\; T_Q^{\mu\nu},
  \label{eq:Ttotal}
\end{equation}
and the Einstein equation $G_{\mu\nu}=\kappa T_{\mu\nu}$ together with the Bianchi identity gives
\begin{equation}
  \nabla_\mu T^{\mu\nu} \;=\; 0
  \label{eq:totalcons}
\end{equation}
exactly, at every order in $\epsilon$. With $\nabla_\mu T_{\rm mag}^{\mu\nu}=F^{\nu\lambda}J_\lambda$, which follows from \eqref{eq:maxwell} and \eqref{eq:Tmag}, Eq.~\eqref{eq:totalcons} is equivalent to
\begin{equation}
   \;\nabla_\mu T_Q^{\mu\nu} \;=\; -\,F^{\nu\lambda}J_\lambda\;
  \label{eq:forcebalance}
\end{equation}
order by order in $\epsilon$. The two constituents are not separately conserved in general. Equation \eqref{eq:forcebalance} says that whatever energy-momentum the electromagnetic field gains through the Lorentz force is what the fermion field loses. The same requirement is why a single charged Dirac particle cannot be treated together with its own Maxwell field \cite{Weinbaum:2026fbp}. The single-particle state obeys the free Dirac equation, its free stress tensor is conserved on its own, and the exchange \eqref{eq:forcebalance} fails. For a general field strength the right-hand side would also contain a dual term $(\ast F)^{\nu}{}_{\sigma}J_m^{\sigma}$, with $J_m$ a magnetic current, arising from the two terms in $\nabla_\mu T_{\rm mag}^{\mu\nu}$ that cancel only by the Bianchi identity. Here it is absent. The charge $q$ is a topological flux rather than a source, $\dd F=0$ holds identically for the background and the perturbation, and $\int_{S^2}F = - 2\pi q \neq 0$ only because $S^2$ is non-contractible. The MMP throat has neither a horizon nor an origin, and the flux threads the wormhole from one mouth to the other.

At order $\epsilon^0$ the condition is satisfied identically,
\begin{equation}
  \nabla_\mu T^{\mu\nu}\big|_{O(\epsilon^0)} \;=\; 0 .
  \label{eq:zerothtotal}
\end{equation}
This follows from the Bianchi identity applied to the background Einstein equation $\bar G_{\mu\nu}=\kappa\bar T_{\mu\nu}$ and holds for any background that solves the field equations. Each constituent is also separately conserved at this order, since the background carries no current,
\begin{equation}
  \bar J^{\mu} \;=\; 0
  \qquad\Longrightarrow\qquad
  \nabla_\mu T_{\rm mag}^{\mu\nu}\big|_{O(\epsilon^0)} = 0,
  \qquad
  \nabla_\mu T_Q^{\mu\nu}\big|_{O(\epsilon^0)} = 0 .
  \label{eq:zerothsplit}
\end{equation}
A static spherically symmetric current could only be a charge density $J^\mu\propto \delta^\mu{}_t$, and the background has no electric field to support one. There is no background Hall current either, since that requires a perturbing field to respond to. Hence $F^{\nu\lambda}J_\lambda$ vanishes at $O(\epsilon^0)$ and \eqref{eq:forcebalance} reduces to $0=0$. This remains true in the throat at $O(\alpha)$, where $\bar T_Q^{\mu\nu}\neq0$ but is divergence-free on its own. The conformal anomaly of MMP's Appendix~F affects the trace of $T_Q$, not its divergence. The consistency condition has no content until first order in $\epsilon$.

Expanding the left-hand side of \eqref{eq:forcebalance} to first order in $\epsilon$, with $\nabla_\mu T_Q^{\mu\nu}=\partial_\mu T_Q^{\mu\nu}+\Gamma^\mu{}_{\mu\lambda}T_Q^{\lambda\nu}+\Gamma^\nu{}_{\mu\lambda}T_Q^{\mu\lambda}$, the variation of each Christoffel factor produces a term that multiplies the background fermion stress tensor,
\begin{equation}
  \delta\!\left(\nabla_\mu T_Q^{\mu\nu}\right)
  \;=\; \nabla_\mu \delta T_Q^{\mu\nu}
  \;+\; \delta\Gamma^{\mu}{}_{\mu\lambda}\,\bar T_Q^{\lambda\nu}
  \;+\; \delta\Gamma^{\nu}{}_{\mu\lambda}\,\bar T_Q^{\mu\lambda},
  \label{eq:bianchiexp}
\end{equation}
where $\nabla_\mu$ on the right is the background connection. With $\delta\Gamma^{\mu}{}_{\mu\lambda}=\tfrac12\partial_\lambda h$ and $h\equiv \bar g^{\sigma\lambda}h_{\sigma\lambda}$ the trace of the metric perturbation, which for the ansatz \eqref{eq:hansatz1}--\eqref{eq:hansatz2} is
\begin{equation}
  h \;=\; \big(-H_0 + H_2 + 2K\big)\,Y(\theta),
  \label{eq:htrace}
\end{equation}
the second term in \eqref{eq:bianchiexp} is $\tfrac12(\partial_\lambda h)\,\bar T_Q^{\lambda\nu}$. It is proportional to the background fermion stress tensor and does not vanish when $\delta T_Q$ is set to zero.

This fixes what each order requires. Where $\bar T_Q^{\mu\nu}=0$, which is the mouth at $O(\alpha^0)$, both correction terms in \eqref{eq:bianchiexp} drop out, and $\delta T_Q^{\mu\nu}=0$ is a consistent truncation provided $\delta(\nabla_\mu T_{\rm mag}^{\mu\nu})=0$. Expanded as in \eqref{eq:bianchiexp},
\begin{equation}
  \delta\big(\nabla_\mu T_{\rm mag}^{\mu\nu}\big)
  \;=\; \nabla_\mu\delta T_{\rm mag}^{\mu\nu}
  \;+\; \delta\Gamma^{\mu}{}_{\mu\lambda}\,\bar T_{\rm mag}^{\lambda\nu}
  \;+\; \delta\Gamma^{\nu}{}_{\mu\lambda}\,\bar T_{\rm mag}^{\mu\lambda} ,
  \label{eq:bianchimag}
\end{equation}
and here the Christoffel terms do not vanish, since the monopole carries stress. They are harmless. For any metric and any $F=\dd A$ the Maxwell stress tensor obeys
\begin{equation}
  \nabla_\mu T_{\rm mag}^{\mu\nu} \;=\; \frac{1}{g_c^{2}}\,\big(\nabla_\mu F^{\mu\sigma}\big)\,F^{\nu}{}_{\sigma},
  \label{eq:magidentity}
\end{equation}
which follows from differentiating \eqref{eq:Tmag} and using $\partial_{[\mu}F_{\nu\sigma]}=0$. Linearising \eqref{eq:magidentity} accounts for the Christoffel terms of \eqref{eq:bianchimag}, and with $\nabla_\mu\bar F^{\mu\sigma}=0$ on the background
\begin{equation}
  \delta\big(\nabla_\mu T_{\rm mag}^{\mu\nu}\big)
  \;=\; \frac{1}{g_c^{2}}\,\delta\big(\nabla_\mu F^{\mu\sigma}\big)\,\bar F^{\nu}{}_{\sigma} .
\end{equation}
The right-hand side is proportional to the perturbed Maxwell equation, which without charged matter is $\delta(\nabla_\mu F^{\mu\sigma})=0$. The Maxwell equations therefore guarantee consistency of the Einstein equations in the mouth at $O(\alpha^0)$, and there the mouth reduces to a closed Einstein--Maxwell problem with no fermionic input (Sec.~\ref{sec:mouthpert}). The fermion stress tensor has no analogue of \eqref{eq:magidentity}. With charged matter the identity \eqref{eq:magidentity}, now with $\nabla_\mu F^{\mu\sigma}=g_c^2J^\sigma$, gives $\nabla_\mu T_{\rm mag}^{\mu\nu}=F^{\nu\lambda}J_\lambda$, and total conservation becomes the force balance \eqref{eq:forcebalance}.

In the throat $\bar T_Q^{\mu\nu}\neq0$, since the Casimir energy holds the throat open. At $O(\alpha)$ the same is true in the mouth, where the conformal anomaly alone gives $\bar T_Q^{\mu\nu}\neq0$ (Sec.~\ref{sec:mouth}). The two correction terms in \eqref{eq:bianchiexp} are then nonzero and of order $\alpha$. Imposing $\delta T_Q^{\mu\nu}=0$ leaves \eqref{eq:forcebalance} violated at $O(\alpha\epsilon)$, and at generic frequency the first-order system has no nonzero solution. The fermion sector must respond. The response is fixed by the two-dimensional conformal Ward identity, and the $\theta$ component of \eqref{eq:forcebalance} then requires an induced Hall current $\delta J^\phi$ (Secs.~\ref{sec:TQ}, \ref{sec:hall} and \ref{sec:mouthpert}). The obstruction would be invisible to anyone who wrote $\delta(\nabla_\mu T_Q^{\mu\nu})=\nabla_\mu\delta T_Q^{\mu\nu}$. Since the background carries no current, the $O(\epsilon)$ force balance \eqref{eq:forcebalance} reads
\begin{equation}
  \nabla_\mu T_Q^{\mu\nu}\big|_{O(\epsilon)}
  \;=\; -\,\bar F^{\nu\lambda}\,\delta J_\lambda ,
  \label{eq:forcebalance1}
\end{equation}
the term $\delta F^{\nu\lambda}\bar J_\lambda$ being absent because $\bar J_\lambda=0$. Section~\ref{sec:hall-force} works out the components of \eqref{eq:forcebalance1}.

\subsection{Harmonic separation}
\label{sec:separation}

Throughout the paper, $\mathcal E_{\mu\nu}$ denotes the residual of the first-order Einstein equation,
\begin{equation}
  \mathcal E_{\mu\nu} \;\equiv\; \delta G_{\mu\nu} \;-\; \kappa\,\delta T_{\mu\nu},
  \qquad
  \delta T_{\mu\nu} = \delta T^{\rm mag}_{\mu\nu} + \delta T^{Q}_{\mu\nu},
  \label{eq:Edef}
\end{equation}
and the content of \eqref{eq:order1} is $\mathcal E_{\mu\nu}=0$. The residual of the first-order Maxwell equation is
\begin{equation}
  \mathcal M^{\nu} \;\equiv\; \delta\big[\partial_\mu\big(\sqrt{-g}\,F^{\mu\nu}\big)\big]
  \;-\; g_c^{2}\,\sqrt{-\bar g}\;\delta J^{\nu},
  \label{eq:Mdef}
\end{equation}
and the first-order Maxwell equation is $\mathcal M^{\nu}=0$. Here $\bar g$ is the background metric, $F_{\mu\nu}=\partial_\mu A_\nu-\partial_\nu A_\mu$, and $\delta J^{\nu}$ is the Hall current of Eq.~\eqref{eq:Jclosed}. The current itself does not depend on $\alpha$ at the order kept. The Maxwell source term $g_c^2\delta J^\nu=4\pi^2\alpha q\,\delta J^\nu$ vanishes at $\alpha=0$ in both regions, and at $O(\alpha)$ only its $\phi$ component is nonzero. In every radial equation $\mathcal M^\phi$ appears divided by $q$, and with $w=4a/q$ the charge drops out.

Every component of $\mathcal E_{\mu\nu}$ is a sum of terms proportional to $Y$, $Y'$ and $Y''$. Using Legendre's equation repeatedly,
\begin{equation}
  Y'' \;=\; -\cot\theta\; Y' \;-\; \lhat\, Y ,
  \qquad
  \lhat \equiv l(l+1),
  \label{eq:legendre}
\end{equation}
all higher derivatives are eliminated. Each tensor component reduces to radial functions times $Y$ and $Y'$ and, in the angular block, $\cot\theta\,Y'$. Parity fixes the pattern,
\begin{align}
  \propto Y:&\qquad \mathcal E_{tt},\; \mathcal E_{tr},\; \mathcal E_{rr},\quad\text{and the angular trace } \mathcal E_{\theta\theta} + \mathcal E_{\phi\phi}/\sin^2\theta,
  \\
  \propto Y':&\qquad \mathcal E_{t\theta},\; \mathcal E_{r\theta},\quad\text{and the $\phi$-component of Maxwell, } \mathcal M^\phi,
  \\
  \text{identically } 0:&\qquad \mathcal E_{t\phi},\;\mathcal E_{r\phi},\;\mathcal E_{\theta\phi},\quad\text{and } \mathcal M^t,\,\mathcal M^r,\,\mathcal M^\theta .
\end{align}
The angular block $\mathcal E_{AB}$ carries two independent pieces. The trace is
\begin{equation}
  \mathcal E_{\theta\theta} + \frac{\mathcal E_{\phi\phi}}{\sin^2\theta} \;=\; \mathcal E^{\rm trace}(r)\,Y(\theta),
  \label{eq:tracepart}
\end{equation}
where the radial function $\mathcal E^{\rm trace}$ is built from all five unknowns and their derivatives, and the coefficient of $Y'$ vanishes identically. Setting $\mathcal E^{\rm trace}$ to zero gives one radial equation. The traceless part is proportional to a single angular function. Using $\mathring\nabla_\theta\mathring\nabla_\theta Y = Y''$, $\mathring\nabla_\phi\mathring\nabla_\phi Y = \sin\theta\cos\theta\,Y'$ and \eqref{eq:legendre},
\begin{equation}
  \mathcal E_{\theta\theta} - \frac{\mathcal E_{\phi\phi}}{\sin^2\theta}
  \;=\; \mathcal E^{\rm tl}(r)\,\big(2\cot\theta\,Y' + \lhat\,Y\big),
  \label{eq:tlcheck}
\end{equation}
which defines the radial coefficient $\mathcal E^{\rm tl}$. The angular factor does not vanish identically for $l\ge2$, and the field equations require $\mathcal E^{\rm tl}=0$. On the ansatz,
\begin{equation}
  \mathcal E^{\rm tl} \;=\; -\tfrac12\left(H_0 - H_2\right),
  \label{eq:tlvalue}
\end{equation}
for any static spherically symmetric background, independently of $A$, $B$ and $R$. Hence
\begin{equation}
  H_0 \;=\; H_2 .
  \label{eq:H0H2}
\end{equation}
The trace \eqref{eq:tracepart} is a second-order equation involving all the unknowns. We do not use it directly. Step~4 below shows that it follows from the constraints already imposed, and it is one of the three linearised Bianchi residuals that must vanish.

We pass to the frequency domain with a Laplace variable,
\begin{equation}
  \left\{H_0,H_1,H_2,K,w\right\}(t,r) \;=\; \left\{H_0,H_1,H_2,K,w\right\}(r)\; e^{p t},
  \qquad p = -i\omega ,
  \label{eq:freq}
\end{equation}
and correspondingly $e^{p\tau}$, $p=-i\Om$, in the throat, where $\Om=\omega\ell$ is the frequency conjugate to the dimensionless time $\tau$. The constraint rules of the reduction divide by $p$, and the reduction assumes $p\neq0$. The static sector is not treated here. Where we set $p=0$ below, we set it in the coefficients of the reduced system. We keep $p$ real in the intermediate steps and set $p\to-i\omega$, $p^2\to-\omega^2$ at the end.

\subsection{The reduction algorithm}
\label{sec:algorithm}

Harmonic separation leaves eight radial equations, $\mathcal E_{tt},\mathcal E_{tr},\mathcal E_{rr},\mathcal E_{t\theta},\mathcal E_{r\theta}$, the angular trace $\mathcal E^{\rm trace}$, the traceless angular combination $\mathcal E^{\rm tl}$, and the $\phi$ component of Maxwell $\mathcal M^\phi$, for five unknown radial functions $H_0,H_1,H_2,K,w$. The eight equations are not independent. The contracted Bianchi identity $\nabla^\mu G_{\mu\nu}\equiv0$ relates them, in the sense made precise in Sec.~\ref{sec:consistency}. The reduction below uses this redundancy in a fixed order, each step using specific equations to eliminate specific unknowns, until a closed system remains for the two functions that carry propagating degrees of freedom. We record the algorithm once in general terms. Sections~\ref{sec:throatmaster} and \ref{sec:mouth} apply it to the throat and to the mouth, with only the background functions $A,B,R$ and the source changing between the two. At every stage each equation is a linear form in the basis elements $K$, $K'$, $w''$, \dots, with coefficients that are functions of the radial coordinate. They are rational at $\alpha^0$, and at $O(\alpha)$ they also contain $\arctan\rho$ and $\log(1+\rho^2)$ in the throat and the redshift function $\Phi_1$ of Sec.~\ref{sec:mouthTQ} in the mouth.

\begin{description}

\item[Step 1 (traceless angular equation).]
The traceless angular equation gives $H_0=H_2$, Eq.~\eqref{eq:H0H2}. This step uses only the one traceless combination and is available before anything else in the system has been touched.

\item[Step 2 (constraints).]
The remaining work is done in the frequency domain, Eq.~\eqref{eq:freq}, where $\partial_t\to p$. After Step~1 has removed $H_0$, three equations involve only first radial derivatives,
\begin{align}
  \mathcal E_{t\theta} &= \mathcal E_{t\theta}\big[H_1,H_1',H_2,K,w\big]=0 ,
  \label{eq:etth_content}\\
  \mathcal E_{r\theta} &= \mathcal E_{r\theta}\big[H_1,H_2,H_2',K',w'\big]=0 ,
  \label{eq:erth_content}\\
  \mathcal E_{tr} &= \mathcal E_{tr}\big[H_1,H_2,K,K'\big]=0.
  \label{eq:etr_content}
\end{align}
Each contains at least one of $H_1'$, $H_2'$ and $K'$. Together they form a $3\times3$ linear system for $(H_1',H_2',K')$ with an invertible coefficient matrix. Solving it gives the derivative rules
\begin{equation}
  H_1' = \mathcal{H}_1\big[H_1,H_2,K,w\big],
  \qquad
  H_2' = \mathcal{H}_2\big[H_1,H_2,K,w'\big],
  \qquad
  K' = \mathcal{K}\big[H_1,H_2,K\big].
  \label{eq:constraints}
\end{equation}
The $H_2$ dependence of $\mathcal E_{tr}$ and of $K'$ comes from the radial derivative of the sphere radius $R$. That derivative vanishes on the AdS$_2\times S^2$ background of the throat at $\alpha=0$, and the rules simplify to \eqref{eq:t-H1p0}--\eqref{eq:t-Kp0}. There $K'$ involves only $H_1$ and $K$, and $w'$ enters only through $H_2'$. In the mouth, where $R=r$, the rule for $K'$ acquires a term $H_2/r$. The full rules for the mouth and for the throat at $O(\alpha)$ are given in Appendix~\ref{app:constraints}. Separately, the $\phi$ component of the perturbed Maxwell equation carries
\begin{equation}
  \mathcal M^\phi = \mathcal M^\phi\big[K,w,w',w''\big] ,
\end{equation}
which contains no metric amplitude other than $K$ once $H_0$ has been eliminated. It is solved for the highest derivative present,
\begin{equation}
  w'' = \mathcal{W}\big[K,w,w'\big] .
  \label{eq:wpp}
\end{equation}
At the end of this step every radial derivative in the system is expressed algebraically in terms of $H_1$, $H_2$, $K$, $w$ and $w'$.

\item[Step 3 (the $rr$ equation).]
After Step~1 the content of $\mathcal E_{rr}$ is
\begin{equation}
  \mathcal E_{rr} = \mathcal E_{rr}\big[H_1,H_2,H_2',K,K',w\big] ,
\end{equation}
which is already first order. Substituting the rules \eqref{eq:constraints} removes $H_2'$ and $K'$ and leaves an algebraic relation, linear in $H_2$,
\begin{equation}
  H_2 = \mathcal H_{\rm alg}\big[H_1,K,w,w'\big] ,
  \label{eq:H2alg}
\end{equation}
where $w'$ enters through the rule for $H_2'$. On the throat background at $\alpha=0$ the $H_1$ and $H_2'$ terms of $\mathcal E_{rr}$ are absent, for the same reason as in Step~2, and only $K'$ needs to be substituted. $H_2$ is therefore never an independent dynamical variable. It is fixed algebraically by the other fields at every $r$.

\item[Step 4 (Bianchi residuals).]
Two field equations remain unused. Both carry second derivatives that Steps~1--3 never touched,
\begin{align}
  \mathcal E_{tt} &= \mathcal E_{tt}\big[H_2,H_2',K,K',K'',w\big] ,
  \\
  \mathcal E^{\rm trace} &= \mathcal E^{\rm trace}\big[H_1,H_1',H_2,H_2',H_2'',K,K',K'',w\big] .
\end{align}
A third condition is internal. Equation \eqref{eq:H2alg} expresses $H_2$ algebraically while \eqref{eq:constraints} gives a rule for $H_2'$, and the two must be compatible. Differentiating \eqref{eq:H2alg} produces $H_1'$, $K'$, $w'$ and $w''$, each of which has an algebraic expression from Step~2. Substituting them leaves a relation with no derivatives, and that relation must reproduce the rule for $H_2'$. The appearance of $w''$ makes this a cross-check between the Einstein and Maxwell sectors, since \eqref{eq:wpp} is used nowhere in the derivation of \eqref{eq:H2alg} or of the rule for $H_2'$. The three residuals are $\mathcal E_{tt}$, $\mathcal E^{\rm trace}$ and this compatibility condition. All three must reduce identically to $0=0$ once Steps~1--3 have been imposed. Steps~1--3 impose six of the eight separated equations, and the contracted Bianchi identity of Sec.~\ref{sec:consistency} makes the rest redundant only if the total source is conserved. The residuals are therefore the force balance \eqref{eq:forcebalance} written in terms of the separated radial equations. They vanish if and only if the source is exactly conserved.

\item[Step 5 (first-order system).]
Once Step~4 has been verified, what survives is a closed, linear, first-order system in the four quantities that were not eliminated,
\begin{equation}
  \frac{\dd}{\dd r}\, \bm{y} \;=\; \mathsf M(r)\,\bm{y},
  \qquad
  \bm{y} \;=\; \left(K,\;H_1,\;w,\;w'\right)^{\!\top} ,
  \label{eq:firstorder}
\end{equation}
with $\mathsf M(r)$ a $4\times4$ matrix of known functions of the background. The variables $H_0$ and $H_2$ are reconstructed afterwards from \eqref{eq:H0H2} and \eqref{eq:H2alg}. The amplitudes $H_0$, $H_1$ and $H_2$ are constrained and non-dynamical. $H_0$ and $H_2$ are fixed algebraically, and $H_1$ is fixed by a first-order equation sourced by the physical fields. The two variables carrying propagating degrees of freedom are $K$ and $w$, one gravitational and one electromagnetic.

\item[Step 6 (master equations).]
Equation \eqref{eq:firstorder} still carries $H_1$ explicitly. Its own equation from \eqref{eq:constraints} eliminates $H_1$ in favour of $K'$, and the system reduces to two coupled second-order equations for $\bm Y\equiv(K,w)^\top$,
\begin{equation}
  \bm{Y}'' \;+\; P(r)\,\bm{Y}' \;+\; Q(r)\,\bm{Y} \;=\; 0 ,
  \label{eq:secondorder}
\end{equation}
with $P,Q$ known $2\times2$ matrices of $r$. Passing to the tortoise coordinate defined by
\begin{equation}
  \frac{\dd x}{\dd r} \;=\; \frac{1}{S(r)} ,
  \qquad S \;\equiv\; \sqrt{\frac{A}{B}} ,
  \label{eq:tortoise}
\end{equation}
the chain rule $\dd/\dd r = S^{-1}\dd/\dd x$ converts \eqref{eq:secondorder} into $\bm Y_{xx}+\tilde P\,\bm Y_x+\tilde Q\,\bm Y=0$ with
\begin{equation}
  \tilde P \;=\; S\,P \;-\; S'\,I_2 ,
  \qquad
  \tilde Q \;=\; S^2\,Q ,
  \label{eq:tortoisePQ}
\end{equation}
where $I_2$ is the $2\times2$ identity matrix. A position-dependent change of variables $\bm Y=\mathcal U\,\tilde{\bm Y}$ removes the first-derivative term. After substituting, the coefficient of $\tilde{\bm Y}_x$ is $2\,\partial_x\mathcal U+\tilde P\,\mathcal U$. It vanishes if
\begin{equation}
  \partial_x\mathcal U \;=\; -\tfrac12\,\tilde P\,\mathcal U ,
  \qquad
  \mathcal U(x) \;=\; \operatorname{P}\exp\!\Big(-\tfrac12\!\int^{x}\!\tilde P\,\dd x'\Big),
  \label{eq:Udef}
\end{equation}
where $\operatorname{P}$ denotes path ordering. Using \eqref{eq:Udef} to eliminate the derivatives of $\mathcal U$ brings the system to coupled Schr\"odinger form,
\begin{equation}
  \partial_x^2\,\tilde{\bm Y} \;+\; \left(\omega^2\,I_2 - V\right) \tilde{\bm Y} \;=\; 0 ,
  \qquad
  V \;=\; -\,p^2 I_2 \;-\; \mathcal U^{-1}\Big(\tilde Q - \tfrac12\,\partial_x\tilde P
  - \tfrac14\,\tilde P^2\Big)\,\mathcal U ,
  \label{eq:schrodinger}
\end{equation}
with $p=-i\omega$ and $\partial_x\tilde P=S\,\tilde P'$. The conjugation by $\mathcal U$ has no scalar analogue. When $\mathcal U$ commutes with the bracket it drops out, and $V$ reduces to
\begin{equation}
  V_{\rm c} \;=\; -\,\tilde Q \;-\; p^2 I_2 \;+\; \tfrac12\,\partial_x\tilde P \;+\; \tfrac14\,\tilde P^2 ,
  \label{eq:schrodinger-commuting}
\end{equation}
which we call the commuting form. It is the result for a single equation carried over to matrices, and it holds whenever $\tilde P$ is proportional to the identity. It does not hold in general. The integration constant in \eqref{eq:Udef} multiplies $\mathcal U$ from the right by a constant matrix. This conjugates $V$ by that matrix, a constant change of basis of $\tilde{\bm Y}$ with no physical content.

Two cases arise in this paper. When $\tilde P$ vanishes, $\mathcal U=I_2$ and \eqref{eq:schrodinger} is exact. This is the exact AdS$_2\times S^2$ throat of Sec.~\ref{sec:alpha0}. When $\tilde P=\alpha\tilde P_1$ is small, the path ordering matters only at $O(\alpha^2)$, and to first order
\begin{equation}
  \mathcal U \;=\; I_2 - \tfrac{\alpha}{2}\,\mathcal G,
  \qquad
  \partial_x\mathcal G=\tilde P_1,
  \qquad
  V \;=\; V_{\rm c} \;-\; \tfrac{\alpha}{2}\big[\,\mathcal G,\tilde Q_0\big] \;+\; O(\alpha^2),
  \label{eq:commutator-general}
\end{equation}
where $\tilde Q_0$ is the $\alpha^0$ part of $\tilde Q$. The commutator is the first-order remnant of the conjugation. It vanishes only when $\mathcal G$ commutes with $\tilde Q_0$. In the throat at $O(\alpha)$ it does not vanish. It is purely off-diagonal in the basis \eqref{eq:Zdef} and contributes to the off-diagonal entries of $V$ (Sec.~\ref{sec:throat-schrodinger} and Appendix~\ref{app:V1a}). Its matrix elements between the two modes of a level that they share cancel against other terms, and on such a level only the Hall current mixes the modes (Sec.~\ref{sec:stab-normalmodes}). When $\tilde P$ is of order one and does not commute with itself at different points, neither the path-ordered exponential nor the conjugation can be done in closed form. This is the case of the mouth. Even when $V$ can be found, it is a $2\times2$ matrix. It separates into two scalar potentials only if a constant basis change diagonalises it at every $x$, as happens for the throat at $\alpha=0$.

\end{description}

At $O(\alpha)$, in the throat and in the mouth alike, the fermion response introduces three extra functions, the Weyl factor $\delta\Upsilon$ and the transport fields $\xi_a$ of Sec.~\ref{sec:response}, defined by the decomposition equations \eqref{eq:deceq_tt}--\eqref{eq:deceq_rr}. The algorithm absorbs them with two sub-steps inserted between Steps~1 and 2, given in Sec.~\ref{sec:throat-reduction}.

\section{The lowest-Landau-level stress tensor in the throat}
\label{sec:TQ}

\subsection{Physical origin of \texorpdfstring{$T^Q$}{TQ}}
\label{sec:throat-origin}

We reduce the four-dimensional fermions to a two-dimensional CFT, identify the state that CFT is in, transport its stress tensor to the curved throat metric with the conformal anomaly, and then decide how much of the resulting tensor to keep.

Consider a single four-dimensional massless Dirac fermion of unit charge in the monopole background \eqref{eq:monopole}. The index theorem reduces it to a two-dimensional problem. A charged fermion on a two-sphere threaded by $q$ units of magnetic flux has exactly $q$ normalisable zero modes of the sphere Dirac operator \cite{Maldacena:2018gjk,Atiyah:1963zz,Wu:1976ge}. These are the lowest Landau level. They form a single multiplet of total angular momentum $j = (q-1)/2$, whose degeneracy $2j+1 = q$ is the index. All other modes have a gap set by the cyclotron energy $\sim\sqrt q/R$. They are never excited. Section~\ref{sec:hll} adds their vacuum polarisation, and their mixing with the lowest level through the perturbation is not included. The construction is a low-energy effective description, valid for excitations with $l \ll \sqrt q$. Each LLL state is a fermion tied to one magnetic field line, free to move along it but unable to move across it. Weyl invariance removes the remaining dependence on $R$. Write the throat metric as a conformal factor times a product,
\begin{equation}
  g \;=\; g_{ab}\,\dd x^a \dd x^b + R^2\,\mathring g
    \;=\; R^2\Big[\underbrace{\tfrac{g_{ab}}{R^2}}_{\textstyle \hat g_{ab}}\dd x^a\dd x^b
    + \mathring g\Big],
  \qquad a,b\in\{\tau,\rho\},
  \label{eq:conformalsplit}
\end{equation}
so that up to an overall factor $R^2$ the geometry is a two-dimensional metric $\hat g$ times a unit sphere. The massless Dirac action in four dimensions is invariant under $g\to e^{2\Upsilon}g$, $\Psi\to e^{-3\Upsilon/2}\Psi$ \cite{Birrell:1982ix}, and the overall factor drops out of the fermion dynamics. This is why $R$ does not appear in the reduced kinetic term and no dilaton coupling is generated at leading order. The same point is made in the reduction of \cite{Freivogel:2026ujn}, eq.~(4.54). The LLL sector is therefore
\begin{equation}
  \text{$q$ massless two-dimensional Dirac fermions on }
  \hat g_{ab} \;=\; \frac{g_{ab}}{R^2},
  \qquad a,b \in \{\tau,\rho\} .
  \label{eq:2dmetric}
\end{equation}
A two-dimensional massless Dirac fermion is two Majorana fermions, each of central charge $1/2$ \cite{DiFrancesco:1997nk}. One Dirac fermion therefore has $c_1=1$, and the $q$ fermions of the LLL sector have
\begin{equation}
  c \;=\; q .
  \label{eq:ceqq}
\end{equation}
This large central charge makes the semiclassical treatment sensible. The one-loop stress tensor is $O(q)$. We assume, as MMP do \cite{Maldacena:2018gjk}, that the corrections to it are suppressed by further powers of $1/q$. This is the requirement $q\gg1$ listed among the assumptions in Sec.~\ref{sec:summary}.

In conformal gauge $\hat g$ is a Weyl factor times flat space,
\begin{equation}
  \hat g = e^{2\bar\Upsilon}\eta,
  \qquad \eta = -\dd\tau^2 + \dd x^2,
  \qquad x = \arctan\rho,
  \qquad \bar\Upsilon = \half\log\!\left(1+\rho^2\right).
  \label{eq:conformalgauge}
\end{equation}
The coordinate $x$ is the tortoise coordinate of the throat. As $\rho$ runs over the real line, $x$ covers the finite interval $(-\pi/2,\pi/2)$, and the whole infinite throat is conformal to a strip of width $\pi$. At $O(\alpha)$ the tortoise coordinate acquires a correction, given in \eqref{eq:t-tortoise}. The Casimir stress is itself of order $\alpha$, and the unperturbed coordinate $x=\arctan\rho$ suffices to compute it.

A stress tensor is a property of a state, and the state is fixed by the global geometry of the field lines. The MMP solution has two mouths, with charges $+q$ and $-q$, held a distance $d$ apart in an asymptotically flat exterior, which we call the ambient region. Field lines leave the $+q$ mouth, cross the ambient region to the $-q$ mouth, and return through the throat. Every field line closes on itself, with total length
\begin{equation}
  L_{\rm loop} \;=\; (\text{throat traversal}) \;+\; (\text{ambient return path}) .
\end{equation}
There are no magnetic monopoles anywhere in the solution, $\dd F = 0$ holds everywhere, and a line of flux can neither begin nor end. The fermion on each field line is therefore quantised on a spatial circle, and the relevant state is the vacuum on that circle. The vacuum of a CFT on a line has vanishing stress tensor. The vacuum on a circle carries Casimir energy, negative for antiperiodic boundary conditions, and this negative energy holds the throat open. This is also why two mouths are needed \cite{Maldacena:2018gjk}. The flux of a single mouth would escape to infinity, the field lines would not close, and there would be no Casimir energy.

The circumference must be measured in the units in which the Casimir stress tensor is computed. The two-dimensional CFT lives on $\hat g_{ab}=g_{ab}/R^2$, and in conformal gauge $\hat g = e^{2\bar\Upsilon}\eta$ with $\eta=-\dd\tau^2+\dd x^2$. The Casimir stress tensor below is computed on the flat metric $\eta$, and the circumference is measured in $\tau$ and $x$,
\begin{equation}
  L_c \;=\; \int_{-\infty}^{\infty}\frac{\dd\rho}{1+\rho^2} \;=\; \pi
  \qquad\text{(in $\tau$, $x$ units).}
  \label{eq:Lc}
\end{equation}
Measured in asymptotic time the same traversal is longer. With $\tau = t/\ell$ it takes
\begin{equation}
  \ell\,L_c \;=\; \pi\ell ,
\end{equation}
which is the throat length quoted by MMP and the origin of the echo period of \cite{Mondal:2025tht}. The factor $\ell$ comes from the matching $\tau=t/\ell$ of the throat time to the asymptotic time, and \eqref{eq:Lc} and $\pi\ell$ measure the same traversal with the two clocks. In the long-throat limit $\ell\gg d$ the ambient return path is negligible against either.

For a two-dimensional CFT on a spatial circle of circumference $L$ the ground-state energy is \cite{Affleck:1986bv,Bloete:1986qm}
\begin{equation}
  E_{\rm Cas} \;=\; \frac{2\pi}{L}\left(\mathsf L_0 + \bar{\mathsf L}_0 - \frac{c}{12}\right) ,
  \label{eq:cylinderE}
\end{equation}
where $c$ is the central charge and $\mathsf L_0$, $\bar{\mathsf L}_0$ are the Virasoro zero modes. The shift $-c/12$ is the Schwarzian term of the map from the plane to the cylinder. It is the same anomaly as \eqref{eq:anomalylaw}, and the same $c$ appears in the trace anomaly $\langle T^a{}_a\rangle = c\hat R/24\pi$. The ground state of the antiperiodic (Neveu--Schwarz) sector has $\mathsf L_0=\bar{\mathsf L}_0 = 0$, hence $E_{\rm Cas} = -\pi c/6L$ and the energy density
\begin{equation}
  \frac{E_{\rm Cas}}{L} \;=\; -\frac{\pi c}{6L^2} .
\end{equation}
The sign is what the wormhole needs, and it belongs to the antiperiodic sector. In the periodic (Ramond) sector the fermion has zero modes, the ground state has $\mathsf L_0\neq0$, and the Casimir energy is not negative. The antiperiodic boundary condition around the closed field line is MMP's choice \cite{Maldacena:2018gjk}, and we adopt it. With $L_c=\pi$ and $c=q$, and converting $t_{xx}$ to $t_{\rho\rho}$ with the Jacobian $(\dd x/\dd\rho)^2 = (1+\rho^2)^{-2}$,
\begin{equation}
  t^{\rm Cas}_{ab} \;=\; -\frac{\pi c}{6 L_c^2}\,\mathrm{diag}\!\left(1,\;\frac{1}{(1+\rho^2)^2}\right)
  \;=\; \begin{pmatrix}- \dfrac{q}{6 \pi} & 0\\[6pt] 0 & - \dfrac{q}{6 \pi \left(1+\rho^{2}\right)^{2}}\end{pmatrix} .
  \label{eq:tcas}
\end{equation}
The two entries are equal in the $x$ coordinate, as tracelessness on flat space requires, $\eta^{ab}t_{ab} = -t_{\tau\tau}+t_{xx}=0$.

Equation \eqref{eq:tcas} is the stress tensor on the flat metric $\eta$. We need it on $\hat g = e^{2\bar\Upsilon}\eta$. In the quantum theory the stress tensor is not Weyl covariant. The nonvanishing trace $\langle T^a{}_a\rangle = c\hat R/24\pi$ expresses this, and its finite form is the transformation law \cite{Birrell:1982ix,Davies:1976ei} (MMP App.~F, eq.~(F.29))
\begin{equation}
  t_{ab}\!\left[e^{2\Upsilon}g_{\rm ref}\right] = t_{ab}\!\left[g_{\rm ref}\right]
  - \frac{c}{12\pi}\Big[\partial_a\Upsilon\,\partial_b\Upsilon
  - \half (g_{\rm ref})_{ab}(\partial\Upsilon)^2
  - \nabla_a\nabla_b\Upsilon + (g_{\rm ref})_{ab}\,\Box\,\Upsilon\Big] .
  \label{eq:anomalylaw}
\end{equation}
In complex coordinates this is the Schwarzian-derivative term in the transformation of $T(z)$. Every term on the right is second order in derivatives of $\Upsilon$ and proportional to $c$, as a one-loop effect must be. Applying \eqref{eq:anomalylaw} with $g_{\rm ref}=\eta$ and $\Upsilon = \bar\Upsilon$ gives the background two-dimensional stress tensor
\begin{equation}
  \bar t_{ab} \;=\; \begin{pmatrix}\dfrac{q\,(\rho^{2} - 2)}{24 \pi} & 0\\[8pt]
  0 & -\,\dfrac{q\,(\rho^{2} + 4)}{24 \pi \left(1+\rho^{2}\right)^{2}}\end{pmatrix} .
  \label{eq:tbar}
\end{equation}
Its traceless part, Eq.~\eqref{eq:thatless} below, is the tensor that MMP display in their eqs.~(5.27) and (5.38) \cite{Maldacena:2018gjk}. Its trace reproduces the anomaly on the unit AdS$_2$ background, where $\hat R = -2$,
\begin{equation}
  \hat g^{ab}\bar t_{ab} \;=\; \frac{c\,\hat R}{24\pi} \;=\; -\frac{c}{12\pi},
  \label{eq:tracevalue}
\end{equation}
which confirms that the anomalous terms in \eqref{eq:anomalylaw} have been applied with the right coefficient. It is conserved, $\hat\nabla^a \bar t_{ab} = 0$. Conservation and tracelessness are independent conditions. The anomaly spoils the second and never the first.

MMP \cite{Maldacena:2018gjk} retain only the two-dimensional traceless part of $\bar t_{ab}$. They discard pieces proportional to the AdS$_2$ and $S^2$ metrics on the grounds that these renormalise the two radii (their Sec.~5.2). For the background this is harmless. A term proportional to the metric shifts the effective cosmological constant in each factor, and one absorbs it by adjusting the radii of the two factors. For the perturbation problem the truncation is not harmless. Decompose the two-dimensional stress tensor into its traceless and trace parts,
\begin{equation}
  \bar t_{ab} \;=\; \bar t^{\rm tl}_{ab} \;+\; \tfrac12\,\hat g_{ab}\,\bar t^{\rm tr} ,
  \qquad
  \bar t^{\rm tr} \;\equiv\; \hat g^{ab}\bar t_{ab},
  \qquad
  \hat g^{ab}\bar t^{\rm tl}_{ab} = 0 .
  \label{eq:tracesplit}
\end{equation}
Evaluating \eqref{eq:tracesplit} on \eqref{eq:tbar} gives the traceless piece
\begin{equation}
  \bar t^{\rm tl}_{ab} \;=\; -\frac{q}{8\pi}\,
  \mathrm{diag}\!\left(1,\;\frac{1}{(1+\rho^2)^2}\right),
  \label{eq:thatless}
\end{equation}
whose components are $\rho$-independent in the flat frame, and which is traceless with respect to $\eta$ as well as $\hat g$. This is the piece MMP retain. Taking the divergence of \eqref{eq:tracesplit},
\begin{equation}
  \hat\nabla^a \bar t_{ab}
  \;=\; \hat\nabla^a \bar t^{\rm tl}_{ab} \;+\; \tfrac12\,\partial_b \bar t^{\rm tr} ,
  \label{eq:divsplit}
\end{equation}
so that the traceless piece is separately conserved if and only if the trace is covariantly constant. On the background the trace is fixed by the anomaly to be proportional to the two-dimensional Ricci scalar, and the unperturbed throat is AdS$_2$ with $\hat R = -2$. Hence
\begin{equation}
  \bar t^{\rm tr} \;=\; \frac{c\,\hat R}{24\pi} \;=\; -\frac{c}{12\pi}
  \qquad\Longrightarrow\qquad
  \partial_b \bar t^{\rm tr} = 0 ,
\end{equation}
and \eqref{eq:divsplit} collapses to $\hat\nabla^a\bar t^{\rm tl}_{ab} = \hat\nabla^a\bar t_{ab} = 0$. Both pieces are conserved on their own, and the trace can be dropped at this order without spoiling conservation. This is MMP's argument, and at $O(\epsilon^0)$ it is correct. At $O(\epsilon)$ it fails, since it rests on $\hat R$ being constant. Perturbing the metric perturbs the two-dimensional curvature, the trace acquires a coordinate dependence,
\begin{equation}
  \delta t^{\rm tr} \;=\; \frac{c\,\delta\hat R}{24\pi} \;\neq\; \text{const},
\end{equation}
and \eqref{eq:divsplit} at first order reads
\begin{equation}
  \delta\big(\hat\nabla^a t^{\rm tl}_{ab}\big)
  \;=\; -\,\tfrac12\,\partial_b\,\delta t^{\rm tr}
  \;=\; -\,\frac{c}{48\pi}\,\partial_b\,\delta\hat R
  \;\neq\; 0 .
  \label{eq:divfail}
\end{equation}
The traceless part is no longer conserved by itself. The trace part is a function of $(\tau,\rho)$ sourced by the perturbation, and \eqref{eq:divfail} says the two pieces exchange energy-momentum at first order. Keeping only $\delta t^{\rm tl}_{ab}$ supplies the Einstein equations with a source whose divergence is nonzero. Its $\tau$ and $\rho$ components are proportional to $\partial_a(\hat\Box\,\delta\Upsilon-2\delta\Upsilon)$. No current can absorb them, and the linearised Bianchi identities then fail (Sec.~\ref{sec:consistency}).

The trace enters at first order by a second route. The sphere pressure in the lift \eqref{eq:liftbg} is $\mathcal P = -\hat g^{ab}t_{ab}/8\pi R^4$, built from the trace. On the background both $\bar t^{\rm tr}$ and $R$ are constant, and $\mathcal P$ is a constant pressure on the sphere. Together with the trace part of $\bar t_{ab}$ it renormalises $r_e$ (Sec.~\ref{sec:bgalpha}). At $O(\epsilon)$ the perturbation makes the sphere radius and the local field strength depend on $\theta$, and even the background value of the trace produces a $\theta$-dependent pressure. This is one of the terms feeding the $\theta$ force balance of Sec.~\ref{sec:hall}, and a traceless truncation discards it. Both failures appear as a failure of Step~4 of Sec.~\ref{sec:algorithm}. With the full tensor all three residuals vanish identically. With $\delta T_Q^{\mu\nu}$ truncated or set to zero they do not (Sec.~\ref{sec:throat-reduction}). We therefore keep the whole tensor, including the two-dimensional trace part and the sphere pressure $\mathcal P = -\hat g^{ab}t_{ab}/8\pi R^4$, and re-solve the $O(\alpha)$ background in Sec.~\ref{sec:bgalpha}.

\subsection{Lift to four dimensions}
\label{sec:lift}

The two-dimensional stress tensor describes the fermions on a single magnetic field line. The Einstein equations require a four-dimensional stress tensor, and the family of two-dimensional CFTs, one for each field line, must be assembled into one object $T^Q_{\mu\nu}$. Three properties of the field lines enter, and we define them for an arbitrary field strength. A field line is a timelike two-dimensional surface tangent to the timelike invariant two-plane of $F^\mu{}_\nu$. In this sector $F\wedge F=0$ and the field is magnetic, and this plane is the kernel of $F$. Its tangent vectors $e_a$ obey
\begin{equation}
  F_{\mu\nu}\,e_a^{\nu} \;=\; 0 ,
  \qquad
  \mathsf g_{ab} \;=\; g_{\mu\nu}\,e_a^{\mu}e_b^{\nu},
  \qquad
  \perp^{\mu\nu} \;=\; g^{\mu\nu} - \mathsf g^{ab}e_a^{\mu}e_b^{\nu},
  \label{eq:kernel}
\end{equation}
with $\mathsf g_{ab}$ the induced metric of the worldsheet and $\perp^{\mu\nu}$ its normal projector. The field strength has no electric part on the worldsheet in this sector, and its magnitude is $|F|$ with $|F|^2=\tfrac12F_{\mu\nu}F^{\mu\nu}$. The density of field lines per unit transverse area is $|F|/2\pi$, and its integral over the sphere is the number $q$ of lowest-Landau-level modes fixed by the index theorem. In an inhomogeneous field the zero modes of the two-dimensional Pauli operator still number the total flux in units of $2\pi$ \cite{Aharonov:1978gb}. The two-dimensional fermions of one line live on the metric
\begin{equation}
  \hat g_{ab} \;=\; \varpi^2\,\mathsf g_{ab},
  \qquad
  \varpi^2 \;=\; \frac{2|F|}{q} .
  \label{eq:linescale}
\end{equation}
The lift is
\begin{equation}
  T_Q^{\mu\nu} \;=\; \frac{|F|}{2\pi q}\Big[\,t^{ab}\,e_a^{\mu}e_b^{\nu} \;-\; \tfrac12\,\varpi^2\big(\hat g^{cd}t_{cd}\big)\perp^{\mu\nu}\Big],
  \label{eq:lift}
\end{equation}
where $t_{ab}$ is the $c=q$ stress tensor of Sec.~\ref{sec:throat-origin} on $\hat g_{ab}$, and its indices are raised with $\mathsf g^{ab}$ in the first term. The prefactor counts the field lines per unit area and divides by $q$, since $t_{ab}$ is the stress tensor of all $q$ modes. On the background the tangent vectors are $e_a=\partial_a$, the field lines sit at fixed $(\theta,\phi)$, $|\bar F|=q/2R^2$ and $\varpi^2=1/R^2$. The metric \eqref{eq:linescale} is then \eqref{eq:2dmetric}, and \eqref{eq:lift} reduces to
\begin{equation}
  T^Q_{ab} = \frac{t_{ab}}{4\pi R^2},
  \qquad
  T^Q_{AB} = \mathcal P\, g_{AB},
  \qquad
  \mathcal P = -\frac{\hat g^{ab}t_{ab}}{8\pi R^4},
  \qquad
  T^Q_{aA} = 0,
  \label{eq:liftbg}
\end{equation}
where $a,b$ run over $(\tau,\rho)$ and $A,B$ over the sphere. The quantity $t_{ab}$ is an energy per unit two-dimensional volume, whereas $T^Q_{ab}$ is an energy per unit four-dimensional volume. Dividing by the area $4\pi R^2$ of the sphere over which the $q$ field lines are spread gives $T^Q_{ab} = t_{ab}/4\pi R^2$. The sphere components are a transverse pressure. They come from the dependence of $\varpi^2$ on the metric through $|F|$, since $\delta\log|F|=-\tfrac12\perp^{\mu\nu}\delta g_{\mu\nu}$ for a magnetic field strength. The pressure is set by the trace of $t_{ab}$, and it makes the lift traceless in four dimensions, $g_{\mu\nu}T_Q^{\mu\nu}=0$, for any field strength and metric. The trace of $t_{ab}$ is nonzero only through the conformal anomaly, and in a classically Weyl-invariant treatment there would be no sphere pressure at all. In Sec.~\ref{sec:hall-action} the lift \eqref{eq:lift} is derived as the metric variation of the effective action of the field lines. It is built from $g_{\mu\nu}$ and $F_{\mu\nu}$ alone and is therefore covariant under diffeomorphisms and invariant under gauge transformations.

At first order in $\epsilon$ the lift \eqref{eq:lift} contains three effects that a lift built on the coordinate planes $(\tau,\rho)$ misses. An azimuthal electric field $\delta F_{a\phi}$ tilts the field lines in $\theta$. In the frame that moves with the field line the transverse electric field vanishes, and the tilt is the $\bm E\times\bm B$ drift of the guiding centres. A change $\delta F_{\theta\phi}$ of the radial magnetic field changes the density of field lines. The same change scales the metric \eqref{eq:linescale} of every line. The fermions stay on their field lines, but the field lines move, and the stress tensor of a moving line has mixed components $T^Q_{aA}$. In the companion sector \cite{Sadhukhan:2026mcu} the perturbation has $\delta F_{\theta\phi}=0$ but carries an electric field along the field lines. There $F\wedge F\neq0$, $F$ has no kernel, and the fermions follow the timelike invariant plane of $F$. The density and the scale of the field lines do not change at $O(\epsilon)$, and \eqref{eq:lift} on that plane reduces to the drift lift used there. The two sectors therefore use one constitutive relation. Here it is the metric variation of the effective action of Sec.~\ref{sec:hall-action} and rests on the assumed scaling \eqref{eq:linescale}. In the companion paper it is stated as an assumption directly.

The scaling $\varpi^2=2|F|/q$ in \eqref{eq:linescale} is an assumption. On the background it is the factor $1/R^2$ of \eqref{eq:2dmetric}, which follows from the Weyl invariance of the massless Dirac action. Away from the background we take the scale of the line metric to follow the local field strength. The lowest Landau level is separated from the next level by the cyclotron gap, which is set by $|F|$, and the two-dimensional description holds below that scale. A cutoff on $\mathsf g_{ab}$ that varies with $|F|$ is equivalent to a uniform cutoff on $\varpi^2\mathsf g_{ab}$. We have not derived \eqref{eq:linescale} from the four-dimensional Dirac reduction. Two results of this paper bear on it, and neither derives it. Section~\ref{sec:response} shows that the $\tau$ and $\rho$ components of the force balance close only with this scaling once the tilt and the density are kept. The lift on the coordinate planes, which has no tilt and a density that ignores the flux, closes as well. Section~\ref{sec:hall-action} shows that the scaling supplies the coupling between flux and curvature that the effective-action derivation of the Hall current requires.

This lift is an approximation. The $q$ lowest-Landau-level states fill the sphere uniformly. Each occupies an area $4\pi R^2/q$ and is localised within an angular patch of size
\begin{equation}
  \Delta\theta_{\rm LLL} \;\sim\; \frac{\ell_B}{R} \;\sim\; \frac{1}{\sqrt q},
  \qquad \ell_B \sim \frac{R}{\sqrt q},
\end{equation}
with $\ell_B$ the magnetic length. A perturbation of multipole $l$ varies on the angular scale $1/l$. Treating $\theta$ as a parameter labelling which field line a mode sits on requires the perturbation to be nearly constant across one wavefunction,
\begin{equation}
  \frac{1}{l} \;\gg\; \frac{1}{\sqrt q}
  \qquad\Longleftrightarrow\qquad
  l \;\ll\; \sqrt q .
  \label{eq:lbound}
\end{equation}
For the fiducial $q=10^{33}$ of \cite{Mondal:2025tht} this permits $l$ up to about $10^{16}$. Without it the assignment of a single two-dimensional metric $\hat g_{ab}(\theta)$ to each mode would not be meaningful. The corrections to \eqref{eq:lift} from the finite width of the states are not computed here. For the charge response of the companion sector the exact LLL states give the factor $1-\lhat/q$ \cite{Sadhukhan:2026mcu}, and we expect corrections of relative order $\lhat/q$ here as well. Multiplying the $O(\alpha)$ response, they enter at $O(\alpha\lhat/q)=O(4\pi^2\lhat\alpha^2/g_c^2)$, beyond the order kept.

$T^Q_{AB}\propto g_{AB}$ has no traceless sphere part. Step~1 of the algorithm uses the traceless angular combination $\mathcal E_{\theta\theta}-\mathcal E_{\phi\phi}/\sin^2\theta$, and a source enters it only through its own anisotropic sphere stress, $T_{\theta\theta}-T_{\phi\phi}/\sin^2\theta$. At background order $\bar T^{Q}_{AB}=\bar{\mathcal P}g_{AB}$ is pure trace. At $O(\epsilon)$ both terms of $\delta T^Q_{AB}=\delta\mathcal P\,g_{AB}+\bar{\mathcal P}\,\delta g_{AB}$ are pure trace as well, since $\delta g_{AB}=R^2 K Y\mathring g_{AB}$. The tangent vectors \eqref{eq:tilt} have sphere components of order $\epsilon$, and they enter $T_Q^{AB}$ only at order $\epsilon^2$. The fermion source therefore drops out of $\mathcal E^{\rm tl}$ entirely, and Step~1 still gives $H_0=H_2$. The fermions have no dynamics on the sphere and cannot generate an anisotropic sphere stress.

On the background the trace is the constant \eqref{eq:tracevalue}, and the background sphere pressure is
\begin{equation}
  \bar{\mathcal P} \;=\; \frac{q}{96\pi^2 r_e^4} .
  \label{eq:Pbar}
\end{equation}
The two-dimensional trace is negative. $\bar{\mathcal P}$ is therefore a positive pressure on the sphere. The full background tensor is
\begin{equation}
  \bar T^{Q}_{\mu\nu} \;=\; \frac{q}{96\pi^2 r_e^2}\;
  \mathrm{diag}\!\left(\rho^2-2,\;\; -\,\frac{\rho^2+4}{(1+\rho^2)^2},\;\; 1,\;\; \sin^2\theta\right) .
  \label{eq:TQzero}
\end{equation}
It is conserved at $O(\epsilon^0)$, and it satisfies
\begin{equation}
  \frac{\kappa}{4\pi r_e^2}\left(\bar t_{\tau\tau} - \half \hat g_{\tau\tau}\,\hat g^{ab}\bar t_{ab}\right)
  = \frac{\kappa\,\bar t^{\rm tl}_{\tau\tau}}{4\pi r_e^2}
  \;=\; -\,\frac{g_c^2}{4\pi^2 q} \;=\; -\,\alpha ,
  \label{eq:MMPmatch}
\end{equation}
using $\kappa = 8g_c^2 r_e^2/q^2$ and $c=q$, which is MMP's normalisation. Equation \eqref{eq:MMPmatch} is an exact identity, and it identifies $\alpha$ as the expansion parameter of the $O(\alpha)$ background. At $O(\epsilon)$ neither factor in $\mathcal P$ stays constant, as shown at the end of Sec.~\ref{sec:throat-origin}.

The parameters $\alpha$ and $\epsilon$ are independent and the calculation is a double expansion in both. The parameter $\alpha = g_c^2/4\pi^2 q$ is a fixed property of the background, measuring the strength of the fermionic backreaction, whereas $\epsilon$ counts powers of the perturbation amplitude. Equation \eqref{eq:MMPmatch} relates background quantities and holds at $O(\epsilon^0)$. The double expansion has four sectors. At $O(\epsilon^0\alpha^0)$ there is the exact AdS$_2\times S^2$ background. At $O(\epsilon^0\alpha^1)$ there is the corrected background $\gamma,\varphi$ of Sec.~\ref{sec:bgalpha}. At $O(\epsilon^1\alpha^0)$ there are the perturbation equations of Sec.~\ref{sec:alpha0}, which close without any fermionic input. At $O(\epsilon^1\alpha^1)$ there is the coupled system of Sec.~\ref{sec:throatmaster}, where the fermion response and the Hall current are required.

\subsection{The \texorpdfstring{$O(\alpha)$}{O(alpha)} background}
\label{sec:bgalpha}

Inserting the ansatz \eqref{eq:bgthroat} into \eqref{eq:order0} with the source \eqref{eq:TQzero} and expanding to first order in $\alpha$ gives three ordinary differential equations. The functions $\gamma$ and $\varphi$ enter \eqref{eq:bgthroat} already multiplied by $\alpha$, and the equations below are the coefficient of $\alpha^1$ in the field equations. The $\tau\tau$, $\rho\rho$ and $\theta\theta$ components read
\begin{align}
  - \rho^{4} \varphi'' - \rho^{3} \varphi' + \rho^{2} \varphi - 2 \rho^{2} \varphi''
    - \frac{\rho^{2}}{3} - \rho \varphi' + \varphi - \varphi'' + \frac{2}{3} &= 0 ,
  \label{eq:bgode_tt}\\[2pt]
  \frac{3 \rho^{3} \varphi' - 3 \rho^{2} \varphi + \rho^{2} + 3 \rho \varphi' - 3 \varphi + 4}
       {3 \left(\rho^{2} + 1\right)^{2}} &= 0 ,
  \label{eq:bgode_rr}\\[2pt]
  \frac{\rho^{2} \varphi''}{2} + \rho \varphi' + 2 \varphi + \frac{\gamma''}{2}
    + \frac{\varphi''}{2} - \frac{1}{3} &= 0 .
  \label{eq:bgode_thth}
\end{align}
Equations \eqref{eq:bgode_tt} and \eqref{eq:bgode_rr} involve $\varphi$ alone, the first at second order and the second at first. $\varphi$ is overdetermined and the compatibility of the two is a check on the source. Equation \eqref{eq:bgode_thth} then determines $\gamma$ by quadrature. The closed-form solution, regular at $\rho=0$ and even in $\rho$, is
\begin{equation}
  \gamma(\rho) = - \frac{4 \rho^{2}}{3} - \rho \left(\rho^{2} + 3\right) \arctan{\rho}
    + \log\!\left(1+\rho^{2}\right),
  \qquad
  \varphi(\rho) = \rho \arctan{\rho} + \frac{4}{3} ,
  \label{eq:bgsol}
\end{equation}
and it solves all three equations. The equations contain $\gamma$ only through $\gamma''$, and a constant $\gamma_0$ added to $\gamma$ is a change of coordinates. At this order the substitution $\tau\to\tau(1+\alpha\gamma_0/2)$, $\rho\to\rho(1-\alpha\gamma_0/2)$ turns the metric with $\gamma$ into the metric with $\gamma+\gamma_0$. We fix the constant by $\gamma(0)=0$. Then $A(0)=r_e^2$, and $\tau$ is the proper time of the static observer at $\rho=0$ in units of $r_e$. Frequencies $\Om$ of the throat are measured by this observer.  We call this choice the clock of the throat. Another constant $\gamma_0$ is another clock, and it changes every first-order shift $\varsigma$ of Sec.~\ref{sec:stab-normalmodes} by $\gamma_0\Om_0^2$. Compared with MMP eqs.~(5.37) and (B.5), $\varphi$ differs by the additive constant $\tfrac13$ and $\gamma$ by the term $-\rho^2/3$. These are the pieces MMP absorbed into the definitions of the two radii. The constant in $\varphi$ renormalises the $S^2$ radius and comes from the two-dimensional trace part. The $\rho^2$ term in $\gamma$ renormalises the AdS$_2$ radius and needs the trace part and the sphere pressure together, which give $-2\rho^2/3$ and $+\rho^2/3$ on their own. Together these two parts are $\alpha/3$ times the Maxwell stress of the monopole, and they take $r_e^2$ to $r_e^2(1+\alpha/3)$ in both radii. A traceless truncation discards both parts. The mouth carries the same renormalisation at $O(\alpha)$ (Sec.~\ref{sec:mouthTQ}). The large-$\rho$ behaviour is unchanged. Expanding \eqref{eq:bgsol},
\begin{equation}
  \varphi(\rho) \;=\; \frac{\pi}{2}\rho + \frac{1}{3} + O(\rho^{-2}),
\end{equation}
so that the physical correction to the sphere radius grows as $\alpha\varphi \sim \tfrac{\pi}{2}\alpha\rho$. This linear growth fixes $\ell$ through the matching to the mouth (MMP eqs.~(5.39)--(5.40)),
\begin{equation}
  \ell \;=\; \frac{4 r_e}{\pi\alpha} ,
  \label{eq:ell}
\end{equation}
and the additive constant $\tfrac13$ does not affect it.

\subsection{Linear response of the LLL sector and the perturbed lift}
\label{sec:response}

We now need $\delta T^Q_{\mu\nu}$. By the discussion around \eqref{eq:bianchiexp} it must be the exact linear response, and its $\tau$, $\rho$ and $\phi$ divergence must vanish with no current to absorb it. We first need the perturbation of the line metric \eqref{eq:linescale}. In Regge--Wheeler gauge $h_{a\theta}=0$, and the tilt changes $\mathsf g_{ab}$ only at $O(\epsilon^2)$. To first order the field strength is $|F|=|F_{\theta\phi}|/R^2\sin\theta$, and \eqref{eq:fpert} gives $\delta F_{\theta\phi}/\bar F_{\theta\phi}=\tfrac12\lhat\,w\,Y$. Hence
\begin{equation}
  |F| \;=\; |\bar F|\big(1-\epsilon\,K_F\,Y\big),
  \qquad
  K_F \;\equiv\; K - \frac{\lhat}{2}\,w ,
  \label{eq:KB}
\end{equation}
and the perturbation of the two-dimensional metric \eqref{eq:linescale} is
\begin{equation}
  \delta \hat g_{ab} \;=\; \frac{h_{ab}}{R^2} \;-\; K_F\,Y\,\hat g_{ab},
  \label{eq:dghat}
\end{equation}
the second term being $\delta\varpi^2\,\mathsf g_{ab}$. The combination $K_F$ is the relative decrease of the field strength. The sphere radius contributes $K$ and the change of the flux contributes $-\lhat w/2$. Any such perturbation can be written as
\begin{equation}
  \delta \hat g_{ab} \;=\; 2\,\delta\Upsilon\;\hat g_{ab} \;+\; \left(\mathcal{L}_\xi \hat g\right)_{ab}.
  \label{eq:decomposition}
\end{equation}
This is a feature of two dimensions. A symmetric rank-two tensor in $D$ dimensions has $D(D+1)/2$ independent components. The right of \eqref{eq:decomposition} supplies one Weyl function $\delta\Upsilon$ and $D$ diffeomorphism functions $\xi_a$, $D+1$ in total. These match only when
\begin{equation}
  \frac{D(D+1)}{2} \;=\; D+1
  \qquad\Longleftrightarrow\qquad
  (D-2)(D+1) = 0
  \qquad\Longleftrightarrow\qquad
  D = 2 .
\end{equation}
In two dimensions $\delta\hat g_{ab}$ has three components $(\tau\tau,\tau\rho,\rho\rho)$ and the right-hand side supplies the three functions $(\delta\Upsilon,\xi_\tau,\xi_\rho)$. For $D>2$ a remainder survives, the transverse-traceless part, which is the graviton. That two-dimensional gravity has no propagating graviton and that \eqref{eq:decomposition} is complete are the same statement. The decomposition is not unique. Adding a conformal Killing vector to $\xi$, with a compensating shift of $\delta\Upsilon$, leaves $\delta\hat g$ unchanged. This freedom is the choice of fermion state, discussed below.

The second term in \eqref{eq:decomposition} is the Lie derivative of the metric along $\xi$. In components,
\begin{equation}
  \left(\mathcal{L}_\xi \hat g\right)_{ab}
  \;=\; \xi^c\,\partial_c \hat g_{ab}
  \;+\; \hat g_{cb}\,\partial_a \xi^c
  \;+\; \hat g_{ac}\,\partial_b \xi^c
  \;=\; \hat\nabla_a \xi_b + \hat\nabla_b \xi_a ,
  \label{eq:liedef}
\end{equation}
and we use the first expression. We take $\xi_a$ with a lower index as the independent function,
\begin{equation}
  \xi^\tau = \hat g^{\tau\tau}\xi_\tau = -\frac{\xi_\tau}{1+\rho^2},
  \qquad
  \xi^\rho = \hat g^{\rho\rho}\xi_\rho = \left(1+\rho^2\right)\xi_\rho .
  \label{eq:xiupper}
\end{equation}
A perturbation of the form \eqref{eq:liedef} is pure gauge from the two-dimensional point of view. Its role is to absorb the part of $\delta\hat g_{ab}$ that is not a Weyl rescaling. The CFT response can then be computed by combining an ordinary tensor transformation with the anomalous Weyl transformation. Every term of \eqref{eq:decomposition} carries the factor $Y(\theta)$. Below we write the coefficients of $Y$ with the same symbols, and from here on $\delta\Upsilon$ and $\xi_a$ denote these coefficients. Writing out the three components with $\hat g$ the unit AdS$_2$ metric gives
\begin{align}
  2 \rho^{3} \xi_\rho + \rho^{2} H_{0} + \rho^{2} K_F + 2 \rho^{2} \delta\Upsilon
    + 2 \rho \xi_\rho + H_{0} + K_F + 2 \delta\Upsilon - 2 \dot\xi_\tau &= 0 ,
  \label{eq:deceq_tt}\\[2pt]
  \rho^{2} r_{e}^{2}\dot\xi_\rho + \rho^{2} r_{e}^{2}\xi_\tau'
    - \rho^{2} H_{1} - 2 \rho r_{e}^{2} \xi_\tau
    + r_{e}^{2}\dot\xi_\rho + r_{e}^{2}\xi_\tau' - H_{1} &= 0 ,
  \label{eq:deceq_tr}\\[2pt]
  2 \rho^{2}\xi_\rho' + 2 \rho \xi_\rho - H_{2} + K_F + 2 \delta\Upsilon + 2\xi_\rho' &= 0 ,
  \label{eq:deceq_rr}
\end{align}
where a dot denotes $\partial_\tau$ and a prime $\partial_\rho$. The $\tau\tau$ component assembles as follows, and the other two follow the same pattern. Write $f \equiv 1+\rho^2$. The background is then $\hat g = \mathrm{diag}(-f,\,1/f)$. From \eqref{eq:dghat} with $h_{\tau\tau} = A H_0 = r_e^2 f H_0$ and $R^2 = r_e^2$,
\begin{equation}
  \delta\hat g_{\tau\tau}
  \;=\; \frac{h_{\tau\tau}}{R^2} - K_F\,\hat g_{\tau\tau}
  \;=\; f H_0 - K_F(-f)
  \;=\; f\left(H_0 + K_F\right).
\end{equation}
The Weyl term is $2\,\delta\Upsilon\,\hat g_{\tau\tau} = -2f\,\delta\Upsilon$. For the Lie term, $\hat g$ is diagonal and depends only on $\rho$, hence only the $\xi^\rho$ term survives in the first term of \eqref{eq:liedef} and only the $\partial_\tau\xi^\tau$ terms survive in the other two,
\begin{equation}
  \left(\mathcal{L}_\xi \hat g\right)_{\tau\tau}
  \;=\; \xi^\rho\,\partial_\rho\!\left(-f\right) + 2\,\hat g_{\tau\tau}\,\partial_\tau \xi^\tau
  \;=\; \left(f\xi_\rho\right)\left(-2\rho\right) + 2\left(-f\right)\left(-\frac{\dot\xi_\tau}{f}\right)
  \;=\; -2\rho f\,\xi_\rho + 2\dot\xi_\tau ,
\end{equation}
using \eqref{eq:xiupper}, $\partial_\rho f = 2\rho$, and $\partial_\tau\xi^\tau = -\dot\xi_\tau/f$. Substituting the three pieces into \eqref{eq:decomposition} gives $f\left(H_0+K_F\right) + 2f\,\delta\Upsilon + 2\rho f\,\xi_\rho - 2\dot\xi_\tau = 0$, and expanding $f = 1+\rho^2$ gives \eqref{eq:deceq_tt} term by term. Only \eqref{eq:deceq_tr} carries a factor of $r_e^2$, since $h_{\tau\rho} = H_1$ is defined without a background metric factor in \eqref{eq:hansatz1} while $h_{\tau\tau}$ and $h_{\rho\rho}$ carry $A$ and $B$. In the throat $H_1$ therefore has dimension length$^2$, while $H_0$, $H_2$, $K$ and $w$ are dimensionless. We use \eqref{eq:deceq_tt}--\eqref{eq:deceq_rr} to trade the three metric amplitudes for the three transport functions. Solving for $H_0,H_1,H_2$ in terms of $(\delta\Upsilon,\xi_\tau,\xi_\rho,K_F)$ gives, for example,
\begin{equation}
  H_0 = - \frac{\left(\rho^{2}+1\right)\left(K_F + 2 \delta\Upsilon\right)
        + 2 \rho\left(\rho^{2}+1\right) \xi_{\rho} - 2 \dot\xi_{\tau}}{\rho^{2} + 1} .
  \label{eq:H0sol}
\end{equation}
$K$ and $w$ are not eliminated. They enter \eqref{eq:dghat} only through the scale $K_F$ of the line metric, and they survive as independent unknowns alongside the transport fields.

The stress tensor of the CFT responds to \eqref{eq:decomposition} through two mechanisms. The generating functional is diffeomorphism invariant, and under $\delta\hat g = \mathcal{L}_\xi\hat g$ every tensor is dragged along, $\delta_\xi t_{ab} = (\mathcal{L}_\xi \bar t)_{ab}$, with no anomaly in this channel. The Weyl part is where the anomaly enters, and the required transformation law is \eqref{eq:anomalylaw} evaluated for an infinitesimal Weyl factor. Setting $\Upsilon = \delta\Upsilon$ and keeping first order, the two terms quadratic in $\partial\Upsilon$ drop out and
\begin{equation}
  \delta_\Upsilon t_{ab}
  \;=\; -\frac{c}{12\pi}\left[-\nabla_a\nabla_b\,\delta\Upsilon + \hat g_{ab}\,\Box\,\delta\Upsilon\right],
\end{equation}
with $\nabla$ and $\Box$ those of the background $\hat g$. Adding the two channels gives the full linear response,
\begin{equation}
  \delta t_{ab} \;=\; \left(\mathcal{L}_\xi \bar t\right)_{ab}
  \;-\;\frac{c}{12\pi}\left[-\nabla_a\nabla_b\,\delta\Upsilon + \hat g_{ab}\,\Box\,\delta\Upsilon\right] .
  \label{eq:response}
\end{equation}
This is the two-dimensional conformal Ward identity, the statement that the generating functional is diffeomorphism invariant and anomalous only under Weyl transformations. Nothing has been fitted and there is no free coefficient. The response still depends on the state of the fermions, which we discuss at the end of this subsection. The factor $c/12\pi$ is the one appearing in the trace anomaly and in the background \eqref{eq:tbar}. With $c=q$ and $L_c = \pi$ the diffeomorphism piece is
\begin{align}
   \left(\mathcal{L}_\xi \bar t\right)_{\tau\tau}
    &= \frac{q}{12\pi}\left[\rho\left(1+\rho^{2}\right)\xi_\rho
       \;-\; \frac{\rho^{2}-2}{1+\rho^{2}}\,\dot\xi_\tau\right],
  \label{eq:lie_tt}\\[4pt]
  \left(\mathcal{L}_\xi \bar t\right)_{\tau\rho}
    &= -\frac{q}{24\pi\left(1+\rho^{2}\right)^{2}}
      \Big[\left(\rho^{2}+4\right)\left(1+\rho^{2}\right)\dot\xi_\rho
      + \left(\rho^{2}-2\right)\left(1+\rho^{2}\right)\xi_\tau'
      - 2\rho\left(\rho^{2}-2\right)\xi_\tau\Big],
  \label{eq:lie_tr}\\[4pt]
  \left(\mathcal{L}_\xi \bar t\right)_{\rho\rho}
    &= -\frac{q}{12\pi\left(1+\rho^{2}\right)}
      \Big[\left(\rho^{2}+4\right)\xi_\rho' \;+\; \rho\,\xi_\rho\Big],
  \label{eq:lie_rr}
\end{align}
and the anomaly piece is
\begin{align}
  \mathcal{A}_{\tau\tau}[\delta\Upsilon]
    &= \frac{q\left(\rho^{2}+1\right)}{12\pi}
      \Big[\left(\rho^{2}+1\right)\delta\Upsilon'' + \rho\,\delta\Upsilon'\Big],
  \label{eq:anom_tt}\\[4pt]
  \mathcal{A}_{\tau\rho}[\delta\Upsilon]
    &= \frac{q}{12\pi\left(\rho^{2}+1\right)}
      \Big[\left(\rho^{2}+1\right)\dot{\delta\Upsilon}' - \rho\,\dot{\delta\Upsilon}\Big],
  \label{eq:anom_tr}\\[4pt]
  \mathcal{A}_{\rho\rho}[\delta\Upsilon]
    &= -\frac{q}{12\pi\left(\rho^{2}+1\right)^{2}}
      \Big[\rho\left(\rho^{2}+1\right)\delta\Upsilon' - \ddot{\delta\Upsilon}\Big],
  \label{eq:anom_rr}
\end{align}
so that $\delta t_{ab} = (\mathcal{L}_\xi\bar t)_{ab} + \mathcal{A}_{ab}[\delta\Upsilon]$. The diffeomorphism piece involves $\xi$ and one derivative. The anomaly piece involves $\delta\Upsilon$ and exactly two derivatives, as a one-loop Weyl variation must, and it is the only place where time derivatives of $\delta\Upsilon$ appear. The consolidated expressions are in Appendix~\ref{app:dTQ}. The conservation law
\begin{equation}
  \hat\nabla^a\!\left(\bar t + \epsilon\,\delta t\right)_{ab} \;=\; 0 + O(\epsilon^2)
  \qquad\text{for arbitrary } \delta\Upsilon,\ \xi_a ,
  \label{eq:ward}
\end{equation}
holds with $\delta\Upsilon$ and $\xi_a$ left as undetermined functions. It tests the structure of \eqref{eq:response} and not a particular solution. Conservation of $\delta T^Q$ is therefore built in, which is what the obstruction of Sec.~\ref{sec:consistency} demands.

Equation \eqref{eq:response} does not fix the stress tensor completely. In the null coordinates $x^\pm=\tau\pm x$ the conservation equation $\hat\nabla^a t_{ab}=0$, together with the fixed trace, leaves one freedom. One may add an arbitrary function $\mathfrak f(x^+)$ to $t_{++}$ and an arbitrary function $\mathfrak g(x^-)$ to $t_{--}$, and the result is still conserved with the same trace. $\mathfrak f$ describes fermions that move freely to the left along the field line and $\mathfrak g$ fermions that move freely to the right. At a fixed frequency every quantity is proportional to $e^{p\tau}$, which forces $\mathfrak f=\mathfrak a_+ e^{px^+}$ and $\mathfrak g=\mathfrak a_- e^{px^-}$ with $\mathfrak a_\pm$ constants. Choosing them is choosing the state of the fermions. The same freedom appears in the reduction. Suppose a solution $(K,w)$ of the master equations has been found at frequency $p$, with $H_1$ fixed by the constraints. Substituting this $K$, $K'$ and $w$ into \eqref{eq:xit} and \eqref{eq:xir} gives two linear first-order equations for $\xi_\tau$ and $\xi_\rho$, with the terms in $K$, $K'$ and $w$ acting as sources. Its general solution is one particular solution plus any solution $\zeta_a$ of the same equations with the sources removed,
\begin{equation}
  \zeta_\tau' = \frac{2\rho}{f}\,\zeta_\tau - p\,\zeta_\rho ,
  \qquad
  \zeta_\rho' = -\,\frac{p}{f^2}\,\zeta_\tau .
  \label{eq:zetaeqs}
\end{equation}
For one and the same perturbation $(K,w)$ the transport fields are determined up to the addition of $\zeta_a$, which contains two free constants. Write $\zeta_\tau=-f\,\mathfrak u$ and use $x=\arctan\rho$, with $\partial_x=f\,\partial_\rho$. Then \eqref{eq:zetaeqs} becomes $\partial_x\mathfrak u=p\,\zeta_\rho$ and $\partial_x\zeta_\rho=p\,\mathfrak u$. The sum $\mathfrak u+\zeta_\rho$ grows like $e^{px}$ and the difference $\mathfrak u-\zeta_\rho$ like $e^{-px}$. Restoring the factor $e^{p\tau}$ gives
\begin{equation}
  \zeta_\tau = -f\big(\mathfrak b_+\,e^{px^+}+\mathfrak b_-\,e^{px^-}\big),
  \qquad
  \zeta_\rho = \mathfrak b_+\,e^{px^+}-\mathfrak b_-\,e^{px^-},
  \label{eq:zeta}
\end{equation}
with $\mathfrak b_+$ and $\mathfrak b_-$ constants. These vectors are conformal Killing vectors of unit AdS$_2$. Adding $\zeta_a$ to $\xi_a$, together with the shift $\delta\Upsilon\to\delta\Upsilon-\tfrac12\hat\nabla_a\zeta^a$, leaves the metric perturbation $\delta\hat g_{ab}$ unchanged and changes the stress tensor by
\begin{equation}
  \delta t_{++} = -\frac{p\,q\left(p^2+4\right)}{12\pi}\mathfrak b_+\,e^{px^+},
  \qquad
  \delta t_{--} = -\frac{p\,q\left(p^2+4\right)}{12\pi}\mathfrak b_-\,e^{px^-},
  \label{eq:freerad}
\end{equation}
with no trace and no mixed component. The constant $\mathfrak b_+$ gives a pure left mover and $\mathfrak b_-$ a pure right mover. The integration constants of the transport equations are the free fermion radiation.\footnote{At a fixed frequency only $\mathfrak f=\mathfrak a_+e^{px^+}$ and $\mathfrak g=\mathfrak a_-e^{px^-}$ survive, which is why the freedom contains exactly two constants. Writing $p=-i\Om$, the vectors \eqref{eq:zeta} with $\mathfrak b_-=0$ or $\mathfrak b_+=0$ are, for even $\Om$, the Virasoro generators $\mathsf L_n$ and $\bar{\mathsf L}_n$ of the circle of circumference $L_c=\pi$, with $n=-\Om/2$. Those with $n=0,\pm1$, at $\Om=0,\mp2$, generate the global conformal group SL(2,$\mathbb R$)$\times$SL(2,$\mathbb R$) of the circle and leave its vacuum invariant. Each of them is a conformal Killing vector of AdS$_2$ and not an isometry. The sum $\mathsf L_0+\bar{\mathsf L}_0$, with $\mathfrak b_+=\mathfrak b_-$, is proportional to the isometry $\partial_\tau$. The other isometries are the combinations with $\mathfrak b_-=-\mathfrak b_+$ at $\Om=\pm1$, which are not periodic around the circle.} The factor $p\left(p^2+4\right)$ vanishes at $\Om=0$ and $\Om=\pm2$. At these frequencies the vectors \eqref{eq:zeta} generate global conformal transformations, which leave the vacuum unchanged and produce no radiation. The metric responds to this radiation through
\begin{equation}
  H_1 = -\,\frac{4\alpha\,p\,r_e^2\left(p^2+4\right)}{3\lhat f}\,\zeta_\rho,
  \qquad
  H_2 = \frac{4\alpha\,p\left(p^2+4\right)}{3\lhat f^2}\,\zeta_\tau,
  \qquad
  K = w = 0 ,
  \label{eq:radresponse}
\end{equation}
which satisfies the equations for $K$, $H_1$ and $w$ and the algebraic relation for $H_2$ to $O(\alpha)$. With the response \eqref{eq:radresponse} the transport equations acquire $O(\alpha)$ terms, among them $H_1$ through the term $K'$ of \eqref{eq:xit}, and their solution is $\xi_a=\zeta_a+\alpha\,\zeta^{(1)}_a$ with
\begin{equation}
  \zeta^{(1)}_\tau = -\,\frac{p^2+4}{3\lhat}\Big[\mathfrak b_+\,\mathsf v_+\,e^{px^+}+\mathfrak b_-\,\mathsf v_-\,e^{px^-}\Big],
  \qquad
  \zeta^{(1)}_\rho = -\,\frac{p^2+4}{3\lhat f}\Big[\mathfrak b_+\,\mathsf v_+\,e^{px^+}-\mathfrak b_-\,\mathsf v_-\,e^{px^-}\Big],
  \qquad
  \mathsf v_\pm = 1+\frac{(\rho\pm p)^2}{p^2+1},
  \label{eq:zeta1}
\end{equation}
up to a solution of \eqref{eq:zetaeqs}, which is a redefinition of $\mathfrak b_\pm$ at $O(\alpha)$. The correction enters $H_1$ and $H_2$ only at $O(\alpha^2)$. At $\Om=\pm1$ the denominator of $\mathsf v_\pm$ vanishes and the correction contains terms linear in $x$. The free radiation changes only $H_1$ and $H_2$ and not $K$ or $w$. This agrees with Sec.~\ref{sec:throat-reduction}, where the master equations for $K$ and $w$ contain neither $\xi_\tau$ nor $\xi_\rho$. In the throat the potentials, the spectrum and the stability analysis do not depend on the state of the fermions at this order. The state enters $H_1$, $H_2$ and the Hall current $J^\phi$ when these are reconstructed from a solution $(K,w)$. In the mouth at $O(\alpha)$ the free radiation also sources $K$ (Sec.~\ref{sec:mouth-master}). The choice of $\mathfrak b_\pm$ is part of the choice of state. We take the state on each line to be invariant under charge conjugation (Sec.~\ref{sec:hall-force}), which does not fix $\mathfrak b_\pm$. Each field line is closed, and periodicity around it restricts $\mathfrak b_\pm$. The free radiation $e^{px^\pm}$ is periodic around a line of conformal length $L_c=\pi$ only if $e^{p\pi}=1$, that is for $\Om\in2\mathbb Z$. At other frequencies periodicity fixes $\mathfrak b_\pm$ once $\delta t_{ab}$ is known along the whole line, through the mouths and the ambient return path, which we do not construct. At $\Om=0,\pm2$ the radiation carries no stress. At $\Om=2n$ with $|n|\ge2$ it is periodic and carries stress, and it is then a normal mode of the fermions on the closed line, with $K=w=0$ in the throat and $H_1$, $H_2$ from \eqref{eq:radresponse}. The levels $\Om_0$ of the closed throat in Sec.~\ref{sec:stab-normalmodes} include these frequencies. Within the throat a throat level does not excite this mode. With $K''$ from \eqref{eq:ads2Kcomp}, the transport equations \eqref{eq:xit}--\eqref{eq:xir} have the local particular solution $\xi_\tau=pK/\lhat$, $\xi_\rho=K'/\lhat$. The transport fields of a throat mode are therefore $\xi_a=\partial_aK/\lhat+\zeta_a$, and a normalisable mode of the closed throat carries no free radiation of its own. A throat level at $\Om_0=2n$ and the mode of the line can couple only through the mouths, where the radiation sources $K$. There the throat mode has decayed to an amplitude of order $\rho_{\rm J}^{-2}$ or smaller, below $10^{-57}$ at the junction of Sec.~\ref{sec:overlap}. This is far below the $O(\alpha)$ shifts of the throat levels, and we neglect the coupling. Two effects are left out at this order. The effect of the induced fields back on the fermions is $O(\alpha^2)$. The excitation of higher Landau levels has a gap of order $\sqrt q/R$, far above any frequency relevant to scattering or to quasinormal modes.

At $O(\epsilon)$ the lift \eqref{eq:lift} varies through the response $\delta t_{ab}$, through the density and the scale of the field lines, and through the tilt. In Regge--Wheeler gauge the tangent vectors are
\begin{equation}
  e_a \;=\; \partial_a + \epsilon\,\partial_a\vartheta\,\partial_\theta ,
  \qquad
  \vartheta \;=\; -\frac{a_\phi}{\bar F_{\theta\phi}} \;=\; \tfrac12\,w\;Y'(\theta),
  \label{eq:tilt}
\end{equation}
which solve \eqref{eq:kernel} at $O(\epsilon)$, since $\partial_a\vartheta=\delta F_{a\phi}/\bar F_{\phi\theta}$. The field line through $\theta$ on the background sits at $\theta+\epsilon \vartheta$. The two-dimensional block gives
\begin{equation}
  \delta T^Q_{ab}
  \;=\; \frac{1}{4\pi r_e^2}\Big[\,\delta t_{ab} \;-\; K_F\,\bar t_{ab}\,\Big]Y(\theta),
  \label{eq:dTQ2d}
\end{equation}
using \eqref{eq:KB} for the prefactor $|F|/2\pi q$. Here $R^2$ appears in two expansions. The background sphere radius is $R^2 = r_e^2\left(1+\alpha\varphi(\rho)\right)$ from \eqref{eq:bgthroat}, and the perturbation contributes a further factor $\left(1+\epsilon K Y\right)$. In the lift we use $R^2 = r_e^2\left(1+\epsilon K Y\right)$ and \eqref{eq:KB} without the $\alpha$-corrected background. The fermion stress tensor is already first order in $\alpha$ once it enters the field equations, since $\kappa\,\bar T^Q_{\theta\theta} = \alpha/3$ and $\kappa\,\bar T^Q_{\tau\tau} = \alpha\left(\rho^{2}-2\right)/3$. Hence $\kappa\,\delta T^Q$ is $O(\alpha\epsilon)$, and the corrected background inside the lift would generate only $O(\alpha^{2}\epsilon)$ terms. The second term in \eqref{eq:dTQ2d} is the background Casimir energy redistributed over a changed density of field lines. The mixed block is the drift of the field lines,
\begin{equation}
  \delta T^Q_{a\theta} \;=\; \big(\bar T^{Q}{}_{a}{}^{c} - \bar{\mathcal P}\,\delta_a{}^c\big)\,u_c ,
  \qquad
  u_c \;=\; \bar g_{\theta\theta}\,\partial_c\vartheta \;=\; \frac{r_e^2}{2}\,\partial_c w\;Y'(\theta),
  \label{eq:driftlift}
\end{equation}
which is the drift lift of \cite{Sadhukhan:2026mcu} with the roles of $\theta$ and $\phi$ exchanged. The sphere block gives
\begin{equation}
  \delta T^Q_{AB} \;=\; \delta\mathcal P\; g_{AB} \;+\; \bar{\mathcal P}\;\delta g_{AB},
  \qquad
  \delta g_{AB} = r_e^2 K Y \mathring g_{AB},
  \label{eq:dTQsphere}
\end{equation}
where $\delta \mathcal P$ receives contributions from $\delta t_{ab}$, from the perturbed inverse metric $\delta\hat g^{ab}$ in the trace, and from the prefactor $|F|\,\varpi^2$ of the pressure term in \eqref{eq:lift}. The components along the field line, $\delta T^Q_{ab}$ with $a,b\in\{\tau,\rho\}$, depend only on $\delta\Upsilon$, $\xi_a$ and $K_F$. The metric amplitudes $H_0$ and $H_2$ enter $\delta T^Q$ only through $\delta\hat g^{ab}$ in $\delta\mathcal P$, and this dependence is only apparent. The trace of the CFT stress tensor is fixed by the anomaly for any metric, $\hat g^{ab}t_{ab}=c\hat R/24\pi$. Its variation is $c\,\delta\hat R/24\pi$, and on the unit AdS$_2$ background $\delta\hat R=4\delta\Upsilon-2\hat\Box\,\delta\Upsilon$. Adding the variation of the prefactor, $\delta\log(|F|\varpi^2)=-2K_FY$, gives $\delta\mathcal P=\bar{\mathcal P}\,[\hat\Box\,\delta\Upsilon-2\delta\Upsilon-2K_F]$, which contains no $H_0$ or $H_2$ (Appendix~\ref{app:dTQ}). The explicit components are collected in Appendix~\ref{app:dTQ}, Eqs.~\eqref{eq:dTQ_tt}--\eqref{eq:dTQ_mix}. Dropping the term in $K_F$ of \eqref{eq:dTQ2d}, which involves no response at all, destroys conservation. The $\tau$ and $\rho$ components of the divergence are then proportional to $\partial_aK_F$, and the $O(\epsilon)$ system fails at Step~4.

The $\tau$, $\rho$ and $\phi$ components of the divergence of the perturbed lift vanish identically in the fields once $h_{ab}$ is expressed through $(\delta\Upsilon,\xi_a,K,w)$ by the decomposition. Once any of the three effects is kept, this holds only when all three are kept together. With $f=1+\rho^2$, the tilt alone, the density alone, and the density with the scale but without the tilt leave
\begin{equation}
  \Big(\nabla_\mu T_Q^{\mu\tau},\;\nabla_\mu T_Q^{\mu\rho}\Big)\Big|_{O(\epsilon)}
  \;=\; \frac{\lhat\,\bar{\mathcal P}}{2r_e^{2}}\,Y(\theta)\times
  \begin{cases}
    \Big(-\dfrac{2\rho^{2}-1}{f^{2}}\,\dot w,\;\;(2\rho^{2}+5)\,w'\Big) & \text{tilt},\\[10pt]
    \Big(\dfrac{\rho^{2}-2}{f^{2}}\,\dot w,\;\;-(\rho^{2}+4)\,w'\Big) & \text{density},\\[10pt]
    \Big(\dfrac{2\rho^{2}-1}{f^{2}}\,\dot w,\;\;-(2\rho^{2}+5)\,w'\Big) & \text{density and scale},
  \end{cases}
  \label{eq:closureres}
\end{equation}
in the throat. The last is minus the first, and the two cancel in the full lift. The tilt together with the density, with the line metric kept at $g_{ab}/R^2$, leaves
\begin{equation}
  \nabla_\mu T_Q^{\mu a}\Big|_{O(\epsilon)} \;=\; \bar{\mathcal P}\,\bar g^{ab}\,\partial_b\!\left(\frac{\delta F_{\theta\phi}}{\bar F_{\theta\phi}}\right) \;=\; \frac{\lhat}{2}\,\bar{\mathcal P}\,\bar g^{ab}\,\partial_b w\;Y(\theta),
  \label{eq:residue}
\end{equation}
and no current can absorb it, since $\bar F^{a\lambda}=0$. The tilt and the density are tied together by the Bianchi identity $\dd F=0$, which carries the density of field lines along with the moving lines. The residue \eqref{eq:residue} is the divergence along the field lines of the transverse pressure, which follows the density of lines while the scale of each line metric stays fixed. Scaling the line metric with $|F|$ removes it exactly. The lift built on the coordinate planes, with $T^Q_{aA}=0$ and with the density and the scale both proportional to $1/R^2$, also passes this test. Its density and scale depend on the metric alone. They miss the change of flux $\delta F_{\theta\phi}$, which the measure $|F_{\theta\phi}|/2\pi$ of the effective action \eqref{eq:Wfact} contains. Both closures pass Step~4 (Sec.~\ref{sec:throat-reduction}). Both also pass the matching of Sec.~\ref{sec:overlap}, since the mouth and the throat then carry the same lift. The choice between them is physical. The measure of \eqref{eq:Wfact} counts the field lines and follows the flux, and the scale of each line metric follows the cyclotron gap. The second part is the assumption \eqref{eq:linescale}. The coordinate planes are the invariant planes of the background field $\bar F$. A lift built on them depends on $\bar F$ separately from the dynamical field $F$ and is not a functional of the fields, as noted for the companion sector \cite{Sadhukhan:2026mcu}. We use it as a comparison and call it the coordinate lift, and \eqref{eq:lift} with the scaling \eqref{eq:linescale} the scaled lift. The Hall current of the coordinate lift follows from its own force balance. The coordinate lift should not be confused with the treatment on the coordinate planes of Sec.~\ref{sec:hall-action}, which adds the Euler term \eqref{eq:lambdafixed} to the effective action.

\section{The induced Hall current}
\label{sec:hall}

\subsection{Force balance and the induced current}
\label{sec:hall-force}

Lifting \eqref{eq:response} with \eqref{eq:lift} and computing the four-dimensional divergence in the perturbed metric, we examine the $O(\epsilon)$ force balance \eqref{eq:forcebalance1} component by component. The only nonvanishing components of the background field strength are $\bar F_{\theta\phi}=-\tfrac q2\sin\theta$ and its antisymmetric partner. For $\nu=\tau$ and $\nu=\rho$ the right-hand side vanishes identically, since $\bar F^{\tau\lambda}=\bar F^{\rho\lambda}=0$ for every $\lambda$. These two components are consistency conditions on the fermion response alone. They are satisfied once $h_{ab}$ is expressed through the Weyl factor $\delta\Upsilon$ and the transport fields $\xi_\tau,\xi_\rho$ by the two-dimensional Ward identity \eqref{eq:ward}. They also need the structure of the lift \eqref{eq:lift}, whose sphere pressure is set by the trace and whose line metric is scaled by $|F|$ (Sec.~\ref{sec:response}). For $\nu=\phi$ the right-hand side is $-\bar F^{\phi\theta}\delta J_\theta$ and the left-hand side vanishes, hence $\delta J_\theta=0$. For $\nu=\theta$ the right-hand side is $-\bar F^{\theta\phi}\delta J_\phi$, and here the left-hand side does not vanish,
\begin{equation}
  \nabla_\mu T_Q^{\mu\theta}\Big|_{O(\epsilon)} \;\neq\; 0 .
  \label{eq:thetaobstruction}
\end{equation}
This single component\footnote{Equation \eqref{eq:ward} does not preclude \eqref{eq:thetaobstruction}. Equation \eqref{eq:ward} is the two-dimensional divergence of the two-dimensional $t_{ab}$ on $\hat g_{ab}$, with indices running only over $(\tau,\rho)$. Equation \eqref{eq:thetaobstruction} is the four-dimensional divergence of the lifted $T^Q_{\mu\nu}$, with free index $\theta$.} fixes the induced current,
\begin{equation}
  \delta J_\phi \;=\; -\,\frac{\nabla_\mu T_Q^{\mu\theta}\big|_{O(\epsilon)}}{\bar F^{\theta\phi}} ,
  \qquad
  \delta J^\phi = g^{\phi\phi}\delta J_\phi \;\propto\; Y'(\theta) ,
  \label{eq:hallcurrent}
\end{equation}
which has the angular structure of the axial gauge perturbation \eqref{eq:hansatz2}. It enters the $\phi$ component of the Maxwell equation \eqref{eq:order1} and sources the mode $w$ we are solving for. The components $\delta J_\tau$ and $\delta J_\rho$ never appear in \eqref{eq:forcebalance1}. They vanish for an independent reason. The lowest-Landau-level fermions live on the two-dimensional $(\tau,\rho)$ block, and a current along those directions would have to be driven by a perturbing gauge field with $(\tau,\rho)$ legs. The gauge perturbation in this sector is purely axial, $a_\phi=a\sin\theta\,Y'(\theta)$ with $a_\tau=a_\rho=0$, and the fermions see no driving field within their own worldsheet. The perturbed two-dimensional metric induces only a stress-tensor response $\delta t_{ab}$ and not a charge density. Charge conjugation reverses the current and leaves the metric unchanged. We take the vacuum on each line to be invariant under it, and the holonomy around each line is unchanged at $O(\epsilon)$ since $a_\tau=a_\rho=0$. A current along the field lines linear in the polar metric perturbation therefore vanishes. In the companion sector the axial components $h_{a\phi}$ do act on the charge of the LLL states, through their kinetic azimuthal momentum and the spin connection \cite{Sadhukhan:2026mcu}, and this sector has no such components.

The asymmetry between the $\theta$ and $\phi$ components has a simple origin. The perturbation changes the sphere radius and the density of field lines, both through $Y(\theta)$, and the Casimir energy per unit area varies from one field line to the next. This is an energy gradient transverse to the magnetic field. The fermions stay on their field lines, and the field lines drift in $\theta$ when $\delta F_{a\phi}\neq0$. The drift carries the momentum flux \eqref{eq:driftlift}. It is fixed by $\delta F_{a\phi}$ and does not balance the gradient. The remaining channel is the Lorentz force. Charges of either sign subject to a transverse force in a magnetic field drift perpendicular to both, in opposite directions along $\phi$, and their currents add. The current comes out proportional to the transverse gradient of the profile,
\begin{equation}
  \delta J^\phi \;\propto\; \frac{Y'(\theta)}{\sin\theta} ,
\end{equation}
and for a $\theta$-independent perturbation the obstruction would vanish identically. The resulting $\delta J^\phi$ is an ordinary electric current of the charged fermions drifting azimuthally. It sources the axial gauge perturbation because of the parity of the radiated multipole, not because of any magnetic charge.

Equation \eqref{eq:hallcurrent} determines the current by division, since the $\theta$ component of \eqref{eq:forcebalance1} contains the single unknown $\delta J_\phi$ multiplied by the known $\bar F^{\theta\phi}$. The response $\delta T_Q$ is fixed by the two-dimensional Ward identity of Sec.~\ref{sec:response} and the lift of Sec.~\ref{sec:lift}, with no freedom to adjust it. Charge conservation holds automatically. Maxwell's equation needs $\nabla_\nu J^\nu=0$, and for an azimuthal current independent of $\phi$ this is $\nabla_\mu J^\mu = (\sqrt{-g})^{-1}\partial_\phi(\sqrt{-g}\,\delta J^\phi)=0$. With the current inserted the source is conserved, its $\tau$, $\rho$ and $\phi$ components by the structure of the lift and its $\theta$ component by \eqref{eq:hallcurrent}. The three Bianchi residuals of Step~4 in Sec.~\ref{sec:algorithm} must then vanish. They do (Sec.~\ref{sec:throat-reduction}), which checks the construction and the reduction. The physical input is the lift itself. Step~4 cannot test it, since every conserved lift passes.

Write $f \equiv 1+\rho^2$, and understand $J^\phi$ from here on as the coefficient of $\epsilon$ in the current $J^\phi$ of \eqref{eq:maxwell}. The $\theta$ component of \eqref{eq:forcebalance1} gives
\begin{equation}
\begin{aligned}
  J^{\phi} \;=\; \frac{1}{48\pi^{2}r_e^{4}\,f^{2}}\,\frac{Y'(\theta)}{\sin\theta}
  \Bigg[\;
  &f^{3}\,\delta\Upsilon'' \;+\; 2\rho f^{2}\,\delta\Upsilon' \;-\; f\,\ddot{\delta\Upsilon}
  \\[2pt]
  +\;&f^{2}\!\left(\rho^{2}+4\right)\xi_\rho' \;+\; 2\rho f^{2}\,\xi_\rho
     \;-\; \left(\rho^{2}-2\right)\dot\xi_\tau
  \\[2pt]
  -\;&\frac{f^{2}}{2}\left(H_{0}-H_{2}+2K_F\right)
  \\[2pt]
  -\;&\frac{f^{2}}{2}\Big[\left(2\rho^{2}+5\right)w'' + 4\rho\,w'\Big] \;+\; \frac{2\rho^{2}-1}{2}\,\ddot w
  \Bigg] ,
\end{aligned}
\label{eq:Jphi_explicit}
\end{equation}
where a prime denotes $\partial_\rho$ and a dot $\partial_\tau$. The four lines are four contributions. The first is the anomalous Weyl response, carrying two derivatives of $\delta\Upsilon$. The second is the Lie drag of the background stress tensor, carrying one derivative of $\xi$. The third has no derivatives and comes from the background Casimir energy redistributed over a changed density of field lines, together with the metric amplitudes $H_0$ and $H_2$. The fourth is the drift of the field lines \eqref{eq:driftlift}, carrying two derivatives of $w$. With $H_0=H_2$ from \eqref{eq:H0H2}, which holds before the current is used, the third line collapses to $-f^{2}K_F$. Once $H_0$ and $H_2$ are eliminated through the decomposition \eqref{eq:deceq_tt}--\eqref{eq:deceq_rr}, the explicit $K_F$ in the third line cancels against the $K_F$ carried by $H_0-H_2$, and the current depends on $\delta\Upsilon$, $\xi_a$ and $w$ alone. The drift then enters only through the combination
\begin{equation}
  \tilde\xi_a \;\equiv\; \xi_a - \tfrac12\,\partial_a w ,
  \label{eq:xitilde}
\end{equation}
and the reduced form is Eq.~\eqref{eq:Jphi_pure} of Appendix~\ref{app:dTQ}. The drift of the field lines acts on the current as a shift of the transport fields. The angular dependence factorises exactly, and the bracket contains no $\theta$. The combination $Y'(\theta)/\sin\theta$ is regular at the poles, since $Y'(\theta) = -\sin\theta\,P_l'(\cos\theta)$, and it is the axial profile carried by $a_\phi = a\sin\theta\,Y'$. Lowering the index with the background metric,
\begin{equation}
  J_\phi \;=\; R^{2}\sin^{2}\theta\;J^{\phi}
  \;=\; r_e^{2}\sin^{2}\theta\;J^{\phi} \;+\; O(\alpha),
\end{equation}
where the $\alpha$ correction to $R^{2}$ is dropped for the reason given in Sec.~\ref{sec:response}. The current enters the Maxwell equation multiplied by $g_c^{2}=4\pi^{2}\alpha q$, and retaining $R^{2}=r_e^{2}(1+\alpha\varphi)$ here would generate only $O(\alpha^{2}\epsilon)$ terms. The same applies to the raising of the index in \eqref{eq:hallcurrent} and to $F^{\theta\phi}$, both evaluated on the exact AdS$_2\times S^2$ background.

The $\theta$ component of the divergence has a closed form for any background of the form \eqref{eq:bgmetric}. The metric is block diagonal, $\sqrt{-g}=\sqrt{-g_2}\,R^2\sin\theta$ with $g_2$ the determinant of the $(t,r)$ block. The $\theta$ dependence enters through $Y(\theta)$ in $h_{ab}$, in $R^2$, in $|F|$ and in $\mathcal P$, and through $Y'(\theta)$ in the mixed components \eqref{eq:driftlift}. With $h_{ab}$ and $\delta\mathcal P$ the coefficients of $Y$ and indices raised by the background,
\begin{equation}
\begin{gathered}
  \nabla_\mu T_Q^{\mu\theta}\Big|_{O(\epsilon)} \;=\; \frac{Y'(\theta)}{R^{2}}\,\mathfrak F ,
  \qquad
  J^{\phi} \;=\; \frac{2}{q}\,\frac{Y'(\theta)}{\sin\theta}\,\mathfrak F ,
  \\[2pt]
  \mathfrak F \;\equiv\; \delta\mathcal P \;+\; \tfrac12\,\bar{\mathcal P}\,\bar g^{ab}h_{ab} \;-\; \tfrac12\,\bar T_Q^{\,ab}h_{ab}
  \;+\; \frac{1}{2R^{2}\sqrt{-g_2}}\,\partial_a\!\Big(\sqrt{-g_2}\,R^{4}\,\mathsf K^{ab}\,\partial_b w\Big),
  \qquad
  \mathsf K^{ab} \;\equiv\; \bar T_Q^{\,ab} - \bar{\mathcal P}\,\bar g^{ab},
\end{gathered}
  \label{eq:Jclosed}
\end{equation}
where the second relation uses $\bar F^{\theta\phi}=-q/(2R^4\sin\theta)$. The matrix $\mathsf K^{ab}$ is the kernel of the drift lift of the companion sector \cite{Sadhukhan:2026mcu}. The term $\tfrac12\bar{\mathcal P}\,\bar g^{ab}h_{ab}$ comes from the $\theta$ dependence of $\sqrt{-g_2}$ acting on the sphere pressure, and the term $-\tfrac12\bar T_Q^{\,ab}h_{ab}$ comes from $\Gamma^\theta_{ab}=-\tfrac12 g^{\theta\theta}\partial_\theta g_{ab}$. On the throat \eqref{eq:Jclosed} reproduces \eqref{eq:Jphi_explicit} identically when $\delta\mathcal P$ is taken from the lift with $\delta\hat g^{ab}$ written through $h_{ab}$. With $\delta\mathcal P$ of \eqref{eq:dP_explicit} it does so once $h_{ab}$ is decomposed by \eqref{eq:deceq_tt}--\eqref{eq:deceq_rr}. It needs no four-dimensional covariant derivative and no angular algebra, and we use it in the mouth in Sec.~\ref{sec:mouthpert}.

\subsection{The Hall current from the effective action}
\label{sec:hall-action}

The current \eqref{eq:Jphi_explicit} was read off from the one component of the force balance that fails to close. We now derive it a second time, forwards, from the effective action of the field lines, with no reference to $T^Q_{\mu\nu}$ at any stage. On the background the action has the factorised form
\begin{equation}
  W \;=\; \int \dd\tau\,\dd\rho\,\dd\theta\,\dd\phi\;
  \frac{|F_{\theta\phi}|}{2\pi}\;\mathfrak L_{2d}\big[\hat g_{ab}(\tau,\rho;\theta)\big]
  \;=\; -\int \dd^{4}x\;\frac{F_{\theta\phi}}{2\pi}\;\mathfrak L_{2d}\big[\hat g_{ab}(\tau,\rho;\theta)\big],
  \qquad
  \int \frac{|F_{\theta\phi}|}{2\pi}\,\dd\theta\,\dd\phi \;=\; q ,
  \label{eq:Wfact}
\end{equation}
where the second form uses $F_{\theta\phi}<0$ for the orientation \eqref{eq:monopole}, a sign that persists at $O(\epsilon)$ since $|\bar F|\gg\epsilon|\delta F|$. Here $|F_{\theta\phi}|/2\pi$ is the local density of magnetic field lines, normalised to the number of lowest-Landau-level modes fixed by the index theorem. The density $\mathfrak L_{2d}$ is the effective Lagrangian of a single two-dimensional fermion on the metric $\hat g_{ab}$ of the field line through $(\theta,\phi)$. The count is carried by the measure, and $\mathfrak L_{2d}$ carries $c_1=1$ rather than $c=q$. The normalisation of the final answer turns on this point. Away from the background the field lines are the surfaces \eqref{eq:kernel}. In the axisymmetric sector $F_{\mu\nu}e_a^\nu=0$ reads $e_a(A_\phi)=0$, and the field lines are the level surfaces of $A_\phi$. The action is the sum over field lines of the effective action $W_{2d}$ of one fermion on the metric \eqref{eq:linescale} of its line,
\begin{equation}
  W \;=\; \int\frac{\dd A_\phi\,\dd\phi}{2\pi}\;W_{2d}\big[\varpi^2\mathsf g_{ab}\big],
  \qquad
  W_{2d}[\hat g] \;=\; \int\dd\tau\,\dd\rho\;\mathfrak L_{2d}[\hat g] ,
  \label{eq:Wcov}
\end{equation}
with the points of each line labelled by $(\tau,\rho)$. At fixed $(\tau,\rho)$ the measure is $\dd A_\phi\,\dd\phi=|F_{\theta\phi}|\,\dd\theta\,\dd\phi$, and on the background \eqref{eq:Wcov} is \eqref{eq:Wfact}. The metric variation of \eqref{eq:Wcov} at fixed $A_\mu$ is the lift \eqref{eq:lift}. The field lines do not move, since the kernel of $F$ does not depend on the metric. The variation of the line metric is $\delta\hat g_{ab}=\varpi^2 e_a^\mu e_b^\nu\,\delta g_{\mu\nu}+\mathsf g_{ab}\,\delta\varpi^2$ with $\delta\log\varpi^2=-\tfrac12\perp^{\mu\nu}\delta g_{\mu\nu}$. With $\delta W_{2d}=\tfrac12\int\sqrt{\hat g}\,\mathfrak t^{ab}\delta\hat g_{ab}$ and $\mathfrak t^{ab}=t^{ab}/q$ this gives \eqref{eq:lift} term by term.

The current follows from the gauge variation. The fermions couple to $A_\nu$ only through $W$. Varying the total action $-\tfrac{1}{4g_c^{2}}\!\int\!\sqrt{-g}\,F^{2} + W$ gives $\nabla_\mu F^{\mu\nu} = -(g_c^{2}/\sqrt{-g})\,\delta W/\delta A_\nu$, and with the convention \eqref{eq:maxwell} the current is $J^\nu = -(\sqrt{-g})^{-1}\,\delta W/\delta A_\nu$. A variation $\delta A_\phi$ at fixed metric does two things. It moves the field line through $(\theta,\tau,\rho)$ to $\theta+\delta \vartheta$ with $\delta \vartheta=-\delta A_\phi/F_{\theta\phi}$. It also changes $|F|$ at a fixed point, by $\delta\log|F|=g^{\phi\phi}\big(F_{\theta\phi}\,g^{\theta\theta}\,\partial_\theta\delta A_\phi+F_{a\phi}\,g^{ab}\,\partial_b\delta A_\phi\big)/|F|^2$. The metric of the displaced line changes by $\delta \vartheta\,\partial_\theta\hat g_{ab}$, through the $\theta$ dependence of the metric along the line, and by $\varpi^2 g_{\theta\theta}\big(\partial_a\vartheta\,\partial_b\delta \vartheta+\partial_b\vartheta\,\partial_a\delta \vartheta\big)$, through the tilt. For the axisymmetric $m=0$ sector $\partial_\phi$ annihilates everything. Integrating by parts in $\theta$ and along the line gives, to first order in $\epsilon$,
\begin{equation}
  J^\phi \;=\; -\,\frac{1}{2\pi\sqrt{-g}}\Big[\frac{\sqrt{\hat g}}{2}\,\bar{\mathfrak t}^{\,ab}\,\partial_\theta\hat g_{ab}
  \;+\;\frac{c_1}{48\pi}\,\partial_\theta\big(\sqrt{\hat g}\,\hat R\big)\Big]
  \;+\;\frac{1}{2\pi\sqrt{-g}}\,\partial_b\Big(\sqrt{\hat g}\,\mathfrak s^{ab}\,\partial_a\vartheta\Big),
  \qquad
  \mathfrak s^{ab}\;\equiv\;\bar{\mathfrak t}^{\,ab}+\tfrac12\,\hat g^{ab}\,\hat g^{cd}\,\bar{\mathfrak t}_{cd} ,
  \label{eq:Jcov}
\end{equation}
with $c_1=1$ the central charge of one fermion, the indices of $\bar{\mathfrak t}$ raised by $\hat g$, the boundary terms at $\theta=0,\pi$ vanishing because $\delta A_\phi \propto \sin\theta\,Y'(\theta)$. Two sign changes compensate each other, $|F_{\theta\phi}|=-F_{\theta\phi}$ in \eqref{eq:Wfact} and $J^\nu=-\delta W/\sqrt{-g}\,\delta A_\nu$ from \eqref{eq:maxwell}. For the opposite orientation, $q<0$ in \eqref{eq:monopole}, one has $|F_{\theta\phi}|=+F_{\theta\phi}$ and the current reverses. This is the behaviour of a Hall drift under reversal of the magnetic field, and it is the only place in the paper where the orientation enters. The three terms of \eqref{eq:Jcov} have separate origins. The first is the displacement of a line through the $\theta$-dependent metric. The second is the change of $|F|$ through $F_{\theta\phi}$ at a fixed point. The trace of the line stress tensor multiplies $\delta\log\varpi^2$, and the trace anomaly $\hat g^{ab}\mathfrak t_{ab}=\hat R/24\pi$ turns this product into the $\theta$ derivative of the Euler density. The third is the tilt, from the displacement of the moving line and from $F_{a\phi}$ in $|F|$. In $\mathfrak s^{ab}$ the second term is the transverse pressure of one line, and $\mathfrak s^{ab}$ is the counterpart for one line of the kernel $\bar T^{Q}{}_a{}^c-\bar{\mathcal P}\delta_a{}^c$ of \eqref{eq:driftlift}.

The current is the transverse gradient of the effective action across the field lines, together with the drift, and its angular profile follows at once. The $\theta$-dependence of $\hat g_{ab}$ enters only through the harmonic,
\begin{equation}
  \hat g_{ab}(\tau,\rho;\theta) \;=\; \hat g^{(0)}_{ab}(\tau,\rho) \;+\; \epsilon\,Y(\theta)\,\delta\hat g_{ab}(\tau,\rho),
  \qquad
  \delta\hat g_{ab} \;=\; \frac{h_{ab}}{R^2}\Big|_{Y=1} - K_F\,\hat g^{(0)}_{ab},
  \label{eq:ghattheta}
\end{equation}
whose components on the unit AdS$_2$ background are, with $f\equiv1+\rho^2$,
\begin{equation}
  \hat g_{\tau\tau} = -f + \epsilon\,f\,(H_0+K_F)\,Y,
  \qquad
  \hat g_{\tau\rho} = \epsilon\,\frac{H_1}{r_e^2}\,Y,
  \qquad
  \hat g_{\rho\rho} = \frac{1}{f} + \epsilon\,\frac{H_2-K_F}{f}\,Y .
  \label{eq:ghatcomponents}
\end{equation}
Neighbouring field lines therefore carry different two-dimensional geometries and hence different effective actions. The factor $Y'(\theta)$ of the first two terms, and with it the axial parity of the current, is derived rather than observed. The third term carries $Y'(\theta)$ through the tilt \eqref{eq:tilt}. That the current sources the axial Maxwell equation and no other is a consequence of \eqref{eq:Jcov}.

The first term of \eqref{eq:Jcov} needs no choice of a local density $\mathfrak L_{2d}$. It uses only the variation of the integrated action of one line, $\delta W_{2d}=\tfrac12\int\sqrt{\hat g}\,\mathfrak t^{ab}\delta\hat g_{ab}$, for a displacement $\delta \vartheta(\tau,\rho)$ that is arbitrary along the line. Hence
\begin{equation}
  \frac{\sqrt{\hat g}}{2}\;\bar{\mathfrak t}^{\,ab}\,\partial_\theta\hat g_{ab}
  \;=\; \epsilon\,Y'(\theta)\;\frac{\sqrt{\hat g^{(0)}}}{2}\;\bar{\mathfrak t}^{\,ab}\,\delta\hat g_{ab}
  \;+\;O(\epsilon^2),
  \label{eq:dthetaellt}
\end{equation}
where $\mathfrak t^{ab}$ has been evaluated on the background because $\partial_\theta\hat g_{ab}$ already carries a factor of $\epsilon$. On unit AdS$_2$ one has $\sqrt{\hat g^{(0)}}=1$ and $\sqrt{-g}=r_e^4\sin\theta$, and the background tensor of one fermion, $\bar{\mathfrak t}_{ab}=\bar t_{ab}/q$ with $\bar t_{ab}$ from \eqref{eq:tbar}, has upper components
\begin{equation}
  \bar{\mathfrak t}^{\,\tau\tau} \;=\; \frac{\rho^2-2}{24\pi f^2},
  \qquad
  \bar{\mathfrak t}^{\,\rho\rho} \;=\; -\frac{\rho^2+4}{24\pi},
  \qquad
  \bar{\mathfrak t}^{\,\tau\rho} \;=\; 0 .
  \label{eq:tbarup}
\end{equation}
The contraction with \eqref{eq:ghatcomponents}, using $\delta\hat g_{\tau\tau}=f(H_0+K_F)$ and $\delta\hat g_{\rho\rho}=(H_2-K_F)/f$, gives
\begin{equation}
  \frac{\sqrt{\hat g}}{2}\;\bar{\mathfrak t}^{\,ab}\,\partial_\theta\hat g_{ab}
  \;=\; \frac{\epsilon\,Y'(\theta)}{48\pi\,f}
  \Big[\,(\rho^2-2)\,H_0 \;-\; (\rho^2+4)\,H_2 \;+\; 2f\,K_F\,\Big] .
  \label{eq:dthetaelltexplicit}
\end{equation}
This contains no derivatives of the metric amplitudes. A contraction of the background stress tensor with the perturbation cannot produce any. One term of \eqref{eq:dthetaelltexplicit} can be checked independently. The coefficient of a single amplitude depends on which amplitudes are held independent, since \eqref{eq:deceq_tt}--\eqref{eq:deceq_rr} relate them. Here, as in \eqref{eq:Jphi_explicit}, they are $\delta\Upsilon$, $\xi_a$, $H_0$, $H_2$, $K_F$ and $w$. Setting $\delta\Upsilon=\xi_a=H_0=H_2=0$ then isolates the pure $K_F$ dependence, which through \eqref{eq:Jcov} gives
\begin{equation}
  J^\phi\Big|_{K_F} \;=\; -\frac{K_F\,Y'(\theta)}{48\pi^{2}r_e^{4}\sin\theta},
  \label{eq:JK}
\end{equation}
in agreement with the geometric term of \eqref{eq:Jphi_explicit}. The test that does not depend on this choice is \eqref{eq:agreement} below. This piece comes from the contraction of $-K_F\hat g_{ab}$ with the trace $\hat g^{ab}\bar{\mathfrak t}_{ab}=-1/12\pi$ and can be obtained a second way. The $\theta$-dependence of $\hat g_{ab}$ through the scale $\varpi^2$ alone is a Weyl factor $\delta\Upsilon=-\tfrac12 K_FY$. The shift in $\mathfrak L_{2d}$ is then fixed by the trace anomaly to be $\delta\mathfrak L_{2d} = \sqrt{\hat g}\,\delta\Upsilon\,\hat g^{ab}\mathfrak t_{ab} = \sqrt{\hat g}\,K_F Y/24\pi$, and substituting into \eqref{eq:Jcov} reproduces \eqref{eq:JK}. A Weyl-invariant treatment would produce no such current.

For the second term no variational identity is used. The Ricci scalar and volume element of the perturbed metric \eqref{eq:ghatcomponents} are computed directly as functions of $(\tau,\rho;\theta)$ to first order in $\epsilon$, and the result is differentiated with respect to $\theta$,
\begin{equation}
  \partial_\theta\!\left[\frac{c_1}{48\pi}\sqrt{\hat g}\,\hat R\right]_{\hat g=\hat g(\theta)} .
  \label{eq:dthetaEuler}
\end{equation}
The background values are $\hat R^{(0)}=-2$ and $\sqrt{\hat g^{(0)}}=1$. The Euler density is constant on the background, and \eqref{eq:dthetaEuler} is $O(\epsilon)$. The derivative with respect to $\theta$ acts only on the harmonic, turning $Y(\theta)$ into $Y'(\theta)$. Equation \eqref{eq:dthetaelltexplicit} is algebraic in the amplitudes, whereas \eqref{eq:dthetaEuler} carries second derivatives with respect to $\tau$ and $\rho$, since $\hat R$ does. The second term supplies the derivative structure that the first lacks. For any two-dimensional metric the first-order variation of the Euler density is
\begin{equation}
  \delta\big(\sqrt{\hat g}\,\hat R\big)
  \;=\; \sqrt{\hat g}\,\Big[\hat\nabla^a\hat\nabla^b\delta\hat g_{ab} - \hat\Box\big(\hat g^{ab}\delta\hat g_{ab}\big)\Big] ,
  \label{eq:dEulergen}
\end{equation}
which follows from $\delta\hat R=-\hat R^{ab}\delta\hat g_{ab}+\hat\nabla^a\hat\nabla^b\delta\hat g_{ab} -\hat\Box(\hat g^{ab}\delta\hat g_{ab})$ together with $\hat R^{ab}=\tfrac12\hat R\hat g^{ab}$ in two dimensions and $\delta\sqrt{\hat g}=\tfrac12\sqrt{\hat g}\,\hat g^{ab}\delta\hat g_{ab}$. The two $\hat R$ terms cancel, leaving a pure divergence, as the topological invariance of $\int\sqrt{\hat g}\hat R$ requires. On the decomposition $\delta\hat g_{ab}=2\delta\Upsilon\,\hat g_{ab}+(\mathcal{L}_\xi\hat g)_{ab}$ the Weyl part gives $\hat g^{ab}\delta\hat g_{ab}=4\delta\Upsilon$ and $\hat\nabla^a\hat\nabla^b(2\delta\Upsilon\hat g_{ab})=2\Box\delta\Upsilon$, hence $-2\Box\delta\Upsilon$. The diffeomorphism part gives $\mathcal{L}_\xi(\sqrt{\hat g}\hat R)=\partial_a(\xi^a\sqrt{\hat g}\hat R)=\hat R\,\nabla_a\xi^a$, which on the background $\hat R=-2$ is $-2\nabla_a\xi^a$. Altogether,
\begin{equation}
  \delta\big(\sqrt{\hat g}\,\hat R\big) \;=\; -\,2\,\big(\Box\delta\Upsilon + \nabla_a\xi^a\big) ,
  \label{eq:dEuler}
\end{equation}
which also follows by direct computation of the Ricci scalar of the perturbed metric.

For the third term, \eqref{eq:tbarup} and $\hat g^{ab}\bar{\mathfrak t}_{ab}=-1/12\pi$ give $\mathfrak s^{\tau\tau}=(2\rho^2-1)/24\pi f^2$ and $\mathfrak s^{\rho\rho}=-(2\rho^2+5)/24\pi$. With the tilt \eqref{eq:tilt},
\begin{equation}
  J^\phi\Big|_{\rm tilt} \;=\; -\,\frac{Y'(\theta)}{96\pi^{2}r_e^{4}\sin\theta}\Big[\left(2\rho^{2}+5\right)w'' + 4\rho\,w' - \frac{2\rho^{2}-1}{f^{2}}\,\ddot w\Big] ,
  \label{eq:Jdrift}
\end{equation}
which is the fourth line of \eqref{eq:Jphi_explicit}. Substituting the three terms into \eqref{eq:Jcov} and expressing the result in $(\delta\Upsilon,\xi,K,w)$ by eliminating $H_0,H_1,H_2$ through \eqref{eq:deceq_tt}--\eqref{eq:deceq_rr} reproduces the current \eqref{eq:Jphi_explicit} of the force balance identically,
\begin{equation}
  J^\phi\Big|_{\text{action}} \;-\; J^\phi\Big|_{\text{force balance}} \;=\; 0 .
  \label{eq:agreement}
\end{equation}
Omitting the second term the agreement fails, and the mismatch is
\begin{equation}
  \frac{Y'(\theta)}{\sin\theta}\;
  \frac{\hat\Box\,\delta\Upsilon + \hat\nabla_a\xi^a}{48\pi^2 r_e^4} .
\end{equation}
Omitting the third, the mismatch is \eqref{eq:Jdrift}. The second term is therefore the entire source of the second derivatives of $\delta\Upsilon$ in the current, through $\hat\Box\delta\Upsilon+\hat\nabla_a\xi^a$, and the third is the entire source of its dependence on the derivatives of $w$. The agreement \eqref{eq:agreement} is a test of the lift and of its algebra. The force balance uses the metric variation of \eqref{eq:Wcov}, and the current is its gauge variation. For a functional invariant under diffeomorphisms and gauge transformations the two variations obey the force balance identically, and \eqref{eq:agreement} is that identity. It confirms that the lift is the metric variation of \eqref{eq:Wcov}. It does not test the scaling \eqref{eq:linescale}, which \eqref{eq:Wcov} contains. The agreement holds for arbitrary $\delta\Upsilon$ and $\xi_a$, and therefore in every state of the fermions.

The factor of $q$ enters the two calculations by different routes and must cancel. In the force-balance route,
\begin{equation}
  \delta J_\phi \;=\; -\,\frac{\nabla_\mu T_Q^{\mu\theta}}{\bar F^{\theta\phi}},
  \qquad
  T_Q \propto c ,
  \qquad
  \bar F^{\theta\phi} = -\frac{q}{2r_e^4\sin\theta} \propto q ,
\end{equation}
so that $J^\phi\propto c/q$, which the index theorem $c=|q|$ collapses to $\mathrm{sign}(q)$, equal to unity for the orientation \eqref{eq:monopole}. In the present route $q$ never appears. The density $\mathfrak L_{2d}$ carries $c_1=1$ and the count is supplied by $|F_{\theta\phi}|/2\pi$. Had $\mathfrak L_{2d}$ been taken with the central charge $q$, \eqref{eq:agreement} would have failed by a factor of $q$. The current is therefore independent of $q$. Doubling the flux doubles the number of field lines. It also doubles the field, which halves the drift of each line under a given force, and the current is unchanged. The vacuum on each line carries no net charge, but particles and antiparticles drift in opposite directions and their currents add (Sec.~\ref{sec:hall-force}).

The coefficient $c_1/48\pi$ of the second term of \eqref{eq:Jcov} is fixed by the trace anomaly, given the scaling \eqref{eq:linescale}. Integrating the anomaly along the Weyl orbit gives the Liouville form of the effective action of one line \cite{Polyakov:1981rd,Birrell:1982ix},
\begin{equation}
  W_{2d}\big[\varpi^2\mathsf g\big] \;=\; W_{2d}[\mathsf g] \;+\; \frac{c_1}{24\pi}\int\dd\tau\,\dd\rho\,\sqrt{\mathsf g}\,\Big(R_{\mathsf g}\log\varpi + \mathsf g^{ab}\,\partial_a\log\varpi\,\partial_b\log\varpi\Big),
  \label{eq:Liouville}
\end{equation}
in the conventions of \eqref{eq:anomalylaw}. With $\varpi^2=2|F|/q$ from \eqref{eq:linescale} the term linear in $\log\varpi$ is
\begin{equation}
  W \;\supset\; \frac{c_1}{48\pi}\int\frac{\dd A_\phi\,\dd\phi}{2\pi}\int\dd\tau\,\dd\rho\,\sqrt{\mathsf g}\,R_{\mathsf g}\,\log\frac{2|F|}{q} ,
  \label{eq:W4d}
\end{equation}
a coupling between the local field strength and the curvature of the field-line worldsheet. Its gauge variation through $|F|$, together with that of the term quadratic in $\log\varpi$ in \eqref{eq:Liouville}, is the second term of \eqref{eq:Jcov}. The two vary together into $\sqrt{\hat g}\hat R$, since $\sqrt{\hat g}\hat R=\sqrt{\mathsf g}\,(R_{\mathsf g}-2\Box_{\mathsf g}\log\varpi)$. A pure Euler term $\lambda_{\rm E}\int\sqrt{\hat g}\,\hat R$ added to $W_{2d}$ is different. It integrates to $4\pi\lambda_{\rm E}\chi$ on every field line, with $\chi$ the Euler characteristic, and it drops out of \eqref{eq:Wcov} for any $\lambda_{\rm E}$, since the field lines are the leaves of \eqref{eq:kernel} and the measure $\dd A_\phi\,\dd\phi$ is constant along each of them. A treatment on the coordinate planes $(\tau,\rho)$ at fixed $(\theta,\phi)$ has no dependence on $F$ through $\varpi^2$. It reproduces the second term of \eqref{eq:Jcov} only if the Euler density is added to $\mathfrak L_{2d}$ by hand with
\begin{equation}
  \lambda_{\rm E} \;=\; \frac{c_1}{48\pi} ,
  \label{eq:lambdafixed}
\end{equation}
and in that treatment the coefficient can only be fixed by requiring agreement with the force balance. In two dimensions in isolation $\lambda_{\rm E}$ is scheme dependent, since it shifts with the logarithm of the renormalisation scale through the term of the heat-kernel expansion proportional to $\int\sqrt{\hat g}\hat R$ \cite{Vassilevich:2003xt}. Here that scale is the cyclotron scale set by $|F|$, and the logarithm is the $\log|F|$ of \eqref{eq:W4d}. The Polyakov action \cite{Polyakov:1981rd} is obtained by integrating the trace anomaly, and a term in the kernel of the Weyl variation is invisible to that procedure. The two-dimensional Einstein--Hilbert term is therefore routinely omitted or absorbed into an inessential constant \cite{Birrell:1982ix}, and anomaly-induced actions are incomplete for this reason \cite{Deser:1993yx}. Fibred over the flux-threaded sphere with the scale \eqref{eq:linescale}, this term is not inert, and its coefficient is fixed by the anomaly. A scale $\varpi^2\propto|F|^n$ would give $n\,c_1/48\pi$.

A coupling of the field strength to a curvature, generated by projection onto a lowest Landau level, is reminiscent of the Wen-Zee term of the quantum Hall literature \cite{Wen:1992ej}. The analogy is structural. The Wen-Zee term couples the gauge field to the spin connection of the same two-dimensional space, whereas \eqref{eq:W4d} couples the field strength on $S^2$ to the curvature of the orthogonal worldsheet. Closer analogues are the responses of lowest-Landau-level states to curvature and to inhomogeneous fields on curved surfaces \cite{Klevtsov:2013iua,Ferrari:2014yba,Can:2014ota}. The methods used to derive the Wen-Zee coefficient from the lowest Landau level \cite{Gromov:2014gta,Bradlyn:2014wla,Abanov:2014ula} appear to be the natural route to deriving the scaling \eqref{eq:linescale} from the four-dimensional reduction directly. Only the variation of $W_{2d}$ and its trace anomaly enter \eqref{eq:Jcov}. We have not derived $W_{2d}$ itself from the reduction of the four-dimensional fermion determinant with its measure, along the lines of \cite{Freivogel:2026ujn} eq.~(4.51).

\section{Master equations in the throat to \texorpdfstring{$O(\alpha)$}{O(alpha)}}
\label{sec:throatmaster}

\subsection{The extended reduction and the master equations}
\label{sec:throat-reduction}

The throat calculation is a double expansion in $\epsilon$ and $\alpha$ (Sec.~\ref{sec:lift}). The $O(\epsilon^1\alpha^0)$ radial equations on exact AdS$_2\times S^2$ contain no fermionic input and are displayed in Appendix~\ref{app:throat-eqs}, Eqs.~\eqref{eq:t-ett}--\eqref{eq:t-mph}. The $O(\epsilon^1\alpha^1)$ equations are the $\alpha$-derivatives of the full residuals, $\partial_\alpha \mathcal E_{\mu\nu}$ and $\partial_\alpha \mathcal M^\phi$, Eqs.~\eqref{eq:t-dett}--\eqref{eq:t-dmph} of the same appendix. Besides the corrected background \eqref{eq:bgsol} they contain the lifted response $\delta T^Q_{\mu\nu}$ of Sec.~\ref{sec:response}, which brings in the Weyl factor $\delta\Upsilon$ and the transport fields $\xi_\tau,\xi_\rho$. The Maxwell equation also contains the Hall current $J^\phi$ of Sec.~\ref{sec:hall}. The sourced and source-free Maxwell equations differ by the expression \eqref{eq:mphdiff} of Appendix~\ref{app:throat-eqs}, which is $-g_c^2\sqrt{-g}\,\delta J^\phi/q$ with $g_c^2=4\pi^2\alpha q$ substituted, and which is proportional to $\alpha$.

The reduction of Sec.~\ref{sec:algorithm} is applied with two additional sub-steps that handle the transport fields. At each step only the terms linear in $\alpha$ are kept.
\begin{description}
\item[Step 1b.] The $\tau\tau$ decomposition equation \eqref{eq:deceq_tt} is algebraic in $\delta\Upsilon$. With $H_0=H_2$ from Step~1 and $\partial_\tau\to p$,
\begin{equation}
  \delta\Upsilon \;=\; -\,\frac{\left(1+\rho^{2}\right)\left(H_2 + K_F\right)
    + 2\rho\left(1+\rho^{2}\right)\xi_\rho - 2p\,\xi_\tau}{2\left(1+\rho^{2}\right)} .
  \label{eq:om1sol}
\end{equation}
\item[Step 1c.] The $\tau\rho$ and $\rho\rho$ decomposition equations \eqref{eq:deceq_tr}, \eqref{eq:deceq_rr} give first-order transport equations for the diffeomorphism parameters. We write $f\equiv1+\rho^2$ and display them at $\alpha^0$, which suffices since they enter the master system only at $O(\alpha)$,
\begin{align}
  \xi_\tau' &= -\,\frac{2p\rho}{\lhat f}\,K \;+\; \frac{2p}{\lhat}\,K' \;+\; \frac{2\rho}{f}\,\xi_\tau
              \;-\; p\,\xi_\rho ,
  \label{eq:xit}\\[4pt]
  \xi_\rho' &= \frac{(\lhat+2)f + 2p^2}{\lhat f^2}\,K \;-\; \frac{2\rho}{\lhat f}\,K'
              \;-\; \frac{p}{f^2}\,\xi_\tau \;-\; \frac{2}{f}\,w .
  \label{eq:xir}
\end{align}
\end{description}
Steps~2 and 3 then use the same three constraint equations and the same $rr$ equation as before. Among the constraint equations only $\mathcal E_{\tau\rho}$ acquires dependence on the transport fields. $\mathcal E_{\tau\theta}$ and $\mathcal E_{\rho\theta}$ receive $O(\alpha)$ corrections to their coefficients and the drift of the field lines, through the mixed components \eqref{eq:driftlift}, which add terms in $w$ and $w'$ and no new fields. The derivative rules for $H_2'$ and $K'$ pick up $\xi_\rho$, $w$ and $w'$ at $O(\alpha)$, with the explicit forms given in Appendix~\ref{app:constraints}, and the algebraic relation for $H_2$ picks up the transport fields through \eqref{eq:om1sol}. The remaining equations $\mathcal E_{\tau\tau}$, $\mathcal E_{\rho\rho}$, $\mathcal E^{\rm trace}$ and $\mathcal M^\phi$ acquire $\delta\Upsilon$ and $\xi_a$ dependence through $\delta T^Q$ and $J^\phi$, and this is what Step~4 tests.

In Step~4 all three residuals vanish identically at $O(\alpha^0)$ and at $O(\alpha^1)$. This confirms that the source is conserved and that the reduction is complete. The linearised Bianchi identities hold when both the stress-tensor response \eqref{eq:response}, lifted by \eqref{eq:lift}, and the Hall current \eqref{eq:Jphi_explicit} are included. If one keeps $\delta T^Q$ but drops $J^\phi$, the residual $\mathcal E_{\tau\tau}$ and the compatibility condition still vanish, but the angular trace does not. After Steps~1--3 it is
\begin{equation}
\begin{aligned}
  \mathcal E^{\rm trace}\Big|_{J^\phi=0} \;=\; \alpha\,\bigg[\,&\frac{\lhat(\lhat+2)f^{2}-4\lhat f-4(p^{2}+1)}{\lhat f^{2}}\,K
  \;+\;\frac{2\rho}{p\,r_e^{2}f}\,H_1\\
  &-\;\frac{\lhat(\lhat+10)f^{2}-6(\lhat+4)f-12p^{2}}{6f^{2}}\,w
  \;-\;\frac{2\rho}{f}\,w'\,\bigg] ,
\end{aligned}
  \label{eq:noHall}
\end{equation}
with $\mathcal E^{\rm trace}$ the radial coefficient of \eqref{eq:tracepart}. The truncation with $J^\phi=0$ is therefore inconsistent. At generic frequency the residual and its derivatives along the $\alpha^0$ system force $K=H_1=w=0$. The same experiment with $\delta T^Q=0$ also fails, as anticipated in Sec.~\ref{sec:consistency}. Step~4 also rejects an incomplete lift. With the tilt and the density kept but the line metric held at $g_{ab}/R^2$, the residue \eqref{eq:residue} leaves $\alpha\lhat f w/6$ in the $\mathcal E_{\tau\tau}$ residual and $-\tfrac{\alpha}{3}\big(w'+\rho w/f\big)$ in the compatibility condition for $H_2$. The coordinate lift of Sec.~\ref{sec:response} passes Step~4, and Step~4 alone does not decide between it and \eqref{eq:lift}. The closed first-order system is six-dimensional, $\bm y=(K,\allowbreak H_1,\allowbreak w,\allowbreak w',\allowbreak \xi_\tau,\allowbreak \xi_\rho)^\top$. The second-order equations for $K$ and $w$ obtained in Step~6 contain neither $\xi_\tau$ nor $\xi_\rho$. The transport fields appear only in intermediate quantities such as \eqref{eq:om1sol} and the algebraic expression for $H_2$, and they cancel from the master system. In the mouth they do not cancel. The equation for $K$ keeps a term proportional to $\alpha p^3$, which is of order $\alpha^3$ in the throat scaling (Sec.~\ref{sec:mouth-master}). The three decomposition equations are local first-order relations, and the reduction never requires inverting a differential operator.

The outcome of Step~6 is the pair of coupled second-order equations \eqref{eq:secondorder} for $\bm Y=(K,w)^\top$ in the coordinate $\rho$,
\begin{equation}
  \bm Y'' + P\,\bm Y' + Q\,\bm Y = 0,
  \qquad
  P = P_0 + \alpha P_1,
  \qquad
  Q = Q_0 + \alpha Q_1 ,
  \label{eq:t-PQ0}
\end{equation}
and, writing $f\equiv1+\rho^2$, $\lhat=l(l+1)$, $p=-i\Om$,
\begin{equation}
  P_0 = \frac{2\rho}{f}\,I_2 ,
  \qquad
  Q_0 = \begin{pmatrix}
  -\,\dfrac{(\lhat+2)f + p^{2}}{f^{2}} & \dfrac{2\lhat}{f}\\[10pt]
  \dfrac{2}{f} & -\,\dfrac{\lhat f + p^{2}}{f^{2}}
  \end{pmatrix} .
  \label{eq:t-P0Q0}
\end{equation}
The matrices $P_0,Q_0$ are the $\alpha^{0}$ terms, evaluated on exact AdS$_2\times S^2$. The background corrections $\gamma,\varphi$ of \eqref{eq:bgsol} enter through $P_1,Q_1$ and are the origin of the $\arctan\rho$ and $\log f$ terms there. $P_1,Q_1$ are not obtained from $P_0,Q_0$ by the substitution $f\to f+\alpha\gamma$, since they also receive the stress-tensor response of Sec.~\ref{sec:response}, the drift of the field lines and the Hall current of Sec.~\ref{sec:hall}. The $O(\alpha)$ matrices are
\begin{equation}
\begin{aligned}
  P_{1,KK} &= \frac{6\lhat(\rho^{2}-1)\arctan\rho - 6\lhat\rho\log f - 8\lhat\rho + 12\rho}{3\lhat f^{2}},
  \qquad
  P_{1,Kw} = \frac{2\rho}{f} + 2\arctan\rho ,
  \\[4pt]
  P_{1,wK} &= \frac{2\rho}{\lhat f^{2}},
  \qquad
  P_{1,ww} = -\,\frac{(3\rho^{4}+9)\arctan\rho + 6\rho\log f + 3\rho^{3} + 14\rho}{3f^{2}} ,
\end{aligned}
\label{eq:t-P1}
\end{equation}
and
\begin{equation}
\begin{aligned}
  Q_{1,KK} &= \frac{1}{3\lhat f^{3}}\Big\{
     -3\lhat(\lhat-2)\rho^{4} - 2(\lhat^{2}-7\lhat+6)\rho^{2} + \lhat^{2} + 8\lhat - 12
     - p^{2}\big(8\lhat\rho^{2} + 12\big)
  \\&\hspace{1.8cm}
     + \big[3\lhat(\lhat+2)f + 6\lhat p^{2}\big]\log f
     + 6\lhat\big[2\rho^{5} - (\lhat-2+p^{2})\rho^{3} - (\lhat+3p^{2})\rho\big]\arctan\rho
  \Big\},
  \\[4pt]
  Q_{1,Kw} &= \frac{\lhat(\lhat-2)\rho^{2} + \lhat^{2} - 18\lhat + 24 - 12\lhat\log f - 12\lhat\rho(\rho^{2}-1)\arctan\rho}{6f^{2}} ,
  \\[4pt]
  Q_{1,wK} &= \frac{\lhat(3\lhat-4)f^{2} + 2\lhat f(5\rho^{2}-9) - 12(p^{2}+f) - 12\lhat f\log f + 24\lhat\rho f\arctan\rho}{6\lhat f^{3}} ,
  \\[4pt]
  Q_{1,ww} &= \frac{1}{12f^{3}}\Big\{-\lhat^{2}f^{2} - 2\lhat f(5\rho^{2}-6) - 4p^{2}(8\rho^{2}-3) + 24f
  \\&\hspace{1.8cm}
     + 12(\lhat f + 2p^{2})\log f - 24\rho\big[\lhat f + p^{2}(\rho^{2}+3)\big]\arctan\rho\Big\} .
\end{aligned}
\label{eq:t-Q1}
\end{equation}
Equations \eqref{eq:t-PQ0}--\eqref{eq:t-Q1} are the complete Zerilli-Moncrief-type system of the MMP throat to first order in $\alpha$. Appendix~\ref{app:throat-eqs} lists the underlying radial equations.

\subsection{The master equations in tortoise form}
\label{sec:throat-master}

The throat tortoise coordinate is defined by \eqref{eq:tortoise}, $\dd x/\dd\rho = \sqrt{B/A}$,
\begin{equation}
  \frac{\dd x}{\dd\rho} \;=\; \frac{1}{f + \alpha\gamma(\rho)},
  \qquad
  x \;=\; \arctan\rho \;-\; \alpha\,\iota(\rho) \;+\; O(\alpha^{2}),
  \qquad
  \iota(\rho) = \int_{0}^{\rho}\frac{\gamma(\rho')}{f(\rho')^{2}}\,\dd\rho' ,
  \label{eq:t-tortoise}
\end{equation}
with $f\equiv1+\rho^{2}$ the unperturbed metric function and $\gamma$ from \eqref{eq:bgsol}. The system \eqref{eq:t-PQ0} becomes
\begin{equation}
  \bm Y_{xx} \;+\; \tilde P\,\bm Y_{x} \;+\; \tilde Q\,\bm Y \;=\; 0 ,
  \qquad
  \bm Y = (K,w)^{\top},
  \qquad
  \tilde P = S\,P - S'\,I_2,
  \quad
  \tilde Q = S^{2}\,Q,
  \quad S = \frac{\dd\rho}{\dd x} = f+\alpha\gamma .
  \label{eq:t-master-tortoise}
\end{equation}
To first order in $\alpha$, with $\lhat = l(l+1)$ and $p=-i\Om$,
\begin{equation}
  \tilde P \;=\; \alpha\,\tilde P_{1} \;+\; O(\alpha^{2}),
  \qquad
  \tilde Q \;=\; \tilde Q_{0} \;+\; \alpha\,\tilde Q_{1} \;+\; O(\alpha^{2}) .
\end{equation}
The first-derivative matrix vanishes identically at $\alpha=0$, $\tilde P_0 = 0$, since $P_0 = 2\rho/f\,I_2 = (\log f)'\,I_2$ is exactly the term the tortoise transformation removes. At first order
\begin{equation}
  \tilde P_{1}
  \;=\; \begin{pmatrix}
  f\arctan\rho + \rho + \dfrac{4\rho}{\lhat f}
  & 2\big(f\arctan\rho + \rho\big)\\[10pt]
  \dfrac{2\rho}{\lhat f} & -\,\dfrac{\rho}{f}
  \end{pmatrix},
  \qquad
  \tilde P_{1} \;=\; f\,P_{1} \;+\; \Big(\frac{2\rho\gamma}{f} - \gamma'\Big) I_2 ,
  \label{eq:t-Ptilde1}
\end{equation}
the second form showing where the background correction $\gamma$ enters. The second row of $\tilde P_1$ comes from the Hall current of Sec.~\ref{sec:hall}. Its entry $\tilde P_{1,ww}=-\rho/f$ comes from the motion of the field lines in that current, through the tilt, the density and the scale of the line metric (Sec.~\ref{sec:lift}). The second row is the counterpart of the drift entry of the first-derivative matrix in the companion sector \cite{Sadhukhan:2026mcu}. The zeroth-order potential matrix is
\begin{equation}
  \tilde Q_{0}
  \;=\; \begin{pmatrix}
  -(\lhat+2)f - p^{2} & 2\lhat f\\[4pt]
  2f & -\lhat f - p^{2}
  \end{pmatrix}
  \;=\; -f\,\mathcal N \;-\; p^{2}I_2,
  \qquad
  \mathcal N \equiv \begin{pmatrix} \lhat+2 & -2\lhat\\ -2 & \lhat\end{pmatrix},
  \label{eq:t-Qtilde0}
\end{equation}
and is diagonalised exactly by the basis change \eqref{eq:Zdef}, $\mathsf T\tilde Q_0\mathsf T^{-1} = -\,\mathrm{diag}\big[(l+1)(l+2)f + p^{2},\;l(l-1)f + p^{2}\big]$, which with $\tilde P_0=0$ is the Poschl-Teller system \eqref{eq:ads2master}. The first-order correction, obtained from \eqref{eq:Qtilde-from-Q} below, is
\begin{align}
  \tilde Q_{1,KK} &= \frac{\lhat\big[(5\lhat+22)\rho^{4} + 6(\lhat+5)\rho^{2} + \lhat + 8\big]
      - 12p^{2} - 12f}{3\lhat f}
      \;-\; (\lhat+2)\log f
  \nonumber\\&\qquad
      \;+\; 2\rho\big[\lhat(\rho^{2}+2) + 2(2\rho^{2}+3)\big]\arctan\rho ,
  \label{eq:t-Qtilde1-KK}\\[6pt]
  \tilde Q_{1,Kw} &= \frac{\lhat(\lhat-34)\rho^{2} + \lhat^{2} - 18\lhat + 24}{6}
      \;+\; 2\lhat\log f \;-\; 2\lhat\rho\big(3\rho^{2}+5\big)\arctan\rho ,
  \label{eq:t-Qtilde1-Kw}\\[6pt]
  \tilde Q_{1,wK} &= \frac{\lhat\big[(3\lhat-26)\rho^{4} + 6(\lhat-8)\rho^{2} + 3\lhat - 22\big]
      - 12p^{2} - 12f}{6\lhat f}
      \;+\; 2\log f \;-\; 4\rho\big(\rho^{2}+2\big)\arctan\rho ,
  \label{eq:t-Qtilde1-wK}\\[6pt]
  \tilde Q_{1,ww} &= \frac{22\lhat\rho^{2} + 12\lhat + 24 - \lhat^{2}f}{12} \;+\; \frac{p^{2}}{f}
      \;-\; \lhat\log f \;+\; 2\lhat\rho\big(\rho^{2}+2\big)\arctan\rho .
  \label{eq:t-Qtilde1-ww}
\end{align}
Equations \eqref{eq:t-master-tortoise}--\eqref{eq:t-Qtilde1-ww} are the same system in the tortoise frame. They follow from the $\rho$-frame matrices \eqref{eq:t-P0Q0}--\eqref{eq:t-Q1}. With $S=f+\alpha\gamma$, expanding $\tilde Q=S^2Q$ to first order gives
\begin{equation}
  \tilde Q_{0} \;=\; f^{2}Q_{0},
  \qquad
  \tilde Q_{1} \;=\; f^{2}Q_{1} + 2f\gamma\,Q_{0} .
  \label{eq:Qtilde-from-Q}
\end{equation}
The frequency enters $\tilde Q_{0}$ only through $-p^{2}I_2$. At $O(\alpha)$ it enters only through the terms $-4p^{2}/\lhat f$, $-2p^{2}/\lhat f$ and $p^{2}/f$ in the $KK$, $wK$ and $ww$ entries. Both $f^{2}Q_{1}$ and $2f\gamma Q_{0}$ contain $p^{2}$ multiplying $\log f$ and $\arctan\rho$, and these terms cancel in \eqref{eq:Qtilde-from-Q}. Unlike the $\alpha^0$ matrices of the mouth (Sec.~\ref{sec:mouth-master}), the $O(\alpha)$ throat matrices are not rational in the radial coordinate. The first-derivative matrix is small here, $\tilde P=O(\alpha)$, whereas in the mouth it is of order one (Sec.~\ref{sec:mouth-master}).

\subsection{Schr\"odinger form}
\label{sec:throat-schrodinger}

Since $\tilde P=\alpha\tilde P_1$, the first-derivative term is removed to first order in $\alpha$ as in \eqref{eq:commutator-general}. We write $\bm Y=\mathcal U\,\tilde{\bm Y}$ with
\begin{equation}
  \mathcal U = I_2 - \tfrac{\alpha}{2}\,\mathcal G,
  \qquad
  \mathcal G = \rho\arctan\rho\begin{pmatrix}1&2\\0&0\end{pmatrix} + \frac{\rho^{2}}{\lhat f}\begin{pmatrix}2&0\\1&0\end{pmatrix} - \frac{\rho^{2}}{2f}\begin{pmatrix}0&0\\0&1\end{pmatrix},
  \label{eq:Xi}
\end{equation}
which solves $\partial_x\mathcal G=\tilde P_1$ with $\mathcal G(0)=0$. Then $\tilde{\bm Y}$ obeys \eqref{eq:schrodinger} with
\begin{equation}
  V \;=\; V_{\rm c} \;-\; \tfrac{\alpha}{2}\big[\,\mathcal G,\tilde Q_0\big] \;+\; O(\alpha^{2}) ,
  \label{eq:commutator}
\end{equation}
and $V_{\rm c}$ the commuting form \eqref{eq:schrodinger-commuting}. At $\alpha=0$ the potential is $-\tilde Q_0-p^2I_2=f\mathcal N$, with $\mathcal N$ the constant matrix of \eqref{eq:t-Qtilde0}. It is diagonalised by the constant matrix
\begin{equation}
  \mathsf T = \begin{pmatrix} 1 & -l \\ 1 & l+1\end{pmatrix},
  \qquad
  \bm{Z} = \mathsf T\,\tilde{\bm Y} = \mathsf T\,\mathcal U^{-1}\bm Y ,
  \label{eq:Zdef}
\end{equation}
so that $\bm Y=\mathcal U\,\mathsf T^{-1}\bm Z$. To first order $\mathcal U^{-1}=I_2+\tfrac{\alpha}{2}\,\mathcal G$, and explicitly
\begin{equation}
\begin{aligned}
  Z_+ &= K-l\,w+\frac{\alpha}{2}\Big[\rho\arctan\rho\,(K+2w)+\frac{(2-l)\,\rho^{2}}{\lhat f}\,K+\frac{l\,\rho^{2}}{2f}\,w\Big],\\
  Z_- &= K+(l+1)\,w+\frac{\alpha}{2}\Big[\rho\arctan\rho\,(K+2w)+\frac{(l+3)\,\rho^{2}}{\lhat f}\,K-\frac{(l+1)\,\rho^{2}}{2f}\,w\Big],
\end{aligned}
\label{eq:Zexplicit}
\end{equation}
up to $O(\alpha^2)$. At $\alpha=0$ these reduce to $Z_+=K-l\,w$ and $Z_-=K+(l+1)\,w$. Writing $V$ in this basis gives
\begin{equation}
  \partial_x^2 \bm Z + \left[\Om^2\left(1 + \alpha V_{1b}\right) - V_0 - \alpha V_{1a}\right]\bm Z = 0,
  \qquad x = \arctan\rho-\alpha\,\iota(\rho) ,
  \label{eq:master}
\end{equation}
with
\begin{equation}
  V_0 \;=\; \begin{pmatrix}\left(l + 1\right) \left(l + 2\right) & 0\\[2pt]
  0 & l\left(l - 1\right)\end{pmatrix}\left(1+\rho^{2}\right).
  \label{eq:V0}
\end{equation}
All potentials are functions of $\rho$, with $\rho(x)$ given by \eqref{eq:t-tortoise}. At $\alpha=0$ one has $1+\rho^2=\sec^2x$ and $V_0=\mathrm{diag}\big[(l+1)(l+2),\,l(l-1)\big]\sec^2x$. The $O(\alpha)$ correction $V_{1a} + p^2 V_{1b}$ splits into a frequency-independent part $V_{1a}$ and a frequency-dependent part $p^2V_{1b}$, and it is exactly quadratic in $p$. With $p^2=-\Om^2$ the $V_{1b}$ term is absorbed into the frequency in \eqref{eq:master}. It is elementary,
\begin{equation}
  V_{1b} \;=\; \frac{1}{(2l+1)\left(1+\rho^{2}\right)}
  \begin{pmatrix} -\,\dfrac{l^{2}+2l-4}{l} & \dfrac{l^{2}-l+4}{l+1}\\[8pt]
  \dfrac{l^{2}+3l+6}{l} & -\,\dfrac{l^{2}-5}{l+1}\end{pmatrix}.
  \label{eq:V1b}
\end{equation}
The four entries of $V_{1a}$ are displayed in Appendix~\ref{app:V1a}, Eqs.~\eqref{eq:V1a00}--\eqref{eq:V1a11}.

The correction is not diagonal in the $\bm Z$ basis. The two modes mix at $O(\alpha)$, and part of the off-diagonal terms comes from the commutator in \eqref{eq:commutator}. All entries of $V_{1a}$ grow at large $\rho$ like $\rho(1+\rho^{2})\arctan\rho\sim\tfrac{\pi}{2}\rho^{3}$,
\begin{equation}
\begin{aligned}
  V_{1a,\,++} &\sim -\,\frac{\pi\,(4l^{3}+16l^{2}+19l+9)}{2\,(2l+1)}\;\rho^{3},
  &\qquad
  V_{1a,\,+-} &\sim -\,\frac{3\pi\,l}{2l+1}\;\rho^{3},
  \\[4pt]
  V_{1a,\,--} &\sim -\,\frac{\pi\,(4l^{3}-4l^{2}-l-2)}{2\,(2l+1)}\;\rho^{3},
  &\qquad
  V_{1a,\,-+} &\sim -\,\frac{3\pi\,(l+1)}{2l+1}\;\rho^{3},
\end{aligned}
\qquad \rho\to\infty .
  \label{eq:V1alarge}
\end{equation}
This is the $\rho^{3}$ growth of the background function $\gamma$ in \eqref{eq:bgsol}, the $O(\alpha)$ background deformation. Relative to $V_0\sim\rho^2$ it is of order $\alpha\rho$, and it is what the mouth solution must absorb in the matching region $1\ll\rho\ll1/\alpha$ (Sec.~\ref{sec:overlap}). At the centre of the throat, for $l=2$,
\begin{equation}
  V_0\big|_{\rho=0} = \begin{pmatrix} 12 & 0\\ 0 & 2\end{pmatrix},
  \qquad
  V_{1a}\big|_{\rho=0} = \begin{pmatrix}-\frac{36}{5} & \frac{11}{5}\\[2pt] \frac{9}{5} & -\frac{29}{30}\end{pmatrix},
  \qquad
  V_{1b}\big|_{\rho=0} = \begin{pmatrix}-\frac{2}{5} & \frac{2}{5}\\[2pt] \frac{8}{5} & \frac{1}{15}\end{pmatrix} .
  \label{eq:Vrho0}
\end{equation}
The commutator vanishes there, since $\mathcal G(0)=0$.

\subsection{The \texorpdfstring{$\alpha\to0$}{alpha->0} limit and the exact AdS\texorpdfstring{$_2\times S^2$}{2 x S2} throat}
\label{sec:alpha0}

Setting $\alpha=0$ in \eqref{eq:t-PQ0}--\eqref{eq:master} reduces the construction to the exact AdS$_2\times S^2$ problem. At $\alpha=0$ a shorter four-variable reduction with no fermion sector gives the same second-order equations \eqref{eq:ads2Kcomp}--\eqref{eq:ads2wcomp} for $K$ and $w$. The background is $A=r_e^2(1+\rho^2)$, $B=r_e^2/(1+\rho^2)$, $R=r_e$. The fermion stress tensor enters the field equations only through $\kappa T^Q$, and by \eqref{eq:MMPmatch} $\kappa T^Q=O(\alpha)$. Both the background source \eqref{eq:TQzero} and its response drop out at this order, and the Bianchi obstruction \eqref{eq:bianchiexp} is absent, as in the mouth at $\alpha^0$ (Sec.~\ref{sec:consistency}). The master system collapses to
\begin{align}
  \Box K - \left(\lhat + 2\right)K + 2\lhat\, w &= 0,
  \label{eq:ads2K}\\
  \Box w - \lhat\, w + 2K &= 0,
  \label{eq:ads2w}
\end{align}
with $\Box$ the unit-AdS$_2$ d'Alembertian, that is, in components
\begin{align}
  K'' &= \frac{1}{1+\rho^2}\left[-2\rho K' + \frac{p^2 K}{1+\rho^2} + (\lhat+2)K - 2\lhat\,w\right],
  \label{eq:ads2Kcomp}\\
  w'' &= \frac{1}{1+\rho^2}\left[-2\rho w' + \frac{p^2 w}{1+\rho^2} + \lhat\,w - 2K\right],
  \label{eq:ads2wcomp}
\end{align}
which is \eqref{eq:t-P0Q0}. The tortoise-frame first-derivative matrix $\tilde P$ vanishes identically, and \eqref{eq:Zdef} diagonalises the system exactly,
\begin{equation}
  \partial_x^2 Z_\pm + \left(\Om^2 - m_\pm^2\sec^2 x\right)Z_\pm = 0,
  \qquad
  m_+^2 = (l+1)(l+2),
  \qquad
  m_-^2 = l(l-1),
  \qquad x=\arctan\rho .
  \label{eq:ads2master}
\end{equation}
These masses agree with the Kaluza--Klein spectrum on AdS$_2\times S^2$ of \cite{Michelson:1999kn}. The mouth reproduces these masses and variables in its near-horizon limit (Sec.~\ref{sec:overlap}). For $l=1$ one finds $m_-^2 = 0$. If $l=0$ is set formally, although the reduction assumes $l\ge2$ (Sec.~\ref{sec:ansatz}), the equation for $K$ becomes $\Box K = 2K$. Its static solution that is even in $\rho$ is $1+\rho\arctan\rho$, up to normalisation. This is the Jackiw--Teitelboim dilaton \cite{Jackiw:1984je,Teitelboim:1983ux} and MMP's correction to the sphere radius, their eq.~(5.37). The function $\varphi$ of \eqref{eq:bgsol} differs from it by the constant $\tfrac13$ of the trace part and obeys $\Box\varphi-2\varphi=-\tfrac23$. The radius $r_e$ cancels from \eqref{eq:ads2master}, since $(\tau,\rho)$ are dimensionless and the $r_e^2$ prefactor is an overall conformal rescaling the sector does not see. It reappears in $\ell = 4r_e/(\pi\alpha)$ when converting $\tau$ to asymptotic time.

The potential $m^2\sec^2x$ is of Poschl-Teller type. Writing $m^2 = s(s-1)$ one has $s = l+2$ for $Z_+$ and $s = l$ for $Z_-$, both integers. The solutions are associated Legendre functions of argument $i\rho$, degree $s-1$ and order $\Om=\omega\ell$, since unit AdS$_2$ is the analytic continuation of the round sphere. For integer $s$ they reduce to $e^{\pm i\Om x}$ times a polynomial of degree $s-1$ in $\tan x$. Each of these two solutions keeps a single phase across the throat. This is the trigonometric counterpart of the reflectionless hyperbolic Poschl-Teller potentials with integer $s$ \cite{Cooper:1994eh}. The levels of the closed throat sit at $\Om=n+s$, $n=0,1,2,\dots$, with unit spacing set by the length $\pi$ of the $x$ interval, and integer $s$ places them at integer $\omega\ell$. The same spacing appears for test scalar and vector fields, where it arises from the two mouth barriers acting as a Fabry--Perot cavity of length $\pi\ell$ in the tortoise coordinate \cite{Mondal:2025tht}. This is the leading-order echo spectrum, and the $O(\alpha)$ corrections \eqref{eq:V1a00}--\eqref{eq:V1a11} and \eqref{eq:V1b} shift it at relative order $\alpha = g_c^2/4\pi^2 q$.

\section{The mouth to \texorpdfstring{$O(\alpha)$}{O(alpha)}}
\label{sec:mouth}

\subsection{The lowest-Landau-level stress tensor and the corrected background}
\label{sec:mouthTQ}

The LLL fermions are not absent from the mouth. They travel along the magnetic field lines through it and out into the ambient region, and that is how the Casimir energy is generated \cite{Maldacena:2018gjk}. Their stress tensor has a Casimir part fixed by the state and an anomaly part that is a local function of the field-line metric, which is not flat in the mouth. The anomaly part gives $\kappa T^Q_{\mu\nu}=O(\alpha)$ there, the same order as in the throat. At $O(\alpha^0)$ the mouth carries no fermion stress tensor, and by Sec.~\ref{sec:consistency} the truncation $\delta T^Q_{\mu\nu}=0$, $J^\nu=0$ is consistent at that order. This section treats the mouth to $O(\alpha)$ with the lift, the response and the Hall current of Secs.~\ref{sec:TQ} to \ref{sec:throatmaster}. The vacuum Einstein--Maxwell mouth, exact in $GM$ and $r_e$, is its $\alpha^0$ part.

The construction of Sec.~\ref{sec:TQ} applies to the mouth without change. As in \eqref{eq:conformalsplit} and \eqref{eq:2dmetric}, the fermions on each field line see the field-line metric $\hat g_{ab}=g_{ab}/R^2$, the $(t,r)$ block of the metric divided by the square of the sphere radius. With the mouth background \eqref{eq:bgmouth}, where $R=r$ and $B=1/A$, it is
\begin{equation}
  \hat g_{ab}\,\dd x^a\dd x^b \;=\; -\frac{A}{r^{2}}\,\dd t^{2} + \frac{\dd r^{2}}{A\,r^{2}} \;=\; \frac{A}{r^{2}}\big(-\dd t^{2}+\dd r_*^{2}\big),
  \qquad
  \frac{\dd r_*}{\dd r}=\frac{1}{A} .
  \label{eq:ghatmouth}
\end{equation}
This is conformal gauge $\hat g=e^{2\bar\Upsilon}\eta$ with $\eta=-\dd t^2+\dd r_*^2$ and $\bar\Upsilon=\tfrac12\log(A/r^2)$. The state is the vacuum of the closed field line, and on $\eta$ its Casimir stress is $-\pi c/6L^2$ with $L\simeq\pi\ell$ the circumference in asymptotic time. The anomalous part of \eqref{eq:anomalylaw} is of order $c/r_e^2$ in the same units. The Casimir part is smaller by $(r_e/\ell)^2=\pi^2\alpha^2/16$, and we drop it. For the null combination $t_{tt}+t_{r_*r_*}$ this ratio is not uniform in $r$, since its anomaly part vanishes at the horizon. Here and below the horizon means the double root $r=r_e$ of $\Delta$ at $GM=r_e$, which the wormhole mouth, with $\delta_M<0$, does not have. What remains is \eqref{eq:anomalylaw} with $g_{\rm ref}=\eta$, a local function of the mouth geometry. We evaluate it on the extremal background $GM=r_e$, where $A=(1-r_e/r)^2$. This is enough, since $\kappa T^Q$ is already $O(\alpha)$ and $GM-r_e$ is $O(\alpha)$ for the wormhole, Eq.~\eqref{eq:mouthmass}. The scalar curvature of \eqref{eq:ghatmouth} is then
\begin{equation}
  \hat R \;=\; -2 \;+\; \frac{12\,r_e}{r} \;-\; \frac{12\,r_e^{2}}{r^{2}} ,
  \label{eq:Rhatmouth}
\end{equation}
which equals $-2$ at $r=r_e$ and as $r\to\infty$ and reaches $+1$ at $r=2r_e$. The field-line metric is close to unit AdS$_2$ at both ends of the mouth. Near $r=r_e$ this is the throat. At large $r$ it is $\hat g\to(-\dd t^2+\dd r^2)/r^2$, unit AdS$_2$ in Poincar\'e coordinates. Lifting with \eqref{eq:lift} gives the background source of the mouth,
\begin{equation}
\begin{gathered}
  \kappa\,T^{Q\,t}{}_{t} \;=\; -\frac{\alpha\,r_e^{2}\,(r^{2}-8rr_e+8r_e^{2})}{3r^{6}},
  \qquad
  \kappa\,T^{Q\,r}{}_{r} \;=\; -\frac{\alpha\,r_e^{2}\,(r-2r_e)^{2}}{3r^{6}},
  \\[4pt]
  \kappa\,T^{Q\,\theta}{}_{\theta} \;=\; \kappa\,T^{Q\,\phi}{}_{\phi} \;=\; \frac{\alpha\,r_e^{2}\,(r^{2}-6rr_e+6r_e^{2})}{3r^{6}} .
\end{gathered}
  \label{eq:TQmouth}
\end{equation}
It is conserved, and its four-dimensional trace vanishes, as the lift guarantees on any background. The Maxwell source is $\kappa T^{{\rm mag}\,t}{}_{t}=\kappa T^{{\rm mag}\,r}{}_{r}=-\kappa T^{{\rm mag}\,\theta}{}_{\theta}=-r_e^{2}/r^{4}$, and the ratio of the two energy densities is
\begin{equation}
  \frac{T^{Q\,t}{}_{t}}{T^{{\rm mag}\,t}{}_{t}} \;=\; \frac{\alpha}{3}\;\frac{r^{2}-8rr_e+8r_e^{2}}{r^{2}} .
  \label{eq:TQratio}
\end{equation}
It tends to $\alpha/3$ at both ends, where \eqref{eq:TQmouth} has the Maxwell form. There the fermions act like an increase of $q^2/g_c^2$ by the factor $1+\alpha/3$. In between they do not. Their energy density is negative for $(4-2\sqrt2)\,r_e<r<(4+2\sqrt2)\,r_e$, and $T^{Q\,r}{}_r\neq T^{Q\,t}{}_t$ away from the two ends. Along a radial null ray the two-dimensional tensor gives $t_{tt}+t_{r_*r_*}=-c\,r_e(r-r_e)^3/6\pi r^6$. This vanishes at the horizon, where $\hat g$ approaches AdS$_2$ in Poincar\'e coordinates and the vacuum of the frame $(t,r_*)$ carries no null energy. A second effect is the vacuum polarisation of the higher Landau levels, which makes the gauge coupling run \cite{Schwinger:1951nm,Dunne:2004nc}. In the mouth it exceeds the $O(\alpha)$ terms of this section, and Sec.~\ref{sec:hll} treats it as a separate small parameter.

The source \eqref{eq:TQmouth} corrects the background. We write the corrected mouth as
\begin{equation}
  \dd s^2 \;=\; -e^{2\alpha\Phi_1(r)}N\,\dd t^2 + \frac{\dd r^2}{N} + r^2\,\mathring g ,
  \qquad
  N \;=\; 1-\frac{2GM}{r}+\frac{r_e^2}{r^2}+\alpha\,N_1(r) ,
  \label{eq:mouthansatz}
\end{equation}
keeping $R=r$ as the radial coordinate. At $O(\alpha)$ the $tt$ and $rr$ components of \eqref{eq:order0} with the source \eqref{eq:TQmouth} are first-order equations,
\begin{equation}
  \big(r\,N_1\big)' \;=\; -\,\frac{r_e^{2}\,(r^{2}-8rr_e+8r_e^{2})}{3r^{4}},
  \qquad
  \Phi_1' \;=\; -\,\frac{2r_e^{3}}{3r^{3}(r-r_e)} .
  \label{eq:mouthODE}
\end{equation}
The integration constant of the first is a shift of $GM$. We fix the second by $\Phi_1(\infty)=0$, which keeps $t$ the asymptotic time. Then
\begin{equation}
  N_1 \;=\; \frac{r_e^{2}}{3r^{2}} - \frac{4r_e^{3}}{3r^{3}} + \frac{8r_e^{4}}{9r^{4}},
  \qquad
  \Phi_1 \;=\; -\frac23\Big[\log\Big(1-\frac{r_e}{r}\Big) + \frac{r_e}{r} + \frac{r_e^{2}}{2r^{2}}\Big],
  \label{eq:mouthbgsol}
\end{equation}
and the $\theta\theta$ component holds identically. Only the linear part $1+2\alpha\Phi_1$ of the factor $e^{2\alpha\Phi_1}$ enters at this order. At large $r$, $N\to1-2GM/r+r_e^2(1+\alpha/3)/r^2$. The charge seen from infinity therefore corresponds to the radius $r_e\sqrt{1+\alpha/3}$ and not to $r_e$. The mass defect of MMP is $O(\alpha^2r_e)$. We fix $GM$ by requiring $N$ and $N'$ to vanish at one point to first order in $\alpha$, which gives
\begin{equation}
  GM \;=\; r_e\Big(1-\frac{\alpha}{18}\Big),
  \qquad
  r_h \;=\; r_e\Big(1+\frac{\alpha}{6}\Big) .
  \label{eq:mouthmass}
\end{equation}
Near $r_e$, with $r=r_e(1+\alpha\hat r)$, one has $N=\alpha^2\hat r(3\hat r-1)/3+O(\alpha^3)$, of the order of the $O(\alpha^2)$ terms that we do not compute. Its minimum sits at $r_h$. Whether $N$ has a double zero there, two simple zeros or none is decided at $O(\alpha^2)$, where the mass defect $\delta_M$ enters. For $\alpha r_e\ll r-r_e\ll r_e$ the two-dimensional curvature of the $(t,r)$ block is $-2/r_h^2+12(r-r_h)/r_e^3$, and continued to $r=r_h$ it is $-2/r_h^2$. The near-horizon geometry is AdS$_2\times S^2$ with both radii equal to $r_h$, as for an extremal Reissner-Nordstrom black hole of radius $r_h$. Both ends of the mouth see the same renormalised radius, $r_h^2=r_e^2(1+\alpha/3)$ at this order. This is the renormalisation of $r_e$ that MMP absorb into the definition of the radii. It comes from the anomaly part of the stress tensor, which is fixed by the local geometry. The Casimir part, which depends on how the field lines close, is dropped. An isolated extremal black hole of the same charge, with the fermions in the same local vacuum, therefore has the same correction at this order. The mass in \eqref{eq:mouthmass} is lower than the renormalised radius by $2\alpha r_e/9$. Between the two ends the fermion energy density lies below $\alpha/3$ times the Maxwell value, by \eqref{eq:TQratio}, and this lowers the mass. Near the horizon $\Phi_1\simeq-\tfrac23\log(r/r_e-1)$. This logarithm changes the normalisation of time relative to the throat. It cancels between $e^{2\alpha\Phi_1}$ and the radial dictionary in the overlap, and it does not enter the matching at first order (Sec.~\ref{sec:overlap}).

The throat sees the same renormalisation. Its $O(\alpha)$ background contains the Casimir part as well as the trace part, and the renormalisation comes from the trace part and the sphere pressure it sets (Sec.~\ref{sec:bgalpha}). At large $\rho$ the background \eqref{eq:bgsol} gives
\begin{equation}
  R \;=\; r_e \;+\; \frac{r_e^{2}}{\ell}\,\rho \;+\; \frac{\alpha\,r_e}{6} \;+\; O\big(\alpha\rho^{-2},\alpha^2\big),
  \qquad
  \frac{A}{r_e^{2}} \;=\; \Big(1-\frac{\alpha}{3}\Big)\rho^{2} \;-\; \frac{\pi\alpha}{2}\,\rho^{3} \;+\; \dots ,
  \label{eq:throatlarge}
\end{equation}
where the dots stand for terms of order $\rho^0$, $\alpha\rho$ and $\alpha\log\rho$. The constant $\alpha r_e/6$ is the shift $r_h-r_e$ of \eqref{eq:mouthmass}, and the factor $1-\alpha/3$ is the AdS$_2$ radius $r_h$. Both come from the trace of the two-dimensional stress tensor, which Sec.~\ref{sec:throat-origin} keeps. The trace enters through the trace part of the source and through the sphere pressure it sets. The corrected mouth reproduces both (Sec.~\ref{sec:overlap}). MMP drop the trace part in both regions. Their throat then matches the uncorrected mouth at this order, and their treatment is consistent. Keeping the trace part in the throat requires keeping it in the mouth.

\subsection{Radial equations, response and reduction}
\label{sec:mouthpert}

The perturbation problem on the corrected mouth has the structure of the throat problem of Secs.~\ref{sec:response}, \ref{sec:hall} and \ref{sec:throatmaster}. The field-line metric is \eqref{eq:linescale}. On the background it is $\hat g_{ab}=g_{ab}/r^2$, and its perturbation is \eqref{eq:dghat} with $R^2=r^2(1+\epsilon KY)$ and $K_F$ of \eqref{eq:KB}. The response is \eqref{eq:response} with $\bar t_{ab}$ the anomaly tensor of Sec.~\ref{sec:mouthTQ}. As in the throat we evaluate the source on the $\alpha^0$ background, since $\kappa T^Q$ is already $O(\alpha)$. The decomposition \eqref{eq:decomposition} reads
\begin{align}
  2r^{5}\,\dot\xi_t \;-\; r\,(r-r_e)^{2}\big(H_0+K_F+2\delta\Upsilon\big) \;+\; 2\,(r-r_e)^{3}(r-2r_e)\,\xi_r &= 0 ,
  \label{eq:m-dec-tt}\\[2pt]
  r^{2}(r-r_e)\big(\dot\xi_r+\xi_t'\big) \;+\; 2r\,(r-2r_e)\,\xi_t \;-\; (r-r_e)\,H_1 &= 0 ,
  \label{eq:m-dec-tr}\\[2pt]
  2\,(r-r_e)^{2}\,\xi_r' \;+\; 2\,(r-r_e)\,\xi_r \;-\; H_2 \;+\; K_F \;+\; 2\delta\Upsilon &= 0 ,
  \label{eq:m-dec-rr}
\end{align}
with a dot for $\partial_t$ and a prime for $\partial_r$. Under the matching \eqref{eq:matching} the transport fields and $H_1$ go over into those of the throat as $\xi_t=\xi_\tau/\ell$, $\xi_r=\ell\,\xi_\rho/r_e^2$ and $H_1=H_1^{\rm throat}/r_e^2$, and \eqref{eq:m-dec-tt}--\eqref{eq:m-dec-rr} become the large-$\rho$ form of \eqref{eq:deceq_tt}--\eqref{eq:deceq_rr}. The Ward identity \eqref{eq:ward} holds on the mouth for arbitrary $\delta\Upsilon$ and $\xi_a$. The lift is \eqref{eq:dTQ2d}--\eqref{eq:dTQsphere} with $r_e^2$ replaced by $r^2$, and its mixed block is \eqref{eq:driftlift} with $u_c=(r^2/2)\,\partial_cw\,Y'(\theta)$. Its $t$, $r$ and $\phi$ divergence vanishes identically once $h_{ab}$ is expressed through \eqref{eq:m-dec-tt}--\eqref{eq:m-dec-rr}, as in the throat. The Hall current follows from the closed form \eqref{eq:Jclosed}. After $H_0$, $H_1$ and $H_2$ are eliminated through \eqref{eq:m-dec-tt}--\eqref{eq:m-dec-rr} the explicit $K_F$ cancels, as in the throat. The drift of the field lines enters only through $\tilde\xi_a$ of \eqref{eq:xitilde}, and
\begin{equation}
\begin{aligned}
  J^{\phi} \;=\; \frac{1}{48\pi^{2}}\,\frac{Y'(\theta)}{\sin\theta}\Bigg[\;
  &\frac{(r-r_e)^{2}}{r^{4}}\,\delta\Upsilon'' + \frac{2r_e(r-r_e)}{r^{5}}\,\delta\Upsilon' - \frac{\ddot{\delta\Upsilon}}{(r-r_e)^{2}} + \frac{2\,(r^{2}-6rr_e+6r_e^{2})}{r^{6}}\,\delta\Upsilon
  \\[2pt]
  +\;&\frac{2\,(r-r_e)^{2}(r^{2}-5rr_e+5r_e^{2})}{r^{6}}\,\tilde\xi_r' + \frac{2r_e(r-r_e)(7r^{2}-25rr_e+20r_e^{2})}{r^{7}}\,\tilde\xi_r
  \\[2pt]
  -\;&\frac{2\,(r^{2}-7rr_e+7r_e^{2})}{r^{2}(r-r_e)^{2}}\,\dot{\tilde\xi}_t\;\Bigg] .
\end{aligned}
\label{eq:Jmouth}
\end{equation}
With the dictionary above each coefficient reduces near the horizon to the large-$\rho$ coefficient of \eqref{eq:Jphi_pure}. As in the throat, the current vanishes for any perturbation that leaves $\hat g_{ab}$ unchanged and does not move the field lines.

We insert the source and the current into \eqref{eq:order1} on the background \eqref{eq:mouthansatz} and keep the first order in $\alpha$. The mass is $GM=r_e(1-\alpha/18)+\delta_M$, with $\delta_M=O(\alpha^2r_e)$ measured here from $r_e(1-\alpha/18)$, hence $GM-r_e=O(\alpha r_e)$. In the terms that already carry a factor $\alpha$ we may therefore set $GM=r_e$, at the cost of $O(\alpha^2)$. The full $GM$, with the shift $-\alpha r_e/18$ and the defect $\delta_M$, enters through the $\alpha^0$ system displayed next, which is exact in $GM$. Near the horizon this needs $r-r_e\gg\alpha r_e$, which holds at the junction of Sec.~\ref{sec:overlap} by a factor $4\times10^{28}$. The reduction then follows the throat. Step~1 gives $H_0=H_2$, since the source has no anisotropic sphere stress. Step~1b solves \eqref{eq:m-dec-tt} for $\delta\Upsilon$. Step~1c turns \eqref{eq:m-dec-tr} and \eqref{eq:m-dec-rr} into transport equations. Steps~2 and 3 are as in Sec.~\ref{sec:algorithm}. In Step~4 all three residuals vanish at $O(\alpha^0)$ for every $l$ and $p$. At $O(\alpha)$ they vanish identically in $r$, $l$ and $p$. Without the Hall current the residuals are nonzero and proportional to $\alpha$, as checked at $(l,p)=(2,\tfrac37)$ and $(3,-\tfrac5{11})$. The mouth at $O(\alpha)$ needs the same two ingredients as the throat.

At $\alpha=0$ the source, the current and the transport fields drop out, since $\kappa T^Q$ is $O(\alpha)$, and the mouth is vacuum Einstein--Maxwell on the background \eqref{eq:bgmouth}, exact in $GM$ and $r_e$. We display this part in full. With \eqref{eq:bgmouth} the $O(\epsilon^0)$ equations \eqref{eq:order0} read
\begin{equation}
  \bar G_{\mu\nu} - \kappa\bar T^{\rm mag}_{\mu\nu}
  \;=\; 0
\end{equation}
identically for every component once \eqref{eq:re} is used, and the zeroth-order Maxwell equations vanish identically for the monopole \eqref{eq:monopole}. Carrying out the expansion and the separation of Sec.~\ref{sec:separation}, and passing to the frequency domain \eqref{eq:freq}, gives the eight radial equations $\mathcal E_{tt},\mathcal E_{tr},\mathcal E_{rr},\mathcal E_{t\theta},\mathcal E_{r\theta},\mathcal E^{\rm trace},\mathcal E^{\rm tl},\mathcal M^\phi$ displayed in Appendix~\ref{app:mouth-eqs}, Eqs.~\eqref{eq:m-ett}--\eqref{eq:m-mph}. We write $\lhat=l(l+1)$ and $w=4a/q$. The equations are polynomial in $r$, $GM$, $r_e$, $\lhat$ and $p$, and $q$ has cancelled. Their content is that listed in \eqref{eq:etth_content}--\eqref{eq:etr_content} and Step~4. The traceless equation is \eqref{eq:m-etl}, and the three constraint equations are \eqref{eq:m-etth}, \eqref{eq:m-erth} and \eqref{eq:m-etr}. From here on we use the abbreviations
\begin{equation}
  \Delta \equiv r^2 - 2GMr + r_e^2 ,
  \qquad
  \mathcal C \equiv (\lhat-2)\,r^2 + 6GM\,r - 4r_e^2 ,
  \qquad
  \mathcal D \equiv 4p^2r^4 - \lhat(\lhat-2)\,r^2 - 4GM\lhat\, r + 2\lhat\, r_e^2 .
  \label{eq:m-abbrev}
\end{equation}

Equation \eqref{eq:m-etl} is algebraic and gives $H_0 = H_2$, Eq.~\eqref{eq:H0H2}. We solve \eqref{eq:m-etth}, \eqref{eq:m-erth} and \eqref{eq:m-etr} for $H_1',H_2',K'$, and $\mathcal M^\phi$ \eqref{eq:m-mph} for $w''$. Substituting into $\mathcal E_{rr}$ \eqref{eq:m-err} gives the algebraic relation \eqref{eq:H2alg},
\begin{equation}
\begin{aligned}
  H_2 \;=\;&
  \frac{r\big[\,2p^2r^5 + (\lhat-2)r^3 + 2GM(3-\lhat)\,r^2 + (\lhat\,r_e^2 - 6G^2M^2)\,r + 2GM\,r_e^2\,\big]}
       {\Delta\,\mathcal C}\;K
  \\[2pt]
  &-\;\frac{2p^2r^4 + GM\lhat\, r - \lhat\, r_e^2}{p\,r\,\mathcal C}\;H_1
  \;-\;\frac{2\lhat\, r_e^2}{\mathcal C}\;w
  \;-\;\frac{4 r_e^2\,\Delta}{r\,\mathcal C}\;w' .
\end{aligned}
\label{eq:m-H2alg}
\end{equation}
At $\alpha^0$ the three residuals of Step~4, $\mathcal E_{tt}$, the angular trace and the consistency of $H_2'$, reduce to zero coefficient by coefficient with no fermionic input. The closed first-order system \eqref{eq:firstorder} is $\bm y'=\mathsf M\bm y$ with $\bm y=(K,H_1,w,w')^{\top}$. Its $4\times4$ matrix $\mathsf M$ is given in Appendix~\ref{app:mouth-M}, Eq.~\eqref{eq:m-M}. It is exact, rational in $r$ and directly integrable numerically.

At $O(\alpha)$ the closed first-order system is six-dimensional, $\bm y=(K,H_1,w,w',\xi_t,\xi_r)$. The two transport equations enter the master system only at $O(\alpha)$, and at $\alpha^0$ they read
\begin{align}
  \xi_t' \;&=\; \frac{H_1}{r^{2}} \;-\; \frac{2\,(r-2r_e)}{r\,(r-r_e)}\,\xi_t \;-\; p\,\xi_r ,
  \label{eq:m-xit}\\[4pt]
  \xi_r' \;&=\; -\,\frac{p\,r^{4}}{(r-r_e)^{4}}\,\xi_t \;-\; \frac{2}{r}\,\xi_r
  \;+\; \frac{r\big[\lhat\,r(r-r_e)^{2}-2(r-r_e)^{3}+2p^{2}r^{5}\big]}{(r-r_e)^{4}\,\mathcal C}\,K
  \nonumber\\&\qquad
  \;-\; \frac{\lhat\,r_e(r-r_e)+2p^{2}r^{4}}{p\,r\,(r-r_e)^{2}\,\mathcal C}\,H_1
  \;-\; \frac{2\lhat\,r_e^{2}}{(r-r_e)^{2}\,\mathcal C}\,w \;-\; \frac{4r_e^{2}}{r\,\mathcal C}\,w' ,
  \label{eq:m-xir}
\end{align}
with $\mathcal C$ of \eqref{eq:m-abbrev} at $GM=r_e$. Their source-free part is solved by the conformal Killing vectors of $\hat g$,
\begin{equation}
  \zeta_t \;=\; -\,\frac{(r-r_e)^{2}}{r^{4}}\Big(\mathfrak b_+\,e^{pr_*}+\mathfrak b_-\,e^{-pr_*}\Big),
  \qquad
  \zeta_r \;=\; \frac{1}{r^{2}}\Big(\mathfrak b_+\,e^{pr_*}-\mathfrak b_-\,e^{-pr_*}\Big),
  \label{eq:m-zeta}
\end{equation}
with $\dd r_*/\dd r=r^2/(r-r_e)^2$. These are the free left- and right-moving fermion radiation of \eqref{eq:zeta}, now along the field lines of the mouth.

\subsection{Master equations}
\label{sec:mouth-master}

At $\alpha^0$, eliminating $H_1$ gives the coupled second-order system \eqref{eq:secondorder} in $r$,
\begin{equation}
  \bm{Y}'' \;+\; P(r)\,\bm{Y}' \;+\; Q(r)\,\bm{Y} \;=\; 0 ,
  \label{eq:secondorderRN}
\end{equation}
whose matrices $P,Q$ follow from the tortoise-frame matrices below through
\begin{equation}
  P \;=\; A^{-1}\,\tilde P \;+\; \frac{A'}{A}\,I_2 ,
  \qquad
  Q \;=\; A^{-2}\,\tilde Q ,
  \qquad
  \frac{A'}{A} = \frac{2\,(GM\,r - r_e^2)}{r\,\Delta} ,
  \label{eq:m-PQfromtilde}
\end{equation}
since here $S=A$ in \eqref{eq:tortoisePQ}. The matrices $P,Q$ themselves are given in Appendix~\ref{app:mouth-M}.

In the tortoise coordinate $\dd r_*/\dd r = 1/A = r^2/\Delta$, the master system for $\bm Y=(K,w)^\top$ is
\begin{equation}
  \bm Y_{r_*r_*} \;+\; \tilde P\,\bm Y_{r_*} \;+\; \tilde Q\,\bm Y \;=\; 0
  \label{eq:m-master}
\end{equation}
with, in terms of \eqref{eq:m-abbrev},
\begin{align}
  \tilde P_{KK} &= -\,\frac{2\lhat\,\Delta\,\mathcal C}{r^3\,\mathcal D},
  &
  \tilde P_{Kw} &= \frac{4 r_e^2\,\Delta\,\big(4p^2r^2 - \lhat^2\big)}{r^3\,\mathcal D},
  &
  \tilde P_{wK} &= \tilde P_{ww} = 0 ,
  \label{eq:m-Ptilde}\\[6pt]
  \tilde Q_{Kw} &= \frac{2\lhat\, r_e^2\,\Delta\,\big(4p^2r^2 - \lhat^2\big)}{r^4\,\mathcal D},
  &
  \tilde Q_{wK} &= \frac{2\Delta}{r^4},
  &
  \tilde Q_{ww} &= -\,\frac{p^2 r^4 + \lhat\,\Delta}{r^4} ,
  \label{eq:m-Qtilde}
\end{align}
and
\begin{equation}
\begin{aligned}
  \tilde Q_{KK} \;=\; \,\frac{1}{r^4\,\mathcal D}\Big[\;
  &\lhat^2\,\Delta\,\big((\lhat-2)r^2 + 4r_e^2\big)
  \;-\;4p^4 r^8
  \\
  &+\;p^2 r^2\Big(\lhat(\lhat-2)\,r^4 + 4GM\lhat\, r^3 - 2\lhat\, r^2 r_e^2 - 48G^2M^2 r^2
  \\
  &\hspace{2.2cm} + 24GM\,r^3 + 88GM\,r\,r_e^2 - 32 r^2 r_e^2 - 32 r_e^4\Big)\Big] .
\end{aligned}
\label{eq:m-QKK}
\end{equation}
Equations \eqref{eq:m-master}--\eqref{eq:m-QKK} are the complete Zerilli-Moncrief-type system of the magnetic Reissner-Nordstrom mouth in the polar-metric and axial-gauge sector. They are exact in $GM$, $r_e$, $l$ and $p$.

The second row of $\tilde P$ vanishes, and $\tilde Q_{wK}$ and $\tilde Q_{ww}$ are elementary. With $p^2=-\omega^2$ the equation for $w$ takes the Regge--Wheeler form \cite{Regge:1957td}
\begin{equation}
  \partial_{r_*}^2 w + \left(\omega^2 - \frac{\lhat\, A}{r^2}\right) w
  \;=\; -\,\frac{2A}{r^2}\,K .
  \label{eq:wRW}
\end{equation}
This is the electromagnetic master equation with potential $\lhat A/r^2$, and the source on the right is the magnetic parity mixing. The equation for $K$ is different. The first row of $\tilde P$ is nonzero, and several entries of $\tilde P$ and $\tilde Q$ contain the frequency dependent factor $\mathcal D$ of \eqref{eq:m-abbrev}. No rescaling of $K$ alone can remove this. The Zerilli-Moncrief variable \cite{Zerilli:1970se,Moncrief:1974am,Martel:2005ir} is not $K$ itself. It is a combination of $K$ and $H_1$ and, in the charged case, of the gauge amplitude. Section~\ref{sec:mouth-RN} constructs it.

At $O(\alpha)$ the same elimination of $H_1$ gives a system that differs from that of the throat. The transport fields do not cancel. The equation for $w$ contains none of them, and the equation for $K$ contains them in a single combination,
\begin{equation}
  K'' + P_{KK}K' + P_{Kw}w' + Q_{KK}K + Q_{Kw}w \;=\; -\,\frac{8\alpha\,p^{3}r_e^{2}\,r^{5}}{3\,(r-r_e)^{4}\,\mathcal D}\;\Pi ,
  \qquad
  \Pi \;\equiv\; \lhat\,r\,\xi_t \;+\; 2p\,(r-r_e)^{2}\,\xi_r ,
  \label{eq:Kchi}
\end{equation}
with $P$ and $Q$ the matrices of \eqref{eq:secondorderRN} extended to $O(\alpha)$ and $\mathcal D$ of \eqref{eq:m-abbrev} at $GM=r_e$. The coefficient vanishes for static perturbations. Under the matching \eqref{eq:matching} it becomes a term of order $\alpha^{3}\rho^{-4}\xi_\tau$ in the $\rho$ equation, two orders beyond the throat calculation. This is why the throat master equations contain no transport fields. In the mouth the term is of order $\alpha$ like the rest of the correction. The state of the fermions therefore enters the gravitational mode in the mouth. The free radiation \eqref{eq:m-zeta} gives $\Pi\neq0$ and sources $K$. Even without it, $\xi_a$ is fixed by $K$, $H_1$, $w$ and $w'$ only through the integration of \eqref{eq:m-xit}--\eqref{eq:m-xir}. The master system of the mouth at $O(\alpha)$ is therefore the pair \eqref{eq:Kchi} and \eqref{eq:m-wrow} together with \eqref{eq:m-xit}--\eqref{eq:m-xir}, where $H_1$ is expressed through $K$, $K'$, $w$ and $w'$ by \eqref{eq:m-Kp} and \eqref{eq:m-H2alg}. This system is of sixth order. Its two extra constants are the amplitudes $\mathfrak b_\pm$ of the free radiation \eqref{eq:m-zeta}, defined relative to a particular solution of \eqref{eq:m-xit}--\eqref{eq:m-xir}. On a closed field line periodicity fixes them away from $\Om=2n$, as in the throat (Sec.~\ref{sec:response}). This needs the line outside the mouth. We do not construct it here. The matching of Sec.~\ref{sec:overlap} does not need it, since the source in \eqref{eq:Kchi} vanishes at $p=0$.

The equation for $w$ keeps a simple form. In the tortoise coordinate of the corrected mouth, $\dd r_*/\dd r=e^{-\alpha\Phi_1}/N$, it reads
\begin{equation}
  w_{r_*r_*} \;+\; \alpha\big(\mathcal F_{wK}\,K_{r_*}+\mathcal F_{ww}\,w_{r_*}\big)
  \;+\; \Big(\frac{2A}{r^{2}}+\alpha\,\mathcal T_{wK}\Big)K
  \;+\; \Big(\omega^{2}-\frac{\lhat\,A}{r^{2}}+\alpha\,\mathcal T_{ww}\Big)w \;=\; 0 ,
  \qquad
  A=e^{2\alpha\Phi_1}N .
  \label{eq:m-wrow}
\end{equation}
At $\alpha=0$ this is \eqref{eq:wRW}. Here $A=e^{2\alpha\Phi_1}N$ is the metric function of \eqref{eq:bgmetric} on the corrected mouth, and the terms in $A$ are the Regge--Wheeler equation on the corrected background. The four fermion coefficients $\mathcal F_{wK}$, $\mathcal F_{ww}$, $\mathcal T_{wK}$ and $\mathcal T_{ww}$ are rational in $r$ and are given in Appendix~\ref{app:mouthalpha}. They vanish at the horizon and fall off at large $r$. The second row of $\tilde P$ is no longer zero. The fermions carry charge, and the electric--magnetic duality that maps the $\alpha^0$ system onto Chandrasekhar's decoupled system (Sec.~\ref{sec:mouth-RN}) does not extend to this order. The equation for $K$ has the same form as the first row of \eqref{eq:m-master} with $O(\alpha)$ corrections in all four coefficients. These are long rational functions of $r$ that also contain $\Phi_1$, and we do not display them in full. Near the horizon they reduce to the first row of \eqref{eq:mouth-nh}, which is all that the matching of Sec.~\ref{sec:overlap} uses for $\tilde Q$, and their size is estimated below.

The $O(\alpha)$ terms are small everywhere in the mouth. Near $r_e$ the largest of them comes from the move of the minimum of $N$ from $r_e$ to $r_h$, which turns $(r-r_e)^2$ into $(r-r_h)^2$ in $A$. Its size relative to the $\alpha^0$ terms is $\alpha r_e/3(r-r_e)$, which is $8\times10^{-30}$ at the fiducial junction of Sec.~\ref{sec:overlap}. The logarithm in $\Phi_1$ contributes only $\alpha\log(r/r_e-1)$. Far from $r_e$ the corrections are of relative order $\alpha$. At $l=2$, $p=0$ and the fiducial $\alpha$ of Sec.~\ref{sec:overlap}, the ratio of each $O(\alpha)$ entry of $\tilde Q$ to its $\alpha^0$ value is $8.4\times10^{-30}$ at $r-r_e=10^{-6}r_e$, $8.2\times10^{-33}$ at $10^{-3}r_e$, and below $5\times10^{-35}$ at $0.1r_e$, $r_e$ and $10r_e$.

\subsection{Chandrasekhar's variables and duality at \texorpdfstring{$\alpha^0$}{alpha0}}
\label{sec:mouth-RN}

At $\alpha^0$ the system has a Schr\"odinger form in suitable combinations of $K$, $H_1$ and $w$. We construct them from the results of Chandrasekhar \cite{Chandrasekhar:1979iz,Chandra} for the electrically charged Reissner-Nordstrom black hole. Its field strength $\bar F\propto\dd t\wedge\dd r$ is parity even, and by the selection rule \eqref{eq:selection} the sectors pair like with like. In each sector he reduced the coupled system to two decoupled Schr\"odinger equations. Define
\begin{equation}
  \mu^2 = (l-1)(l+2),\qquad
  \Delta = r^2 - 2GMr + r_e^2,\qquad
  \beta_{1,2} = 3GM \pm \sqrt{9G^2M^2 + 4r_e^2\mu^2},
  \label{eq:chandra-def}
\end{equation}
so that $\beta_1+\beta_2 = 6GM$ and $\beta_1 \beta_2 = -4r_e^2\mu^2$. The master equations are
\begin{equation}
  \frac{\dd^2 Z^{(\pm)}_i}{\dd r_*^2} + \big(\omega^2 - V^{(\pm)}_i\big)Z^{(\pm)}_i = 0,
  \qquad
  V^{(\pm)}_i = \pm \beta_i\, \frac{\dd \Theta_i}{\dd r_*} + \beta_i^2 \Theta_i^2 + \mu^2(\mu^2+2)\Theta_i,
  \qquad
  \Theta_i = \frac{\Delta}{r^3(\mu^2 r + \beta_i)} ,
  \label{eq:chandra-V}
\end{equation}
for $i=1,2$, with $\dd r_*/\dd r = r^2/\Delta$. Chandrasekhar writes $q_i$ for the roots $\beta_i$ and builds his $i$-th pair of potentials from the other root $q_j$ with $j\neq i$ \cite{Chandrasekhar:1979iz}. We label each pair by the root it contains. Our $V^{(\pm)}_1$ and $Z_1$ are therefore his $V^{(\pm)}_2$ and $Z_2$, and the reverse. We write $\beta_i$ and $\Theta_i$ to avoid a clash with the flux $q$ and with $f=1+\rho^2$. The potential $V^{(+)}_i$ governs the polar sector and $V^{(-)}_i$ the axial sector. The two are isospectral, being related by Chandrasekhar's transformation \cite{Chandra}. As $r_e\to0$ they reduce to the Schwarzschild potentials. The potential $V^{(+)}_1$ becomes the Zerilli potential
\begin{equation}
  V_Z = \left(1-\frac{2GM}{r}\right)
  \frac{2n_{\rm Z}^2(n_{\rm Z}+1)r^3 + 6n_{\rm Z}^2 GM r^2 + 18 n_{\rm Z}\,G^2M^2 r + 18 G^3M^3}{r^3\left(n_{\rm Z}r+3GM\right)^2},
  \qquad n_{\rm Z} = \tfrac{\mu^2}{2},
\end{equation}
and $V^{(+)}_2$ becomes $\lhat\,(1-2GM/r)/r^2$, the Regge--Wheeler potential for an electromagnetic field.

Source-free Einstein--Maxwell theory is invariant under the duality rotation $F\to\star F$ \cite{Deser:1976iy,Pereniguez:2023wxf}. The stress tensor \eqref{eq:Tmag} does not change and the metric is untouched. Duality therefore maps the electric black hole to the magnetic one with the same metric and the same $r_e$. The Hodge star contains $\varepsilon_{\theta\phi}$, which is parity odd. Duality thus turns an even parity gauge perturbation into an odd parity one and leaves the metric perturbation alone. Chandrasekhar's polar sector is mapped to the sector studied here, with polar metric and axial gauge perturbations, and the magnetic mouth obeys the decoupled equations \eqref{eq:chandra-V} with $V^{(+)}_i$. The magnetic and electric black holes have the same quasinormal spectrum \cite{Pereniguez:2023wxf,DeFelice:2023rra}. Our amplitude $w=4a/q$ is the dual image of Chandrasekhar's polar gauge amplitude. The duality holds only without charged matter. It holds in the mouth at $O(\alpha^0)$, where $T^Q$ and $J^\nu$ vanish. It fails in the throat, where the fermions carry electric charge, and in the mouth at $O(\alpha)$ (Sec.~\ref{sec:mouth-master}). The one-loop action of the higher Landau levels breaks it as well (Sec.~\ref{sec:hll-mouth}). Duality also fixes the master equations but not the normalisation of the master variables. We therefore construct the variables directly.

Define the combination
\begin{equation}
  \mathcal Z \;\equiv\; r^{2}K \;-\; \frac{\Delta}{p\,r}\,H_1 ,
  \qquad p=-i\omega ,
  \label{eq:Xdef}
\end{equation}
which is the Zerilli-Moncrief combination of the metric amplitudes in Regge--Wheeler gauge \cite{Martel:2005ir}. It has dimension length$^2$, since $K$ and $H_1$ are dimensionless in the mouth. The two variables, of dimension length,
\begin{equation}
  Z_i \;=\; \frac{\mathcal Z}{\mu^{2}r + \beta_i} \;+\; \frac{2r_e^{2}}{\beta_i}\,w ,
  \qquad i=1,2,\;
  \label{eq:Zi-Kw}
\end{equation}
satisfy the polar equations
\begin{equation}
  \frac{\dd^2 Z_i}{\dd r_*^2} + \big(\omega^2 - V^{(+)}_i\big)Z_i = 0
  \label{eq:Zi-eq}
\end{equation}
exactly, for every solution of the first-order system \eqref{eq:m-M}. We found \eqref{eq:Zi-Kw} by asking which linear combination of $K$, $H_1$, $w$ and $w'$, with coefficients rational in $r$, obeys \eqref{eq:Zi-eq}. Among combinations whose coefficients are cubic polynomials in $r$ divided by $r(\mu^2r+\beta_i)$, the answer is unique up to an overall constant and contains no $w'$ term. We checked this uniqueness for $l=2$ at $GM=r_e$ and $GM\simeq0.77\,r_e$, and for $l=3$ at $GM=r_e$ and $GM\simeq1.13\,r_e$. We have not shown it in general. Equation \eqref{eq:Zi-eq} holds identically in $r$, $p$, $GM$ and $l$, for both roots $\beta_{1,2}$. To show this we eliminate $r_e^2=\beta(\beta-6GM)/4\mu^2$ in favour of $\beta$, which makes every coefficient rational. The coefficient of $w$ can also be written $2r_e^2/\beta_i=(\beta_i-6GM)/2\mu^2=-\beta_j/2\mu^2$ with $j\neq i$. The relations invert to
\begin{equation}
  w \;=\; \frac{2\big[(\mu^{2}r+\beta_1)Z_1 - (\mu^{2}r+\beta_2)Z_2\big]}{r\,(\beta_1-\beta_2)},
  \qquad
  \mathcal Z \;=\; (\mu^{2}r+\beta_1)\Big(Z_1 - \frac{2r_e^{2}}{\beta_1}\,w\Big),
  \label{eq:Kw-Zi}
\end{equation}
with $\beta_1-\beta_2 = 2\sqrt{9G^2M^2+4r_e^2\mu^2}$. The amplitudes $K$ and $H_1$ then follow separately from $\mathcal Z$ and $\dd\mathcal Z/\dd r$, using the first two rows of \eqref{eq:m-M}. Since $\mathcal Z$ contains $H_1$, the change of variables is not a transformation of $(K,w)$ alone. The system has no simple Schr\"odinger form in the variables $(K,w)$ and acquires one only after $K$ is traded for the combination \eqref{eq:Xdef}.

The uncharged limit connects \eqref{eq:Zi-Kw} to known results. For $r_e\to0$ one has $\beta_1\to6GM$ and $2r_e^2/\beta_1\to0$, hence $Z_1\to\big[r^{2}K - (r-2GM)H_1/p\big]/(\mu^{2}r+6GM)$, which is the Zerilli-Moncrief variable up to normalisation \cite{Zerilli:1970se,Moncrief:1974am}. For the second variable $\beta_2\to0$ and $2r_e^{2}/\beta_2\to-3GM/\mu^{2}$. The combination $-(\mu^{2}/3GM)\,Z_2$ then tends to $w - \mathcal Z/(3GM\,r)$, which obeys the source-free electromagnetic Regge--Wheeler equation. In terms of the physical amplitude $a=qw/4$ this is $a - q\mathcal Z/(12GM\,r)$, which reduces to $a$ as $q\to0$. At fixed frequency one can integrate either the first-order form \eqref{eq:m-M} or the decoupled form \eqref{eq:Zi-eq}, in the second case reconstructing $K$ and $w$ through \eqref{eq:Kw-Zi}. The matrix of \eqref{eq:m-M} contains $1/p$, and in the time domain the decoupled form applies directly.

\subsection{Matching the mouth and the throat in the overlap}
\label{sec:overlap}

The extremal near-horizon limit connects \eqref{eq:Zi-Kw} to the throat. At $GM=r_e$ one has $\sqrt{9+4\mu^{2}}=2l+1$, hence $\beta_1 = 2r_e(l+2)$ and $\beta_2=-2r_e(l-1)$. At $r=r_e$ this gives $\mu^{2}r_e+\beta_1 = r_e(l+1)(l+2)$ and $\mu^{2}r_e+\beta_2 = r_e\,l(l-1)$. We eliminate $H_1$ through the first row of \eqref{eq:m-M} and apply the scaling $r=r_e+\rho\,r_e^2/\ell$, $p_{\rm mouth}=p/\ell$ at fixed $\rho$. At leading order the derivative terms drop out, and
\begin{equation}
  Z_1 \;\to\; \frac{r_e}{(l+1)(l+2)}\,\big[K + (l+1)\,w\big] = \frac{r_e}{(l+1)(l+2)}\,Z_-,
  \qquad
  Z_2 \;\to\; \frac{r_e}{l(l-1)}\,\big[K - l\,w\big] = \frac{r_e}{l(l-1)}\,Z_+ ,
  \label{eq:Zi-nh}
\end{equation}
with $Z_\pm$ the throat variables \eqref{eq:Zdef}. The superscript of $V^{(\pm)}_i$ labels the parity, and the subscript of $Z_\pm$ and $m_\pm$ labels the throat mass. The masses pair up the same way, $V^{(+)}_1\to m_-^2=l(l-1)$ and $V^{(+)}_2\to m_+^2=(l+1)(l+2)$, as in \eqref{eq:nhmasses}. The variable $Z_1$, which tends to the Zerilli-Moncrief variable as $r_e\to0$, thus continues into the lighter throat mode $Z_-$, and $Z_2$ continues into $Z_+$. The corrections to \eqref{eq:Zi-nh} are of relative order $r_e\rho/\ell=\tfrac{\pi}{4}\alpha\rho$, and they mix in the other throat mode. This is the counterpart, at the level of the variables, of the $O(\alpha)$ mixing found in Sec.~\ref{sec:throat-schrodinger}. The Chandrasekhar variables of the mouth continue into the Poschl-Teller variables of the throat.

The same limit can be taken in the equations. Set $GM = r_e$, write $r = r_e + y$, and expand for small $y$. The coefficients of the second-order system \eqref{eq:secondorder} develop poles at $y=0$. The leading ones go like $1/y$ in $P$ and like $1/y^2$ and $p^2/y^4$ in $Q$. The terms of $Q$ in $1/y^3$ and $1/y$ are suppressed by at least one power of $\alpha$ under the matching used below and drop out at $\alpha^0$. Keeping the leading terms gives
\begin{align}
  K_{yy} &= -\frac{2}{y}K_y + \frac{p^2 r_e^4}{y^4}K + \frac{1}{y^2}\left[(\lhat+2)K - 2\lhat\, w\right],
  \label{eq:nhK}\\
  w_{yy} &= -\frac{2}{y}w_y + \frac{p^2 r_e^4}{y^4}w + \frac{1}{y^2}\left[\lhat\, w - 2K\right].
  \label{eq:nhw}
\end{align}
These are the throat equations of Sec.~\ref{sec:alpha0} at $\alpha=0$ under the MMP matching \eqref{eq:matching}, $\rho = \ell\,(r-r_e)/r_e^2$ and $\tau = t/\ell$, in the regime $1 \ll \rho$ where $1+\rho^2 \to \rho^2$. Here $p$ is the mouth Laplace variable, with $p_{\rm mouth} = p_{\rm throat}/\ell$. The same limit applied to Chandrasekhar's potentials \eqref{eq:chandra-V} gives
\begin{equation}
  V^{(\pm)}_1 \;\longrightarrow\; \frac{y^2}{r_e^4}\,m_-^2,
  \qquad V^{(\pm)}_2 \;\longrightarrow\; \frac{y^2}{r_e^4}\,m_+^2,
  \label{eq:nhmasses}
\end{equation}
which are the two masses of the throat, Sec.~\ref{sec:alpha0}. They are the eigenvalues of the matrix that multiplies $(K,w)/y^2$ in \eqref{eq:nhK}--\eqref{eq:nhw}, which is the matrix $\mathcal N$ of \eqref{eq:t-Qtilde0}. A different choice of master variables recombines $K$ and $w$ linearly and changes this matrix by a similarity transformation, which leaves its eigenvalues unchanged. Any decoupling must therefore reproduce these two masses. With $r_*=-r_e^2/y=\ell(x-\pi/2)$ the potentials \eqref{eq:nhmasses} become $m_-^2/r_*^2$ and $m_+^2/r_*^2$, which is the large-$\rho$ form of $m^2\sec^2x$ in \eqref{eq:ads2master} once $\Om=\omega\ell$.

The mouth system \eqref{eq:m-master}, with its $O(\alpha)$ part from Sec.~\ref{sec:mouth-master}, and the throat system \eqref{eq:t-master-tortoise} are written in the same variables $\bm Y=(K,w)^\top$ and the same form. Their coefficient matrices can be compared in the overlap region $1\ll\rho$, $\alpha\rho\ll1$. Near extremality $\dd r_*/\dd r = r^2/\Delta\to r_e^2/y^2$, hence $r_*=-r_e^2/y$, while $x=\arctan\rho\to\pi/2-1/\rho$. With the MMP matching \eqref{eq:matching} and $\ell=4r_e/\pi\alpha$ these give
\begin{equation}
  r_* \;=\; \ell\Big(x-\frac{\pi}{2}\Big),
  \qquad
  \dd r_* \;=\; \ell\,\dd x,
  \qquad
  p_{\rm mouth} \;=\; \frac{p}{\ell},
  \label{eq:overlap-dict}
\end{equation}
and the two equations are the same equation if and only if
\begin{equation}
  \ell\,\tilde P^{\rm mouth} \;=\; \tilde P^{\rm throat},
  \qquad
  \ell^{2}\,\tilde Q^{\rm mouth} \;=\; \tilde Q^{\rm throat}
  \label{eq:overlap-condition}
\end{equation}
in the overlap.

Both calculations are first order in $\alpha$, but they hold different variables fixed. The throat expands in $\alpha$ at fixed $\rho$. The mouth expands in $\alpha$ at fixed $y=r-r_e$. In the overlap $\rho=4y/(\pi\alpha r_e)$, and a throat term becomes
\begin{equation}
  \alpha^{n_\alpha}\rho^{n_\rho} \;=\; \Big(\frac{4y}{\pi r_e}\Big)^{n_\rho}\,\alpha^{\,n_\alpha-n_\rho} .
  \label{eq:overlap-orders}
\end{equation}
At fixed $y$ this term is of order $\alpha^{n_\alpha-n_\rho}$. We count this power from the leading term of the same matrix, which is $\rho^2\propto\alpha^{-2}$ for $\tilde Q$ and $\alpha\rho^2\propto\alpha^{-1}$ for $\ell\tilde P$. We call the result the mouth order $n_{\rm m}$. It is $n_{\rm m}=n_\alpha-n_\rho+2$ for $\tilde Q$ and $n_{\rm m}=n_\alpha-n_\rho+1$ for $\ell\tilde P$. The throat calculation gives every term with $n_\alpha\le1$. The mouth calculation gives every term with $n_{\rm m}\le1$. Both give the terms with $n_\alpha\le1$ and $n_{\rm m}\le1$. In $\tilde Q$ these are $\rho^2$ and $\alpha\rho^3$ with $n_{\rm m}=0$, and $\rho$ and $\alpha\rho^2$ with $n_{\rm m}=1$. In $\ell\tilde P$ they are $\rho$ and $\alpha\rho^2$ with $n_{\rm m}=0$, and $\rho^0$ and $\alpha\rho$ with $n_{\rm m}=1$. The terms with $n_{\rm m}=0$ test the $\alpha^0$ mouth, and the terms with $n_{\rm m}=1$ test the $O(\alpha)$ mouth.

The terms of mouth order 0 come from the $\alpha^0$ mouth. Substituting $GM=r_e$, $r=r_e+\rho\,r_e^{2}/\ell$ and \eqref{eq:overlap-dict} into \eqref{eq:m-Ptilde}--\eqref{eq:m-QKK} and expanding to first order in $\alpha$ at fixed $\rho$ gives
\begin{equation}
  \ell\,\tilde P^{\rm mouth}
  = \begin{pmatrix}\tfrac{\pi\alpha}{2}\rho^{2} & \pi\alpha\rho^{2}\\ 0 & 0\end{pmatrix},
  \quad
  \ell^{2}\tilde Q^{\rm mouth}
  = \begin{pmatrix}
  -(\lhat+2)\rho^{2} - p^{2} + \pi\alpha(\lhat+4)\rho^{3} & 2\lhat\rho^{2} - 3\pi\alpha\lhat\rho^{3}\\[4pt]
  2\rho^{2} - 2\pi\alpha\rho^{3} & -\lhat\rho^{2} - p^{2} + \pi\alpha\lhat\rho^{3}
  \end{pmatrix}.
  \label{eq:overlap-mouth}
\end{equation}
These are the large-$\rho$ limits of \eqref{eq:t-Ptilde1}--\eqref{eq:t-Qtilde1-ww}. With $\arctan\rho\to\pi/2$ and $\log f$ subleading, $\alpha\tilde P_{1}$ reproduces the first matrix, its $wK$ and $ww$ entries falling off as $2\alpha/\lhat\rho$ and $-\alpha/\rho$. Then $\tilde Q_{0}$ with $f\to\rho^{2}$ reproduces the $\alpha^{0}$ terms, and the $\arctan\rho$ terms of $\alpha\tilde Q_{1}$ reproduce the four $\rho^{3}$ coefficients $\pi\alpha(\lhat+4)$, $-3\pi\alpha\lhat$, $-2\pi\alpha$ and $\pi\alpha\lhat$. The $\rho^3$ growth of the $O(\alpha)$ throat coefficients is thus the near-horizon tail of the Reissner-Nordstrom exterior, and with it the linear growth $\alpha\varphi\sim\tfrac{\pi}{2}\alpha\rho$ that fixes $\ell$ in \eqref{eq:ell}.

The terms of mouth order 1 need the corrected mouth and a dictionary correct to $O(\alpha)$. We write both metrics in the gauge of the throat, $AB=r_e^4$, with $t=\ell\tau$. For the mouth this fixes
\begin{equation}
  \frac{\dd r}{\dd\rho} \;=\; \frac{r_e^{2}}{\ell}\,e^{-\alpha\Phi_1(r)} ,
  \label{eq:overlap-dict1}
\end{equation}
and we fix the integration constant by $R=r$. The inversion of \eqref{eq:overlap-dict1} then gives
\begin{equation}
  \frac{\ell y}{r_e^{2}} \;=\; \rho \;+\; \frac{2}{3\pi} \;+\; \frac{\alpha\rho}{3}\Big[1+2\log\frac{\pi\alpha\rho}{4}\Big] \;+\; \dots ,
  \label{eq:overlap-sigma}
\end{equation}
and \eqref{eq:overlap-condition} holds with $p_{\rm mouth}=p/\ell$ as before. The constant $2/3\pi$ is the shift $r_h-r_e=\alpha r_e/6$ in throat units. The logarithm comes from $\Phi_1\simeq-\tfrac23\log(y/r_e)$ and cancels against the redshift $e^{2\alpha\Phi_1}$ in every compared coefficient. A different normalisation of time, a rescaling of $\ell$ by a factor $1+O(\alpha)$, changes no compared coefficient. The background is the first test. With $GM=r_e(1+\alpha\,\mathfrak m)$ the expansion of $\ell^2A^{\rm mouth}/r_e^2$ contains the term $-16(18\mathfrak m+1)/9\pi^2\alpha$, which has no counterpart in the throat. It vanishes only for $\mathfrak m=-1/18$, the value at which $N$ and $N'$ vanish together at $O(\alpha)$, which \eqref{eq:mouthmass} already imposes. The matching reproduces it. The remaining terms agree with \eqref{eq:throatlarge} at $\rho^2$, $\rho$, $\alpha\rho^3$ and $\alpha\rho^2$, and $R$ agrees including the constant $\alpha r_e/6$.

The master matrices agree at every compared order, for all $l$. The terms of mouth order 1 are zero on both sides except the $\alpha\rho^2$ terms of $\tilde Q$, which are
\begin{equation}
  \tilde Q\big|_{\alpha\rho^{2}} \;=\; \begin{pmatrix} -\tfrac{\lhat+2}{3} & \tfrac{\lhat(\lhat+2)}{6}\\[4pt] \tfrac{3\lhat-2}{6} & -\tfrac{\lhat(\lhat+2)}{12}\end{pmatrix}
  \label{eq:overlap-arho2}
\end{equation}
in the throat and in the corrected mouth alike. In the throat these numbers carry the trace part of the source, its response, the density and the drift of the field lines, and the Hall current. In the mouth they come from the same pieces, through $N_1$, $\Phi_1$ and the terms of Sec.~\ref{sec:mouth-master}. The uncorrected mouth fails this test. On the MMP dictionary $r-r_e=\rho\,r_e^2/\ell$ every $\alpha\rho^2$ term of the uncorrected mouth vanishes, against $-1/3$ in $A$ and \eqref{eq:overlap-arho2} in $\tilde Q$, and $R$ misses the constant $\alpha r_e/6$. A constant shift of the dictionary does not help, since it moves the $\rho$ terms of $A$ and $\tilde Q$ away from zero. The $\alpha\rho^2$ terms of the throat are therefore a test of the $O(\alpha)$ mouth, and only the corrected mouth passes it.

The mouth side of \eqref{eq:overlap-arho2} follows from the expansion of the mouth near the horizon, Eqs.~\eqref{eq:mouth-nh} and \eqref{eq:mouth-nh-DC} of Appendix~\ref{app:mouthalpha}. On the dictionary \eqref{eq:overlap-sigma} the $O(\alpha)$ term linear in $y$ cancels the $\rho$ term that the constant $2/3\pi$ produces, and the logarithms cancel. Collecting the $\alpha\rho^2$ terms,
\begin{equation}
  \ell^{2}\,\tilde Q^{\rm mouth}\big|_{\alpha\rho^{2}} \;=\; -\tfrac23\,\mathcal N \;+\; \tfrac12\,\mathsf Q_3 \;+\; \mathsf Q_2 ,
  \label{eq:mouth-arho2}
\end{equation}
with $\mathcal N$ the matrix of \eqref{eq:t-Qtilde0} and $\mathsf Q_3$, $\mathsf Q_2$ from \eqref{eq:mouth-nh-DC}. This is \eqref{eq:overlap-arho2} entry by entry.

We have also evaluated both sides numerically at $r_{\rm J}=r_e(1+\delta_{\rm J})$ for the fiducial parameters of \cite{Mondal:2025tht}, $q=10^{33}$, $\ell_p=1.6\times10^{-33}\,{\rm cm}$ for the Planck length, $g_c=1$ and $\delta_{\rm J}=10^{-6}$, for which
\begin{equation}
  r_e = 2.8359\,{\rm cm},\quad
  \ell = 1.4255\times10^{35}\,{\rm cm},\quad
  \alpha = 2.5330\times10^{-35},\quad
  \rho_{\rm J} = 5.0265\times10^{28},\quad
  \alpha\rho_{\rm J} = 1.2732\times10^{-6},
  \label{eq:overlap-params}
\end{equation}
with $\delta_M=-5.61\times10^{-70}\,{\rm cm}$ the mass defect of MMP's eq.~(5.42). The comparison uses the $\alpha^0$ mouth, whose extremal mass is $r_e$, and there $\delta_M$ is measured from $r_e$. The two sides agree entry by entry (Table~\ref{tab:junction}).
\begin{table}[htbp]
\caption{Coefficients of the master systems of the mouth and of the throat at the fiducial junction $r_{\rm J}=r_e(1+10^{-6})$. The mouth column is $\ell\tilde P$ and $\ell^2\tilde Q$ of the $\alpha^0$ mouth at $GM=r_e+\delta_M$. The throat column is $\tilde P$ and $\tilde Q$ of the $O(\alpha)$ throat at $\rho_{\rm J}$. The last three columns give the relative differences, and the absolute differences for $\tilde P_{wK}$ and $\tilde P_{ww}$.}
\label{tab:junction}
\begin{tabular}{lrrccc}
\hline\hline
 & \multicolumn{2}{c}{value at $l=2$, $\Om=1$}
 & \multicolumn{3}{c}{relative difference}\\
\cline{2-3}\cline{4-6}
 & \multicolumn{1}{c}{mouth} & \multicolumn{1}{c}{throat} & $l=2$ & $l=3$ & $l=4$ \\
\hline
\multicolumn{6}{l}{$\tilde P$ matrix, mouth side multiplied by $\ell$}\\
$\tilde P_{KK}$ & $1.0053070\times10^{23}$ & $1.0053096\times10^{23}$
  & $2.7\times10^{-6}$ & $2.8\times10^{-6}$ & $2.9\times10^{-6}$\\
$\tilde P_{Kw}$ & $2.0106092\times10^{23}$ & $2.0106193\times10^{23}$
  & $5.0\times10^{-6}$ & $5.0\times10^{-6}$ & $5.0\times10^{-6}$\\
$\tilde P_{wK}$ & $0$ & $1.68\times10^{-64}$
  & $1.7\times10^{-64}$ & $8.4\times10^{-65}$ & $5.0\times10^{-65}$\\
$\tilde P_{ww}$ & $0$ & $-5.04\times10^{-64}$
  & $5.0\times10^{-64}$ & $5.0\times10^{-64}$ & $5.0\times10^{-64}$\\[4pt]
\multicolumn{6}{l}{$\tilde Q$, mouth side multiplied by $\ell^{2}$}\\
$\tilde Q_{KK}$ & $-2.0212849\times10^{58}$ & $-2.0212849\times10^{58}$
  & $1.6\times10^{-11}$ & $1.3\times10^{-11}$ & $1.2\times10^{-11}$\\
$\tilde Q_{Kw}$ & $\phantom{-}3.0319243\times10^{58}$ & $3.0319243\times10^{58}$
  & $2.1\times10^{-11}$ & $2.1\times10^{-11}$ & $2.1\times10^{-11}$\\
$\tilde Q_{wK}$ & $\phantom{-}5.0532172\times10^{57}$ & $5.0532172\times10^{57}$
  & $1.0\times10^{-11}$ & $1.0\times10^{-11}$ & $1.0\times10^{-11}$\\
$\tilde Q_{ww}$ & $-1.5159652\times10^{58}$ & $-1.5159652\times10^{58}$
  & $1.0\times10^{-11}$ & $1.0\times10^{-11}$ & $1.0\times10^{-11}$\\
\hline\hline
\end{tabular}
\end{table}
The mouth entries are quoted after multiplication by the factor $\ell$ or $\ell^{2}$ required by \eqref{eq:overlap-condition}. The mouth side is the $\alpha^0$ system at $GM=r_e+\delta_M$. The relative differences are unchanged to the digits shown for $\Om=0$ and $\Om=5$, since the frequency enters only through $p^{2}$ against $\rho^{2}\sim10^{57}$. The entries $\tilde P_{wK}$ and $\tilde P_{ww}$ vanish identically in the $\alpha^0$ mouth. In the throat at $l=2$ they are $2\alpha\rho/(\lhat f)\simeq1.7\times10^{-64}$ and $-\alpha\rho/f\simeq-5.0\times10^{-64}$, and only the absolute differences are meaningful there. The two accuracies are as expected. The matrix $\tilde Q$ starts at $O(\rho^{2})$ and its $O(\alpha)$ term is $O(\alpha\rho^{3})$. The first neglected term is $O(\alpha^{2}\rho^{4})$ and the relative mismatch scales as $(\alpha\rho)^{2}$. The matrix $\tilde P$ is itself $O(\alpha\rho^{2})$, and its relative mismatch scales as $\alpha\rho$. Moving the junction through $\delta_{\rm J}=10^{-9},\dots,10^{-2}$ confirms both. The ratios ${\rm rel.}\,\tilde Q_{KK}/(\alpha\rho)^2$ and ${\rm rel.}\,\tilde P_{KK}/\alpha\rho$ stay at $9.77$ and $2.09$ for $l=2$, at $8.21$ and $2.23$ for $l=3$ and at $7.46$ and $2.28$ for $l=4$. They change only when $\alpha\rho$ approaches $10^{-2}$, where the $O(\alpha^{2})$ terms begin to be felt. The junction value $\delta_{\rm J}=10^{-6}$ used in the echo computation of \cite{Mondal:2025tht} sits inside the overlap window. The $O(\alpha)$ terms of the mouth do not change the table. At this junction they are of relative size $8\times10^{-30}$ (Sec.~\ref{sec:mouth-master}), far below the relative differences of about $10(\alpha\rho_{\rm J})^2\simeq1.6\times10^{-11}$ that the neglected throat terms produce in Table~\ref{tab:junction}. The numerical test checks the terms of mouth order 0, and the expansion above checks those of mouth order 1.

Repeating the comparison at $l=2$, $\Om=1$ and $\delta_{\rm J}=10^{-6}$ with the $\alpha^{0}$ throat matrices alone, that is with $\tilde P=0$ and $\tilde Q=\tilde Q_{0}$, degrades the agreement from $O((\alpha\rho)^{2})$ to $O(\alpha\rho)$,
\begin{equation}
\begin{aligned}
  \text{rel.\ diff.\ with }O(\alpha):&\quad
  \tilde Q_{KK}:1.6\times10^{-11},\quad \tilde Q_{Kw}:2.1\times10^{-11},\quad
  \tilde Q_{wK},\tilde Q_{ww}:1.0\times10^{-11},\\
  \text{rel.\ diff.\ at }\alpha^{0}\text{ only}:&\quad
  \tilde Q_{KK}:5.0\times10^{-6},\quad \tilde Q_{Kw}:6.0\times10^{-6},\quad
  \tilde Q_{wK},\tilde Q_{ww}:4.0\times10^{-6},
\end{aligned}
\end{equation}
an improvement by a factor of $3\times10^{5}$. Equation \eqref{eq:overlap-condition} is a consistency check on the coefficients and not a junction condition. Matching a solution still requires continuity of $(K,w,K',w')$ across the overlap. In the mouth at $O(\alpha)$ the transport fields enter the equation for $K$ and the system is of sixth order, while in the throat they are fixed by $(K,w)$ up to the constants $\mathfrak b_\pm$. How they continue across the junction is not discussed here. The mass defect $\delta_M$ is $O(\alpha^2r_e)$. Its largest effect is on $\Delta=(r-r_e)^2-2\delta_M r$, which it changes at this junction by the relative amount $2|\delta_M|/(r_e\delta_{\rm J}^2)\simeq4\times10^{-58}$. Setting it to zero changes none of the digits above. The working precision of this comparison is discussed in Appendix~\ref{app:numerics}.

\section{The null energy condition and linear stability}
\label{sec:stability}

The throat is held open by the negative Casimir energy of the lowest-Landau-level fermions. Two questions arise once the perturbation is switched on. Does the null energy condition remain violated, and can the perturbation grow? The second is the substantive one. In semiclassical gravity the linear response of the quantum stress tensor is part of the stability problem \cite{Anderson:2002fk}, and here it is the response of Sec.~\ref{sec:response}.

\subsection{The null energy condition}
\label{sec:NEC0}

A radial null vector in the $(\tau,\rho)$ block of the throat is
\begin{equation}
  k^a \;=\; k_0\left(1,\;f\right),
  \qquad f \equiv 1+\rho^2,
  \qquad \hat g_{ab}k^ak^b = -f k_0^2 + \frac{f^2k_0^2}{f} = 0 ,
\end{equation}
with $k^A=0$ on the sphere. Only the two-dimensional block of $T^Q$ contributes. Using \eqref{eq:tbar} and the lift \eqref{eq:lift},
\begin{equation}
  T_{\mu\nu}k^\mu k^\nu
  \;=\; \frac{\bar t_{ab}k^ak^b}{4\pi r_e^2}
  \;=\; -\,\frac{q\,k_0^2}{16\pi^2 r_e^2} \;<\; 0 ,
  \label{eq:NECbackground}
\end{equation}
independent of $\rho$. The violation is uniform along the throat. The magnetic field is NEC-marginal, since $T^{\rm mag}_{\mu\nu}k^\mu k^\nu = 0$ for radial null vectors, and the entire violation comes from the fermions. Within $\bar t_{ab}$ it comes entirely from the traceless part. The trace piece is proportional to $\hat g_{ab}$ and contracts to zero against a null vector. The trace part of $\bar t_{ab}$, which Sec.~\ref{sec:throat-origin} showed must be retained for the perturbation problem, plays no role in supporting the throat.

At $O(\epsilon)$ both the stress tensor and the null vector change. The null vector must remain null in the perturbed metric. With its component $k^\tau=k_0$ held fixed, this fixes
\begin{equation}
  k^\rho \;=\; f k_0\left(1 + \epsilon\,k_1 Y(\theta)\right),
  \qquad
  k_1 = -\,\frac{r_e^2\left(H_0+H_2\right) + 2H_1}{2r_e^2} .
\end{equation}
Computing $\delta\!\left(T^Q_{ab}k^ak^b\right)$ with this correction included gives an expression linear in the perturbation fields and proportional to $Y(\theta)e^{p\tau}$. For an oscillatory mode, $p=-i\Om$ with $\Om$ real, it averages to zero over a period. The time-averaged null energy condition along the throat is unchanged at $O(\epsilon)$. Instantaneously the contraction oscillates about the negative background value \eqref{eq:NECbackground} with amplitude $O(\epsilon)$, and it stays negative for all sufficiently small $\epsilon$. This says only that a small perturbation is small. The question with content is whether modes with $\mathrm{Re}\,p>0$ exist.

\subsection{The stability criterion and the throat at \texorpdfstring{$\alpha=0$}{alpha=0}}
\label{sec:criterion}

A growing mode is a solution with $p$ real and positive, equivalently $\omega^2<0$, of the master system in Schr\"odinger form,
\begin{equation}
  -\,\partial_x^2 \bm{Z} \;+\; V\bm{Z} \;=\; \omega^2 \bm{Z} .
\end{equation}
Here $\omega$ is the Laplace frequency of whichever region is considered. In the mouth $x=r_*$ and the eigenvalue is $\omega^2$. In the throat $x$ is the tortoise coordinate \eqref{eq:t-tortoise}, equal to $\arctan\rho$ at $\alpha=0$, and the eigenvalue is $\Om^2=\omega^2\ell^2$, as in \eqref{eq:master}. Only the sign of the eigenvalue enters the argument, and it is the same in both. Contracting with $\bm Z^{\!\top}$ and integrating over the domain,
\begin{equation}
  \int \left|\partial_x \bm{Z}\right|^2 \dd x
  \;+\; \int \bm{Z}^{\!\top} V \bm{Z}\,\dd x
  \;=\; \omega^2 \int \left|\bm{Z}\right|^2 \dd x .
  \label{eq:quadform}
\end{equation}
If the symmetric part of $V$ is positive definite on the whole domain, both terms on the left are positive. Hence $\omega^2>0$ and no growing mode exists. Only the symmetric part enters, since $\bm Z^{\!\top}V\bm Z = \bm Z^{\!\top}V_{\rm sym}\bm Z$ with $V_{\rm sym}=\tfrac12\left(V+V^{\!\top}\right)$. This is the argument that establishes stability of Schwarzschild from positivity of the Regge--Wheeler and Zerilli potentials \cite{Regge:1957td,Vishveshwara:1970cc,Zerilli:1970se,Wald:1979lth}, and of Reissner-Nordstrom from Moncrief's potentials \cite{Moncrief:1974ng,Moncrief:1974am,Moncrief:1974gw}. We apply it region by region.

In the throat at $\alpha=0$ the result is exact. In the tortoise coordinate $x=\arctan\rho\in\left(-\pi/2,\pi/2\right)$ the reduction of Sec.~\ref{sec:alpha0} gives $\tilde P=0$ and a potential matrix that is diagonal in the basis \eqref{eq:Zdef}, $Z_+ = K - l\,w$ and $Z_- = K + (l+1)\,w$, with
\begin{equation}
  V_+ = (l+1)(l+2)\sec^2 x ,
  \qquad
  V_- = l(l-1)\sec^2 x ,
  \label{eq:Vthroat}
\end{equation}
and no dependence on $p$. Both potentials are strictly positive for $l\ge2$. They are bounded below by $l(l-1)\ge2$ at $x=0$ and diverge at the ends of the interval. By \eqref{eq:quadform} no growing mode can originate in the $\alpha=0$ throat. The zero of $V_-$ at $l=1$ lies outside the range $l\ge2$ of this paper, where the $l=1$ sector carries a single physical degree of freedom (Sec.~\ref{sec:ansatz}).

\subsection{The throat at \texorpdfstring{$O(\alpha)$}{O(alpha)}}
\label{sec:stab-throat1}

Write $V=V_0+\alpha(V_{1a}+p^2V_{1b})$ from \eqref{eq:master}. The correction is not symmetric and depends on $p^2$. Its symmetric part decides whether a mode can grow as a pure exponential. Its antisymmetric part decides whether the frequency can become complex, which would allow a mode that grows while it oscillates. To separate the two, write a mode as $\bm Z(x)\,e^{p\tau}$, set $\Om^2=-p^2$ and move the $p^2V_{1b}$ term of $V$ to the right. The throat equation becomes
\begin{equation}
  -\bm Z''+\big(V_0+\alpha V_{1a}\big)\bm Z=\Om^2\big(1+\alpha V_{1b}\big)\bm Z ,
  \label{eq:throatOm}
\end{equation}
where a prime is $\partial_x$. The mode grows if $\mathrm{Re}\,p>0$. Multiply by $\bm Z^{\dagger}$, integrate over $x$ and drop the boundary terms, which vanish for a normalisable mode. This gives $\Om^2=\mathfrak e/\mathfrak n$ with
\begin{equation}
  \mathfrak e=\int\dd x\,\Big[|\bm Z'|^2+\bm Z^{\dagger}\big(V_0+\alpha V_{1a}\big)\bm Z\Big],
  \qquad
  \mathfrak n=\int\dd x\;\bm Z^{\dagger}\big(1+\alpha V_{1b}\big)\bm Z .
  \label{eq:EN}
\end{equation}
For a real matrix $\mathsf G$ the real part of $\bm Z^{\dagger}\mathsf G\bm Z$ comes only from the symmetric part of $\mathsf G$ and the imaginary part only from the antisymmetric part. The real part of $\mathfrak e$ contains $|\bm Z'|^2$ and $(V_0+\alpha V_{1a})_{\rm sym}$, and the real part of $\mathfrak n$ contains $(1+\alpha V_{1b})_{\rm sym}$. The imaginary parts of $\mathfrak e$ and $\mathfrak n$ contain only the antisymmetric parts of $V_{1a}$ and $V_{1b}$, and they are of order $\alpha$. To first order in $\alpha$,
\begin{equation}
  \mathrm{Re}\,\Om^2=\frac{\mathrm{Re}\,\mathfrak e}{\mathrm{Re}\,\mathfrak n}+O(\alpha^2),
  \qquad
  \mathrm{Im}\,\Om^2=\frac{\mathrm{Im}\,\mathfrak e\,\mathrm{Re}\,\mathfrak n-\mathrm{Re}\,\mathfrak e\,\mathrm{Im}\,\mathfrak n}{|\mathfrak n|^2}=O(\alpha).
  \label{eq:realp}
\end{equation}

The sign of $\mathrm{Re}\,\Om^2$ rests on two matrices. The first is $(V_0+\alpha V_{1a})_{\rm sym}$, which is the symmetric part $V_{\rm sym}=V_0+\frac{\alpha}{2}\big(V_{1a}+V_{1a}^{\!\top}\big)$ of the full potential at $p=0$. The unperturbed potential is bounded below by $l(l-1)\ge2$, and the ratio of $\alpha V_{1a}$ to $V_0$ grows only as $\alpha\rho$, from \eqref{eq:V1alarge} against $V_0\sim\rho^2$. Positivity can therefore be lost only where $\alpha\rho\sim1$. The relation between $x$ and $\rho$ does not enter here. The criterion is pointwise in the potential, and the scan runs over $\rho$. We evaluated the minimum eigenvalue of $V_{\rm sym}$ numerically (Appendix~\ref{app:numerics}) over $0\le\alpha\rho\le0.1$ for $l=2,3,4$, for $p=0,1,3$, and for $\alpha=10^{-6},10^{-3},10^{-2}$. It stays within a few per cent of $l(l-1)$ in every case. A search up to $\alpha\rho=1$ (Appendix~\ref{app:numerics}) finds that positivity is first lost at $\alpha\rho\approx0.24$ for $l=2$, and at $0.27$ and $0.28$ for $l=3$ and $4$, for every $p$ and $\alpha$ in the scan. The higher Landau levels move these points outwards (Sec.~\ref{sec:hll-throat}). The threshold does not move with $p$, since the term $\alpha p^2V_{1b}$ falls off as $1/\rho^2$ while $V_0$ grows as $\rho^2$. For the fiducial parameters the junction sits at $\rho_{\rm J}\simeq5\times10^{28}$ and $\alpha\simeq2.5\times10^{-35}$, hence $\alpha\rho_{\rm J}\sim10^{-6}$, five orders of magnitude below the point where positivity is lost. The second matrix is $(1+\alpha V_{1b})_{\rm sym}$. It is positive since $\alpha V_{1b}$ is small compared with the identity, and its smallest eigenvalue is $0.988$ at $\alpha=10^{-2}$. Hence $\mathrm{Re}\,\mathfrak e>0$, $\mathrm{Re}\,\mathfrak n>0$ and $\mathrm{Re}\,\Om^2>0$, up to the region $\alpha\rho>0.1$ that the scan does not cover. There the $O(\alpha)$ form of $V$ itself fails. A normalisable mode with $\bm Z\sim\cos^sx$ gives that region a weight of relative order $(10\alpha)^{2s-1}$ in $\mathfrak e$.

$\Om^2$ is therefore never real and negative, and there is no mode with $p$ real and positive. The argument covers every value of $p$ at once, since $p$ enters \eqref{eq:throatOm} only through $\Om^2$. The scan at $p=0$ is all that is needed, and the scans at $p=1$ and $p=3$ are a consistency check. Positivity does not exclude a complex $\Om^2$. For complex $\Om^2$ with $\mathrm{Re}\,\Om^2>0$ one root $p$ has $\mathrm{Re}\,p\simeq|\mathrm{Im}\,\Om^2|/2\sqrt{\mathrm{Re}\,\Om^2}$. Such a mode would grow while oscillating, at a rate of order $\alpha$. Whether $\mathrm{Im}\,\Om^2$ vanishes is decided by the antisymmetric parts in \eqref{eq:realp}, not by the symmetric part. Section~\ref{sec:stab-normalmodes} shows that $\mathrm{Im}\,\Om^2=0$ at first order in $\alpha$ on every level with the scaled lift, and on the levels examined with the coordinate lift.

\subsection{Normal modes of the closed throat at \texorpdfstring{$O(\alpha)$}{O(alpha)}}
\label{sec:stab-normalmodes}

We follow the normal modes of the throat to first order in $\alpha$. At $\alpha=0$ the two channels of \eqref{eq:ads2master} are Poschl-Teller problems $-\partial_x^2Z+s(s-1)\sec^2x\,Z=\Om^2Z$ on $x\in(-\pi/2,\pi/2)$, with $s=l+2$ for $Z_+$ and $s=l$ for $Z_-$. With $Z$ normalisable at both ends \cite{Ishibashi:2004wx} the eigenfunctions and levels are
\begin{equation}
  \psi_n(x)=\cos^{s}x\;C^{(s)}_n(\sin x),\qquad \Om_0=n+s,\qquad n=0,1,2,\dots,
  \label{eq:PTmodes}
\end{equation}
with $C^{(s)}_n$ the Gegenbauer polynomials. The $Z_+$ levels are $\Om_0=l+2,l+3,\dots$ and the $Z_-$ levels are $\Om_0=l,l+1,\dots$. Every level of $Z_+$ coincides with a level of $Z_-$. The mode $Z_+$ with quantum number $n$ and the mode $Z_-$ with quantum number $n+2$ share $\Om_0=n+l+2$, and only the two lowest $Z_-$ levels are non-degenerate. On a shared level the antisymmetric part of $V_{1a}-\Om_0^2V_{1b}$ acts at first order, since no gap separates the two modes. This holds at $\eta_{\rm H}=0$ (Sec.~\ref{sec:hll-throat}). These modes live deep inside the throat. At the junction $\rho_{\rm J}\simeq5\times10^{28}$ their weight is $\psi_n^2\sim\rho_{\rm J}^{-2s}$, of order $10^{-115}$ or smaller, and the normalisable condition at the ends of the interval is an excellent idealisation. We expect transmission through the mouths to remove energy from these modes and to add a negative imaginary part to $\Om$. We have not shown this with a flux identity, which the $O(\alpha)$ problem would need, since its potential depends on frequency and is not symmetric.

At $O(\alpha)$ the potentials in \eqref{eq:throatOm} are functions of $\rho$, and $\rho$ is related to $x$ by \eqref{eq:t-tortoise}. Inverting that relation to first order gives $\rho=\tan x+\alpha\,\iota\sec^2x$, and hence $1+\rho^2=\sec^2x\,(1+2\alpha\rho\,\iota)$. In $V_{1a}$ and $V_{1b}$, which already carry a factor $\alpha$, one may set $\rho=\tan x$. In $V_0$ one may not. The throat equation becomes
\begin{equation}
  -\partial_x^2\bm Z+\Big[V_0+\alpha\big(V_{1a}+2\rho\,\iota\,V_0\big)\Big]\bm Z
  \;=\;\Om^2\big(1+\alpha V_{1b}\big)\bm Z ,
  \qquad \rho=\tan x ,
  \label{eq:xform}
\end{equation}
with $V_0=\mathrm{diag}\big[(l+1)(l+2),\,l(l-1)\big]\sec^2x$ and $\iota$ from \eqref{eq:t-tortoise}. The term $2\alpha\rho\,\iota\,V_0$ is diagonal and comes entirely from the difference between $x$ and $\arctan\rho$. We call it the remap term. Omitting it changes the diagonal shifts below by amounts between $2$ and $107$, comparable to the splittings. The frequency-dependent matrix $V_{1b}$ of \eqref{eq:V1b} has
\begin{equation}
  \det V_{1b} \;=\; -\,\frac{4}{\lhat\,f^{2}} ,
  \label{eq:V1bmatrix}
\end{equation}
and rank two. Both off-diagonal entries are positive for every $l$, and their product is positive.

At $\alpha=0$ any linear combination of the modes $\psi_j$ of a level solves \eqref{eq:xform}. Expanding $\Om^2=\Om_0^2+\alpha\,\varsigma+O(\alpha^2)$ and projecting the $O(\alpha)$ equation onto $\psi_i$ removes the unknown first-order correction to the mode, since $-\partial_x^2+V_0$ is self-adjoint and $(-\partial_x^2+V_0-\Om_0^2)\psi_i=0$. The coefficients of the combination then form an eigenvector, with eigenvalue $\varsigma$, of the matrix
\begin{equation}
  \mathcal S_{ij}=\frac{\int_{-\pi/2}^{\pi/2}\dd x\;\psi_i\Big[\big(V_{1a}-\Om_0^2V_{1b}\big)_{\mathsf c_i\mathsf c_j}
  +\delta_{\mathsf c_i\mathsf c_j}\,2\rho\,\iota\,V_{0,\mathsf c_i}\Big]\psi_j}{\int_{-\pi/2}^{\pi/2}\dd x\;\psi_i^2} ,
  \label{eq:Smatrix}
\end{equation}
where $i,j$ run over the modes of the level and $\mathsf c_i\in\{+,-\}$ is the channel of mode $i$. The shift $\varsigma$ is an eigenvalue of $\mathcal S$, as in degenerate first-order perturbation theory \cite{Kato:1995}. The integrals converge, since the integrands grow at most like $\cos^{-3}x$, up to a logarithm, against $\psi_i^2\sim\cos^{2s}x$ with $s\ge2$. The region $\alpha\rho\gtrsim1$, where the $O(\alpha)$ form of $V$ and the expansion of $\rho(x)$ both fail, contributes only at relative order $\alpha^{2s-2}$. The $O(\alpha)$ tortoise coordinate is unbounded, since $\iota\to-\tfrac{\pi}{2}\log\rho$ at large $\rho$, and its range is not $(-\pi/2,\pi/2)$. We keep the interval and the normalisable condition of the $\alpha=0$ problem. The boundary condition at $\alpha\rho\sim1$, where the throat opens into the mouths, is not analysed. For $l=2$ the two non-degenerate levels give $\varsigma=-\tfrac{194}{15}+8\log2=-7.39$ at $\Om_0=2$ and $\varsigma=-\tfrac{381}{10}+18\log2=-25.62$ at $\Om_0=3$. On the first three shared levels, with the modes ordered as $(Z_+,Z_-)$,
\begin{equation}
\begin{aligned}
  \Om_0=4:&\quad \mathcal S=\begin{pmatrix}-\tfrac{748}{21}+32\log2 & \tfrac{48}{35}\\[2pt] \tfrac{9}{5} & -\tfrac{1172}{15}+32\log2\end{pmatrix},\\
  \Om_0=5:&\quad \mathcal S=\begin{pmatrix}-\tfrac{6415}{84}+50\log2 & \tfrac{10}{7}\\[2pt] \tfrac{15}{2} & -\tfrac{805}{6}+50\log2\end{pmatrix},\\
  \Om_0=6:&\quad \mathcal S=\begin{pmatrix}-\tfrac{923}{7}+72\log2 & \tfrac{22}{15}\\[2pt] \tfrac{99}{5} & -\tfrac{7248}{35}+72\log2\end{pmatrix},
\end{aligned}
\label{eq:Sl2}
\end{equation}
with eigenvalues $\varsigma=-13.38,\,-56.01$ at $\Om_0=4$, $\varsigma=-41.53,\,-99.69$ at $\Om_0=5$ and $\varsigma=-81.57,\,-157.56$ at $\Om_0=6$. Every diagonal entry is a rational number plus $2\Om_0^2\log2$, on every level (Appendix~\ref{app:numerics}). The logarithmic part is the same for $Z_+$ and $Z_-$ and moves the two modes of a shared level together. The difference $\mathcal S_{++}-\mathcal S_{--}$ of the diagonal entries is rational, $\tfrac{1488}{35}$, $\tfrac{4855}{84}$ and $\tfrac{2633}{35}$ on the three levels. The off-diagonal entries depend on the normalisation of \eqref{eq:PTmodes}, but their product does not. It is $\tfrac{432}{175}$, $\tfrac{75}{7}$ and $\tfrac{726}{25}$ on these three levels. A positive product makes the two eigenvalues of $\mathcal S$ real whatever its diagonal entries, and the two modes of a shared level repel. On every shared level the product is
\begin{equation}
  \mathcal S_{+-}\mathcal S_{-+}=\frac{\mu^2}{4(2l+1)^2}\,\frac{\Om_0^2(\Om_0-l)(\Om_0-l-1)(\Om_0+2l+1)^2}{(\Om_0+l)(\Om_0+l+1)},
  \label{eq:prodS}
\end{equation}
with $\mu^2=(l-1)(l+2)$ of \eqref{eq:chandra-def}. The product is positive for every $\Om_0\ge l+2$ (Appendix~\ref{app:offdiag}). Every first-order shift of the closed throat is therefore real. For $l=2$ and $\Om_0\le6$ a direct numerical solution of the $O(\alpha)$ eigenvalue problem at $\alpha=10^{-3}$, $3\times10^{-4}$ and $10^{-4}$ reproduces these values to $O(\alpha)$ and finds no complex eigenvalue. Appendix~\ref{app:numerics} describes the quadrature, the identification of the closed forms and the direct solution.

The off-diagonal entries have a simple structure. On a shared level the two modes have the same $\Om_0$. Up to terms whose matrix elements between them vanish, both off-diagonal entries of $V_{1a}-\Om_0^2V_{1b}$ are then constant multiples of
\begin{equation}
  V_{\rm mix}=\lhat-2\big(\Om_0^2-1\big)\cos^2x .
  \label{eq:Vmix}
\end{equation}
The constants are $l(l-1)/[2(l+1)(2l+1)]$ for $\mathcal S_{+-}$ and $(l+1)(l+2)/[2l(2l+1)]$ for $\mathcal S_{-+}$ (Appendix~\ref{app:offdiag}). The mixing comes entirely from the Hall current of Sec.~\ref{sec:hall}. The current enters only the equation for $w$. With the scaled lift its terms in $\tilde P_1$ and $V_{1b}$ are $(2/\lhat)(\rho/f)\,\partial_xK_F$ and $(2/\lhat)K_F/f$, with $K_F$ the relative decrease of $|F|$ in \eqref{eq:KB}. We repeated the reduction with the source of the current removed from the Maxwell equation, solving the same equations for the second-order system. The constant multiples of $V_{\rm mix}$ then vanish identically for both lifts, and only terms with vanishing matrix elements remain. The remaining equations of Step~4 fail without the current. The comparison therefore identifies the part of the potential carried by the current and does not define a closure without it. The other terms of the potential, the commutator \eqref{eq:commutator} among them, contribute to the off-diagonal entries, but their contributions cancel on every shared level.

With $\Om=\Om_0+\alpha\varsigma/2\Om_0$, a negative shift lowers the normal frequency. Here $\Om$ refers to the clock of the static observer at $\rho=0$. A constant $\gamma_0$ added to $\gamma$ changes the clock and adds $\gamma_0\Om_0^2$ to every $\varsigma$ (Sec.~\ref{sec:bgalpha}). The term $2\Om_0^2\log2$ of every diagonal entry has this form, and with $\gamma(0)=-2\log2$ every diagonal entry would be rational. The splittings, the off-diagonal entries of $\mathcal S$, the absence of imaginary parts and the differences between the two parity sectors on the same background do not depend on this choice. The frequency measured at infinity follows from $\Om$ through the redshift between the centre of the throat and infinity, which at $O(\alpha)$ needs $\ell$ to relative order $\alpha$. The matching of Sec.~\ref{sec:overlap} does not fix it. On every level computed $|\varsigma|<7.2\,\Om_0^2$, for the eigenvalues and for the diagonal entries of $\mathcal S$. On these levels, for $l=2$ and $3$, the ratio $|\varsigma|/\Om_0^2$ grows roughly like $2\log\Om_0$ (Sec.~\ref{sec:hll-throat}). The frequency shift is then below $4\alpha\Om_0$, and first-order perturbation theory needs $\alpha\Om_0\log\Om_0\ll1$, which keeps it below the unit level spacing. For $l=2$ and $l=3$ all the shifts we computed are negative, and for this clock the $O(\alpha)$ corrections lower every one of these normal frequencies. For larger $l$ the upper mode of the lowest shared levels moves up. At $l=4$ the shifts on the shared level $\Om_0=6$ are $\varsigma=20.52$ and $-85.03$, and at $l=6$ the upper mode has $\varsigma=130.67$, $107.49$ and $73.83$ on the three lowest shared levels. The non-degenerate $Z_-$ levels move down for every $l$ examined. The corrections also lift the degeneracy between $Z_+$ and $Z_-$. On a shared level the two modes, degenerate at $\alpha=0$, move to $\Om^2=\Om_0^2+\alpha\varsigma_1$ and $\Om_0^2+\alpha\varsigma_2$, and the gap $\alpha(\varsigma_1-\varsigma_2)$ between them is $42.6\,\alpha$, $58.2\,\alpha$ and $76.0\,\alpha$ on the shared levels of $l=2$. This splitting is the leading imprint of the $O(\alpha)$ backreaction on the level structure of the throat. The higher Landau levels split these levels by a larger amount, of order $\eta_{\rm H}$ (Sec.~\ref{sec:hll-throat}). The non-symmetric part of the $O(\alpha)$ potential does not produce oscillatory growth at first order on any of these levels.

The shifts depend on the scaling \eqref{eq:linescale} of the line metric. The coordinate lift (Sec.~\ref{sec:response}) has $P_1-\delta_{\rm c}P_1$ and $Q_1-\delta_{\rm c}Q_1$ in place of the matrices \eqref{eq:t-P1} and \eqref{eq:t-Q1}, with
\begin{equation}
  \delta_{\rm c}P_1 \;=\; -\,\frac{\rho}{f^{2}}\begin{pmatrix}0&0\\0&1\end{pmatrix},
  \qquad
  \delta_{\rm c}Q_1 \;=\; \begin{pmatrix}
  -\dfrac{\lhat}{3f} & \dfrac{\lhat(\lhat+2)}{6f}\\[10pt]
  \dfrac{(\lhat-4)f-6}{6f^{2}} & \dfrac{2\lhat f(\rho^{2}+4)-\lhat^{2}f^{2}+12p^{2}}{12f^{3}}
  \end{pmatrix}.
  \label{eq:liftdiff}
\end{equation}
Table~\ref{tab:lifts} compares the eigenvalues of $\mathcal S$ for the two lifts. With the coordinate lift the diagonal entries of $\mathcal S$ are again rational numbers plus $2\Om_0^2\log2$. On every shared level the product of the off-diagonal entries is negative and equal to $-4/\mu^2$ times \eqref{eq:prodS}. Entry by entry, $\mathcal S_{+-}$ is multiplied by $-2/(l-1)$ and $\mathcal S_{-+}$ by $2/(l+2)$. With this lift the Hall current responds to $K$ instead of $K_F$, and its terms are those above with $K_F$ replaced by $K$ (Appendix~\ref{app:offdiag}). On the levels computed the eigenvalues stay real, since the diagonal entries are far apart. These levels go up to $\Om_0=20$ for $l=2$, $\Om_0=15$ for $l=3$ and $\Om_0=l+5$ for $l=4,5,6$. On them every shift is negative, including the upper modes that move up with the scaled lift for $l\ge4$. Signs refer to the clock of Sec.~\ref{sec:bgalpha}. The diagonal entries change by up to 58 per cent at $l=2$, and the upper mode of the shared level $\Om_0=4$ moves from $-13.38$ to $-21.27$. The gaps between the two shifts of each shared level of $l=2$ decrease by 4.5 to 7.5 per cent. With the coordinate lift the positivity thresholds of Sec.~\ref{sec:stab-throat1} are the same at $\alpha=10^{-6}$ and lie below those of the scaled lift by at most $0.002$ at $\alpha=10^{-2}$. They are set by the $\rho^3$ terms of the corrected background, which both lifts share. Three results therefore rest on the scaling. They are the values of the shifts, the sign of the products, and the coincidence of the products with the companion sector (Sec.~\ref{sec:summary}). The shifts are real on every level with the scaled lift and on the levels examined with the coordinate lift.
\begin{table}[htbp]
\caption{First-order shifts of the normal frequencies of the closed throat for the two lifts of Sec.~\ref{sec:response}. The scaled lift is the lift \eqref{eq:lift} with the metric of each field line scaled by the local field strength, Eq.~\eqref{eq:linescale}. The coordinate lift is the lift on the coordinate planes $(\tau,\rho)$. A level with unperturbed frequency $\Om_0$ moves to $\Om^2=\Om_0^2+\alpha\varsigma$. For $\Om_0=l$ and $\Om_0=l+1$ only $Z_-$ has a mode, and $\varsigma$ is a single number. For $\Om_0\ge l+2$ the level is shared by the mode of $Z_+$ with quantum number $n=\Om_0-l-2$ and the mode of $Z_-$ with $n=\Om_0-l$. The two modes mix, and the two values of $\varsigma$ are the eigenvalues of the $2\times2$ matrix $\mathcal S$ of Eq.~\eqref{eq:Smatrix}, the larger one first. With the higher Landau levels of Sec.~\ref{sec:hll} the two modes no longer mix at first order, and the shifts are the diagonal entries of $\mathcal S$, within 3.2 per cent of these values. The last two columns give the product $\mathcal S_{+-}\mathcal S_{-+}$ of its off-diagonal entries, which does not depend on the normalisation of the modes. It is \eqref{eq:prodS} for the scaled lift and $-4/\mu^2$ times \eqref{eq:prodS} for the coordinate lift. A positive product makes the eigenvalues real for any diagonal entries. With a negative product they are real here since the diagonal entries are far apart. The frequencies refer to the clock of Sec.~\ref{sec:bgalpha}.}
\label{tab:lifts}
\centering
\setlength{\tabcolsep}{6pt}
\begin{tabular}{rrlrrrr}
\hline\hline
 & & & \multicolumn{2}{c}{$\varsigma$} & \multicolumn{2}{c}{$\mathcal S_{+-}\mathcal S_{-+}$}\\
\cline{4-5}\cline{6-7}
$l$ & $\Om_0$ & modes & scaled lift & coordinate lift & scaled lift & coordinate lift\\
\hline
2 & 2 & $Z_-$ & $-7.39$ & $-8.59$ &  & \\
2 & 3 & $Z_-$ & $-25.62$ & $-28.32$ &  & \\
2 & 4 & $Z_+$, $Z_-$ & $-13.38$, $-56.01$ & $-21.27$, $-60.69$ & $432/175$ & $-432/175$\\
2 & 5 & $Z_+$, $Z_-$ & $-41.53$, $-99.69$ & $-52.62$, $-106.81$ & $75/7$ & $-75/7$\\
2 & 6 & $Z_+$, $Z_-$ & $-81.57$, $-157.56$ & $-96.41$, $-167.57$ & $726/25$ & $-726/25$\\
3 & 3 & $Z_-$ & $-13.49$ & $-17.44$ &  & \\
3 & 4 & $Z_-$ & $-36.14$ & $-42.54$ &  & \\
3 & 5 & $Z_+$, $Z_-$ & $-2.69$, $-70.09$ & $-22.44$, $-79.40$ & $250/49$ & $-100/49$\\
3 & 6 & $Z_+$, $Z_-$ & $-31.26$, $-116.19$ & $-56.47$, $-128.86$ & $1014/49$ & $-2028/245$\\
4 & 4 & $Z_-$ & $-20.11$ & $-30.90$ &  & \\
4 & 5 & $Z_-$ & $-47.13$ & $-62.01$ &  & \\
4 & 6 & $Z_+$, $Z_-$ & $20.52$, $-85.03$ & $-21.52$, $-104.45$ & $90/11$ & $-20/11$\\
4 & 7 & $Z_+$, $Z_-$ & $-7.78$, $-134.51$ & $-58.58$, $-158.91$ & $3136/99$ & $-6272/891$\\
6 & 8 & $Z_+$, $Z_-$ & $130.67$, $-103.57$ & $-8.24$, $-175.55$ & $2688/169$ & $-1344/845$\\
\hline\hline
\end{tabular}
\end{table}
\subsection{The mouth}
\label{sec:stab-mouth}

In the mouth we use Chandrasekhar's decoupled variables $Z_i$ of \eqref{eq:Zi-Kw}, which obey the Schr\"odinger equations \eqref{eq:Zi-eq} with the potentials $V^{(+)}_i$ of \eqref{eq:chandra-V}. These are isospectral to the axial potentials, which have the simpler closed form
\begin{equation}
  V_i^{(-)} = \frac{\Delta}{r^5}\left[\left(\mu^2+2\right)r - \beta_i + \frac{4r_e^2}{r}\right].
  \label{eq:chandra-mouth}
\end{equation}
For $V^{(+)}_i$ we checked positivity on the mouth numerically for $l=2,\dots,10$, at extremality and on a horizonless mouth. At extremality the ratio $r^4V^{(+)}_i/(r-r_e)^2$ lies between its value at $r=r_e$, the near-horizon mass of \eqref{eq:nhmasses}, and its value $\lhat$ at infinity (Appendix~\ref{app:numerics}). The potentials $V^{(-)}_i$ are positive on the mouth for every $l\ge2$, and the argument is analytic. On the wormhole mouth $GM=r_e+\delta_M$ with $\delta_M<0$, hence $\Delta=(r-r_e)^2-2\delta_M r>0$ everywhere and there is no horizon. The domain is $r\ge r_{\rm J}$, with $r_{\rm J}$ the junction radius with the throat and $r_{\rm J} - r_e\ll r_e$. The bracket in $V_1^{(-)}$ evaluated at exact extremality and $r=r_e$ is $r_e\,l(l-1)\ge0$, and its minimum over $r$ lies at $r^2=4r_e^2/(\mu^2+2)<r_e^2$ for $l\ge2$, outside the domain. The bracket therefore increases monotonically outward. The two descriptions join at the junction. In the extremal near-horizon limit the potentials reduce to \eqref{eq:nhmasses}, with the masses of \eqref{eq:Vthroat}, and the variables $Z_i$ reduce to $Z_\mp$ (Sec.~\ref{sec:overlap}).

At $O(\alpha)$ the mouth equations are those of Sec.~\ref{sec:mouth-master}. The duality fails at this order, and we have not constructed decoupled variables for it. The mouth then has no Schr\"odinger form at this order, and \eqref{eq:quadform} does not apply to it as written. The $O(\alpha)$ terms, including the transport coupling \eqref{eq:Kchi}, are small everywhere in the mouth (Sec.~\ref{sec:mouth-master}). A growing mode at first order in $\alpha$ would have to come from these terms, acting on an $\alpha^0$ problem whose quadratic form is positive on the whole mouth. Near the junction the dominant correction is the move of the minimum of $N$ to $r_h$, which replaces $y$ by $y-\alpha r_e/6$ in the potentials \eqref{eq:nhmasses} and leaves them positive. The expansion itself requires $y\gg\alpha r_e$, which holds at the junction by a factor $4\times10^{28}$ (Sec.~\ref{sec:mouthpert}). We therefore expect no growing mode in the mouth at $O(\alpha)$. This is an estimate from the size of the corrections and not a proof. With the higher Landau levels the mouth is a problem of nonlinear electrodynamics. For $l=2,\dots,10$ and the values of $\eta_{\rm H}$ we scanned its master potential is positive, and the mouth has no growing mode at $O(\eta_{\rm H})$ (Sec.~\ref{sec:hll-mouth}).

\section{The higher Landau levels}
\label{sec:hll}

The previous sections keep only the lowest Landau level. The higher levels are gapped by the cyclotron energy and are never excited here, but their vacuum polarisation is not suppressed by $1/q$. For a massless Dirac fermion of unit charge it is the one-loop Euler-Heisenberg action \cite{Heisenberg:1936nmg,Schwinger:1951nm,Dunne:2004nc} with the lowest level removed, and its logarithm is the running of the gauge coupling. We include it as a second small parameter,
\begin{equation}
  \eta_{\rm H}\;\equiv\;\frac{g_c^2}{12\pi^2},
  \qquad
  \frac{\eta_{\rm H}}{\alpha}\;=\;\frac q3 ,
  \label{eq:etaH}
\end{equation}
to first order. At $g_c=1$ one has $\eta_{\rm H}=8.4\times10^{-3}$, larger than $\alpha$ by $3\times10^{32}$ at the fiducial parameters. To first order the two corrections add. Each is sourced on the Einstein--Maxwell solution, and the stress tensor of the lowest level and the one-loop action enter the background and the perturbation equations linearly. The first term that couples them is of order $\alpha\eta_{\rm H}$, and we do not compute it. The results of the previous sections therefore stand, and the effects of this section add to them. They also change one conclusion. The levels shared by $Z_+$ and $Z_-$ are split at $O(\eta_{\rm H})$. On the levels examined this pushes the mixing of Sec.~\ref{sec:stab-normalmodes} to second order (Sec.~\ref{sec:hll-throat}).

\subsection{The one-loop action}
\label{sec:hll-action}

We write the Lagrangian of the gauge field as $\mathcal L(\mathcal Q,\mathcal R)$, a function of the invariants $\mathcal Q=-\tfrac14F_{\mu\nu}F^{\mu\nu}$ and $\mathcal R=-\tfrac14F_{\mu\nu}\tilde F^{\mu\nu}$, with $\tilde F^{\mu\nu}=\tfrac12\varepsilon^{\mu\nu\rho\sigma}F_{\rho\sigma}$. Maxwell theory is $\mathcal L=\mathcal Q/g_c^2$. The stress tensor and the field equation become
\begin{equation}
  T^{\rm mag}_{\mu\nu}=\mathcal L_{\mathcal Q}\,F_{\mu\lambda}F_\nu{}^\lambda+g_{\mu\nu}\big(\mathcal L-\mathcal R\,\mathcal L_{\mathcal R}\big),
  \qquad
  \nabla_\mu\big(\mathcal L_{\mathcal Q}F^{\mu\nu}+\mathcal L_{\mathcal R}\tilde F^{\mu\nu}\big)=J^\nu ,
  \label{eq:nled}
\end{equation}
with subscripts denoting derivatives. For $\mathcal L=\mathcal Q/g_c^2$ they are \eqref{eq:Tmag} and \eqref{eq:maxwell}. On the magnetic background $\mathcal R=0$ and $\mathcal Q=-\tfrac12\bar B^2$, with $\bar B=|\bar F|=q/2R^2$ the field strength of Sec.~\ref{sec:lift} in an orthonormal frame. It is not the metric function $B$ of \eqref{eq:bgmetric}. The linearised equations need $\mathcal L_{\mathcal Q}$, $\mathcal L_{\mathcal Q\mathcal Q}$ and $\mathcal L_{\mathcal R\mathcal R}$ there, and the mixed derivative vanishes, since $\mathcal L$ is even in $\mathcal R$. For the levels $n\ge1$ of one massless Dirac fermion (Appendix~\ref{app:EH})
\begin{equation}
  \mathcal L=-\tfrac12\bar B^2\mathsf h(\bar B),
  \qquad
  \mathsf h(\bar B)=\frac{1}{g_c^2}\Big[1-\eta_{\rm H}\log\frac{\bar B}{B_0}\Big],
  \label{eq:EHcoef1}
\end{equation}
\begin{equation}
  \varepsilon_\perp\equiv g_c^2\mathcal L_{\mathcal Q}=1-\eta_{\rm H}\log\frac{\bar B}{B_0}-\frac{\eta_{\rm H}}2 ,
  \qquad
  g_c^2\bar B^2\mathcal L_{\mathcal Q\mathcal Q}=\eta_{\rm H},
  \qquad
  g_c^2\bar B^2\mathcal L_{\mathcal R\mathcal R}=c_\parallel\,\eta_{\rm H},
  \label{eq:EHcoef2}
\end{equation}
with $c_\parallel=\gamma_{\rm E}-\tfrac12+12\,\zeta'(-1)=-1.9078\ldots$, $\gamma_{\rm E}$ Euler's constant and $\zeta$ the Riemann zeta function. We define $g_c$ by $\mathsf h(B_0)=1/g_c^2$ at the field $B_0=q/2r_e^2$ of the throat, with $r_e$ the radius of its sphere. Then \eqref{eq:re} holds as it stands, and $g_c$ is the coupling at the scale $\sqrt q/r_e$ used by MMP \cite{Maldacena:2018gjk}. The logarithm in $\mathsf h$ is the running of $1/g_c^2$ with the beta function of one Dirac fermion. A change of scheme adds a multiple of $\mathcal Q$ to $\mathcal L$, which is absorbed into $g_c$, and $c_\parallel$ does not depend on the scheme. An electric field across $\bar B$ enters $\mathcal L$ as $\varepsilon_\perp E^2/2g_c^2$, and an electric field along $\bar B$ as $\varepsilon_\parallel E^2/2g_c^2$ with $\varepsilon_\parallel=\varepsilon_\perp+c_\parallel\eta_{\rm H}$. The polar-metric sector has $\delta\mathcal R=0$ at $O(\epsilon)$, and only $\mathcal L_{\mathcal Q}$ and $\mathcal L_{\mathcal Q\mathcal Q}$ enter. The constant $c_\parallel$ enters the companion sector \cite{Sadhukhan:2026mcu}.

The one-loop action is local. It holds for fields that vary slowly on the magnetic length $\bar B^{-1/2}\sim R/\sqrt q$ and for frequencies below the gap $\sqrt q/R$ of the higher levels. The first corrections are of relative order $1/q$ and $l(l+1)/q$. With $\eta_{\rm H}/q=\alpha/3$ they are of the order of the effects of the lowest level. They contain couplings between the field strength and the curvature, derivatives of the field along the sphere and the discreteness of the Landau spectrum on the sphere. We do not compute them. We also neglect the transitions of the lowest level to the higher ones induced by the perturbation, and we have not estimated them.

\subsection{The backgrounds}
\label{sec:hll-bg}

In the throat $\bar B=B_0$, and the $\epsilon^0$ equations keep AdS$_2\times S^2$,
\begin{equation}
  \dd s^2=R_A^2\Big[-(1+\rho^2)\,\dd\tau_A^2+\frac{\dd\rho^2}{1+\rho^2}\Big]+r_e^2\,\mathring g,
  \qquad
  R_A^2=\frac{r_e^2}{1-\eta_{\rm H}} .
  \label{eq:hllthroat}
\end{equation}
The sphere radius is fixed by the energy density $\tfrac12\bar B^2\mathsf h$ and stays $r_e$. The AdS$_2$ radius is fixed by $\bar B^2(\mathcal L_{\mathcal Q}-\tfrac12\mathsf h)$ and grows. The difference $1/r_e^2-1/R_A^2$ is $-\kappa/2$ times the trace of the stress tensor, $4(\mathcal L-\mathcal Q\mathcal L_{\mathcal Q})=-\bar B^2/12\pi^2$, the trace anomaly of the higher levels. It plays the role that the trace part of the lowest level plays in Sec.~\ref{sec:bgalpha}.

In the mouth $\bar B=q/2r^2$ and $\log(\bar B/B_0)=-2\log(r/r_e)$. The energy density $\tfrac12\bar B^2\mathsf h$ exceeds its Maxwell value at large $r$, where the coupling is weaker. With $B=1/A$ and $R=r$ the $\epsilon^0$ equations are solved by
\begin{equation}
  A=1-\frac{2GM}{r}+\frac{r_e^2}{r^2}\Big[1+2\eta_{\rm H}\Big(1+\log\frac{r}{r_e}\Big)\Big]
  \label{eq:hllmouth}
\end{equation}
for every $GM$, exactly for the one-loop action. The mouth is extremal at $GM=r_e(1+\eta_{\rm H})$. There $A$ has a double zero at $r=r_e$ and $A\simeq(r-r_e)^2/R_A^2$ near it, with $R_A$ of \eqref{eq:hllthroat}. The near-horizon region is the throat. With the shift of \eqref{eq:mouthmass} the extremal mass is $r_e(1+\eta_{\rm H}-\alpha/18)$ at first order, and the wormhole mouth lies below it by $|\delta_M|$. The charge radius seen from far away grows slowly with $r$, as $r_e[1+2\eta_{\rm H}(1+\log(r/r_e))]^{1/2}$. We keep \eqref{eq:hllmouth} and the potentials below without expanding in $\eta_{\rm H}\log(r/r_e)$, which grows at large $r$. The one-loop coupling decreases there, and the higher loops stay small.

Up to terms of order $\alpha\eta_{\rm H}$ the throat and the mouth with both corrections are
\begin{equation}
\begin{aligned}
  \text{throat:}&\quad \dd s^2=R_A^2\Big[-(1+\rho^2+\alpha\gamma)\,\dd\tau_A^2+\frac{\dd\rho^2}{1+\rho^2+\alpha\gamma}\Big]+r_e^2(1+\alpha\varphi)\,\mathring g,\\
  \text{mouth:}&\quad \dd s^2=-(1+2\alpha\Phi_1)N\,\dd t^2+\frac{\dd r^2}{N}+r^2\mathring g,\\
  &\quad N=1-\frac{2GM}{r}+\frac{r_e^2}{r^2}\Big[1+2\eta_{\rm H}\Big(1+\log\frac{r}{r_e}\Big)\Big]+\alpha N_1,
\end{aligned}
\label{eq:bgboth}
\end{equation}
with $R_A$ of \eqref{eq:hllthroat}, $\gamma$ and $\varphi$ of \eqref{eq:bgsol}, $N_1$ and $\Phi_1$ of \eqref{eq:mouthbgsol}, and $GM=r_e(1+\eta_{\rm H}-\alpha/18)+\delta_M$.

\subsection{The throat}
\label{sec:hll-throat}

The reduction of Sec.~\ref{sec:algorithm} runs as before with the field equations \eqref{eq:nled}, and the residuals of Step~4 vanish identically. In the tortoise coordinate $x=\arctan\rho$ the first-derivative matrix vanishes and the potential is $(1+\rho^2)\,\mathcal N_{\rm H}$ for the time $\tau_A$ of \eqref{eq:hllthroat}, with the constant matrix
\begin{equation}
  \mathcal N_{\rm H}=\frac{1}{1-\eta_{\rm H}}
  \begin{pmatrix}
    \lhat+2-2\eta_{\rm H} & -\lhat\,(2-\eta_{\rm H})\\[4pt]
    -\dfrac{2(2-3\eta_{\rm H})}{2-\eta_{\rm H}} & \dfrac{\lhat\,(2-3\eta_{\rm H})}{2-\eta_{\rm H}}
  \end{pmatrix}
  \label{eq:NH}
\end{equation}
in the basis $(K,w)$. At $\eta_{\rm H}=0$ it is the matrix $\mathcal N$ of \eqref{eq:t-Qtilde0}. The system is again two Poschl-Teller problems. Their masses $m_\pm^2R_A^2$ are the eigenvalues of \eqref{eq:NH}, with sum $2(2\lhat+2-\eta_{\rm H})/(2-\eta_{\rm H})$ and product $\lhat(\lhat-2)(2-3\eta_{\rm H})/[(2-\eta_{\rm H})(1-\eta_{\rm H})^2]$. To first order
\begin{equation}
  m_-^2R_A^2=l(l-1)\Big[1+\eta_{\rm H}\,\frac{l+1}{2l+1}\Big],
  \qquad
  m_+^2R_A^2=(l+1)(l+2)\Big[1+\eta_{\rm H}\,\frac{l}{2l+1}\Big],
  \label{eq:mH}
\end{equation}
and the decoupled variables are
\begin{equation}
  Z_+=K-\Big[l-\eta_{\rm H}\,\frac{(l-1)l(l+1)}{2(2l+1)}\Big]w,
  \qquad
  Z_-=K+\Big[l+1+\eta_{\rm H}\,\frac{l(l+1)(l+2)}{2(2l+1)}\Big]w .
  \label{eq:ZH}
\end{equation}
Both masses grow in units of $R_A$, and the potentials stay positive. In units of $r_e$ they decrease, $m_-^2r_e^2=l(l-1)[1-\eta_{\rm H}l/(2l+1)]$ and $m_+^2r_e^2=(l+1)(l+2)[1-\eta_{\rm H}(l+1)/(2l+1)]$. The throat has no growing mode at this order. With the $O(\alpha)$ terms of Sec.~\ref{sec:stab-throat1} the potential becomes $V_0+\eta_{\rm H}V_\eta+\alpha(V_{1a}+p^2V_{1b})$, with $\eta_{\rm H}V_\eta$ the $O(\eta_{\rm H})$ part of $f\,\mathcal N_{\rm H}$. A constant change of basis leaves the kinetic term $-\partial_x^2$ unchanged. In the time $\tau_A$ and in the eigenbasis of \eqref{eq:NH} the $O(\eta_{\rm H})$ part is $(1+\rho^2)$ times a positive diagonal matrix, with entries $\eta_{\rm H}l(l+1)(l+2)/(2l+1)$ and $\eta_{\rm H}l(l^2-1)/(2l+1)$, while $V_{1a}$ and $V_{1b}$ change only at order $\alpha\eta_{\rm H}$. The smallest eigenvalue of the symmetric part therefore cannot drop below its value at $\eta_{\rm H}=0$, and a scan with both corrections confirms this (Appendix~\ref{app:numerics}). The point where positivity is first lost moves outwards. In the eigenbasis of \eqref{eq:NH} it moves for $l=2$ from $\alpha\rho=0.241$ to $0.242$ at $\eta_{\rm H}=1/12\pi^2$ and to $0.251$ at $\eta_{\rm H}=0.1$. The offsets $s_\pm$, with $s(s-1)=m^2R_A^2$, are no longer integers, and the property of Sec.~\ref{sec:alpha0} that each solution keeps a single phase across the throat is lost at $O(\eta_{\rm H})$.

In the time $\tau_A$ the levels of the closed throat are $\Om_A=n+s_\pm$, with
\begin{equation}
  s_-=l+\eta_{\rm H}\,\frac{(l-1)l(l+1)}{(2l-1)(2l+1)},
  \qquad
  s_+=l+2+\eta_{\rm H}\,\frac{l(l+1)(l+2)}{(2l+1)(2l+3)} .
  \label{eq:sH}
\end{equation}
The clock of Sec.~\ref{sec:bgalpha} is the proper time of the static observer at $\rho=0$ in units of $r_e$, which is $R_A\tau_A/r_e$. On this clock $\Om=\Om_A\sqrt{1-\eta_{\rm H}}$, and a level moves to
\begin{equation}
  \Om^2=\Om_0^2+\alpha\varsigma+\eta_{\rm H}\varsigma_{\rm H},
  \qquad
  \varsigma_{\rm H}=2\Om_0\,\frac{\partial s}{\partial\eta_{\rm H}}-\Om_0^2 .
  \label{eq:varsigmaH}
\end{equation}
The term $-\Om_0^2$ is the change from $\tau_A$ to the clock. For $l=2$ one finds $\varsigma_{\rm H}=-\tfrac{12}5$ and $-\tfrac{33}5$ on the levels $\Om_0=2$ and $3$ of $Z_-$. On every level $\Om_0\ge l>2\,\partial s_-/\partial\eta_{\rm H}$ for $Z_-$ and $\Om_0\ge l+2>2\,\partial s_+/\partial\eta_{\rm H}$ for $Z_+$. Every $\varsigma_{\rm H}$ is therefore negative, and on this clock the higher Landau levels lower all normal frequencies of the throat. On a shared level $\Om_0=n+l+2$ the $Z_+$ mode with quantum number $n$ and the $Z_-$ mode with $n+2$ separate by
\begin{equation}
  \Om_{Z_+}-\Om_{Z_-}=\eta_{\rm H}\,\frac{l(l+1)}{(2l-1)(2l+3)},
  \label{eq:splitH}
\end{equation}
which is $2\eta_{\rm H}/7$ for $l=2$ on every shared level and does not depend on the clock. The perturbation \eqref{eq:NH} does not mix the two modes. In the basis of \eqref{eq:Zdef} its first-order part is a constant matrix times $\sec^2x$. The overlap of $\cos^{l+2}x\,C^{(l+2)}_n(\sin x)$ with $\sec^2x\cos^lx\,C^{(l)}_{n+2}(\sin x)$ vanishes, since $C^{(l)}_{n+2}$ is orthogonal to every polynomial of lower degree with the weight $(1-u^2)^{l-1/2}$ in $u=\sin x$.

On the levels we computed the splitting \eqref{eq:splitH} exceeds the entries $\alpha\mathcal S_{ij}$ of \eqref{eq:Smatrix} by a factor of order $q$. Degenerate perturbation theory in $\alpha$ then starts from split levels. The $O(\alpha)$ shift of each mode is its diagonal entry, $\mathcal S_{++}$ or $\mathcal S_{--}$, a matrix element of a real potential between real modes, and it is real. The off-diagonal entries enter at order $\alpha^2/\eta_{\rm H}=3\alpha/q$, beyond the order kept. On the levels of Table~\ref{tab:lifts} the diagonal entries differ from the eigenvalues listed there by at most $0.41$ in absolute value and by at most 3.1 per cent of the entry. The values are those of \eqref{eq:Sl2} for $l=2$. The positive products of the off-diagonal entries, and the repulsion of Sec.~\ref{sec:stab-normalmodes}, describe the throat at $\eta_{\rm H}=0$. With $\eta_{\rm H}\neq0$ the splitting exceeds the $O(\alpha)$ shifts as long as $\alpha|\varsigma|\ll\eta_{\rm H}\Om_0/2$. The ratio $|\varsigma|/\Om_0^2$ grows slowly with $\Om_0$. For $l=2$ it rises from $3.5$ at $\Om_0=4$ to $6.8$ at $\Om_0=20$ with the scaled lift, and to $7.1$ with the coordinate lift, roughly as $2\log\Om_0$. For $l=3$ it grows at the same rate and stays below $5.5$ up to $\Om_0=15$. If this growth continues, the splitting exceeds the $O(\alpha)$ shifts for $\Om_0\log\Om_0\ll q/12$. That range contains every level below the gap $\sqrt q$ of the higher Landau levels, where the one-loop action applies. Under this assumption no level below the gap is degenerate at first order, and oscillatory growth cannot appear there at first order. With both corrections a shared level carries the first-order matrix $\alpha\mathcal S+\eta_{\rm H}\,{\rm diag}(\varsigma_{{\rm H}+},\varsigma_{{\rm H}-})$, since the higher levels do not mix the two modes. Its discriminant is
\begin{equation}
  \big[\alpha(\mathcal S_{++}-\mathcal S_{--})+\eta_{\rm H}(\varsigma_{{\rm H}+}-\varsigma_{{\rm H}-})\big]^2+4\alpha^2\mathcal S_{+-}\mathcal S_{-+} .
  \label{eq:discH}
\end{equation}
Here $\eta_{\rm H}(\varsigma_{{\rm H}+}-\varsigma_{{\rm H}-})$ is $2\Om_0$ times \eqref{eq:splitH} and is positive. With the scaled lift the last term of \eqref{eq:discH} is positive on every shared level by \eqref{eq:prodS}, and the shifts are real for every ratio of the two parameters. With the coordinate lift it is negative. On every shared level examined in Sec.~\ref{sec:stab-normalmodes} one has $\mathcal S_{++}>\mathcal S_{--}$ for both lifts, and \eqref{eq:discH} is positive at $\eta_{\rm H}=0$. It then grows with $\eta_{\rm H}/\alpha$, and on these levels the shifts of the coordinate lift are also real for every ratio.

\subsection{The mouth}
\label{sec:hll-mouth}

On the background \eqref{eq:hllmouth} the reduction gives a closed first-order system for $(K,H_1,w,w')$ as in \eqref{eq:m-M}. The residuals of Step~4 vanish identically in $l$, $p$, $\eta_{\rm H}$, $GM$ and $r_e$. The family \eqref{eq:hllmouth} obeys $A'=(1-A)/r-r_e^2[1+2\eta_{\rm H}\log(r/r_e)]/r^3$ for every $GM$, and this equation stands in for $A$ in the reduction. At $GM=r_e(1+\eta_{\rm H})$ the near-horizon limit reproduces the throat equations with the matrix \eqref{eq:NH} identically in $l$ and $\eta_{\rm H}$. The one-loop action breaks the duality of Sec.~\ref{sec:mouth-RN}, and Chandrasekhar's variables no longer decouple the system. At the formal point $m_S^2=0$ of the axial sector its throat masses agree with those of this one only if $c_\parallel=1$, that is $\mathcal L_{\mathcal R\mathcal R}=\mathcal L_{\mathcal Q\mathcal Q}$ on the background \cite{Sadhukhan:2026mcu}.

Nomura, Yoshida and Soda found master variables for this sector for any $\mathcal L(\mathcal Q,\mathcal R)$ on a magnetic background \cite{Nomura:2020tpc}. In our notation they are
\begin{equation}
  \mathcal X_{\rm g}=\frac{\mu\,\lhat\,\mathcal Z}{r\,\mathcal Y},
  \qquad
  \mathcal X_{\rm e}=r_e\lhat\sqrt{\varepsilon_\perp}\,\Big(w-\frac{2\mathcal Z}{r^2\mathcal Y}\Big),
  \qquad
  \mathcal Y=rA'-2A+\lhat ,
  \label{eq:NYSvars}
\end{equation}
with $\mathcal Z$ of \eqref{eq:Xdef} and $\Delta$ there replaced by $r^2A$, $\mu^2=(l-1)(l+2)$, $\varepsilon_\perp=1-\eta_{\rm H}/2+2\eta_{\rm H}\log(r/r_e)$ and a prime denoting $\partial_r$. At $\eta_{\rm H}=0$ one has $r^2\mathcal Y=\mu^2r^2+6GMr-4r_e^2=(\mu^2r+\beta_1)(\mu^2r+\beta_2)/\mu^2$, and with $Z_1$ and $Z_2$ of \eqref{eq:Zi-Kw}
\begin{equation}
  \mathcal X_{\rm g}=\frac{\mu\lhat\,(\beta_1Z_1-\beta_2Z_2)}{\beta_1-\beta_2},
  \qquad
  \mathcal X_{\rm e}=\frac{2\mu^2r_e\lhat\,(Z_1-Z_2)}{\beta_1-\beta_2} .
  \label{eq:NYStoZ}
\end{equation}
The coefficients do not depend on $r$. For every solution of the first-order system
\begin{equation}
  \begin{aligned}
  -\partial_t^2\mathcal X_{\rm g}+\frac1r\,\partial_{r_*}\Big(r^2\,\partial_{r_*}\frac{\mathcal X_{\rm g}}{r}\Big)-A\big(V_{11}\mathcal X_{\rm g}+V_{12}\mathcal X_{\rm e}\big)&=0,\\
  -\partial_t^2\mathcal X_{\rm e}+\frac{1}{\sqrt{\varepsilon_\perp}}\,\partial_{r_*}\Big(\varepsilon_\perp\,\partial_{r_*}\frac{\mathcal X_{\rm e}}{\sqrt{\varepsilon_\perp}}\Big)-A\big(V_{12}\mathcal X_{\rm g}+V_{22}\mathcal X_{\rm e}\big)&=0,
  \end{aligned}
  \label{eq:NYSeqs}
\end{equation}
with $\dd r_*/\dd r=1/A$ and the symmetric matrix $V$ of \eqref{eq:NYSV}. We verified \eqref{eq:NYSeqs} identically in $r$, $l$, $p$, $\eta_{\rm H}$, $GM$ and $r_e$, which checks our reduction and their master equations against each other. At $\eta_{\rm H}\neq0$ the potentials of \eqref{eq:NYSeqs} written in Schr\"odinger form do not commute at different radii, and no constant combination of $\mathcal X_{\rm g}$ and $\mathcal X_{\rm e}$ decouples them. The system \eqref{eq:NYSeqs} conserves the energy
\begin{equation}
  \mathfrak E=\int\dd r_*\Big[|\partial_t\mathcal X_{\rm g}|^2+|\partial_t\mathcal X_{\rm e}|^2+r^2\Big|\partial_{r_*}\frac{\mathcal X_{\rm g}}{r}\Big|^2+\varepsilon_\perp\Big|\partial_{r_*}\frac{\mathcal X_{\rm e}}{\sqrt{\varepsilon_\perp}}\Big|^2+A\,(\mathcal X_{\rm g}^*,\mathcal X_{\rm e}^*)\,V\,(\mathcal X_{\rm g},\mathcal X_{\rm e})^{\!\top}\Big].
  \label{eq:NYSenergy}
\end{equation}
Its kinetic terms are positive for $\varepsilon_\perp>0$, and $A>0$ on the wormhole mouth. A positive definite $V$ therefore excludes a growing mode. We evaluated the smallest eigenvalue of $r^2V$ for $r_e<r\le10^4r_e$ at $GM=r_e(1+\eta_{\rm H})$, for $l=2,\dots,10$ and $\eta_{\rm H}=0$, $1/12\pi^2$, $0.05$ and $0.1$ (Appendix~\ref{app:numerics}). It is positive everywhere and smallest at $r=r_e$, where $R_A^2V$ has the eigenvalues of \eqref{eq:NH}, whose first-order form is \eqref{eq:mH}. For $l=2$ and $\eta_{\rm H}=1/12\pi^2$ its minimum is $1.993$, against $2$ at $\eta_{\rm H}=0$. For the values of $l$ and $\eta_{\rm H}$ scanned, the mouth of this sector has no growing mode with the higher Landau levels included, within the same region-by-region limits as in Sec.~\ref{sec:stab-mouth}. At $\eta_{\rm H}=0$ the scan repeats the positivity of the Einstein--Maxwell mouth in other variables.

\section{Conclusions and discussion}
\label{sec:summary}

We have studied linear perturbations of the MMP wormhole in the sector that pairs polar metric perturbations with axial gauge perturbations, the sector that carries the polar gravitational degree of freedom. On a magnetic background the parity sectors pair crosswise, as the selection rule \eqref{eq:selection} shows. In both regions we reduced the linearised Einstein--Maxwell--fermion system to closed master equations, Eqs.~\eqref{eq:m-master}--\eqref{eq:m-QKK} for the mouth and Eqs.~\eqref{eq:t-PQ0}--\eqref{eq:master} for the throat. At $O(\alpha^0)$ the mouth problem is Einstein--Maxwell theory. The axial gauge variable satisfies a Regge--Wheeler equation sourced by the gravitational variable, Eq.~\eqref{eq:wRW}. Electric--magnetic duality maps the mouth onto Chandrasekhar's polar sector of the Reissner-Nordstrom black hole, and we gave his variables in terms of our amplitudes, Eq.~\eqref{eq:Zi-Kw}. The near-horizon limit of the mouth reproduces the throat equations, and the Chandrasekhar variables continue into the Poschl-Teller variables of the throat.

The throat is where the new physics lies. Its background carries the Casimir stress tensor of the lowest Landau level. The linearised Bianchi identities hold at $O(\alpha)$ only when two ingredients are included together. The first is the exact response of the LLL stress tensor, Eq.~\eqref{eq:response}, lifted on the field lines of the full field strength with its full trace and sphere pressure, Eq.~\eqref{eq:lift}. MMP's traceless truncation is harmless for the background and inconsistent for the perturbation problem. The lift moves with the field lines. An azimuthal electric field tilts them, which gives the mixed components \eqref{eq:driftlift}, and a perturbed radial magnetic field changes their density. Once the tilt and the density are kept, the $\tau$ and $\rho$ force balance closes only if the metric of each line is scaled by the local field strength, Eq.~\eqref{eq:linescale}. On the timelike invariant plane of $F$ the same constitutive relation is the drift lift of the companion sector \cite{Sadhukhan:2026mcu}, where the density of field lines does not change and the relation is stated as an assumption. The second ingredient is the induced Hall current, Eq.~\eqref{eq:Jphi_explicit}. If either is dropped the first-order system has no nonzero solution at generic frequency. The current is an azimuthal drift of the charged fermions. The transverse gradient of the Casimir energy across the field lines and the drift of the field lines drive it, and it sources the axial gauge mode itself. With both, the source is conserved by construction, and Step~4 of the reduction confirms it. Consistency does not select the lift, since the coordinate lift passes Step~4 and the matching as well. We also derived the Hall current from the gauge variation of the effective action of the field lines, Eq.~\eqref{eq:Wcov}. The two derivations agree term by term, as they must for one functional. Through the trace anomaly, the scaling \eqref{eq:linescale} gives a coupling between the field strength and the curvature of the field-line worldsheet with coefficient $c_1/48\pi$, Eq.~\eqref{eq:W4d}, whose structure resembles a Wen-Zee term. A treatment on the coordinate planes must add the Euler term \eqref{eq:lambdafixed} by hand to reproduce it, and there the coefficient can only be fixed by consistency. The density of one fermion carries $c_1=1$ and the count $q$ is carried by the flux density, and the Hall current of a lowest Landau level does not depend on the number of flux quanta. The state of the fermions enters through two constants, the free left- and right-moving radiation along the field lines. In the throat these affect $H_1$, $H_2$ and the Hall current but not the master equations.

The fermions also act in the mouth. There the Casimir part of their stress tensor is $O(\alpha^3)$, but the anomaly part is $O(\alpha)$, the same order as in the throat, Eq.~\eqref{eq:TQmouth}. At both ends of the mouth it has the Maxwell form with ratio $\alpha/3$. The charge radius is renormalised to $r_e\sqrt{1+\alpha/3}$ at infinity and near $r_e$ alike, and the mass of the extremal mouth is $GM=r_e(1-\alpha/18)$, Eq.~\eqref{eq:mouthmass}. This is the renormalisation that MMP absorb into the radii. The perturbed mouth needs the same response and Hall current as the throat, Eq.~\eqref{eq:Jmouth}. With them the Bianchi identities hold at $O(\alpha)$ for every $l$ and $p$, and without the current they fail. The corrected mouth matches the $O(\alpha)$ throat in the overlap at every order that both calculations determine, including the $\alpha\rho^2$ terms \eqref{eq:overlap-arho2}. These terms carry the trace part of the source, its response, the density and drift of the field lines and the Hall current. The uncorrected mouth fails this test, and the matching reproduces the mass $GM=r_e(1-\alpha/18)$ at which $N$ and $N'$ vanish together at $O(\alpha)$. One effect has no counterpart in the throat. In the mouth the transport fields, and with them the state of the fermions, enter the equation for $K$ at $O(\alpha)$, Eq.~\eqref{eq:Kchi}, and the mouth problem becomes of sixth order. Under the matching they become terms of order $\alpha^3\rho^{-4}$ in the throat equations, two orders beyond the throat calculation. The comparison of Table~\ref{tab:junction} uses the $\alpha^0$ mouth and does not contain them. How the transport fields continue across the junction is left open. The running of $g_c$ is larger than these $O(\alpha)$ terms of the mouth, and Sec.~\ref{sec:hll} treats it.

At $\alpha=0$ the throat modes decouple into two Poschl-Teller problems with integer squared masses $l(l-1)$ and $(l+1)(l+2)$, Eq.~\eqref{eq:ads2master}. The potentials are the trigonometric counterparts of reflectionless Poschl-Teller potentials, and the levels of the closed throat sit at integer $\omega\ell$, the leading-order echo spectrum. The $O(\alpha)$ corrections are given in closed form. They are not symmetric, they depend on frequency, they couple the two modes, in part through the commutator \eqref{eq:commutator} that the scalar reduction formula omits, and they carry $\arctan\rho$ and $\log(1+\rho^2)$ from the corrected background. Stability rests on two results of Sec.~\ref{sec:stability}. The symmetric part of the master potential is positive in the mouths at $\alpha^0$ and in the throat for $\alpha\rho\le0.1$ at $l=2,3,4$, the range we scanned. In the throat it first fails at $\alpha\rho\approx0.24$, far beyond the junction value $\alpha\rho_{\rm J}\sim10^{-6}$. In the mouths at $O(\alpha)$ there is no Schr\"odinger form, and we estimate that the corrections are far too small to change the sign. Within these limits no mode grows as a pure exponential, for any frequency. The antisymmetric part of the $O(\alpha)$ potential could make the frequency complex and allow a mode that grows while it oscillates. A first-order analysis of the normal modes shows that it does not. With the scaled lift the shifts are real on every level. The frequencies refer to the static observer at the centre of the throat, and a different clock adds the same multiple of $\Om_0^2$ to every shift of a level. Their values at infinity need $\ell$ to relative order $\alpha$, which the matching does not fix. On the levels shared by $Z_+$ and $Z_-$ the off-diagonal entries of the first-order matrix have a positive product, Eq.~\eqref{eq:prodS}, and at $\eta_{\rm H}=0$ the two modes repel. On a shared level only the Hall current mixes them. For this clock the corrections lower every normal frequency we computed for $l=2$ and $l=3$, and for larger $l$ they raise the upper mode of the lowest shared levels. On every level the diagonal shift is a rational number plus $2\Om_0^2\log2$, once the remap term of Eq.~\eqref{eq:xform} is kept. These values and the sign of the products rest on the scaling \eqref{eq:linescale}. The coordinate lift gives negative products and lowers every level computed, and its shifts are real on the levels computed (Table~\ref{tab:lifts}). Within these limits, region by region and at $O(\alpha)$, we find no growing mode for $l=2,3,4$ in this sector. Stability of the wormhole as a whole is not shown. The null-energy contraction along radial null rays is unchanged on average at $O(\epsilon)$.

The higher Landau levels add the one-loop Euler-Heisenberg action of the gauge field (Sec.~\ref{sec:hll}). Its parameter $\eta_{\rm H}=g_c^2/12\pi^2$ exceeds $\alpha$ by the factor $q/3$. In the throat the background stays AdS$_2\times S^2$, and the trace anomaly of the higher levels enlarges the AdS$_2$ radius to $r_e/\sqrt{1-\eta_{\rm H}}$. Both masses grow in units of that radius, Eq.~\eqref{eq:mH}, and on the clock of the static observer every normal frequency decreases. The master equations stay two Poschl-Teller problems, whose offsets are no longer integers. The levels shared by $Z_+$ and $Z_-$ split by $\eta_{\rm H}l(l+1)/[(2l-1)(2l+3)]$ in $\Om$, Eq.~\eqref{eq:splitH}, far more than the $O(\alpha)$ entries of $\mathcal S$. On the levels examined the $O(\alpha)$ shift of each mode is then a diagonal entry of $\mathcal S$ and is real, and the mixing of the two modes becomes a second-order effect. The same holds on every level below the gap of the higher levels if the slow growth of the shifts with $\Om_0$ continues. The two corrections add at first order. Together they keep the symmetric part of the throat potential positive on the window $\alpha\rho\le0.1$. With the scaled lift they keep the shifts of every shared level real for every ratio of $\eta_{\rm H}$ to $\alpha$, and with the coordinate lift on the levels examined. In the mouth the extremal mass moves to $r_e(1+\eta_{\rm H})$ and the charge radius runs logarithmically. The reduction closes, the near-horizon limit reproduces the throat, and the master variables of Nomura, Yoshida and Soda give a conserved energy whose potential is positive for the values of $l$ and $\eta_{\rm H}$ we scanned. For these values, region by region and on the levels examined, the higher levels add no growing mode in this sector at first order in $\eta_{\rm H}$.

These results rest on the following assumptions. We assumed $q\gg1$. Corrections to the one-loop fermion stress tensor are then suppressed by powers of $1/q$. We kept only the lowest Landau level as a dynamical sector, since the higher levels have masses of order $\sqrt q/r_e$ (Sec.~\ref{sec:throat-origin}). In the mouth we also dropped the Casimir part of its stress tensor, which is $O(\alpha^3)$ there. Its null component exceeds the anomaly part only very close to $r_e$, inside the throat. The higher Landau levels enter through the one-loop Euler-Heisenberg action of one massless Dirac fermion, with $g_c$ defined at the field of the throat. We keep it to first order in $\eta_{\rm H}$ and do not compute the cross terms of order $\alpha\eta_{\rm H}$ or the two-loop terms. The action holds below the gap $\sqrt q/R$ of the higher levels. Its derivative corrections and the discreteness of the Landau spectrum on the sphere are of relative order $1/q$ and $l(l+1)/q$, comparable to the effects of the lowest level, and we have not computed them. The transitions of the lowest level to the higher ones induced by the perturbation are neglected and not estimated. We neglected the ambient part of each closed field line, which gives corrections of order $d/\ell$, and each circle then has conformal circumference $L_c=\pi$. Each mouth is treated as an isolated static mouth. The field of the other mouth and the orbital motion of MMP's Sec.~6 are neglected, which requires $r\ll d$. MMP keep the mouths apart by a circular orbit of frequency $\Omega_{\rm orb}=2\sqrt{r_e/d^3}$ and require $\Omega_{\rm orb}\ell\ll1$ (their eqs.~(6.50) and (6.51)). The separation must then satisfy $(4r_e\ell^2)^{1/3}\ll d\ll\ell$. At the fiducial parameters $(4r_e\ell^2)^{1/3}=6\times10^{23}$\,cm and $\ell=1.4\times10^{35}$\,cm. The Unruh-like temperature of the orbit is of order $\Omega_{\rm orb}$ and also lies below the level spacing $1/\ell$ of the throat. The lift to four dimensions treats $\theta$ as a label for each field line and holds for $l\ll\sqrt q$, Eq.~\eqref{eq:lbound}. Corrections to the lift from the finite width of the LLL states are not computed, and we expect them at relative order $l(l+1)/q$. The scaling \eqref{eq:linescale} of the line metric with the local field strength is assumed, and the first-order spectrum depends on it (Table~\ref{tab:lifts}). We have not derived it from the four-dimensional Dirac reduction. The state of each closed field line is the antiperiodic vacuum with vanishing holonomy, which is invariant under charge conjugation. We did not fix the free radiation, which periodicity around the closed line fixes away from $\Om=2n$. At $\Om_0=2n$, $n\ge2$, we neglected the coupling of the throat levels to the modes of the line through the mouths. There the throat modes are of order $\rho_{\rm J}^{-2}$ or smaller, and the coupling is far below the $O(\alpha)$ shifts. The calculation is first order in $\epsilon$, in $\alpha$ and in $\eta_{\rm H}$, and the effect of the induced fields back on the fermions is of order $\alpha^2$. The stability statement has further limits. It is made region by region and not with a single master variable for the whole wormhole. We use $Z_\pm$ in the throat, Chandrasekhar's variables in the mouth and the variables of Nomura, Yoshida and Soda in the mouth with the higher levels. The scan with the higher levels covers $l\le10$, $r\le10^4r_e$ and the extremal mass $GM=r_e(1+\eta_{\rm H})$. In the throat the positivity of the symmetric part is shown for $\alpha\rho\le0.1$ and $l=2,3,4$. In the mouth at $O(\alpha)$ it rests on the size of the corrections and not on decoupled variables. The exclusion of oscillatory growth rests on first-order perturbation theory and on the closed throat with the interval of the $\alpha=0$ problem. With the scaled lift the first-order matrix has real eigenvalues on every level. First-order perturbation theory needs $\alpha|\varsigma|\ll\Om_0$, which on the levels computed amounts to $\alpha\Om_0\log\Om_0\ll1$. With the coordinate lift the reality rests on the levels examined, those up to $\Om_0=20$ for $l=2$, up to $\Om_0=15$ for $l=3$ and the lowest four shared levels for $l=4,5,6$. The exclusion of oscillatory growth is a statement at first order in $\alpha$ and not a theorem at finite $\alpha$. Transmission through the mouths, which we expect to give the levels a negative imaginary part, is not included. The $l=0$ sector, which controls the length of the throat, and the $l=1$ sector, which needs its own gauge fixing, are not treated.

We have treated one of the two parity sectors. The other sector, which couples axial metric perturbations to polar gauge perturbations, is treated in \cite{Sadhukhan:2026mcu}. There the polar gauge field is a two-dimensional gauge field on the worldsheet of every field line, and the lowest Landau level responds through its charge with the exact Schwinger current. The electric field along the field lines acquires the screening mass $m_S^2=\alpha q^2$ of MMP's Appendix~A, which is not small and survives in the mouth. In the throat, and in the mouth at $O(\alpha)$, the Casimir stress tensor is assumed to follow the tilted worldsheets. Force balance then fixes a current across the field lines, and charge conservation fixes a polarisation current along them, up to currents that are conserved by themselves. These are the counterparts of the Hall current found here. A neutral mode stays exactly massless. Its zero-energy solution is a persistent current along the field lines. The finite-element runs there find no bound state at their resolution. The channel sits at the threshold for one, and its stability is left open. In the closed throat the massive channels have no growing normal mode at $\alpha=0$. At $O(\alpha)$ their first-order shifts are real on every level at $m_S^2=0$, and for the light massive channel at the physical $m_S^2$ on every level below those of the screened electric field. The isospectrality of the two parities, exact for Reissner-Nordstrom, does not survive. At the physical $m_S^2$ the Schwinger response breaks it already at $\alpha^0$. At $O(\alpha)$ and $m_S^2=0$ part of the difference is fixed by the Einstein--Maxwell mouths, and the rest comes from the fermion response. The one-loop action of the higher Landau levels breaks it too, since $c_\parallel\neq1$. The point $m_S^2=0$ is formal, since $m_S^2=\alpha q^2$. There the products of the off-diagonal entries of the first-order shift matrices of the two sectors coincide on every shared level, while the diagonal shifts differ. In the variables $K/\lhat$ and $w-2K/\lhat=-2K_F/\lhat$ the $\alpha=0$ throat of this sector is that of the companion sector in its gravitational and electric variables. In these variables the Hall current found here has the same friction and frequency-dependent terms as the drift of the field lines in the companion sector, which acts on the electric flux along the lines. Together with the reductions of the off-diagonal entries in both sectors, this makes the two sectors mix the modes of a shared level with the same strength \cite{Sadhukhan:2026mcu}. The coincidence needs the scaling \eqref{eq:linescale}, with which the current follows the change of $|F|$. With the coordinate lift the current follows $K$ alone, and the products differ by the factor $-4/\mu^2$. Open directions are a single master variable valid across the whole wormhole, the normal modes of Sec.~\ref{sec:stab-normalmodes} beyond first order in $\alpha$ and, for the coordinate lift, on all levels, and the $l=0$ sector. Others are a first-principles derivation of the scaling \eqref{eq:linescale} from the lowest Landau level in four dimensions, and a time-domain computation of gravitational-wave echoes. That computation would use the decoupled mouth equations \eqref{eq:Zi-eq} and the throat system \eqref{eq:master} with its $O(\alpha)$ terms, and it would extend the test-field results of \cite{Mondal:2025tht} to the full coupled problem. At the fiducial parameters the $O(\alpha)$ terms of the mouth can be dropped there. The length $\pi\ell$ of the throat scales with $\ell=16\pi^{3/2}\ell_pq^2/g_c^3$. At the fiducial parameters the echoes repeat after about $3\times10^{25}$\,s, twice the light crossing time of the throat.

\appendix
\renewcommand{\thesubsection}{\thesection.\arabic{subsection}}
\section{The fermion response and the Hall current in components}
\label{app:dTQ}

This appendix collects the explicit first-order fermion stress tensor and Hall current on the throat, which the main text gives in covariant form, Eqs.~\eqref{eq:response}, \eqref{eq:dTQ2d} and \eqref{eq:dTQsphere}, and grouped by contribution, Eq.~\eqref{eq:Jphi_explicit}. Throughout, $f\equiv1+\rho^2$, a dot denotes $\partial_\tau$ and a prime $\partial_\rho$, and every quantity displayed is the coefficient of $\epsilon$. The fields $\delta\Upsilon$, $\xi_\tau$, $\xi_\rho$, $K$ and $w$ are functions of $(\tau,\rho)$, with $\xi_a$ carrying a lower index as in Sec.~\ref{sec:response} and $K_F=K-\tfrac12\lhat w$ from \eqref{eq:KB}, and the metric amplitudes $H_0,H_1,H_2$ have been eliminated in their favour through the decomposition \eqref{eq:deceq_tt}--\eqref{eq:deceq_rr}. The central charge is $c=q$.

Evaluating \eqref{eq:response} on the unit AdS$_2$ background with $\bar t_{ab}$ of \eqref{eq:tbar}, the response of one field line is $\delta t_{ab}\,Y(\theta)$ with
\begin{align}
  \delta t_{\tau\tau} &= \frac{q}{12\pi}\Big[\,f^{2}\,\delta\Upsilon'' + \rho f\,\delta\Upsilon'
     + \rho f\,\xi_\rho - \frac{\rho^{2}-2}{f}\,\dot\xi_\tau\,\Big],
  \label{eq:dt_tt}\\[4pt]
  \delta t_{\tau\rho} &= \frac{q}{12\pi}\Big[\,\dot{\delta\Upsilon}{}' - \frac{\rho}{f}\,\dot{\delta\Upsilon}
     + \frac{\rho\,(\rho^{2}-2)}{f^{2}}\,\xi_\tau - \frac{\rho^{2}-2}{2f}\,\xi_\tau'
     - \frac{\rho^{2}+4}{2f}\,\dot\xi_\rho\,\Big],
  \label{eq:dt_tr}\\[4pt]
  \delta t_{\rho\rho} &= \frac{q}{12\pi}\Big[\,\frac{1}{f^{2}}\,\ddot{\delta\Upsilon}
     - \frac{\rho}{f}\,\delta\Upsilon' - \frac{\rho}{f}\,\xi_\rho
     - \frac{\rho^{2}+4}{f}\,\xi_\rho'\,\Big].
  \label{eq:dt_rr}
\end{align}

Inserting \eqref{eq:dt_tt}--\eqref{eq:dt_rr} into the perturbed lift, \eqref{eq:dTQ2d} and \eqref{eq:dTQsphere}, gives $\delta T^Q_{\mu\nu}\,Y(\theta)$ with the $(\tau,\rho)$ block
\begin{align}
  \delta T^Q_{\tau\tau} &= \frac{q}{96\pi^{2}r_e^{2}}\Big[\,2f^{2}\,\delta\Upsilon'' + 2\rho f\,\delta\Upsilon'
     + 2\rho f\,\xi_\rho - \frac{2(\rho^{2}-2)}{f}\,\dot\xi_\tau - (\rho^{2}-2)\,K_F\,\Big],
  \label{eq:dTQ_tt}\\[4pt]
  \delta T^Q_{\tau\rho} &= \frac{q}{96\pi^{2}r_e^{2}}\Big[\,2\,\dot{\delta\Upsilon}{}' - \frac{2\rho}{f}\,\dot{\delta\Upsilon}
     + \frac{2\rho\,(\rho^{2}-2)}{f^{2}}\,\xi_\tau - \frac{\rho^{2}-2}{f}\,\xi_\tau'
     - \frac{\rho^{2}+4}{f}\,\dot\xi_\rho\,\Big],
  \label{eq:dTQ_tr}\\[4pt]
  \delta T^Q_{\rho\rho} &= \frac{q}{96\pi^{2}r_e^{2}}\Big[\,\frac{2}{f^{2}}\,\ddot{\delta\Upsilon}
     - \frac{2\rho}{f}\,\delta\Upsilon' - \frac{2\rho}{f}\,\xi_\rho
     - \frac{2(\rho^{2}+4)}{f}\,\xi_\rho' + \frac{\rho^{2}+4}{f^{2}}\,K_F\,\Big],
  \label{eq:dTQ_rr}
\end{align}
and the sphere block
\begin{equation}
  \delta T^Q_{\theta\theta} \;=\; \frac{\delta T^Q_{\phi\phi}}{\sin^{2}\theta}
  \;=\; \frac{q}{96\pi^{2}r_e^{2}}\Big[\,f\,\delta\Upsilon'' + 2\rho\,\delta\Upsilon'
     - \frac{1}{f}\,\ddot{\delta\Upsilon} - 2\,\delta\Upsilon - 2K_F + K\,\Big] .
  \label{eq:dTQ_thth}
\end{equation}
The mixed block \eqref{eq:driftlift} multiplies $Y'(\theta)$ instead of $Y(\theta)$,
\begin{equation}
  \delta T^Q_{\tau\theta} \;=\; -\,\frac{q}{192\pi^{2}r_e^{2}}\,\frac{2\rho^{2}-1}{f}\,\dot w\;Y'(\theta),
  \qquad
  \delta T^Q_{\rho\theta} \;=\; -\,\frac{q}{192\pi^{2}r_e^{2}}\,\frac{2\rho^{2}+5}{f}\,w'\;Y'(\theta) .
  \label{eq:dTQ_mix}
\end{equation}
The $(\tau,\rho)$ block satisfies $\delta T^Q_{ab}=(\delta t_{ab}-K_F\bar t_{ab})/4\pi r_e^{2}$ exactly, the $K_F$ terms being the background Casimir tensor redistributed over the changed density of field lines. Writing the sphere block as $\delta T^Q_{AB}=\delta\mathcal P\,g_{AB}+\bar{\mathcal P}\,\delta g_{AB}$, the perturbed pressure takes the covariant form
\begin{equation}
  \delta\mathcal P \;=\; \bar{\mathcal P}\,\Big[\,\hat\Box\,\delta\Upsilon - 2\,\delta\Upsilon - 2K_F\,\Big],
  \qquad
  \bar{\mathcal P}=\frac{q}{96\pi^{2}r_e^{4}},
  \qquad
  \hat\Box\,u = -\frac{\ddot u}{f} + \big(f\,u'\big)' ,
  \label{eq:dP_explicit}
\end{equation}
which is what \eqref{eq:lift} requires, $\delta\mathcal P$ being fixed by the variation of the anomalous trace, $\delta\hat R = 4\,\delta\Upsilon - 2\hat\Box\,\delta\Upsilon$ on the unit background, together with $\delta\log\big(|F|\,\varpi^2\big)=-2K_FY$. The components \eqref{eq:dTQ_tt}--\eqref{eq:dTQ_mix} are conserved at $O(\epsilon)$ in the $\tau$, $\rho$ and $\phi$ directions identically in the fields. Only the $\theta$ component of the divergence is nonzero, and it is balanced by the Lorentz force of the Hall current.

With $H_0$ and $H_2$ eliminated, the current \eqref{eq:Jphi_explicit} depends on the response fields and on $w$ alone. With $\tilde\xi_a$ of \eqref{eq:xitilde},
\begin{equation}
  J^\phi \;=\; \frac{1}{48\pi^{2}r_e^{4}}\,\frac{Y'(\theta)}{\sin\theta}
  \Big[\,f\,\delta\Upsilon'' + 2\rho\,\delta\Upsilon' - \frac{1}{f}\,\ddot{\delta\Upsilon}
     + 2\,\delta\Upsilon
     + \left(2\rho^{2}+5\right)\tilde\xi_\rho' + 4\rho\,\tilde\xi_\rho
     - \frac{2\rho^{2}-1}{f^{2}}\,\dot{\tilde\xi}_\tau\,\Big],
  \label{eq:Jphi_pure}
\end{equation}
or, in terms of the unit-AdS$_2$ Laplacian and divergence with $\tilde\xi^a=\hat g^{ab}\tilde\xi_b$,
\begin{equation}
  J^\phi \;=\; \frac{1}{48\pi^{2}r_e^{4}}\,\frac{Y'(\theta)}{\sin\theta}
  \Big[\,\hat\Box\,\delta\Upsilon + 2\,\delta\Upsilon + 2\,\hat\nabla_a\tilde\xi^a
     + 3\Big(\tilde\xi_\rho' + \frac{\dot{\tilde\xi}_\tau}{f^{2}}\Big)\Big] .
  \label{eq:Jphi_grouped}
\end{equation}
Substituting the decomposition \eqref{eq:deceq_tt}--\eqref{eq:deceq_rr} into \eqref{eq:Jphi_explicit} reproduces \eqref{eq:Jphi_pure} identically. The explicit $K_F$ of \eqref{eq:Jphi_explicit} cancels against the $K_F$ carried by $H_0-H_2$. Hence the Hall current is driven entirely by the Weyl factor, the transport fields and the drift of the field lines, and it vanishes for a perturbation that leaves $\hat g_{ab}$ unchanged and does not move the field lines.

\section{Radial equations of the throat}
\label{app:throat-eqs}

All expressions are in throat coordinates $(\tau,\rho)$ with $p=-i\Om$, $w=4a/q$, primes denoting $\dd/\dd\rho$, $\lhat=l(l+1)$ and $f=1+\rho^2$.

\subsection{The \texorpdfstring{$O(\epsilon\,\alpha^0)$}{O(epsilon alpha0)} equations on exact AdS\texorpdfstring{$_2\times S^2$}{2 x S2}}

On the background \eqref{eq:bgthroat} at $\alpha=0$, with no fermionic source,
\begin{equation}\label{eq:t-ett}
  \mathcal E_{\tau\tau}\big|_{\alpha^0}:\quad
  \tfrac{f}{2}\Big[\lhat\,H_{2} + (\lhat+2)\,K - 2\rho\,K' - 2f\,K'' - 2\lhat\,w\Big] = 0 ,
\end{equation}
\begin{equation}\label{eq:t-etr}
  \mathcal E_{\tau\rho}\big|_{\alpha^0}:\quad
  \lhat f\,H_{1} - 2p\,r_{e}^{2}\big(f\,K' - \rho\,K\big) = 0 ,
\end{equation}
\begin{equation}\label{eq:t-err}
  \mathcal E_{\rho\rho}\big|_{\alpha^0}:\quad
  \lhat f\big(H_{0} + 2w\big) + 2\rho f\,K' - \big[(\lhat+2)f + 2p^{2}\big]K = 0 ,
\end{equation}
\begin{equation}\label{eq:t-etth}
  \mathcal E_{\tau\theta}\big|_{\alpha^0}:\quad
  f\,H_{1}' + 2\rho\,H_{1} - p\,r_{e}^{2}\big(H_{2} + K - 2w\big) = 0 ,
\end{equation}
\begin{equation}\label{eq:t-erth}
  \mathcal E_{\rho\theta}\big|_{\alpha^0}:\quad
  r_{e}^{2}\Big[f\big(H_{0}' - K' + 2w'\big) + \rho\big(H_{0} + H_{2}\big)\Big] - p\,H_{1} = 0 ,
\end{equation}
\begin{multline}\label{eq:t-etrace}
  \mathcal E^{\rm trace}\big|_{\alpha^0}:\quad
  r_{e}^{2}\Big\{2f^{2}\big(K'' - H_{0}''\big) - 2\rho f\big(3H_{0}' + H_{2}' - 2K'\big) + \lhat f\big(H_{0} - 4w\big)
  \\ - \big[(\lhat+4)f + 2p^{2}\big]H_{2} + 2\big(4f - p^{2}\big)K\Big\} + 4p\big(f\,H_{1}' + \rho\,H_{1}\big) = 0 ,
\end{multline}
\begin{equation}\label{eq:t-etl}
  \mathcal E^{\rm tl}\big|_{\alpha^0}:\quad -\tfrac12\big(H_{0} - H_{2}\big) = 0 ,
\end{equation}
\begin{equation}\label{eq:t-mph}
  \mathcal M^\phi\big|_{\alpha^0}:\quad
  f^{2}\,w'' + 2\rho f\,w' - \big(\lhat f + p^{2}\big)w + f\big(H_{0} - H_{2} + 2K\big) = 0 .
\end{equation}
Here $\mathcal E_{ab}$ and $\mathcal M^\phi$ are the residuals \eqref{eq:Edef} and \eqref{eq:Mdef}, evaluated on exact AdS$_2\times S^2$ with no fermionic source, and each is multiplied by a factor that clears denominators. In the order shown the factors are $1$, $2r_e^2f$, $2f^2$, $2r_e^2$, $2r_e^2f$, $2r_e^2f$, $1$ and $4f$, with $\mathcal M^\phi$ divided by $q$ first. Their reduction gives \eqref{eq:t-P0Q0} and, in Schr\"odinger form, \eqref{eq:ads2master}.

\subsection{The \texorpdfstring{$O(\epsilon\,\alpha^1)$}{O(epsilon alpha1)} equations}

The $\alpha$-derivatives of the full residuals at $\alpha=0$ are given below. They include the corrected background \eqref{eq:bgsol} and the lifted response $\delta T^Q_{\mu\nu}$ of Sec.~\ref{sec:response}, which enters through $\delta\Upsilon$, $\xi_\tau$ and $\xi_\rho$, and through the density, the scale and the drift of the field lines, which add terms in $w$, $w'$ and $w''$. The Maxwell equation also includes the Hall current $J^\phi$. Here $\partial_\alpha \mathcal E_{ab}$ and $\partial_\alpha \mathcal M^\phi$ are the $\alpha$-derivatives at $\alpha=0$ of the residuals \eqref{eq:Edef} and \eqref{eq:Mdef}. The only factor applied to each residual is the prefactor shown on its left-hand side, which clears the denominators. In \eqref{eq:t-detrace} the sphere pressure enters with $\delta\hat g^{ab}$ written through $h_{ab}$. With $\delta\mathcal P$ of \eqref{eq:dP_explicit} its right side changes by terms that vanish once $h_{ab}$ is decomposed by \eqref{eq:deceq_tt}--\eqref{eq:deceq_rr}. All of them are linear in the perturbation fields, and the coefficient of every field is collected. The $\arctan\rho$ and $\log f$ pieces are grouped. Several equations share the combination
\begin{equation}\label{eq:t-Ssrc}
  \mathcal J \equiv 2p^{2}f\,\delta\Upsilon - 4\rho f^{2}\,\delta\Upsilon' - 2f^{3}\,\delta\Upsilon''
  + 2p(\rho^{2}-2)\,\xi_{\tau} - 4\rho f^{2}\,\xi_{\rho} - 2f^{2}(\rho^{2}+4)\,\xi_{\rho}' ,
\end{equation}
and the Maxwell equation also contains the drift of the field lines,
\begin{equation}\label{eq:t-Sdrift}
  \mathcal J_{\rm d} \equiv f^{2}\big[(2\rho^{2}+5)\,w'' + 4\rho\,w'\big] - p^{2}(2\rho^{2}-1)\,w .
\end{equation}
With these, the equations read
\begin{multline}\label{eq:t-dett}
  6f\,\partial_\alpha \mathcal E_{\tau\tau} = -2f(\rho^{2}-2)\,H_{0}
  + 3f^{2}\big(\rho + f\arctan\rho\big)H_{2}'
  \\ + f\Big\{-2\big[2\lhat(2\rho^{2}+1) - 3(\rho^{2}+2)\big] + 3\lhat\log f
  - 6\rho\big[(\lhat-1)\rho^{2} + 2\lhat - 1\big]\arctan\rho\Big\}H_{2}
  \\ + f\Big\{-2\big[2\lhat(2\rho^{2}+1) + 15\rho^{2} + 14\big] + 3(\lhat+2)\log f
  - 6\rho\big[(\lhat+4)\rho^{2} + 2\lhat + 6\big]\arctan\rho\Big\}K
  \\ + 2\rho f\big[5\rho^{2} + 1 - 3\log f + 3\rho(\rho^{2}+3)\arctan\rho\big]K'
  + 4f^{2}\big[4\rho^{2} - 3\log f + 3\rho(\rho^{2}+3)\arctan\rho\big]K''
  \\ + \lhat f\big[23\rho^{2} + 18 - 6\log f + 6\rho(3\rho^{2}+5)\arctan\rho\big]w
  \\ - 4f^{3}\,\delta\Upsilon'' - 4\rho f^{2}\,\delta\Upsilon' + 4p(\rho^{2}-2)\,\xi_{\tau} - 4\rho f^{2}\,\xi_{\rho} ,
\end{multline}
\begin{multline}\label{eq:t-detr}
  6r_{e}^{2}f^{2}\,\partial_\alpha \mathcal E_{\tau\rho} = -f\big[2(2\lhat f + \rho^{2} - 2) + 3\lhat\rho f\arctan\rho\big]H_{1}
  + 3p\,r_{e}^{2}f\big(\rho + f\arctan\rho\big)H_{2}
  \\ - 2p\,r_{e}^{2}\big[\rho(3\rho^{2}+7) + 3\rho\log f + 3(\rho^{4}+\rho^{2}+2)\arctan\rho\big]K
  + 4p\,r_{e}^{2}f\big(\rho\,\delta\Upsilon - f\,\delta\Upsilon'\big)
  \\ - 4r_{e}^{2}\rho(\rho^{2}-2)\,\xi_{\tau} + 2r_{e}^{2}f(\rho^{2}-2)\,\xi_{\tau}'
  + 2p\,r_{e}^{2}f(\rho^{2}+4)\,\xi_{\rho} ,
\end{multline}
\begin{multline}\label{eq:t-derr}
  6r_{e}^{2}f^{3}\,\partial_\alpha \mathcal E_{\rho\rho} = -\lhat\,r_{e}^{2}f\big(4 + 3\log f - 6\rho\arctan\rho\big)H_{0}
  - 3r_{e}^{2}f^{2}\big(\rho + f\arctan\rho\big)H_{0}'
  \\ + 6pf\big(\rho + f\arctan\rho\big)H_{1}
  - 2r_{e}^{2}f^{2}\big(4 + 3\rho\arctan\rho\big)H_{2}
  \\ + r_{e}^{2}\Big\{2\big[(2\lhat + 7\rho^{2} + 8)f - 8p^{2}\rho^{2}\big] + 3\big[(\lhat+2)f + 4p^{2}\big]\log f
  - 6\rho\big[(\lhat - 2\rho^{2})f + 2p^{2}(\rho^{2}+3)\big]\arctan\rho\Big\}K
  \\ - 2r_{e}^{2}f\big[4\rho + 3\rho\log f - 3(\rho^{2}-1)\arctan\rho\big]K'
  - \lhat\,r_{e}^{2}f\big[7\rho^{2} + 12 + 6\log f + 6\rho(\rho^{2}-1)\arctan\rho\big]w
  \\ - 4p^{2}r_{e}^{2}f\,\delta\Upsilon + 4r_{e}^{2}\rho f^{2}\big(\delta\Upsilon' + \xi_{\rho}\big)
  + 4r_{e}^{2}f^{2}(\rho^{2}+4)\,\xi_{\rho}' ,
\end{multline}
\begin{multline}\label{eq:t-detth}
  6r_{e}^{2}\,\partial_\alpha \mathcal E_{\tau\theta} = -\big(11\rho + 9f\arctan\rho\big)H_{1}
  - \big[4\rho^{2} - 3\log f + 3\rho(\rho^{2}+3)\arctan\rho\big]H_{1}'
  \\ - p\,r_{e}^{2}\Big[2\big(4 + 3\rho\arctan\rho\big) - \frac{2\rho^{2}-1}{f}\Big]w ,
\end{multline}
\begin{multline}\label{eq:t-derth}
  6r_{e}^{2}f^{2}\,\partial_\alpha \mathcal E_{\rho\theta} = -r_{e}^{2}\big[\rho(3\rho^{2}+7) + 3\rho\log f + 3(\rho^{4}+\rho^{2}+2)\arctan\rho\big]H_{0}
  \\ - p\big[4\rho^{2} - 3\log f + 3\rho(\rho^{2}+3)\arctan\rho\big]H_{1}
  - r_{e}^{2}\big[4\rho + 3\rho\log f - 3(\rho^{2}-1)\arctan\rho\big]H_{2}
  \\ - 3r_{e}^{2}f\big(2\rho^{2} + 1 + 2\rho f\arctan\rho\big)w' ,
\end{multline}
\begin{multline}\label{eq:t-detrace}
  3r_{e}^{2}f^{2}\,\partial_\alpha \mathcal E^{\rm trace} = r_{e}^{2}f^{2}\big(4 + 3\log f - 6\rho\arctan\rho\big)\big(K'' - H_{0}''\big)
  + 3r_{e}^{2}f^{2}\big[\rho + (\rho^{2}+4)\arctan\rho\big]H_{0}'
  \\ - r_{e}^{2}f(\rho^{2}-2)\,H_{0}
  + 2pf^{2}\big(4 + 3\rho\arctan\rho\big)H_{1}'
  + 2p\big[4\rho^{3} - 3\rho\log f + 3(\rho^{4}+2\rho^{2}-1)\arctan\rho\big]H_{1}
  \\ + 3r_{e}^{2}f^{2}\arctan\rho\,\big(H_{2}' - 2K'\big)
  + r_{e}^{2}\Big\{(7\rho^{2}+10)f - 4p^{2}(2\rho^{2}+1) + 3p^{2}\log f
  + 6\rho\big[f^{2} - p^{2}(\rho^{2}+2)\big]\arctan\rho\Big\}H_{2}
  \\ - r_{e}^{2}\Big\{2\big[7f^{2} + 2p^{2}(2\rho^{2}+1)\big] - 3p^{2}\log f
  + 6\rho\big[2f^{2} + p^{2}(\rho^{2}+2)\big]\arctan\rho\Big\}K
  \\ + \lhat\,r_{e}^{2}f^{2}\big(7 + 6\rho\arctan\rho\big)w + r_{e}^{2}\,\mathcal J ,
\end{multline}
\begin{multline}\label{eq:t-dmph}
  24f^{2}\,\partial_\alpha \mathcal M^\phi = -f^{2}\big(7 + 6\rho\arctan\rho\big)\big(H_{0} - H_{2} + 2K\big)
  \\ + 2\Big\{4\big(\lhat f^{2} - p^{2}\rho^{2}\big) + 3p^{2}\log f
  + 3\rho\big[\lhat f^{2} - p^{2}(\rho^{2}+3)\big]\arctan\rho\Big\}w
  \\ - 2f^{2}\big(11\rho + 9f\arctan\rho\big)w'
  - 2f^{2}\big[4\rho^{2} - 3\log f + 3\rho(\rho^{2}+3)\arctan\rho\big]w''
  - \lhat f^{2}w + \mathcal J + \mathcal J_{\rm d} ,
\end{multline}
\begin{equation}\label{eq:mphdiff}
  24f^{2}\big(\mathcal M^\phi_{\rm sourced} - \mathcal M^\phi_{\rm free}\big)
  = \alpha\big[f^{2}\big(H_{0} - H_{2} + 2K_F\big) + \mathcal J + \mathcal J_{\rm d}\big] .
\end{equation}
The traceless combination $\mathcal E^{\rm tl}=-\tfrac12(H_0-H_2)$ does not depend on $\alpha$, see Eq.~\eqref{eq:tlvalue}. Equation~\eqref{eq:mphdiff} is the Hall-current contribution alone. It equals $-g_c^2\sqrt{-g}\,\delta J^\phi/q$ with $g_c^2=4\pi^2\alpha q$ substituted. It is proportional to $\alpha$. Equation~\eqref{eq:t-dmph} already includes it.

\section{The \texorpdfstring{$O(\alpha)$}{O(alpha)} throat potential \texorpdfstring{$V_{1a}$}{V1a}}
\label{app:V1a}

The frequency-independent part of the $O(\alpha)$ correction in \eqref{eq:master}, in the $(Z_+,Z_-)$ basis \eqref{eq:Zdef}, is, with $f=1+\rho^2$,
\begin{align}
  V_{1a,\,++} &= \frac{1}{12l(2l+1)}\Big[\,
    l\,(l^{5} + 10l^{4} - 25l^{3} - 174l^{2} - 216l - 100)\,f
  \nonumber\\&\qquad
    + 2(13l^{4} + 41l^{3} + 47l^{2} + 46l + 12) - \frac{12(l^{2}+2l-4)}{f}
  \nonumber\\&\qquad
    - 12l\,(4l^{3} + 16l^{2} + 19l + 9)\,\rho f\arctan\rho
    + 12l(l+1)(l+2)(2l+1)\,\big(\log f - 2\rho\arctan\rho\big)\Big],
  \label{eq:V1a00}\\[6pt]
  V_{1a,\,+-} &= \frac{1}{12(l+1)(2l+1)}\Big[\,
    -l(l-1)(l^{4} - 5l^{2} - 12l - 26)\,f + 6l(l+1)(l^{2}-l+4)
  \nonumber\\&\qquad
    + \frac{12(l^{2}-l+4)}{f} - 72\,l(l+1)\,\rho f\arctan\rho\Big],
  \label{eq:V1a01}\\[6pt]
  V_{1a,\,-+} &= \frac{1}{12l(2l+1)}\Big[\,
    -(l+1)(l+2)(l^{4} + 4l^{3} + l^{2} + 6l - 18)\,f + 6l(l+1)(l^{2}+3l+6)
  \nonumber\\&\qquad
    + \frac{12(l^{2}+3l+6)}{f} - 72\,l(l+1)\,\rho f\arctan\rho\Big],
  \label{eq:V1a10}\\[6pt]
  V_{1a,\,--} &= \frac{1}{12(l+1)(2l+1)}\Big[\,
    (l+1)(l^{5} - 5l^{4} - 55l^{3} + 49l^{2} + 22l + 24)\,f
  \nonumber\\&\qquad
    + 2(13l^{4} + 11l^{3} + 2l^{2} - 23l - 15) - \frac{12(l^{2}-5)}{f}
  \nonumber\\&\qquad
    - 12(l+1)(4l^{3} - 4l^{2} - l - 2)\,\rho f\arctan\rho
    + 12l(l-1)(l+1)(2l+1)\,\big(\log f - 2\rho\arctan\rho\big)\Big].
  \label{eq:V1a11}
\end{align}

The two diagonal entries share the combination $\log f - 2\rho\arctan\rho$. The terms in $1/f$ are the entries of $V_{1b}$ in \eqref{eq:V1b}. The off-diagonal entries include the commutator term \eqref{eq:commutator} that the scalar reduction formula omits. Appendix~\ref{app:offdiag} reduces them on the levels shared by $Z_+$ and $Z_-$. Like $V_0$ and $V_{1b}$, $V_{1a}$ goes into itself under $l\to-1-l$ with $Z_+$ and $Z_-$ exchanged, since the map leaves $\lhat$ fixed and exchanges the rows of $\mathsf T$ in \eqref{eq:Zdef}.

\section{Off-diagonal entries of the shift matrix}
\label{app:offdiag}

This appendix derives the closed form \eqref{eq:prodS} and the structure of the off-diagonal entries stated in Sec.~\ref{sec:stab-normalmodes}. On a level shared by $Z_+$ and $Z_-$ let $\psi_+$ be the mode \eqref{eq:PTmodes} of $Z_+$ with index $n$ and $\psi_-$ the mode of $Z_-$ with index $n+2$, both at $\Om_0=n+l+2$. We write $\|\psi\|^2=\int_{-\pi/2}^{\pi/2}\psi^2\,\dd x$ and $\Lambda_{\mathsf c}=-\partial_x^2+V_{0,\mathsf c}-\Om_0^2$, with $V_{0,\mathsf c}=m_{\mathsf c}^2\sec^2x$, $m_+^2=(l+1)(l+2)$ and $m_-^2=l(l-1)$. Then $\Lambda_+\psi_+=\Lambda_-\psi_-=0$. All integrals run over $(-\pi/2,\pi/2)$.

Let the functions $\mathsf a_1(x)$ and $\mathsf a_0(x)$ obey $2\mathsf a_0'=(V_{0,+}-V_{0,-})\,\mathsf a_1-\mathsf a_1''$. The operator $\Lambda_+(\mathsf a_1\partial_x+\mathsf a_0)-(\mathsf a_1\partial_x+\mathsf a_0+2\mathsf a_1')\Lambda_-$ then contains no derivative. It is the multiplication by
\begin{equation}
  \mathcal O=-\mathsf a_0''+(V_{0,+}-V_{0,-})\,\mathsf a_0-2\mathsf a_1'\big(V_{0,-}-\Om_0^2\big)-\mathsf a_1\,\partial_xV_{0,-} .
  \label{eq:Fab}
\end{equation}
Its matrix element between $\psi_+$ and $\psi_-$ vanishes. The second term of the operator gives zero since $\Lambda_-\psi_-=0$. In the first, $\Lambda_+$ is moved onto $\psi_+$ by two integrations by parts. The boundary terms vanish when $\mathsf a_1$ and $\mathsf a_0$ grow at most like $\sec x$, since $\psi_+\sim\cos^{l+2}x$ and $\psi_-\sim\cos^lx$. Terms of the form \eqref{eq:Fab} are removed by the change of variables $Z_+\to Z_++\alpha(\mathsf a_1\partial_x+\mathsf a_0)Z_-$ together with a multiple of the equation of $Z_-$, and they do not affect $\mathcal S$. Two choices matter. With $\mathsf a_1=0$ and $\mathsf a_0=1$ one has $\mathcal O=(4l+2)\sec^2x$. This is the orthogonality used in Sec.~\ref{sec:hll-throat}. With
\begin{equation}
  \mathsf a_1=-\tfrac12\big(x+\sin x\cos x\big),
  \qquad
  \mathsf a_0=-\tfrac12\big[(2l+1)\,x\tan x+\sin^2x\big]
  \label{eq:abchoice}
\end{equation}
one has $\mathcal O=V_{\rm mix}-3\lhat\,x\tan x\sec^2x+l(l-1)\sec^2x$, with $V_{\rm mix}$ of \eqref{eq:Vmix}. With $\rho=\tan x$, $f=\sec^2x$ and $\arctan\rho=x$ this gives
\begin{equation}
  \int\psi_+\,f\,\psi_-\,\dd x=0,
  \qquad
  \int\psi_+\big(V_{\rm mix}-3\lhat\,\rho f\arctan\rho\big)\psi_-\,\dd x=0 .
  \label{eq:nullpair}
\end{equation}

In the basis $(K,w)$ let $\mathsf R$ be the matrix that maps $(K,w)$ to $(0,\,w-2K/\lhat)=(0,\,-2K_F/\lhat)$, with $K_F$ of \eqref{eq:KB}. It is a projector of rank one, onto the perturbations with $K=0$ along those with $K_F=0$, which leave $|F|$ unchanged. In this basis $V_{1b}$ of \eqref{eq:V1b} is $-\mathsf R/f+4\,{\rm diag}(1,0)/\lhat f$. In the basis $Z_\pm$ of \eqref{eq:Zdef}
\begin{equation}
  \mathsf R=\frac{1}{2l+1}\begin{pmatrix} l+2 & -\dfrac{l(l-1)}{l+1}\\[8pt] -\dfrac{(l+1)(l+2)}{l} & l-1\end{pmatrix},
  \qquad
  {\rm diag}(1,0)\;\to\;\frac{1}{2l+1}\begin{pmatrix} l+1 & l\\ l+1 & l\end{pmatrix}.
  \label{eq:RZ}
\end{equation}
For the scaled lift the off-diagonal entries of $V_{1a}-\Om_0^2V_{1b}$ obey, identically in $\rho$, $l$ and $\Om_0$,
\begin{equation}
\begin{aligned}
  \big(V_{1a}-\Om_0^2V_{1b}\big)_{+-}&=\frac{l(l-1)}{2(l+1)(2l+1)}\,V_{\rm mix}+\frac{2}{(l+1)(2l+1)}\big(V_{\rm mix}-3\lhat\,\rho f\arctan\rho\big)\\
  &\qquad-\frac{l(l-1)(l^4-5l^2-12l-26)}{12(l+1)(2l+1)}\,f,\\[4pt]
  \big(V_{1a}-\Om_0^2V_{1b}\big)_{-+}&=\frac{(l+1)(l+2)}{2l(2l+1)}\,V_{\rm mix}+\frac{2}{l(2l+1)}\big(V_{\rm mix}-3\lhat\,\rho f\arctan\rho\big)\\
  &\qquad-\frac{(l+1)(l+2)(l^4+4l^3+l^2+6l-18)}{12l(2l+1)}\,f .
\end{aligned}
  \label{eq:offdiagred}
\end{equation}
The first coefficients are $-\tfrac12\mathsf R_{+-}$ and $-\tfrac12\mathsf R_{-+}$. The second are $2/\lhat$ times the off-diagonal entries of ${\rm diag}(1,0)$ in \eqref{eq:RZ}. They pair the entry $4/\lhat f$ of $V_{1b}$ with the terms in $\rho f\arctan\rho$ of \eqref{eq:V1a01} and \eqref{eq:V1a10}. By \eqref{eq:nullpair} only the first terms contribute to $\mathcal S$. Since $\mathsf R$ has rank one, $\mathsf R_{+-}\mathsf R_{-+}=\mathsf R_{++}\mathsf R_{--}=(l-1)(l+2)/(2l+1)^2$. The second row of $V_{1b}$, like that of $\tilde P_1$, comes from the Hall current. Without its source the first terms of \eqref{eq:offdiagred} are absent, for both lifts, while the second terms are unchanged and the terms in $f$ change. For the coordinate lift $\mathsf R$ is replaced by the matrix $\mathsf R_{\rm c}$ that maps $(K,w)$ to $(0,-2K/\lhat)$. The second terms of \eqref{eq:offdiagred} are the same, the terms in $f$ change, and the first coefficients become $-\tfrac12\mathsf R_{{\rm c},+-}=-l/[(l+1)(2l+1)]$ and $-\tfrac12\mathsf R_{{\rm c},-+}=(l+1)/[l(2l+1)]$. The matrix $\mathsf R_{\rm c}$ is nilpotent, and $\mathsf R_{{\rm c},+-}\mathsf R_{{\rm c},-+}=-4/(2l+1)^2$. This gives the factor $-4/\mu^2$ and the entry factors $-2/(l-1)$ and $2/(l+2)$ of Sec.~\ref{sec:stab-normalmodes}.

The matrix element of $V_{\rm mix}$ follows from the Gegenbauer polynomials. With $u=\sin x$ one has $\psi_+=(1-u^2)^{(l+2)/2}C^{(l+2)}_n(u)$, $\psi_-=(1-u^2)^{l/2}C^{(l)}_{n+2}(u)$ and $\dd x=\dd u/\sqrt{1-u^2}$. The relation $(m+\lambda)\,C^{(\lambda)}_m=\lambda\big(C^{(\lambda+1)}_m-C^{(\lambda+1)}_{m-2}\big)$ \cite[Eq.~22.7.23]{Abramowitz:1964} expresses $C^{(l)}_{n+2}$ through $C^{(l+1)}_{n+2}$ and $C^{(l+1)}_n$, and through $C^{(l+2)}_{n+2}$, $C^{(l+2)}_n$ and $C^{(l+2)}_{n-2}$. Summed over $m=n,n-2,\dots$ it expresses $C^{(l+2)}_n$ through $C^{(l+1)}_{n-2j}$ with $j\ge0$. Orthogonality with the weights $(1-u^2)^{l+1/2}$ and $(1-u^2)^{l+3/2}$ then gives
\begin{equation}
\begin{aligned}
  \int\psi_+\psi_-\,\dd x&=-\frac{l\,(\Om_0-1)}{(l+1)\,\Om_0}\,\big\|\cos^{l+1}x\,C^{(l+1)}_n(\sin x)\big\|^2,\\
  \int\psi_+\cos^2x\,\psi_-\,\dd x&=-\frac{2\lhat}{\Om_0^2-1}\,\|\psi_+\|^2 .
\end{aligned}
  \label{eq:gegover}
\end{equation}
With $\|\cos^sx\,C^{(s)}_n(\sin x)\|^2=2^{1-2s}\pi\,\Gamma(n+2s)/[n!\,(n+s)\,\Gamma(s)^2]$ \cite[Eq.~22.2.3]{Abramowitz:1964} the factor $\Om_0^2-1$ cancels, and
\begin{equation}
  \int\psi_+V_{\rm mix}\,\psi_-\,\dd x=\frac{l\,(\Om_0-1)(\Om_0+2l+1)}{l+1}\,\big\|\cos^{l+1}x\,C^{(l+1)}_n(\sin x)\big\|^2 .
  \label{eq:Vmixelem}
\end{equation}
For the modes \eqref{eq:PTmodes} the entries of $\mathcal S$ are then
\begin{equation}
\begin{aligned}
  \mathcal S_{+-}&=\frac{2l^2(l-1)\,\Om_0(\Om_0+2l+1)}{(2l+1)(\Om_0+l)(\Om_0+l+1)},\\
  \mathcal S_{-+}&=\frac{(l+2)\,\Om_0(\Om_0-l)(\Om_0-l-1)(\Om_0+2l+1)}{8l^2(2l+1)} ,
\end{aligned}
  \label{eq:Spolar}
\end{equation}
and their product is \eqref{eq:prodS}. These are the values of \eqref{eq:Sl2} and Table~\ref{tab:lifts}. They also reproduce the products identified by PSLQ on the highest levels of Appendix~\ref{app:numerics}, for example $\tfrac{4200}{13}$ and $\tfrac{1530000}{253}$ at $l=2$ and $\Om_0=10$ and $20$, and $\tfrac{1996500}{931}$ at $l=3$ and $\Om_0=15$.

\section{Radial equations and reduction matrices of the mouth}
\label{app:mouth-eqs}

All expressions in this appendix belong to the mouth at $O(\alpha^0)$. They are exact in $GM$, $r_e$, $l$ and $p$, with $\lhat=l(l+1)$, and are given in mouth coordinates $(t,r)$ with $H_i'\equiv\dd H_i/\dd r$. Throughout, $p=-i\omega$ and $w=4a/q$.

\subsection{The eight radial equations}

After harmonic separation (Sec.~\ref{sec:separation}) and passage to the frequency domain \eqref{eq:freq}, the first-order Einstein and Maxwell equations on the background \eqref{eq:bgmouth} are
\begin{align}
\mathcal E_{tt}:\quad & \Delta\Big[-2r^{2}\Delta\,K'' + 2r\Delta\,H_{2}'
  - 2r\big(3r^{2}-5GMr+2r_{e}^{2}\big)K'
\nonumber\\ &\qquad
  + \big((\lhat+2)r^{2}-2r_{e}^{2}\big)H_{2}
  + \big((\lhat-2)r^{2}+4r_{e}^{2}\big)K
  - 2\lhat\,r_{e}^{2}\,w\Big] = 0 ,
\label{eq:m-ett}\\[6pt]
\mathcal E_{tr}:\quad & \lhat\,\Delta\,H_{1} + 2pr\,\Delta\big(H_{2}-rK'\big)
  - 2pr\big(r^{2}-3GMr+2r_{e}^{2}\big)K = 0 ,
\label{eq:m-etr}\\[6pt]
\mathcal E_{rr}:\quad & \Delta\Big[-2r\Delta\,H_{0}' + 2r^{2}(r-GM)\,K'
  + \lhat\,r^{2}H_{0} + 4pr^{3}H_{1} - 2\big(r^{2}-r_{e}^{2}\big)H_{2}
\nonumber\\ &\qquad
  - \big((\lhat-2)r^{2}+4r_{e}^{2}\big)K + 2\lhat\,r_{e}^{2}\,w\Big]
  - 2p^{2}r^{6}K = 0 ,
\label{eq:m-err}\\[6pt]
\mathcal E_{t\theta}:\quad & r\Delta\,H_{1}' + 2\big(GMr-r_{e}^{2}\big)H_{1}
  - pr^{3}\big(H_{2}+K\big) + 2pr\,r_{e}^{2}\,w = 0 ,
\label{eq:m-etth}\\[6pt]
\mathcal E_{r\theta}:\quad & \Delta\big[r^{2}\big(H_{0}'-K'\big) + 2r_{e}^{2}\,w'\big]
  - r\big(r^{2}-3GMr+2r_{e}^{2}\big)H_{0}
\nonumber\\ &\qquad
  + r^{2}(r-GM)\,H_{2} - pr^{4}H_{1} = 0 ,
\label{eq:m-erth}\\[6pt]
\mathcal E^{\rm trace}:\quad & \Delta\Big[2r^{2}\Delta\big(K''-H_{0}''\big)
  - 2r\big(r^{2}+GMr-2r_{e}^{2}\big)H_{0}'
  + 2r^{2}(r-GM)\big(2K'-H_{2}'\big)
\nonumber\\ &\qquad
  + 4pr^{4}H_{1}' + \lhat\,r^{2}H_{0} - \big(\lhat\,r^{2}+4r_{e}^{2}\big)H_{2}
  + 8r_{e}^{2}K - 4\lhat\,r_{e}^{2}\,w\Big]
\nonumber\\ &{}
  + 4pr^{4}(r-GM)\,H_{1} - 2p^{2}r^{6}\big(H_{2}+K\big) = 0 ,
\label{eq:m-etrace}\\[6pt]
\mathcal E^{\rm tl}:\quad & -\tfrac12\big(H_{0}-H_{2}\big) = 0 ,
\label{eq:m-etl}\\[6pt]
\mathcal M^\phi:\quad & r\Delta^{2}\,w'' + 2\big(GMr-r_{e}^{2}\big)\Delta\,w'
  - r\big(\lhat\,\Delta+p^{2}r^{4}\big)w
  + r\Delta\big(H_{0}-H_{2}+2K\big) = 0 .
\label{eq:m-mph}
\end{align}
Here $\mathcal E_{ab}$ and $\mathcal M^\phi$ are the residuals \eqref{eq:Edef} and \eqref{eq:Mdef}. Each is the coefficient of its angular harmonic, $Y$ or $Y'$ as listed in Sec.~\ref{sec:separation}, and the harmonic itself is not written. $\mathcal E^{\rm trace}$ and $\mathcal E^{\rm tl}$ are defined by \eqref{eq:tracepart} and \eqref{eq:tlcheck}, and $\Delta$ by \eqref{eq:m-abbrev}. Each equation has been multiplied by an $r$-dependent factor that clears denominators.

\subsection{The first-order matrix and the second-order coefficients}
\label{app:mouth-M}

With $\Delta$, $\mathcal C$ and $\mathcal D$ as in \eqref{eq:m-abbrev}, the matrix $\mathsf M$ of the closed first-order system \eqref{eq:firstorder}, $\bm y'=\mathsf M\bm y$, $\bm y=(K,H_1,w,w')^\top$, is
\begin{align}
  \mathsf M_{11} &= \frac{2p^2r^6 + GM(\lhat-6)\,r^3 + 12G^2M^2r^2 - (\lhat-8)\,r^2r_e^2 - 22GM\,r\,r_e^2 + 8r_e^4}
                {r\,\Delta\,\mathcal C},
  \nonumber\\
  \mathsf M_{12} &= -\,\frac{\mathcal D}{2p\,r^2\,\mathcal C},
  \qquad
  \mathsf M_{13} = -\,\frac{2\lhat\, r_e^2}{r\,\mathcal C},
  \qquad
  \mathsf M_{14} = -\,\frac{4r_e^2\,\Delta}{r^2\,\mathcal C},
  \nonumber\\
  \mathsf M_{21} &= \frac{2p\,r^2\Big[p^2r^6 + (\lhat-2)\,r^4 + 2GM(4-\lhat)\,r^3 + (\lhat-3)\,r^2r_e^2 - 9G^2M^2r^2 + 8GM\,r\,r_e^2 - 2r_e^4\Big]}
                {\Delta^2\,\mathcal C},
  \nonumber\\
  \mathsf M_{22} &= -\,\frac{2p^2r^6 + GM(3\lhat-4)\,r^3 + 12G^2M^2r^2 - (3\lhat-4)\,r^2r_e^2 - 20GM\,r\,r_e^2 + 8r_e^4}
                {r\,\Delta\,\mathcal C},
  \nonumber\\
  \mathsf M_{23} &= -\,\frac{4p\,r_e^2\big[(\lhat-1)\,r^2 + 3GM\,r - 2r_e^2\big]}{\Delta\,\mathcal C},
  \qquad
  \mathsf M_{24} = -\,\frac{4p\,r\,r_e^2}{\mathcal C},
  \nonumber\\
  \mathsf M_{31} &= \mathsf M_{32} = \mathsf M_{33} = 0,
  \qquad
  \mathsf M_{34} = 1,
  \nonumber\\
  \mathsf M_{41} &= -\,\frac{2}{\Delta},
  \qquad
  \mathsf M_{42} = 0,
  \qquad
  \mathsf M_{43} = \frac{p^2r^4 + \lhat\,\Delta}{\Delta^2},
  \qquad
  \mathsf M_{44} = -\,\frac{2\,(GM\,r - r_e^2)}{r\,\Delta} .
\label{eq:m-M}
\end{align}
The first two rows are the derivative rules \eqref{eq:constraints} after substitution of \eqref{eq:m-H2alg}. The last row is the Maxwell equation \eqref{eq:m-mph} solved for $w''$.

For reference, the matrices $P,Q$ of the second-order system \eqref{eq:secondorderRN} in the coordinate $r$, related to the tortoise-frame matrices \eqref{eq:m-Ptilde}--\eqref{eq:m-QKK} by \eqref{eq:m-PQfromtilde}, are, entry by entry, with $\Delta$, $\mathcal C$ and $\mathcal D$ as in \eqref{eq:m-abbrev},
\begin{equation}
\begin{aligned}
  P_{KK} &= \frac{2\,(GMr - r_e^{2})}{r\,\Delta} - \frac{2\lhat\,\mathcal C}{r\,\mathcal D},
  &\qquad
  P_{Kw} &= \frac{4r_e^{2}\,\big(4p^{2}r^{2} - \lhat^{2}\big)}{r\,\mathcal D},
  \\[4pt]
  P_{wK} &= 0,
  &\qquad
  P_{ww} &= \frac{2\,(GMr - r_e^{2})}{r\,\Delta},
\end{aligned}
\end{equation}
\begin{equation}
\begin{aligned}
  Q_{KK} &= -\frac{p^{2}r^{4}}{\Delta^{2}}
  + \frac{8p^{2}r^{2}\big(3GMr - 4r_e^{2}\big) + \lhat^{2}\big[(\lhat-2)\,r^{2} + 4r_e^{2}\big]}{\Delta\,\mathcal D},
  \\[4pt]
  Q_{Kw} &= \frac{2\lhat\,r_e^{2}\,\big(4p^{2}r^{2} - \lhat^{2}\big)}{\Delta\,\mathcal D},
  \qquad
  Q_{wK} = \frac{2}{\Delta},
  \qquad
  Q_{ww} = -\frac{p^{2}r^{4} + \lhat\,\Delta}{\Delta^{2}} .
\end{aligned}
\end{equation}

\section{The \texorpdfstring{$O(\alpha)$}{O(alpha)} terms of the mouth}
\label{app:mouthalpha}

This appendix gives the four fermion coefficients of \eqref{eq:m-wrow} and the expansion of the $O(\alpha)$ mouth near the horizon. They are the parts of the $O(\alpha)$ coefficients that do not come from $A=e^{2\alpha\Phi_1}N$. The function $\Phi_1$ enters the equation for $w$ only through $A$. With $y\equiv r-r_e$, $\lhat=l(l+1)$, $p=-i\omega$ and $\mathcal D$ of \eqref{eq:m-abbrev} at $GM=r_e$,
\begin{align}
  \mathcal F_{wK} &= -\,\frac{y\,\big(8p^{2}r^{4}r_e - 3\lhat^{2}r^{2}y + 6\lhat\,r^{3} - 24\lhat\,r^{2}r_e + 34\lhat\,r\,r_e^{2} - 16\lhat\,r_e^{3}\big)}{3r^{3}\,\mathcal D} ,
  \label{eq:wrow-PwK}\\[4pt]
  \mathcal F_{ww} &= \frac{r_e\,y^{2}\big[\,5\,(r-2r_e)\,\mathcal D + 4\lhat\,r_e\,y\,(3r-5r_e)\big]}{3r^{5}\,\mathcal D} ,
  \label{eq:wrow-Pww}\\[4pt]
  \mathcal T_{wK} &= -\,\frac{y}{3r^{6}\,\mathcal D}\Big[\lhat^{3}r^{4}y - 5\lhat^{2}r^{2}r_e\,y\,(2r-3r_e) + 2\lhat\,p^{2}r^{6}(r-3r_e)
  \nonumber\\&\hspace{2.2cm}
  - 4\lhat\,y^{2}\big(r^{3}-5r^{2}r_e+6r\,r_e^{2}-r_e^{3}\big) - 4p^{2}r^{4}y\,\big(2r^{2}-17r\,r_e+22r_e^{2}\big)\Big]
  \nonumber\\&\quad
  + \frac{y^{2}\big[(\lhat-4)\,r^{2} + 20\,r\,r_e - 20\,r_e^{2}\big]}{6r^{6}} ,
  \label{eq:wrow-SwK}\\[4pt]
  \mathcal T_{ww} &= -\,\frac{\lhat\,r_e^{2}\,y^{2}\big(12p^{2}r^{4} - 3\lhat^{2}r^{2} + 4\lhat\,r_e\,r - 4\lhat\,r_e^{2}\big)}{3r^{6}\,\mathcal D}
  \nonumber\\&\quad
  + \frac{y\big[\,8p^{2}r^{4}r_e - \lhat(\lhat-2)\,r^{3} + \lhat(\lhat-10)\,r^{2}r_e + 16\lhat\,r\,r_e^{2} - 8\lhat\,r_e^{3}\big]}{12r^{6}} .
  \label{eq:wrow-Sww}
\end{align}
The term $5(r-2r_e)\mathcal D$ in $\mathcal F_{ww}$ and the last terms of $\mathcal T_{wK}$ and $\mathcal T_{ww}$ come from the motion of the field lines, through the tilt, the density and the scale of the line metric (Sec.~\ref{sec:lift}). The coefficient $\mathcal F_{wK}$ does not depend on them. All four vanish at the horizon. For $p\neq0$, $\mathcal F_{wK}$, $\mathcal T_{wK}$ and $\mathcal T_{ww}$ vanish linearly in $y$ and $\mathcal F_{ww}$ quadratically. At $p=0$ all four vanish quadratically.

Near the horizon the whole $O(\alpha)$ mouth, the equation for $K$ included, takes a simple form. Write $\hat y=y/r_e$ and set $p=0$, since the frequency does not enter the orders compared in Sec.~\ref{sec:overlap}. The $\alpha^0$ mouth at $GM=r_e$ and the $O(\alpha)$ part of the corrected mouth, which contains the fermion terms, $N_1$, $\Phi_1$ and the mass shift $GM=r_e(1-\alpha/18)$, give
\begin{equation}
  r_e^{2}\,\tilde Q^{\rm mouth} \;=\; -\mathcal N \hat y^{2} \;+\; \mathsf Q_3\, \hat y^{3}
  \;+\; \alpha\Big[\tfrac13\,\mathcal N \hat y \;+\; \Big(\mathsf Q_2 + \tfrac43\,\mathcal N\log \hat y\Big)\hat y^{2}\Big] \;+\; \dots ,
  \label{eq:mouth-nh}
\end{equation}
where $\mathcal N$ is the matrix of \eqref{eq:t-Qtilde0}, the dots stand for higher orders in $\hat y$ and $\alpha$, and
\begin{equation}
  \mathsf Q_3 \;=\; \begin{pmatrix} 4(\lhat+4) & -12\lhat\\ -8 & 4\lhat\end{pmatrix},
  \qquad
  \mathsf Q_2 \;=\; \begin{pmatrix} -\tfrac{5\lhat+22}{3} & \tfrac{\lhat(\lhat+30)}{6}\\[4pt] \tfrac{3\lhat+14}{6} & -\tfrac{\lhat(\lhat+18)}{12}\end{pmatrix} .
  \label{eq:mouth-nh-DC}
\end{equation}
The subscripts of $\mathsf Q_3$ and $\mathsf Q_2$ are the powers of $\hat y$ they multiply. The $O(\alpha)$ part has no term of order $\hat y^0$, since $N$ and $N'$ vanish together at $r_h$ to $O(\alpha)$. On the dictionary $\ell^2/r_e^2=16/\pi^2\alpha^2$ and $\hat y=\tfrac{\pi\alpha}{4}\,\ell y/r_e^2$, with $\ell y/r_e^2$ from \eqref{eq:overlap-sigma}. The term $-\mathcal N \hat y^2$ becomes $-\mathcal N(\ell y/r_e^2)^2$, which contains $-\tfrac{4}{3\pi}\mathcal N\rho$ and $-\tfrac23\mathcal N\alpha\rho^2\big[1+2\log\tfrac{\pi\alpha\rho}{4}\big]$. The term $\tfrac13\alpha\mathcal N \hat y$ gives $\tfrac{4}{3\pi}\mathcal N\rho$ and cancels the first. The logarithm of the $\alpha \hat y^2$ term gives $\tfrac43\mathcal N\alpha\rho^2\log\tfrac{\pi\alpha\rho}{4}$ and cancels the logarithm of the second. The term $\mathsf Q_3 \hat y^3$ becomes $\tfrac{\pi\alpha}{4}\mathsf Q_3\big(\rho^3+\tfrac{2}{\pi}\rho^2\big)$ plus higher orders, and its $\rho^3$ part is the $\alpha\rho^3$ part of \eqref{eq:overlap-mouth}. The $\alpha\rho^2$ terms that remain are \eqref{eq:mouth-arho2}.

\section{The constraint rules}
\label{app:constraints}

This appendix gives the derivative rules \eqref{eq:constraints} explicitly. They are obtained by solving $\mathcal E_{t\theta}$, $\mathcal E_{r\theta}$ and $\mathcal E_{tr}$ for $H_1'$, $H_2'$ and $K'$ after imposing $H_0=H_2$. The Laplace variable is $p$, and $\lhat=l(l+1)$.

In the mouth, with $\Delta$ as in \eqref{eq:m-abbrev} and a prime denoting $\dd/\dd r$,
\begin{align}
  H_1' &= \frac{2\big(r_e^2-GM r\big)H_1 + p\,r^3\big(H_2+K\big) - 2p\,r\,r_e^2\,w}{r\,\Delta} ,
  \label{eq:m-H1p}\\[4pt]
  H_2' &= \frac{\lhat\,\Delta+2p^2r^4}{2p\,r^2\Delta}\,H_1
     + \frac{r^2-4GMr+3r_e^2}{r\,\Delta}\,H_2
     - \frac{r^2-3GMr+2r_e^2}{r\,\Delta}\,K
     - \frac{2r_e^2}{r^2}\,w' ,
  \label{eq:m-H2p}\\[4pt]
  K' &= \frac{\lhat}{2p\,r^2}\,H_1 + \frac{1}{r}\,H_2 - \frac{r^2-3GMr+2r_e^2}{r\,\Delta}\,K .
  \label{eq:m-Kp}
\end{align}

In the throat a prime denotes $\dd/\dd\rho$, $p$ is the throat Laplace variable, and $f=1+\rho^2$. Each rule is split as $H_1'=H_1'{}_{(0)}+\alpha\,H_1'{}_{(1)}+O(\alpha^2)$, and likewise for $H_2'$ and $K'$. At $\alpha=0$,
\begin{align}
  H_1'{}_{(0)} &= -\frac{2\rho}{f}\,H_1 + \frac{p\,r_e^2}{f}\,\big(H_2+K-2w\big) ,
  \label{eq:t-H1p0}\\[4pt]
  H_2'{}_{(0)} &= \frac{\lhat f+2p^2}{2p\,r_e^2\,f}\,H_1 - \frac{2\rho}{f}\,H_2 + \frac{\rho}{f}\,K - 2\,w' ,
  \label{eq:t-H2p0}\\[4pt]
  K'_{(0)} &= \frac{\lhat}{2p\,r_e^2}\,H_1 + \frac{\rho}{f}\,K .
  \label{eq:t-Kp0}
\end{align}
These are the near-horizon limits of \eqref{eq:m-H1p}--\eqref{eq:m-Kp}, except that the term $H_2/r$ in $K'$ is absent since the sphere radius is constant at this order. The first-order corrections are
\begin{align}
    H_1'{}_{(1)} &= \frac{1}{3f^{2}}\Big\{\big[\rho\,(3\rho^{2}+11) + 6\rho\log f + 3(\rho^{4}+3)\arctan\rho\big]H_1
     - 3p\,r_e^{2}\,\gamma\,\big(H_2+K\big)
  \nonumber\\&\qquad
     + p\,r_e^{2}\big(9 - 2\rho^{2} + 6\log f - 12\rho\arctan\rho\big)\,w\Big\},
  \label{eq:t-H1p1}\\[6pt]
  H_2'{}_{(1)} &= -\Big[\frac{\lhat\,(2+3\rho\arctan\rho)}{6p\,r_e^{2}}
       + \frac{p\,(f+3\gamma)}{3r_e^{2}f^{2}}\Big]H_1
     + \frac{\rho\,(7\rho^{2}+23) + 12\rho\log f + 3(3\rho^{4}+2\rho^{2}+7)\arctan\rho}{6f^{2}}\,H_2
  \nonumber\\&\quad
     - \frac{\mathcal B}{3f^{2}}\,K + \Big[2\big(1+\rho\arctan\rho\big) - \frac{\lhat f + 4\rho^{2} + 10}{6f}\Big]w' + \frac{\lhat\rho}{6f}\,w + \frac{2(p^{2}+4)}{3f}\,\xi_\rho ,
  \label{eq:t-H2p1}\\[6pt]
  K'_{(1)} &= -\Big[\frac{\lhat\,(2+3\rho\arctan\rho)}{6p\,r_e^{2}} + \frac{p}{3r_e^{2}f}\Big]H_1
     + \frac{\rho + 3f\arctan\rho}{6f}\,H_2
     - \frac{\mathcal B}{3f^{2}}\,K - \frac{\lhat+4}{6}\,w' + \frac{\lhat\rho}{6f}\,w + \frac{2(p^{2}+4)}{3f}\,\xi_\rho ,
  \label{eq:t-Kp1}
\end{align}
with $\gamma=-\tfrac43\rho^2-\rho(\rho^2+3)\arctan\rho+\log f$ the background function of \eqref{eq:bgsol} and
\begin{equation}
  \mathcal B \equiv 2\rho(\rho^{2}+3) + 3\rho\log f + 3(\rho^{4}+\rho^{2}+2)\arctan\rho .
\end{equation}
$H_2'$ and $K'$ share the same $K$ coefficient, the same $w$ coefficient and the same $\lhat$ term in the $H_1$ coefficient. The transport field $\xi_\rho$ enters only at $O(\alpha)$, through $\mathcal E_{\tau\rho}$. It enters $H_2'$ and $K'$ with the same coefficient. Its own derivative rule, and that of $\xi_\tau$, are \eqref{eq:xit}--\eqref{eq:xir}.

\section{The one-loop action of the higher Landau levels}
\label{app:EH}

In this appendix $B$ is the background field $\bar B$ of an orthonormal frame and $E$ an electric field along it. For parallel $\bm E$ and $\bm B$ the one-loop Lagrangian of a Dirac fermion of unit charge has Schwinger's proper-time form \cite{Schwinger:1951nm,Dunne:2004nc}
\begin{equation}
  \mathcal L_1=-\frac{1}{8\pi^2}\int_0^\infty\frac{\dd s}{s^3}\Big[(Bs\coth Bs)(Es\cot Es)-1-\frac{s^2}{3}\big(B^2-E^2\big)e^{-\mu_{\rm R}^2s}\Big],
  \label{eq:propertime}
\end{equation}
where the last term is the counterterm that defines the coupling at the scale $\mu_{\rm R}$. The factor $Bs\coth Bs=Bs+\varphi_{\rm L}(Bs)$, with $\varphi_{\rm L}(u)=2u/(e^{2u}-1)=\sum_{n\ge1}2u\,e^{-2nu}$, separates the Landau levels. The term $n$ of the sum is the level of energy squared $2nB$, and the term $Bs$ is the lowest level. Its part at $E=0$ is scaleless and is dropped. Its dependence on $E$ is the vacuum polarisation of the massless two-dimensional fermions of the lowest level, which is infrared divergent. In the companion sector it is the Schwinger response, and in this sector it does not enter. Without the lowest level,
\begin{equation}
  \mathcal L_{\rm H}=-\frac{1}{8\pi^2}\int_0^\infty\frac{\dd s}{s^3}\Big[\varphi_{\rm L}(Bs)\,(Es\cot Es)+Bs-1-\frac{s^2}{3}\big(B^2-E^2\big)e^{-\mu_{\rm R}^2s}\Big],
  \label{eq:LH}
\end{equation}
where the term $Bs$ restores the scaleless piece and makes the integral converge at $s\to0$. At $E=0$ the integral is $-B^2I_0/8\pi^2+(B^2/24\pi^2)\log(B/\mu_{\rm R}^2)$, with
\begin{equation}
  I_0=\int_0^\infty\frac{\dd u}{u^3}\Big[\varphi_{\rm L}(u)-1+u-\frac{u^2}3e^{-u}\Big]=\frac13-4\zeta'(-1)-\frac13\log2 .
\end{equation}
Adding the tree term $-B^2/2g_{\mu_{\rm R}}^2$ and defining $g_c$ by $\mathsf h(B_0)=1/g_c^2$ removes $\mu_{\rm R}$ and $I_0$ and gives $\mathsf h$ of \eqref{eq:EHcoef1}. On the magnetic background $\mathcal L(\mathcal Q)=\mathcal Q\,\mathsf h\big(\sqrt{-2\mathcal Q}\big)$, which gives $\mathcal L_{\mathcal Q}=\mathsf h-1/24\pi^2$ and $\mathcal L_{\mathcal Q\mathcal Q}=1/12\pi^2B^2$, the first two relations of \eqref{eq:EHcoef2}. For $\bm E\parallel\bm B$ one has $\mathcal R=EB$ and $\mathcal Q=\tfrac12(E^2-B^2)$, and the $E^2$ term of $\mathcal L$ is $\tfrac12(\mathcal L_{\mathcal Q}+B^2\mathcal L_{\mathcal R\mathcal R})E^2$. From $Es\cot Es=1-E^2s^2/3+\dots$ the $E^2$ term of \eqref{eq:LH} is
\begin{equation}
  \frac{E^2}{24\pi^2}\int_0^\infty\frac{\dd s}{s}\big[\varphi_{\rm L}(Bs)-e^{-\mu_{\rm R}^2s}\big]=\frac{E^2}{24\pi^2}\Big[-C_\varphi-\log\frac{B}{\mu_{\rm R}^2}\Big],
\end{equation}
with
\begin{equation}
  C_\varphi=\int_0^\infty\frac{\dd u}{u}\big[e^{-u}-\varphi_{\rm L}(u)\big]=\log2-\gamma_{\rm E}.
\end{equation}
With the tree term $E^2/2g_{\mu_{\rm R}}^2$ and the same $g_c$ the scale drops out again, and
\begin{equation}
  B^2\mathcal L_{\mathcal R\mathcal R}=\frac{1}{12\pi^2}\Big(\frac12-3I_0-C_\varphi\Big)=\frac{c_\parallel}{12\pi^2},
\end{equation}
which is the third relation of \eqref{eq:EHcoef2}. The trace $4(\mathcal L-\mathcal Q\mathcal L_{\mathcal Q})=-B^2/12\pi^2$ is the trace anomaly of one Dirac fermion. We evaluated $I_0$ and $C_\varphi$ by quadrature to 35 digits and identified the closed forms. A second derivation sums the levels as two-dimensional massive fermions with a covariant zeta-function regularisation. It gives the same $\mathsf h$, $\mathcal L_{\mathcal Q\mathcal Q}$ and $c_\parallel$.

The potential of the master equations \eqref{eq:NYSeqs} of the mouth is \cite{Nomura:2020tpc}
\begin{equation}
\begin{aligned}
  V_{11}&=-\frac{\mu^2}{r^2}+\frac{2\mu^2(\mu^2+2-2A)}{r^2\mathcal Y}+\frac{2\mu^4A}{r^2\mathcal Y^2}+\frac{8r_e^2\varepsilon_\perp\mu^2A}{r^4\mathcal Y^2},\\
  V_{12}&=2\mu r_e\sqrt{\varepsilon_\perp}\,\Big[\frac1{r^3}\Big(1-\frac{2(\mu^2+2-A)}{\mathcal Y}-\frac{2\mu^2A}{\mathcal Y^2}\Big)+\frac{2\eta_{\rm H}A}{\varepsilon_\perp r^3\mathcal Y}-\frac{8r_e^2\varepsilon_\perp A}{r^5\mathcal Y^2}\Big],\\
  V_{22}&=\frac{\lhat}{r^2}\Big(1-\frac{\eta_{\rm H}}{\varepsilon_\perp}\Big)-\frac{4r_e^2\varepsilon_\perp}{r^4}+\frac{8r_e^2\varepsilon_\perp}{r^4\mathcal Y}\Big(\lhat+\frac{\mu^2A}{\mathcal Y}-\frac{2\eta_{\rm H}A}{\varepsilon_\perp}\Big)+\frac{32r_e^4\varepsilon_\perp^2A}{r^6\mathcal Y^2},
\end{aligned}
\label{eq:NYSV}
\end{equation}
written with \eqref{eq:EHcoef1}, \eqref{eq:EHcoef2}, $\kappa q^2/8g_c^2=r_e^2$ and the variables \eqref{eq:NYSvars}. Here $A$ is \eqref{eq:hllmouth}, $\mathcal Y=rA'-2A+\lhat$ and $\varepsilon_\perp=1-\eta_{\rm H}/2+2\eta_{\rm H}\log(r/r_e)$. At $r=r_e$ and $GM=r_e(1+\eta_{\rm H})$ one has $A=0$ and $\mathcal Y=\lhat$, and $R_A^2V$ has the eigenvalues of \eqref{eq:NH}, whose first-order form is \eqref{eq:mH}.

\section{Numerical methods}
\label{app:numerics}

This appendix describes the numerical methods behind the results of Secs.~\ref{sec:mouth-master}, \ref{sec:overlap}, \ref{sec:stability} and \ref{sec:hll}. The other results are exact, apart from the closed forms of $I_0$ and $C_\varphi$ in Appendix~\ref{app:EH}, which we identified from 35-digit quadrature.

The comparison at the junction in Sec.~\ref{sec:overlap} was computed in 120-digit arithmetic, at which $GM=r_e+\delta_M$ is resolved. The constants \eqref{eq:overlap-params} follow from \eqref{eq:re}, \eqref{eq:alpha} and \eqref{eq:ell} with $\kappa=8\pi\ell_p^2$, and the mass defect is $\delta_M=-r_e^3/2\ell^2$ (MMP eq.~(5.42)). The mouth side is the $\alpha^0$ system \eqref{eq:m-Ptilde}--\eqref{eq:m-QKK} at $GM=r_e+\delta_M$ and $p_{\rm mouth}=-i\Om/\ell$. The throat side is \eqref{eq:t-master-tortoise} with $\tilde P=\alpha\tilde P_1$ and $\tilde Q=\tilde Q_0+\alpha\tilde Q_1$ at $\rho_{\rm J}=\ell(r_{\rm J}-r_e)/r_e^2$. Near the junction $\Delta$ is smaller than $r^2$ by twelve orders of magnitude, and the relative differences in the table are of order $10^{-11}$. With 60 digits, at which $GM$ rounds to $r_e$ since $|\delta_M|/r_e\simeq2\times10^{-70}$, these differences change by less than $10^{-38}$ of their size, and with 30 digits by about $10^{-8}$. The scan over the junction position repeats the evaluation at $\delta_{\rm J}=10^{-9},10^{-8},\dots,10^{-2}$. The size of the $O(\alpha)$ mouth terms at the junction, $|\alpha\tilde Q_1/\tilde Q_0|=8.4\times10^{-30}$ for every entry, was evaluated in 50-digit arithmetic with $\Phi_1$ from \eqref{eq:mouthbgsol} and with $GM=r_e(1-\alpha/18)$. The same evaluation at $l=2$ and $p=0$, at $r-r_e=10^{-3}r_e$, $0.1\,r_e$, $r_e$ and $10\,r_e$, gives the other sizes quoted in Sec.~\ref{sec:mouth-master}.

For Sec.~\ref{sec:stab-throat1} the matrices $V_0$, $V_{1a}$ and $V_{1b}$ are evaluated in double precision from \eqref{eq:V0}, Appendix~\ref{app:V1a} and \eqref{eq:V1b}, with $p$ real. The smallest eigenvalue of $V_{\rm sym}$ is taken on 2000 equally spaced points of $0\le\alpha\rho\le0.1$, for $l=2,3,4$, $p=0,1,3$ and $\alpha=10^{-6},10^{-3},10^{-2}$. Over the whole scan $\min{\rm eig}\,V_{\rm sym}/l(l-1)$ stays above $0.987$, with the lowest value at $l=4$, $p=3$ and $\alpha=10^{-2}$. The threshold is the first zero of the smallest eigenvalue as a function of $\alpha\rho$. We bracket it on 300 points of $0.02\le\alpha\rho\le1$ and refine it to $10^{-7}$ in $\alpha\rho$ by Brent's method \cite{Brent:1971}. It is the same for $p=0,1,3,10,30$. At $\alpha=10^{-6}$ it is $0.241$, $0.269$ and $0.282$ for $l=2,3,4$, and at $\alpha=10^{-2}$ it is larger by at most $0.003$. The same scan for the coordinate lift of Sec.~\ref{sec:response} gives the same thresholds at $\alpha=10^{-6}$. At $\alpha=10^{-2}$ its thresholds lie below those of the scaled lift by at most $0.002$. The smallest eigenvalue of $(1+\alpha V_{1b})_{\rm sym}$ is taken on 4000 equally spaced points of $0\le\rho\le10$, and its minimum is $0.988$ at $l=2$ and $\alpha=10^{-2}$. For Sec.~\ref{sec:hll-throat} the scan is repeated with $\eta_{\rm H}=1/12\pi^2$, $0.05$ and $0.1$ added, for both lifts, in the eigenbasis of \eqref{eq:NH} and in the basis $Z_\pm$ of $\eta_{\rm H}=0$. With $\eta_{\rm H}>0$ the smallest ratio is never below its value at $\eta_{\rm H}=0$. At $\alpha=10^{-6}$ and $\eta_{\rm H}=1/12\pi^2$ the thresholds in the eigenbasis of \eqref{eq:NH} are $0.242$, $0.270$ and $0.283$ for $l=2,3,4$. The discriminant \eqref{eq:discH} uses the matrices $\mathcal S$ of the levels examined in Sec.~\ref{sec:stab-normalmodes}, computed as above.

The matrices $\mathcal S$ of \eqref{eq:Smatrix} are computed for both lifts by tanh-sinh quadrature \cite{Takahasi:1974} on $[-\pi/2,0]$ and $[0,\pi/2]$ in 30-digit arithmetic, with the modes \eqref{eq:PTmodes} evaluated from the Gegenbauer polynomials. The nodes of this rule cluster double-exponentially at the ends, where the factor $\cos^{2s}x$ of the modes meets the growth of $V_{1a}$. The remap term contains $\iota(\tan x)$, which is itself an integral, and we reduce it to a single quadrature by parts. Let mode $i$ have channel $\mathsf c$, index $n$ and $s$ as in \eqref{eq:PTmodes}, and write $V_{0,\mathsf c}=m_{\mathsf c}^2\sec^2x$ with $m_+^2=(l+1)(l+2)$ and $m_-^2=l(l-1)$. Then
\begin{equation}
  \int_{-\pi/2}^{\pi/2}\dd x\;2\rho\,\iota\,V_{0,\mathsf c}\,\psi_i^2
  \;=\; -2\int_0^{\pi/2}\dd x\;U_i(\sin x)\,\gamma(\tan x)\cos^2x ,
  \qquad
  U_i(u) \;=\; -2m_{\mathsf c}^2\int_u^1\dd v\;v\,(1-v^2)^{s-2}\big[C^{(s)}_n(v)\big]^2 ,
  \label{eq:iotaparts}
\end{equation}
with $\rho=\tan x$. The identity follows from $\dd\iota/\dd x=\gamma\cos^2x$, which is \eqref{eq:t-tortoise} at leading order, and from the evenness of the integrand on the left. The boundary terms vanish, since $\iota(0)=0$ and $U_i$ vanishes as $\cos^{2s-2}x$ at $x=\pi/2$ while $\iota$ grows only as $\log\rho$. The polynomial $U_i$ is integrated exactly. For the mode $n=2$ of $Z_-$ at $l=2$, \eqref{eq:iotaparts} agrees with a nested quadrature of its left side to twelve digits.

The closed forms in \eqref{eq:Sl2}, the shifts quoted for $l=2$, the diagonal entries of both lifts and the products of Table~\ref{tab:lifts} were identified with the PSLQ integer-relation algorithm \cite{Ferguson:1999}. Each diagonal entry was tested against the basis $\{1,\log2\}$, and each off-diagonal entry and each product against $\{1\}$, with tolerance $10^{-22}$ and integer coefficients up to $10^{12}$. For the diagonal entries the largest coefficient found is about $1.1\times10^7$ and the largest residual $8\times10^{-25}$, two orders of magnitude below the tolerance, and the coefficient of $\log2$ is $2\Om_0^2$ in every case. As a check we also computed $\mathcal S$ after the change of variables $x_0=\arctan\rho$, the tortoise coordinate at $\alpha=0$. With $\hat\gamma=\gamma/f$ one has $\partial_x=(1+\alpha\hat\gamma)\partial_{x_0}$ exactly, and with $\bm Z=(1-\tfrac{\alpha}{2}\hat\gamma)\bm W$ Eq.~\eqref{eq:throatOm} becomes, to first order in $\alpha$,
\begin{equation}
  -\partial_{x_0}^2\bm W+\Big[V_0+\alpha\Big(V_{1a}-2\hat\gamma V_0+\tfrac12\partial_{x_0}^2\hat\gamma\Big)\Big]\bm W
  \;=\;\Om^2\big[1+\alpha\big(V_{1b}-2\hat\gamma\big)\big]\bm W ,
  \label{eq:yform}
\end{equation}
where $V_0=\mathrm{diag}(m_+^2,m_-^2)\sec^2x_0$ holds exactly. The diagonal terms in $\hat\gamma$ take the place of the remap term of \eqref{eq:xform}. Projecting \eqref{eq:yform} on the modes \eqref{eq:PTmodes} in $x_0$, in 40-digit arithmetic, gives the same closed forms for $l=2$ and $\Om_0=2,\dots,6$.

The growth law of Sec.~\ref{sec:hll-throat} and the statements of Sec.~\ref{sec:stab-normalmodes} on the higher levels use the levels up to $\Om_0=20$ for $l=2$ and up to $\Om_0=15$ for $l=3$. On these levels $\mathcal S$ is computed for both lifts in 40-digit arithmetic, with $[-\pi/2,\pi/2]$ cut into sixteen subintervals of equal length. PSLQ then uses tolerance $10^{-30}$ and integer coefficients up to $10^{20}$. It identifies every entry except the diagonal entry of $Z_+$ at $l=2$ and $\Om_0=20$, for both lifts. Recomputed in 60-digit arithmetic on 48 subintervals, these two entries are identified with tolerance $10^{-45}$, and their rational parts have the denominator $22309287$. Every product of the off-diagonal entries is rational and equals \eqref{eq:prodS} for the scaled lift and $-4/\mu^2$ times \eqref{eq:prodS} for the coordinate lift. On every one of these levels $\mathcal S_{++}>\mathcal S_{--}$, and the discriminant \eqref{eq:discH} is positive at $\eta_{\rm H}=0$.

The form found by PSLQ holds on every level. For integer $s$ the Gegenbauer polynomials have rational coefficients and parity $(-1)^n$. Hence $\psi_i^2$ is $\cos^{2s}x$ times a polynomial in $z=\sin^2x$ with rational coefficients. For both lifts the diagonal entries of $V_{1a}$ are $m_{\mathsf c}^2\log f$ plus an even rational function of $\rho$ plus $\rho\arctan\rho$ times another even rational function. Those of $V_{1b}$ are even rational functions of $\rho$. Every denominator is a power of $f=1/(1-z)$ times a polynomial in $l$ that does not vanish for $l\ge2$. With $\rho=\tan x$ and $\arctan\rho=x$, all integrals in \eqref{eq:Smatrix} and \eqref{eq:iotaparts} reduce to
\begin{equation}
  \begin{gathered}
  \int_{-\pi/2}^{\pi/2}\!\sin^{2a}x\,\dd x=\pi\,\mathfrak c_a,\qquad
  \int_{-\pi/2}^{\pi/2}\!x\sin^{2a+1}x\cos x\,\dd x=\frac{\pi(1-\mathfrak c_{a+1})}{2a+2},\\
  \int_{-\pi/2}^{\pi/2}\!\sin^{2a}x\log\cos x\,\dd x=-\pi\,\mathfrak c_a\Big(\log2+\frac{\mathfrak h_a}{2}\Big),
  \end{gathered}
  \label{eq:threeints}
\end{equation}
with $\mathfrak c_a=(2a)!/(4^aa!^2)$ and $\mathfrak h_a$ the harmonic number. The third follows from $\int_0^{\pi/2}\log\cos x\,\dd x=-\tfrac\pi2\log2$ and the recurrence obtained by integrating $\partial_x(\sin^{2a+1}x\cos x\log\cos x)$. Only the logarithms produce $\log2$. The term $m_{\mathsf c}^2\log f=-2m_{\mathsf c}^2\log\cos x$ of $V_{1a}$ contributes $2m_{\mathsf c}^2\log2$. In \eqref{eq:iotaparts} the term $\log f$ of $\gamma$ contributes $-4\log2\int_0^{\pi/2}U_i\cos^2x\,\dd x$, divided by the norm. The derivative of $U_i(\sin x)$ is $\psi_i^2\,\partial_xV_{0,\mathsf c}$. An integration by parts with $\int_0^x\cos^2x'\,\dd x'$ then turns $-4\int_0^{\pi/2}U_i\cos^2x\,\dd x$ into $\int x\,\psi_i^2\,\partial_xV_{0,\mathsf c}\,\dd x+2m_{\mathsf c}^2\int\tan^2x\,\psi_i^2\,\dd x$ over $(-\pi/2,\pi/2)$. The virial identity $\int x\,\psi_i^2\,\partial_xV_{0,\mathsf c}\,\dd x=2\int(\partial_x\psi_i)^2\,\dd x$ of the Poschl-Teller problem makes the remap contribution $2(\Om_0^2-m_{\mathsf c}^2)\log2$. Every diagonal entry is therefore a rational number plus $2\Om_0^2\log2$. The off-diagonal entries contain no $\log f$ and no remap term, and they are rational. Appendix~\ref{app:offdiag} gives them in closed form. Evaluated exactly, \eqref{eq:threeints} reproduces \eqref{eq:Sl2}, Table~\ref{tab:lifts} and the PSLQ values on the highest levels computed, including the entries with the denominator $22309287$.

The direct solution quoted in Sec.~\ref{sec:stab-normalmodes} discretises the throat equation by second-order finite differences on $30000$ interior nodes, uniform in $x$. Dirichlet conditions are imposed at $\alpha\rho=\pm 0.1$, where the modes have decayed to $\cos^sx\sim(10\alpha)^s$. We solve two forms of the equation. Form (a) is \eqref{eq:xform}, with $\iota$ obtained from $\dd\iota/\dd x=\gamma\cos^2x$ by the trapezoidal rule. Form (b) does not expand the remap. It takes $\rho(x)$ from $\dd\rho/\dd x=f+\alpha\gamma$, integrated by the explicit eighth-order Runge--Kutta method of Prince and Dormand \cite{Prince:1981} with relative tolerance $10^{-12}$, and evaluates $V_0=\mathrm{diag}(m_+^2,m_-^2)\,f$, $V_{1a}$ and $V_{1b}$ at that $\rho$. In both forms $1+\alpha V_{1b}$ is inverted node by node, and the eigenvalues $\Om^2$ nearest $\Om_0^2$ are found by shift-and-invert Arnoldi iteration \cite{Lehoucq:1998} with shift $\Om_0^2$. We form $\delta_{\rm num}(\alpha)=(\Om^2-\Om_0^2)/\alpha$ at $\alpha=10^{-3}$, $3\times10^{-4}$ and $10^{-4}$. We extrapolate linearly in $\alpha$ from the two smaller values. The extrapolation
\begin{equation}
  \varsigma \;\simeq\; \tfrac12\Big[3\,\delta_{\rm num}\big(10^{-4}\big)-\delta_{\rm num}\big(3\times10^{-4}\big)\Big]
  \label{eq:extrap}
\end{equation}
reproduces the eigenvalues of $\mathcal S$ for $l=2$ and $\Om_0=2,\dots,6$ to $0.1$ per cent in form (a) and to $0.02$ per cent in form (b). Every computed $\Om^2$ is real. A third discretisation, uniform in $x_0$ with $12000$ nodes and walls at $\alpha\rho=0.05$, solves the exact transform of \eqref{eq:throatOm} to $x_0$,
\begin{equation}
  -\partial_{x_0}\big(\mathsf k\,\partial_{x_0}\bm Z\big)+\mathsf k^{-1}\big(V_0+\alpha V_{1a}\big)\bm Z \;=\; \Om^2\mathsf k^{-1}\big(1+\alpha V_{1b}\big)\bm Z,
  \qquad \mathsf k=1+\alpha\hat\gamma ,
  \label{eq:yexact}
\end{equation}
as a generalised eigenvalue problem. It gives the same shifts on the shared levels $\Om_0=4,5,6$ of $l=2$. Setting $\mathsf k=1$ in \eqref{eq:yexact} reproduces the eigenvalues of $\mathcal S$ computed without the remap term.

For Sec.~\ref{sec:stab-mouth} the potentials \eqref{eq:chandra-V} and \eqref{eq:chandra-mouth} are evaluated from their closed forms. At extremality we write the ratio $r^4V^{(\pm)}_i/(r-r_e)^2$ as a rational function of $r$ for each $l=2,\dots,10$ and evaluate it on $2\times10^5$ points uniform in $r_e/r$ between $10^{-6}$ and $1-10^{-6}$. For both $i$, both signs and every $l$ it lies between its two limits, the mass $m_\mp^2$ of \eqref{eq:nhmasses} at $r=r_e$ and $\lhat$ as $r\to\infty$. On a horizonless mouth with $\delta_M=-10^{-3}r_e$ the four potentials are positive on $8000$ points of $r_e(1+10^{-4})\le r\le200\,r_e$ for $l=2,\dots,10$.

For Sec.~\ref{sec:hll-mouth} the potential \eqref{eq:NYSV} is evaluated in double precision at $GM=r_e(1+\eta_{\rm H})$ on 4000 points with $r-r_e$ spaced logarithmically between $10^{-8}r_e$ and $10^4r_e$, for $l=2,\dots,10$. The smallest eigenvalue of $r^2V$ sits at the first point in every case. At $\eta_{\rm H}=1/12\pi^2$ it is $1.993$, $5.978$ and $11.954$ for $l=2,3,4$, and at $\eta_{\rm H}=0.1$ it is $1.907$, $5.700$ and $11.373$. Step~4 of the mouth and the identities \eqref{eq:NYSeqs} hold identically in $l$, $p$, $\eta_{\rm H}$, $r_e$ and $r$, with $A$ a function subject to its first-order equation. The terms in $\log(r/r_e)$ cancel separately, since $\log r$ is transcendental over the rational functions of $r$. The same reduction with the explicit $A$ was repeated at $(l,p,\eta_{\rm H},GM/r_e)=(2,\tfrac37,\tfrac1{97},\tfrac{98}{97})$, $(3,-\tfrac5{11},\tfrac2{31},\tfrac{398}{403})$ and $(5,\tfrac73,\tfrac17,\tfrac{134}{105})$, with $r_e=1$ at the first two points and $r_e=\tfrac32$ at the third, and gives the same first-order system. The diagonal entries $\mathcal S_{++}$ and $\mathcal S_{--}$ quoted in Sec.~\ref{sec:hll-throat} come from the quadrature described above.

\bibliography{MMP_perturbations}

\end{document}